\documentclass[11pt,a4paper]{article}
 
\usepackage{amsmath,amsthm,amssymb}

\usepackage{cite}

\pdfoutput=1

\usepackage{graphics,graphicx}
\usepackage{epsfig}
\usepackage{multicol}
\usepackage{color}
\makeatletter
\@addtoreset{equation}{section}
\makeatother

\usepackage{braket}
\usepackage{bm}
\usepackage{graphicx}

\usepackage{subcaption}
\usepackage{relsize}

\usepackage{soul}

\usepackage{lineno}

\begin{document}

\begin{center}
\LARGE{\bf Quantum reference frames, revisited}
\end{center}

\begin{center}
Matthew J. Lake${}^{a,b,c,d,e}$\footnote{matthewjlake@narit.or.th} 
and Marek Miller${}^{f}$\footnote{mlm@math.ku.dk} 
\end{center}
\begin{center}
\emph{$^{a}$National Astronomical Research Institute of Thailand, \\ 260 Moo 4, T. Donkaew,  A. Maerim, Chiang Mai 50180, Thailand \\}
\emph{$^{b}$Department of Physics and Materials Science, \\ Faculty of Science, Chiang Mai University, \\ 239 Huaykaew Road, T. Suthep, A. Muang, Chiang Mai 50200, Thailand \\}
\emph{$^{c}$ School of Physics, Sun Yat-Sen University, Guangzhou 510275, China} \\
\emph{$^{d}$Department of Physics, Babe\c s-Bolyai University, \\ Mihail Kog\u alniceanu Street 1, 400084 Cluj-Napoca, Romania \\}
\emph{$^{e}$Office of Research Administration, Chiang Mai University, \\ 239 Huaykaew Rd, T. Suthep, A. Muang, Chiang Mai 50200, Thailand \\}
\emph{$^{f}$Centre for the Mathematics of Quantum Theory, University of Copenhagen, Universitetsparken 5, DK-2100 Copenhagen, Denmark \\}
\vspace{0.1cm}
\end{center}


\abstract{
The topic of quantum reference frames (QRFs) has attracted a great deal of attention in the recent literature. 
Potentially, the correct description of such frames is important for both the technological applications of quantum mechanics and for its foundations, including the search for a future theory of quantum gravity. 
In this letter, we point out potential inconsistencies in the mainstream approach to this subject and propose an alternative definition that avoids these problems. 
Crucially, we reject the notion that transformations between QRFs can be represented by unitary operators and explain the clear physical reasons for this.
An experimental protocol, capable of empirically distinguishing between competing definitions of the term, is also proposed. 
The implications of the new model, for uncertainty relations, spacetime symmetries, gauge symmetries, the quantisation of gravity, and other foundational issues are discussed, and possible directions for future work in this field are considered. 
} 

\tableofcontents 

\section{Introduction} \label{Sec.1}

The kinematic description of all physical systems is based on the mathematical construction of a reference frame.
Only relative to a frame of reference can the motion of a physical system be represented analytically.
In classical mechanics,
a family of equivalent reference frames serves as a model of the geometry of space,
and admissible transformations of coordinate systems account for the symmetries of the laws of motion.  
The reference frame becomes an idealised physical observer, 
capable of obtaining information about the motion of the system it observes.

Quantum mechanics describes how the probability amplitude associated with a physical system evolves against the fixed three-dimensional Euclidean background,
understood as a classical frame of reference.
If we were to promote the reference frame to the role of an \emph{quantum} observer,
we would need to account for the quantum character of the interaction between a system in a superposition of states and the background geometry. 
We would then face the problem of how to build a fully quantum theory of reference frames and their transformations. 
These \emph{quantum reference frames} (QRF),
if defined consistently,
would be our point of departure into a much needed theory of motion taking place against a quantum superposition of background geometries.

The study of QRFs was initiated by Aharonov and Susskind \cite{Aharonov:1967zza,Aharonov:1967zz}, 
who analysed the relation between reference frames in quantum mechanics and the superselection rules
(see also \cite{Aharonov:1984zz,Bartlett:2006tzx}). 
They showed that, for quantum systems entangled with a laboratory frame of reference, 
the superselection rules need no longer hold locally.
A quantum observable, 
say the total number of photons, 
can be conserved for any system-laboratory composite state, 
and, at the same time, 
if the two subsystems share information about their relative phase, 
the measurement results need not be constrained by the superselection rule. 
The authors conclude that from the point of view of a quantum observer in a genuine supersposition of states, 
the susperselection rules are no longer something inevitable.

Later, the problem of building a theory of quantum frames of reference appeared naturally in many branches of physics.
It has been studied in the context of quantum gravity \cite{Rovelli:1990pi,1603.04583,1803.03523v1} and quantum information theory \cite{quant-ph/0602069v2,Angelo:2011,Kitaev:2003zj,arXiv:0711.0043v2,Palmer:2013zza},
among others.
From early on, 
the topic has been connected to the relational approach to quantum theory \cite{Poulin:2005dn},
a framework in which only relative observables between subsystems are considered physical.

The systematic study of QRFs was initiated by Giacomini, Castro-Ruiz and Brukner (GCB) \cite{Giacomini:2017zju}.
In a series of follow-up papers
\cite{Vanrietvelde:2018pgb,Vanrietvelde:2018dit,Hohn:2018toe,Hohn:2018iwn,Krumm:2020fws,Ballesteros:2020lgl,QRF_Bell_Test:2021,Giacomini:2021gei,Castro-Ruiz:2021vnq,delaHamette:2021iwx,Cepollaro:2021ccc,AliAhmad:2021adn,Hoehn:2021flk,Carrozza:2021gju,delaHamette:2021oex,delaHamette:2021piz,Giacomini:2022hco,Overstreet:2022zgq,Kabel:2022cje,Apadula:2022pxk,Amelino-Camelia:2022dsj,Wang:2023koz,Hoehn:2023axh,Hoehn:2023ehz,Kabel:2023jve} their approach was expanded to provide a theoretical exposition of QRFs within the framework of the relational quantum mechanics. 
Taking as a starting point the question: 
\emph{Can a quantum system be considered as a reference frame?}, 
GCB developed a method to quantise transformations between reference frames and to describe genuine ``superpositions of coordinate transformations'' \cite{Giacomini:2017zju}.
Consequently, they arrived at a version of quantum theory for which entanglement, and quantum superposition in general, are frame-dependent properties of quantum systems. 
Following an operational approach, these authors developed a framework, which we will call the GCB formalism \cite{Giacomini:2017zju}, in which they can assign a state to any external quantum system with arbitrary precision --
in contrast with previous formulations \cite{Aharonov:1984zz,quant-ph/0602069v2,Kitaev:2003zj,arXiv:0711.0043v2,Palmer:2013zza,Poulin:2005dn}, where the lack of a shared reference frame leads to a situation in which an observer in a quantum superposition can assign only a noisy quantum state to the observed system.  
Starting from the assumption that all physical systems are quantum, they build a relational theory in which only quantum systems can represent genuine physical observers (i.e., reference frames).
The resulting transformations that effect the transition between QRFs
``takes the states of all systems external to the initial QRF as input, and outputs the states of all systems external to the final QRF'' \cite{Giacomini:2017zju}.
As these transformations are proved to be unitary, 
they preserve the probability of measurement outcomes.

In this paper, 
we develop an alternative theory of quantum reference frames. 
Our staring point is the detailed analysis, 
given below,
of what we believe are potential inconsistencies hidden in the GCB formalism.
In the next subsection, we present a strong argument for why unitary operators cannot serve as a model of transformations between QRFs. 
Later, we will outline an alternative definition that is free from these inconsistencies and propose an empirical protocol to distinguish between the two competing approaches.
Unlike the GCB formalism, our take on quantum reference frames never eschews the classical background geometry, and thus remains consistent with canonical quantum mechanics.
In the latter part of our work we will apply our theory to the problem of quantum uncertainty relations and the quantisation of gravity.

We begin our discussion by establishing a basic definition of degrees of freedom of an multi-particle system and presenting the main points of the GCB formalism.

\subsection{How many degrees of freedom describe an $N$-particle state?} \label{Sec.1.1}

The logical structure of the GCB formalism \cite{Giacomini:2017zju}, and of later follow-up works \cite{Vanrietvelde:2018pgb,Vanrietvelde:2018dit,Hohn:2018toe,Hohn:2018iwn,Krumm:2020fws,Ballesteros:2020lgl,QRF_Bell_Test:2021,Giacomini:2021gei,delaHamette:2021iwx,Cepollaro:2021ccc,Castro-Ruiz:2021vnq,AliAhmad:2021adn,Hoehn:2021flk,Carrozza:2021gju,delaHamette:2021oex,delaHamette:2021piz,Giacomini:2022hco,Overstreet:2022zgq,Kabel:2022cje,Apadula:2022pxk,Amelino-Camelia:2022dsj,Kabel:2023jve,Hoehn:2023ehz,Hoehn:2023axh,Wang:2023koz}, is based on the following two assumptions: 
(i) that if an observer $A$ sees a quantum particle $B$ as being in a superposition of states then $B$ perceives $A$ as being in an equivalent superposition of states (related by a coordinate transformation), and 
(ii) that classical observers and, hence, classical frames of reference (CRFs), do not exist. 
The first of these assumptions is illustrated in Fig. 1, for the scenario in which $B$ exists in a superposition of two highly localised position states. 
The second assumption is based on the observation that all physical observers must be embodied as material systems, and, therefore, that they must ultimately be quantum mechanical in nature. 

Combining assumptions (i) and (ii), we are led to the conclusion that $A$ is a quantum system and, hence, that a two-particle system is described by $3$ independent degrees of freedom \cite{Muller_Group_Website}. 
This is sufficient to describe either $B$'s superposition in three-dimensional space, as perceived by $A$, or, equivalently, $A$'s superposition in three-dimensional space, as perceived by $B$, but it is not sufficient to describe both, from the perspective of an external CRF. 
Applying the same logic to a three-particle system, the formalism treats the composite state of the subsystems $A$, $B$ and $C$ as a function of $6$ independent variables \cite{Giacomini:2017zju}. 
A general $N$-particle system is then described by a maximum of $3(N-1)$ independent degrees of freedom.
\footnote{In later papers, for example \cite{Vanrietvelde:2018dit}, it was argued that additional constraints can be placed on the classical Lagrangian of the system, to remove all `non-relational' degrees of freedom {\it prior} to quantisation, reducing this number even further. We consider the legitimacy of this procedure, also, in the following analysis.} 

\subsubsection{Classical mechanics and ignorable coordinates} \label{Sec.1.1.1}

This `relational' quantum theory, in which only the relative positions and momenta between physical subsystems are considered as physically meaningful, also appears to fit well with the corresponding classical theory. 
In classical mechanics, the presence of ignorable coordinates is well established and the correct means of their analysis are well understood; a generalised coordinate is ignorable if it does not appear in the Hamiltonian, but its canonically conjugate momentum does \cite{Classical_Mechanics_Kibble}. 

In a classical $N$-particle system, which is not subject to an external potential, the total potential energy is a function of the $N-1$ independent inter-particle displacements, and, therefore, of $3(N-1)$ independent degrees of freedom. 
However, the canonical Hamiltonian, which describes the total energy of the system from the perspective of an arbitrary CRF, $O$, contains $N$ momentum terms (one for each particle). 
It follows that the 3 components of one of these momenta are ignorable and it is often convenient to perform a canonical coordinate transformation, in order to transform the equations of motion (EOM) into a separable form. 

There is no unique way to do this, but one of the most common techniques it to transform the canonical displacements, $({\bf r}_{0},{\bf r}_{1}, \dots {\bf r}_{N-1})$, which label the particle positions relative to the centre of $O$'s coordinate system, such that $({\bf r}_{0},{\bf r}_{1}, \dots {\bf r}_{N-1}) \mapsto ({\bf x}_{0},{\bf x}_{1}, \dots {\bf x}_{N-1})$, where ${\bf x}_{0}$ denotes the position of the centre-of-mass, relative to $O$, and ${\bf x}_{i} = {\bf r}_{i} - {\bf r}_{0}$ is the displacement of the $i^{\rm th}$ particle, relative to the $0^{\rm th}$ particle, where the latter may be chosen arbitrarily. 
Hamilton's equations for the centre-of-mass coordinate then take the form of a free-particle EOM, which describes the inertial motion of the total mass of the system, $M = \sum_{I=0}^{N-1}m_{I}$, relative to $O$. 
Since there is no way to define or detect absolute inertial motion, which therefore has no physical meaning, it is clear that the degrees of freedom in this EOM are, indeed, ignorable at the classical level\cite{Classical_Mechanics_Kibble}. 

\subsubsection{Relational quantum theories and the GCB formalism} \label{Sec.1.1.2}

It is obvious that these results are relevant to the assumptions of relational physical theories, including extensions of the GCB approach \cite{Giacomini:2017zju, Vanrietvelde:2018pgb,Vanrietvelde:2018dit,Hohn:2018toe,Hohn:2018iwn,Krumm:2020fws,Ballesteros:2020lgl,QRF_Bell_Test:2021,Giacomini:2021gei,delaHamette:2021iwx,Cepollaro:2021ccc,Castro-Ruiz:2021vnq,AliAhmad:2021adn,Hoehn:2021flk,Carrozza:2021gju,delaHamette:2021oex,delaHamette:2021piz,Giacomini:2022hco,Overstreet:2022zgq,Kabel:2022cje,Apadula:2022pxk,Amelino-Camelia:2022dsj,Kabel:2023jve,Hoehn:2023ehz,Hoehn:2023axh,Wang:2023koz}, as well as the relational quantum mechanics originally introduced by Rovelli \cite{Rovelli:1995fv,Zych:2018nao,Hoehn:2019fsy,Hoehn:2020epv,Robson:2023hux}.
\footnote{See \cite{Lawrence:2022uvk,Drezet:2022zkc,Cavalcanti:2023rtl,Lawrence:2022pwu,Lawrence:2023gbl} for objections to the relational interpretation formulated in \cite{Rovelli:1995fv}, by other authors, as well as counter-claims by its proponents.} 
If we do not admit the validity of $O$'s perspective, since this does not represent the perspective of a physical observer, embodied as a material system, then we may regard the $N-1$ relative displacements, $\left\{{\bf x}_{I}\right\}_{I=1}^{N-1}$, as representing the view of the $N$-particle system from the perspective of the $0^{\rm th}$ particle. 
By quantising only these degrees of freedom, we obtain a relational quantum theory, which is consistent (e.g., for $N=3$) with the formalism presented in \cite{Giacomini:2017zju}. 

\begin{figure}[h] \label{Fig.1}
\begin{center}
\includegraphics[width=12cm]{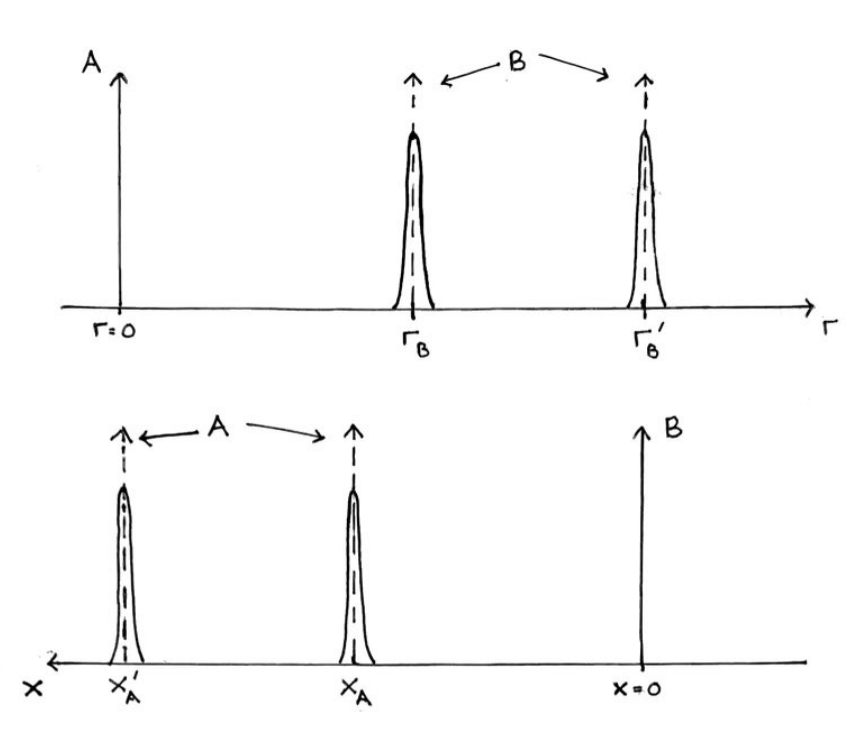}
\caption{According to the GCB formalism \cite{Giacomini:2017zju}, and related works \cite{Vanrietvelde:2018pgb,Vanrietvelde:2018dit,Hohn:2018toe,Hohn:2018iwn,Krumm:2020fws,Ballesteros:2020lgl,QRF_Bell_Test:2021,Giacomini:2021gei,delaHamette:2021iwx,Cepollaro:2021ccc,Castro-Ruiz:2021vnq,AliAhmad:2021adn,Hoehn:2021flk,Carrozza:2021gju,delaHamette:2021oex,delaHamette:2021piz,Giacomini:2022hco,Overstreet:2022zgq,Kabel:2022cje,Apadula:2022pxk,Amelino-Camelia:2022dsj,Kabel:2023jve,Hoehn:2023ehz,Hoehn:2023axh,Wang:2023koz}, when a quantum observer $A$ observes a quantum particle $B$ as being in a superposition of states, $B$ perceives $A$ as being in an equivalent superposition of states, related by a coordinate transformation \cite{Muller_Group_Website}. In this example, $B$ is in a superposition of two highly localised position states, centred on the values ${\bf r}_{B}$ and ${\bf r}'_{B}$ in $A$'s coordinate system (Fig. 1(a), above); $B$ then perceives $A$ as being in a superposition of states, centred on the values ${\bf x}_{A}=-{\bf r}_{B}$ and ${\bf x}'_{A}=-{\bf r}'_{B}$ (Fig. 1(b), below). By contrast, we contend that this picture holds only when either $A$ or $B$ is effectively classical, and, therefore, not in a superposition of states. When both $A$ and $B$ are quantum mechanical in nature, we require $3 \times 2 = 6$ independent degrees of freedom, to describe {\it both} their configurations in three-dimensional space.}
\end{center}
\end{figure}

\subsubsection{Canonical quantum mechanics - can we safely ignore non-relational degrees of freedom?} \label{Sec.1.1.3}

Despite these rather neat-and-tidy results, the assumptions of the GCB formalism may be questioned. 
Here, we raise two primary objections and show how they combine to suggest a different view of quantum reference frames, yielding an alternative formalism, and an alternative definition of the term. 
We limit our discussion, in this section, to general considerations, and present more detailed criticisms in Sec. \ref{Sec.2}.

First, let us state, explicitly, that we {\it partially} agree with the two main assumptions on which relational quantum theories, including the formalism developed in \cite{Giacomini:2017zju}, are based. 
More specifically, we agree (a) that all physical observers are embodied as material systems, and (b) that classical material systems do not exist. 
{\it Hence, we agree that all physical observers are embodied as material quantum systems.}

However, we do {\it not} agree that these results combine to imply the non-existence of CRFs. 
In both classical mechanics, and canonical quantum mechanics (QM), the classical background geometry is physical, and influences the dynamics of material systems \cite{Frankel:1997ec,Nakahara:2003nw}. 
Furthermore, according to the Erlangen Program in mathematics, there is a formal equivalence between classical spaces and their symmetries \cite{ErlangenProgram_Klein_1872,ErlangenProgram_EMS_2015,Zuber:2013rha,Kisil:2010,Horwood:2004uh,ErlangenProgram_Encycolpedia_Srpinger,Lev:2020igj,Goenner:2015}. 

The background geometry of classical Newtonian mechanics, and of non-relativisitc quantum theory, is three-dimensional Euclidean space, $\mathbb{E}^{3}$. 
Formally, this is equivalent to the Euclidean group of symmetries, denoted ${\rm E}^3$, which is comprised of three subgroups: translations, rotations about arbitrary spatial points, and reflections through arbitrary two-dimensional planes. 
These define the equivalence of all CRFs that are related by translations, rotations or reflections, and these transformations also form subgroups of the group of Galilean symmetries \cite{Jones:1998}. 

When one adds material dynamical systems to the physical background geometry, defined as scalar, vector, or tensor functions of the spatial coordinates, one obtains the dual momentum space, which is given by the tangent bundle to the physical space \cite{Frankel:1997ec,Nakahara:2003nw}. 
For Euclidean geometry, the tangent spaces at all points are also Euclidean, so that the momentum space is isomorphic to the physical geometry. 
A translation in momentum space is then equivalent to a Galilean velocity boost, yielding the group of Galilean transformations, ${\rm G}^{3}$, as the symmetry group of physical systems in $\mathbb{E}^{3}$ \cite{Ibragimov:2015,Giachetta:2011}. 

This leads us to an important conclusion. 
Put simply, we may say that `because the physical background space exists, CRFs also exist'. 
There is a one-to-one correspondence between the two, as specified by the Erlangen program \cite{ErlangenProgram_Klein_1872,ErlangenProgram_EMS_2015,Zuber:2013rha,Kisil:2010,Horwood:2004uh,ErlangenProgram_Encycolpedia_Srpinger,Lev:2020igj,Goenner:2015}. 
It follows that we are free to describe the dynamics of a physical system, whether classical or quantum, from the perspective of an arbitrary CRF. 
Such an `objective' description is valid, at least mathematically, despite the fact that no physical observer can inhabit such a frame. 
This is our first point of departure from the GCB formalism: instead of omitting CRFs from the description of physical reality, altogether, we admit their validity, due to the assumed physical existence of the classical background space, within the formalism of canonical quantum theory \cite{Jones:1998,Giachetta:2011}. 
We are then prompted to ask an important physical question, namely: how does the imperfect information available to a real physical observer, embodied as a material quantum system, differ from the perfect information available to an idealised CRF? 

The answer to this question will determine how we define, in a mathematically precise way, the intuitive concept of a QRF, where the latter is assumed to be associated, in {\it some} way, with a superposition of CRFs. 
At this point, it is instructive to return to the scenario depicted in Figs. 1(a) and 1(b), which illustrate the basic assumptions underlying relational quantum theories \cite{Muller_Group_Website}. 
Having established the objective existence of CRFs, at least formally, within the framework of canonical quantum mechanics, we note that the centre-of-mass coordinates of massive macroscopic observers {\it effectively} inhabit such frames.
\footnote{If the positional uncertainty of a macroscopic observer is effectively negligible, $\Delta x \simeq 0$, then its momentum uncertainty will be very large. However, this need not have devastating practical effects. Due to its very large mass, its velocity uncertainty can be negligible, also; $\Delta v = \Delta p/M \simeq 0$. This point was made, and indeed utilised, in the original QRF papers by Aharonov and Susskind \cite{Aharonov:1967zza,Aharonov:1967zz}. We are free to rewrite all momentum space wave functions as functions of velocity space, $\tilde{\psi}({\bf v}) \equiv \tilde{\psi}({\bf p})$, and it is possible for both $|\psi({\bf x})|^2$ and $|\tilde{\psi}({\bf v})|^2$ to adopt approximately Dirac delta configurations, simultaneously. When $M$ is large enough for both $\Delta x$ and $\Delta v$ to be unmeasurable, given the accuracies of currently available experiments, an observer can be said to inhabit a CRF for all practical purposes. (See Sec. \ref{Sec.3} for further details.)} 
Let us suppose, therefore, that the observer $A$ is an effectively classical system, whereas $B$ is unambiguously quantum, and in a superposition of two highly localised position states. 
It is obvious that $A$ will perceive these accurately, in the sense that any position measurements she makes on $B$'s state will return values closely clustered around ${\bf r}_{B}$ and ${\bf r}'_{B}$. 
But how will $B$ perceive $A$? 

It seems obvious, to us, that $B$ cannot but perceive $A$ as being in a superposition of position states, tightly clustered around the values ${\bf x}_{A} = -{\bf r}_{B}$ and ${\bf x}'_{A} = -{\bf r}'_{B}$. 
This point is crucial. 
We expect this result when $A$ is classical and $B$ is quantum, but the proponents of relational quantum theories claim that it holds true, even when both $A$ {\it and} $B$ are quantum mechanical in nature. 
This is our second point of departure from the GCB formalism, and related works \cite{Giacomini:2017zju,Vanrietvelde:2018pgb,Vanrietvelde:2018dit,Hohn:2018toe,Hohn:2018iwn,Krumm:2020fws,Ballesteros:2020lgl,QRF_Bell_Test:2021,Giacomini:2021gei,delaHamette:2021iwx,Cepollaro:2021ccc,Castro-Ruiz:2021vnq,AliAhmad:2021adn,Hoehn:2021flk,Carrozza:2021gju,delaHamette:2021oex,delaHamette:2021piz,Giacomini:2022hco,Overstreet:2022zgq,Kabel:2022cje,Apadula:2022pxk,Amelino-Camelia:2022dsj,Kabel:2023jve,Hoehn:2023ehz,Hoehn:2023axh,Wang:2023koz,Muller_Group_Website}. 

If such formalisms offer the correct description of physical reality, this begs the question, what happens if $A$ is classical? 
This is met with a terse response from relational theories; CRFs are not physical, so the question is invalid. 
We disagree. 
As already stated above, $3$ degrees of freedom is sufficient to describe either $B$'s superposition in three-dimensional space, or $A$'s, but not both. 
We require $2  \times 3 = 6$ degrees of freedom to describe a situation in which $A$ and $B$ exist as either independent, or entangled, quantum states. 
This fact may not be immediately apparent, to either $A$ or $B$ individually, but it must certainly be so to an idealised external observer, represented by an arbitrary CRF, $O$. 
This scenario is depicted in Fig. 2. 

\begin{figure}[h] \label{Fig.2}
\begin{center}
\includegraphics[width=10cm]{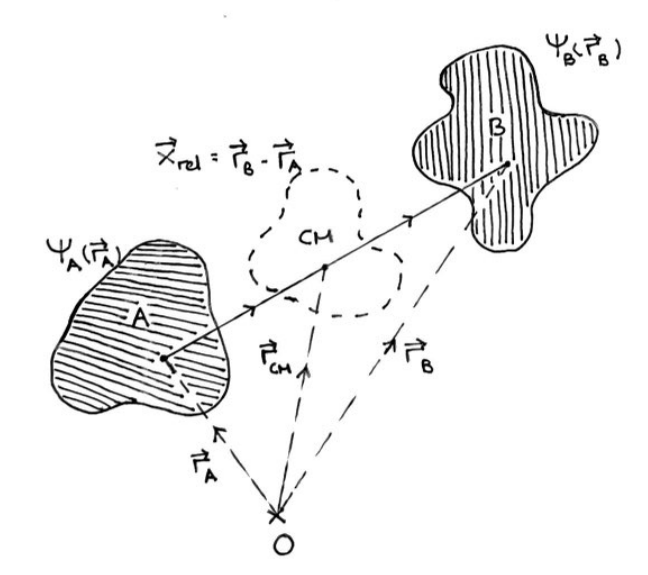}
\caption{Because the massive particles $A$ and $B$ each exist in a superposition of position eigenstates, the centre-of-mass of the joint $AB$ system also exists in a superposition of position eigenstates. Ignoring the 3 degrees of freedom associated with the {\it effective} wave function of this virtual entity is equivalent to ignoring 3 physical degrees of freedom, associated with either $A$ or $B$. This leads to incorrect physical predictions, and, specifically, to incorrect expressions for the uncertainty relations associated with the only {\it directly} measurable uncertainties, $\Delta x_{\rm rel}^{i}$ and $\Delta p_{{\rm rel} j}$. (See Sec. \ref{Sec.3} for details.) Note that, in this diagram, we have labelled the centre-of-mass coordinate as ${\bf r}_{\rm CM}$, not ${\bf x}_{\rm CM}$, to emphasise the fact that this variable denotes its position with respect to the CRF $O$. The shaded regions represent the volumes of space in which the wave functions of the physical particles, $\psi_{A}({\bf r}_{A})$ and $\psi_{B}({\bf r}_{B})$, are nonzero, whereas the dashed closed curve represents the region in which the effective wave function associated with the centre-of-mass coordinate, $\phi_{\rm CM}({\bf r}_{\rm CM}) \equiv \phi_{\rm CM}({\bf x}_{\rm CM})$, is nonzero.}
\end{center}
\end{figure}

What we require, therefore, is a map from an arbitrary CRF $O$ to a QRF, embodied as a material system, by either $A$ or $B$. 
How might a QRF, defined in this way, differ from the purely relational description? 
To answer this question, let us return to the $N$-particle coordinate transformation, discussed above, and limit our attention to the simplest non-trivial example, $N=2$. 
Thus, we have $({\bf r}_{A},{\bf r}_{B}) \mapsto ({\bf x}_{\rm CM},{\bf x}_{\rm rel})$, where ${\bf x}_{\rm CM} = (m_{B}{\bf r}_{B} + m_{A}{\bf r}_{A})/M$ is the centre-of-mass, $M = m_{B}+m_{B}$ is the total mass, and ${\bf x}_{\rm rel} = {\bf r}_{B} - {\bf r}_{A}$ is the relative displacement between the quantum observer $A$ and the observed particle $B$. 
It is well known that, in this case, the canonical Hamiltonian may be rewritten as \cite{U_Texas_Notes:2023,Michigan_State_U_Notes:2009}
\begin{eqnarray} \label{canonical_Hamiltonian_N=2}
&&\hat{H}_{AB}(\hat{{\bf r}},\hat{\boldsymbol{\kappa}}) = \frac{\hat{\kappa}_{A}^2}{2m_{A}} + \frac{\hat{\kappa}_{B}^2}{2m_{B}} + \hat{V}(|\hat{{\bf r}}_{B}-\hat{{\bf r}}_{A}|)
\nonumber\\
&\equiv& \hat{H}_{AB}(\hat{{\bf x}},\hat{{\bf p}}) = \frac{\hat{p}_{\rm CM}^2}{2M} + \frac{\hat{p}_{\rm rel}^2}{2\mu} + \hat{V}(|\hat{{\bf x}}_{\rm rel}|) \, , 
\end{eqnarray}
where ${\bf p}_{\rm CM} = \boldsymbol{\kappa}_{B} + \boldsymbol{\kappa}_{A}$ is canonically conjugate to the centre-of-mass coordinate, at the classical level, ${\bf p}_{\rm rel} = (m_{A}\boldsymbol{\kappa}_{B}-m_{B}\boldsymbol{\kappa}_{A})/M$ is conjugate to the relative displacement, and $\mu = m_{B}m_{A}/M$ is the reduced mass. 
In the quantum regime, the corresponding conditions are $[x_{\rm CM}^{i},p_{{\rm CM}j}] = i\hbar \, \delta^{i}{}_{j}$ and $[x_{\rm rel}^{i},p_{{\rm rel}j}] = i\hbar \, \delta^{i}{}_{j}$ \cite{U_Texas_Notes:2023,Michigan_State_U_Notes:2009}.

The Schr{\" o}dinger equation for the composite $AB$ system therefore separates into a free particle equation for the {\it effective} wave function associated with the centre-of-mass coordinate, plus an EOM with an interaction potential for the wave function associated with the relative displacement.
\footnote{The centre-of-mass and relative displacements are virtual entities, not physical particles, but the total Hilbert space of the system can be partitioned into subspaces in different ways, thanks to the separability of the Hamiltonian, as $\mathcal{H}_{AB} = \mathcal{H}_{A} \otimes \mathcal{H}_{B} = \mathcal{H}_{\rm CM} \otimes \mathcal{H}_{\rm rel}$. The first partition contains states of the form $\ket{\psi_{A}}_{A} \otimes \ket{\psi_{B}}_{B}$, where $\psi_{A}({\bf r}_{A})$ ($\tilde{\psi}_{A}(\boldsymbol{\kappa}_{A})$) and $\psi_{B}({\bf r}_{B})$ ($\tilde{\psi}_{B}(\boldsymbol{\kappa}_{B})$) are the position (momentum) space wave functions of the material particles $A$ and $B$, respectively. Similarly, the second contains states of the form $\ket{\phi_{\rm CM}}_{\rm CM} \otimes \ket{\phi_{\rm rel}}_{\rm rel}$, where $\phi_{\rm CM}({\bf x}_{\rm CM})$ ($\tilde{\phi}_{\rm CM}({\bf p}_{\rm CM})$) and $\phi_{\rm rel}({\bf x}_{\rm rel})$ ($\tilde{\phi}_{\rm rel}({\bf p}_{\rm rel})$) represent superpositions of ${\bf x}_{\rm CM}$ (${\bf p}_{\rm CM}$) and ${\bf x}_{\rm rel}$ (${\bf p}_{\rm rel}$) eigenstates. Of course, both partitions also contain entangled states, representing more general superpositions of the relevant variables. In general, a physical state that is separable in one basis, $\ket{{\bf r}_A}_{A} \otimes \ket{{\bf r}_B}_{B}$ ($\ket{\boldsymbol{\kappa}_A}_{A} \otimes \ket{\boldsymbol{\kappa}_B}_{B}$), will be entangled with respect to the other, $\ket{{\bf x}_{\rm CM}}_{\rm CM} \otimes \ket{{\bf x}_{\rm rel}}_{\rm rel}$ ($\ket{{\bf p}_{\rm CM}}_{\rm CM} \otimes \ket{{\bf p}_{\rm rel}}_{\rm rel}$).} 
In general, solutions to the latter include a sequence of discrete bound states, $\ket{n,l,m}_{\rm rel}$, where $n$, $l$ and $m$ are the energy, total angular momentum and magnetic angular momentum quantum numbers, respectively, plus a continuum of free states, $\ket{{\bf p}_{\rm rel}}_{\rm rel}$. 
The eigenstates of the centre-of-mass, however, are simply free states, $\ket{{\bf p}_{\rm CM}}_{\rm CM}$. 
But can these be safely ignored?

We claim that they cannot. 
Unlike the classical case, the EOM for $\hat{{\bf x}}_{\rm CM}$ and $\hat{{\bf p}}_{\rm CM}$, which are obtained from the Heisenberg picture, do {\it not} define inertial motion of the centre-of-mass. 
Instead, the closest we can get to this condition is 
\begin{eqnarray} \label{non-classical_CoM}
M\frac{{\rm d}}{{\rm d}t}\braket{\hat{{\bf x}}_{\rm CM}} = \braket{\hat{{\bf p}}_{\rm CM}} \, , \quad \frac{{\rm d}}{{\rm d}t}\braket{\hat{{\bf p}}_{\rm CM}} = 0 \, ; \quad \frac{{\rm d}^2}{{\rm d}t^2}\braket{\hat{{\bf x}}_{\rm CM}} = 0 \, ,
\end{eqnarray}
which follows from the Eherenfest theorem \cite{Griffiths:2017}. 
Thus, although the expectation value of the centre-of-mass undergoes (undetectable) inertial motion, the centre-of-mass itself exists in a superposition of position eigenstates, distributed within a volume of three-dimensional space.  
It cannot be otherwise, since both massive particles, $A$ and $B$, exist in superpositions, and this scenario is depicted in Fig. 2.

Furthermore, it is straightforward to show that the nonzero variance of the centre-of-mass coordinate has {\it physical} consequences, for the bipartite quantum system, which cannot be ignored. 
This differs from the classical scenario, in which ${\bf x}_{\rm CM}$ is ignorable because its inertial motion is not observable, by means of measurements performed from within either $A$ or $B$'s frames of reference. 
Specifically, it is easy to show that 
\begin{eqnarray} \label{variance_rel}
&&(\Delta x_{\rm rel}^{i})^2 = (\Delta r_{A}^{i})^2 + (\Delta r_{B}^{i})^2 \, , 
\nonumber\\ 
&&(\Delta p_{{\rm rel} j})^2 = (\Delta \kappa_{Aj})^2 + (\Delta \kappa_{Bj})^2 \, , 
\end{eqnarray}
and
\begin{eqnarray} \label{variance_CM}
&&(\Delta x_{\rm CM}^{i})^2 = \left(\frac{m_{A}}{M}\right)^2(\Delta r_{A}^{i})^2 + \left(\frac{m_{B}}{M}\right)^2(\Delta r_{B}^{i})^2 \, , 
\nonumber\\
&&(\Delta p_{{\rm CM} j})^2 = \left(\frac{m_{B}}{M}\right)^2(\Delta \kappa_{Aj})^2 + \left(\frac{m_{A}}{M}\right)^2(\Delta \kappa_{Bj})^2 \, , 
\end{eqnarray}
where 
\begin{eqnarray} \label{canonical_HUP_A/B}
\Delta r_{A}^{i} \, \Delta \kappa_{Aj} \geq \frac{\hbar}{2} \, \delta^{i}{}_{j} \, , \quad \Delta r_{B}^{i} \, \Delta \kappa_{Bj} \geq \frac{\hbar}{2} \, \delta^{i}{}_{j} \, , 
\end{eqnarray}
are the canonical Heisenberg uncertainty relations for $A$ and $B$, respectively. 
It follows that $\Delta x_{\rm rel}^{i}$ and $\Delta x_{\rm CM}^{i}$ are not independent quantities, since they have a functional dependence on one another, and, likewise, neither are $\Delta p_{{\rm rel} j}$ and $\Delta p_{{\rm CM} j}$. 
Thus, by ignoring the degrees of freedom associated with the centre-of-mass, ${\bf x}_{\rm CM}$ or ${\bf p}_{\rm CM}$, we discard {\it physical} information. 
This information is required, in order to compute the correct uncertainty relations, for measurements of the {\it relative} displacement and momentum between the two particles.
\footnote{Note that, here, we do not dispute the fact that ${\bf x}_{\rm rel}$ or ${\bf p}_{\rm rel}$ are the only physically measurable quantities. In this respect, we are in complete agreement with the proponents of relational quantum theories, \cite{Giacomini:2017zju,Vanrietvelde:2018pgb,Vanrietvelde:2018dit,Hohn:2018toe,Hohn:2018iwn,Krumm:2020fws,Ballesteros:2020lgl,QRF_Bell_Test:2021,Giacomini:2021gei,delaHamette:2021iwx,Cepollaro:2021ccc,Castro-Ruiz:2021vnq,AliAhmad:2021adn,Hoehn:2021flk,Carrozza:2021gju,delaHamette:2021oex,delaHamette:2021piz,Giacomini:2022hco,Overstreet:2022zgq,Kabel:2022cje,Apadula:2022pxk,Amelino-Camelia:2022dsj,Kabel:2023jve,Hoehn:2023ehz,Hoehn:2023axh,Wang:2023koz,Muller_Group_Website} and \cite{Rovelli:1995fv,Zych:2018nao,Hoehn:2019fsy,Hoehn:2020epv,Robson:2023hux}. It follows, immediately, that only $\Delta x_{\rm rel}^{i}$ and $\Delta p_{{\rm rel} j}$ can be computed, directly, from the statistical spreads of repeated measurements, whereas  $\Delta x_{\rm CM}^{i}$ and $\Delta p_{{\rm CM} j}$ are `unobservable' in this sense. This does not make them unphysical. When each is nonzero, this alters the form of the uncertainty relations obeyed by the {\it measurable} quantities $\Delta x_{\rm rel}^{i}$ and $\Delta p_{{\rm rel} j}$, in a way that is not captured by the GCB formalism \cite{Giacomini:2017zju}, or related models \cite{Vanrietvelde:2018pgb,Vanrietvelde:2018dit,Hohn:2018toe,Hohn:2018iwn,Krumm:2020fws,Ballesteros:2020lgl,QRF_Bell_Test:2021,Giacomini:2021gei,delaHamette:2021iwx,Cepollaro:2021ccc,Castro-Ruiz:2021vnq,AliAhmad:2021adn,Hoehn:2021flk,Carrozza:2021gju,delaHamette:2021oex,delaHamette:2021piz,Giacomini:2022hco,Overstreet:2022zgq,Kabel:2022cje,Apadula:2022pxk,Amelino-Camelia:2022dsj,Kabel:2023jve,Hoehn:2023ehz,Hoehn:2023axh,Wang:2023koz,Muller_Group_Website}. (See Sec. \ref{Sec.3} for further details.)}
 
In Sec. \ref{Sec.3}, we show, explicitly, that generalised variances like Eqs. (\ref{variance_rel}) lead to generalised uncertainty relations (GURs) of the form
\begin{eqnarray} \label{GURs}
(\Delta x_{\rm rel}^{i})^2 \, (\Delta p_{{\rm rel} j})^2 \geq \dots \geq \hbar^2 \, (\delta^{i}{}_{j})^2 \, , 
\end{eqnarray}
where the dots in the middle of Eqs. (\ref{GURs}) represent a sum of terms, in either $(\Delta x_{\rm rel}^{i})^2$ or $(\Delta p_{{\rm rel} j})^2$, that is generically greater than or equal to the Schr{\" o}dinger-Robertson bound on the far right-hand side \cite{Rae:2002,Isham:1995}. 
We stress that such non-Heisenberg statistics are an {\it unavoidable} consequence of the fact that both the `observer' and the `observed' are embodied as material quantum systems, which each exist in a superposition of position, or, equivalently, momentum eigenstates. 
By ignoring degrees of freedom that are ignorable in the classical limit, but {\it not} in the quantum regime, relational quantum theories, such as the GCB formalism \cite{Giacomini:2017zju}, are unable to capture this behaviour. 

The structure of this paper is then as follows. 
In Sec. \ref{Sec.2}, we present additional objections to the formalism developed in \cite{Giacomini:2017zju}, in greater detail. 
These are related, but not equivalent to the two fundamental criticisms presented above. 
In Sec. \ref{Sec.3}, we develop an alternative QRF formalism and outline its main features. 
In brief, these are: (i) the unavoidable existence of generalised (non-Heisenberg) uncertainty relations, for all relative variables, due to the quantum nature of both the `observer' and the `observed', and (ii) the non-unitary nature of CRF-to-QRF and QRF-to-QRF transitions. 
The second feature is intimately related to the first and follows directly from the fact that, a priori, there is no relation between the wave function of the observer and the wave functions of the observed systems. 
Section \ref{Sec.4} highlights various advantages of the new model, in comparison to the issues raised in Sec. \ref{Sec.2} concerning relational quantum theories \cite{Giacomini:2017zju,Vanrietvelde:2018pgb,Vanrietvelde:2018dit,Hohn:2018toe,Hohn:2018iwn,Krumm:2020fws,Ballesteros:2020lgl,QRF_Bell_Test:2021,Giacomini:2021gei,delaHamette:2021iwx,Cepollaro:2021ccc,Castro-Ruiz:2021vnq,AliAhmad:2021adn,Hoehn:2021flk,Carrozza:2021gju,delaHamette:2021oex,delaHamette:2021piz,Giacomini:2022hco,Overstreet:2022zgq,Kabel:2022cje,Apadula:2022pxk,Amelino-Camelia:2022dsj,Kabel:2023jve,Hoehn:2023ehz,Hoehn:2023axh,Wang:2023koz,Muller_Group_Website,Rovelli:1995fv,Zych:2018nao,Hoehn:2019fsy,Hoehn:2020epv,Robson:2023hux}. 
In Sec. \ref{Sec.5}, we apply the new QRF formalism to the problem of quantum gravity, or, more specifically, quantum {\it geometry}. 
As a toy model, we replace the classical spatial points of the Euclidean background geometry with QRFs representing superpositions of spatial coordinates. 
This allows us to make contact with models of phenomenological quantum gravity and we are able to derive the three most widely studied GURs, known as the generalised uncertainty principle (GUP) \cite{Maggiore:1993rv,Adler:1999bu,Scardigli:1999jh}, the extended uncertainty principle (EUP), and the extended generalised uncertainty principle (EGUP) \cite{Bolen:2004sq,Park:2007az,Bambi:2007ty}, within this framework. 
These derivations differ from those presented in the bulk of the existing literature, which are based on implementing modified commutation relations \cite{Kempf:1996ss,Hossenfelder:2012jw,Tawfik:2015rva,Tawfik:2014zca,Bosso:2023aht}, and naturally avoid various pathologies associated with this approach \cite{Lake:2018zeg,Lake:2019nmn,Lake:2019oaz,Lake:2020rwc,Lake:2021beh,Lake:2020chb,Lake:2021gbu,Lake:2022hzr,Lake:2023lvh,Lake:2023uoi}. 
In this way, we are able to unify the GUR and QRF paradigms, by modelling the quantum background space, in which material systems `live', as a non-material QRF. 
The implications for this unification, for spacetime symmetries, gauge symmetries, and the quantisation of gravity, are also discussed. 
A brief summary of our conclusions, and a discussion of prospects for future work in this field, are given in Sec. \ref{Sec.6}. 

\section{Objections to the GCB formalism} \label{Sec.2}

\subsection{The spatial background as a CRF} \label{Sec.2.1}

In the GCB formalism \cite{Giacomini:2017zju}, bipartite systems, in which one subsystem acts as the `observer' and the other as the `observed', are described by a single three-variable wave function. 
Similarly, in tripartite systems, in which one quantum subsystem is prescribed to act as a reference frame (a QRF), separable states are described by two, independent, three-variable wave functions. 
Of course, the generation of entangled states means that it is not possible, in general, to associate a complex function of three variables, in a unique way, with an individual subspace of the total Hilbert space, and the notion of a single-particle wave function breaks down in this case. 
Nonetheless, in the analysis that follows, we will use the term `wave function' as a kind of linguistic shorthand for the phrase `three independent quantum mechanical degrees of freedom'. 
Thus, we will speak, somewhat lazily, about systems of $N$ particles being described by `$N$ wave functions' or `$N-1$ wave functions', etc., in different quantum formalisms. 

In this sense, the formalism developed in \cite{Giacomini:2017zju} for tripartite states naturally generalises to $N$-body systems, which are described by $N-1$ wave functions, or $3(N-1)$ degrees of freedom. 
This represents a radical break with canonical quantum theory, in which the dynamics of $N$-body systems are described with respect to {\it classical} observers, and are characterised by $N$ wave functions, or, equivalently, by $3N$ independent variables.  

As discussed in Sec. \ref{Sec.1}, the rationale for this omission, of one whole wave function per $N$-partite system, arises from the combination of two assumptions. 
The first assumption is that a QRF (by definition) is not permitted to obtain any information about its own internal degrees of freedom \cite{Giacomini:2017zju}.
\footnote{In the Introduction, we phrased this slightly differently, as the assumption that `if $A$ perceives $B$ as being in a superposition of states, then $B$ perceives $A$ as being in an equivalent superposition of states'. The key word here is {\it equivalent}. In the GCB formalism, this equivalence is given a precise meaning, and implies that the two descriptions are related by a unitary transformation \cite{Giacomini:2017zju}. However, as explained in Sec. \ref{Sec.1.1}, it may be argued that this picture holds only when either $A$ or $B$ is effectively classical, so that the total system can be described using only three quantum mechanical variables. The alternative is to assume that, although both $A$ and $B$ are quantum mechanical in nature, the designated observer has no physical access to their own `internal' degrees of freedom, which are therefore omitted from their description of the composite state. This assumption is, therefore, equivalent to the one stated above.}
The second assumption is that, since all {\it physical} observers must be embodied as quantum systems, classical frames of reference do not exist \cite{Giacomini:2017zju}. 
Both assumptions lead to the same conclusion, regarding the degrees of freedom required to describe a quantum $N$-particle state, and accepting them implies that one can safely remove one wave function, per $N$-partite system, from the description of physical reality; as opposed to, for example, performing a partial trace over the degrees of freedom associated with a single subsystem, selected as the `observer'. 

In this section, we show that the assumptions on which the GCB formalism is based are inconsistent with other aspects of the quantum mechanical formalism, which are present in the canonical theory and which are tacitly assumed to hold, also, in their model. 
The most important of these is the Born rule. 
In both canonical quantum mechanics and the formalism developed in \cite{Giacomini:2017zju} the probability of observing a particle $B$, at a given coordinate ${\bf x}_{B}$, is ${\rm d}^{3}P({\bf x}_{B}|\Phi_{B}({\bf x}_{B})) = |\Phi_{B}({\bf x}_{B})|^2{\rm d}^{3}{\rm x}_{B}$, for some normalised wave function, $\Phi_{B}({\bf x}_{B})$. 
In order for this interpretation to hold, $\Phi_{B}$ must be defined, regardless of which coordinate system we choose to employ, as a complex scalar in three-dimensional Euclidean space. 
(In particular, this remains true, even if the coordinate ${\bf x}_{B}$ denotes the {\it relative} distance between two quantum subsystems.)
Analogous remarks also hold in the momentum space representation, in which the relevant probability is ${\rm d}^{3}P({\bf p}_{B}|\tilde{\Phi}_{B}({\bf p}_{B})) = |\tilde{\Phi}_{B}({\bf p}_{B})|^2{\rm d}^{3}{\rm p}_{B}$. 
The momentum space manifold is given by the tangent bundle associated with the physical space manifold, which, for Euclidean space, is isomorphic to the original physical space, $\mathbb{E}^3$ \cite{Frankel:1997ec,Nakahara:2003nw}. 

This demonstrates, unambiguously, that the {\it classical} spatial background exists, at least formally, even in the GCB model \cite{Giacomini:2017zju}.
\footnote{We do not, here, address the question of whether this is the correct, or most complete, description of the spacetime background. We only note that, in the standard non-relativistic approximation, and when the spatial background is not endowed with quantum degrees of freedom of its own, classical Euclidean space forms an essential part of the mathematical formalism that describes the properties and dynamics of quantum matter \cite{Frankel:1997ec,Nakahara:2003nw,Jones:1998,Giachetta:2011}.} 
Without it, we cannot employ the Born rule, and we have no way of assigning probabilities to the possible outcomes of physical measurements. 
Next, we need only note that any spatial point in $\mathbb{E}^3$ may be regarded, at a given time $t$, as the origin of a set of coordinate axes, which determine a frame of reference. 
In short, we have reached the conclusion that {\it because} the classical spatial background exists, classical frames of reference {\it also} exist, as already noted in Sec. \ref{Sec.1.1}. 

Furthermore, because classical frames of reference exist, we must be free to describe the properties of an $N$-partite quantum system, at least mathematically, from the perspective of an arbitrary CRF. 
This does not, however, alter that the fact that no {\it physical} observer can exist as a CRF. 
What we require, therefore, is a map from an arbitrary CRF, $O$, to a given QRF, $A$, within the $N$-partite system. 
In other words, we must start by describing quantum systems from the perspective of an idealised classical observer and then ask how the complete information available to $O$, about an $N$-partite state, differs from the information that is available, to a subsystem $A$, about the remaining $N-1$ subsystems.   

In Sec. \ref{Sec.3}, we explicitly construct such a map, and show that it must be {\it non-unitary}. 
Before concluding this section, however, we again stress the logical necessity of such a construction, in relation to the Born rule. 
The logic is as follows; (i) the Born rule is required, even in QRF formalisms, (ii) the Born rule only makes sense if the classical spatial background exists, (iii) if the classical spatial background exists, then classical frames of reference also exist, (iv) if classical frames of reference exist, then we are free to describe an $N$-partite quantum system from the perspective of an arbitrary CRF, (v) to construct a model of a QRF, we therefore require a map from an arbitrary CRF to the perspective of a quantum subsystem, which acts as an `observer' of the remaining $N-1$ subsystems.  

\subsection{A minimally bipartite universe? (Or, do wave functions objectively exist?)} \label{Sec.2.2}

The logical conclusion of the GCB formalism \cite{Giacomini:2017zju}, in which $N$-partite systems are described by $N-1$ wave functions, is that it is impossible for physical science to describe a Universe that consists of a single quantum mechanical particle in empty space. 
We find this very strange. 
Although such a Universe is clearly unrealistic, and does not correspond to our own, we see no fundamental reason why it should be logically impossible, and, therefore, physically inadmissible. 

Furthermore, as established in Sec. \ref{Sec.2.1}, classical frames of reference {\it must} exist in QRF models, as a direct consequence of the classicality of the spatial background, and we must surely be at liberty to describe a one-particle Universe from the perspective of an arbitrary spatial point. 
Put another way, the GCB formalism implies that the Universe is minimally bipartite, and that there is no such thing as an $N$-wave function description of an $N$-partite quantum system. 
This viewpoint is inconsistent with the existence of the classical background space, which, in turn, {\it requires} that $N$ quantum mechanical particles be described by $3N$ independent degrees of freedom.

The fundamental question at issue, therefore, is `Do wave functions objectively exist?'. 
In the canonical theory, and in the alternative QRF formalism proposed in Sec. \ref{Sec.3}, the answer is `Yes, wave functions exist, objectively, in physical three-dimensional space'. 
(See \cite{Stoica:2021owy} for an elegant discussion of this point.) 
We therefore require one wave function, $\psi_{I}(x,y,z)$, per particle in an $N$-particle system, with $I \in \left\{1,2, \dots N\right\}$. 
In the GCB formalism \cite{Giacomini:2017zju}, by contrast, there is no way to assign a given wave function, $\psi_{I}$, to a given particle $m_{I}$, even if all particles in the system are {\it distinguishable}. 

To see why this is so, let us consider, as an example, a tripartite system consisting of an electron, a proton, and a neutron. 
In the formalism developed in \cite{Giacomini:2017zju}, one of these particles is designated as the `observer' and the composite state of the remaining two particles is described by a functional of two wave functions, $F[\psi\phi]$. 
To begin, we assume that the observer (i.e., the QRF) is the electron, so that $F[\psi\phi]$ describes the composite state of the proton and the neutron. 
According to the GCB model, the composite state of the electron and the proton, as seen by the neutron, is given by a unitary transformation of this functional, $\hat{U}F[\psi\phi]$. 
In other words, the physical state of the composite electron-proton system, as seen by the neutron, is characterised by the {\it same} two wave functions that characterise the state of the proton-neutron system, as seen by the electron. 

It is therefore impossible to assign the existence of a specific wave function to a specific particle, in {\it any} basis, even if all particles within the system are distinguishable (or should be so) by virtue of their intrinsic quantum numbers, including mass, electric charge, spin, and any additional charges derived from gauge symmetries \cite{Frankel:1997ec,Nakahara:2003nw}. 
We argue that this departure from canonical quantum theory is not logically necessary. 
In Sec. \ref{Sec.3}, we show how to construct a QRF by performing (among other transformations) a partial trace over the degrees of freedom associated with a single subsystem, $A$, of an $N$-partite state. 
This renders detailed information about $A$'s state {\it inaccessible to $A$}, but does not remove it entirely from the description of physical reality. 
It is, for example, still accessible to other subsystems, $B \neq A$. 
This description is compatible with the objective existence of wave functions, in which one three-variable wave function is associated with each quantum mechanical particle in the classical three-dimensional background space \cite{Stoica:2021owy}. 

\subsection{The ambiguity of transformations between QRFs} \label{Sec.2.3}

As shown in Sec. \ref{Sec.2.2}, the assumption that transformations between QRFs can be modelled by unitary transformations implies that individual wave functions cannot be associated, objectively, with the existence of individual particles. 
Furthermore, there is an inherent ambiguity in the definition of such a transition, in the GCB formalism \cite{Giacomini:2017zju}, since {\it at least two inequivalent operators} are required to complete its description. 

For a tripartite system, consisting of the subsystems Alice, Bob and Charlie, the transition QRF $C$ $\rightarrow$ QRF $A$ is associated with the operators 
\begin{eqnarray} \label{S_x}
\hat{S}_{x}^{(C \rightarrow A)} := \hat{\mathcal{P}}_{C \leftrightarrow A} \exp\left(\frac{i}{\hbar}{\bf x}_{A} \, . \, {\bf p}_{B}\right) \, , 
\end{eqnarray}
and 
\begin{eqnarray} \label{S_p}
\hat{S}_{p}^{(C \rightarrow A)} := \hat{\mathcal{P}}_{C \leftrightarrow A} \exp\left(-\frac{i}{\hbar}{\bf p}_{A} \, . \, {\bf x}_{B}\right) \, ,
\end{eqnarray}
where $\hat{\mathcal{P}}_{C \leftrightarrow A}$ is the parity swap operator \cite{Giacomini:2017zju}. 
These act on the composite state of $A$ and $B$, as viewed from subsystem $C$'s perspective, which is described by only two wave functions, $\ket{\psi_{A}}_{A} \otimes \ket{\phi_{B}}_{B}$, and which is assumed to to be separable, here, for simplicity. 
The former, $\hat{S}_x^{(C \rightarrow A)} $ (\ref{S_x}), implements the coordinate transformation 
\begin{eqnarray} \label{S_x_coord_transf}
{\bf q}_{A} \mapsto {\bf x}_{A} = -{\bf q}_{A} \, , \quad {\bf q}_{B} \mapsto {\bf x}_{B} = {\bf q}_{B} - {\bf q}_{C} \, , 
\nonumber\\
\boldsymbol{\pi}_{A} \mapsto {\bf p}_{A} = -(\boldsymbol{\pi}_{B} + \boldsymbol{\pi}_{C}) \, , \quad \boldsymbol{\pi}_{B} \mapsto {\bf p}_{B} = \boldsymbol{\pi}_{B} \, , 
\end{eqnarray}
where $({\bf q}_{I},\boldsymbol{\pi}_{I})$ denote the phase space coordinates of the subsystem $I \in \left\{A,B\right\}$, as seen by $C$, and $({\bf x}_{J},{\bf p}_{J})$ denote the coordinates of the subsystem $J \in \left\{B,C\right\}$, as seen by $A$, whereas the latter, $\hat{S}_p^{(C \rightarrow A)} $ (\ref{S_p}), implements the transformation 
\begin{eqnarray} \label{S_p_coord_transf}
{\bf q}_{A} \mapsto {\bf x}_{A} = -({\bf q}_{B} + {\bf q}_{C}) \, , \quad {\bf q}_{B} \mapsto {\bf x}_{B} = {\bf q}_{B} \, , 
\nonumber\\
\boldsymbol{\pi}_{A} \mapsto {\bf x}_{A} = -\boldsymbol{\pi}_{A} \, , \quad \boldsymbol{\pi}_{B} \mapsto {\bf p}_{B} = \boldsymbol{\pi}_{B} - \boldsymbol{\pi}_{C} \, .
\end{eqnarray}

However, only the transformations on the first line of Eqs. (\ref{S_x_coord_transf}) and on the second line of Eqs. (\ref{S_p_coord_transf}) make intuitive sense, according to a canonical reference frame diagram; see, for example, Fig. 6, below, or the figures provided in \cite{Giacomini:2017zju}. 
According to GCB this problem is solved by applying the operator $\hat{S}_x^{(C \rightarrow A)}$ to the composite state of Alice and Bob, in the position space representation, and applying, instead, the operator $\hat{S}_p^{(C \rightarrow A)}$, to the same state, when using the momentum space representation. 

In other words, $\hat{S}_x^{(C \rightarrow A)}$ is supposed to transform Charlie's view of Alice and Bob, in position space, into Alice's view of Bob and Charlie (also in position space) whereas $\hat{S}_p^{(C \rightarrow A)}$ transforms Charlie's view of Alice and Bob, in momentum space, into Alice's view of Bob and Charlie's system (also in momentum space). 
In the GCB formalism, therefore, one must first choose a particular representation for the state of the composite system, viewed from the perspective of a given QRF, before one considers how this state changes when transforming to the perspective of a different subsystem. 
Different representations, chosen initially, imply different transformations, and therefore lead to different final states \cite{Giacomini:2017zju}. 

An obvious objection to this construction comes when one considers, for example, the energy eigenstate basis. 
Let us suppose, for simplicity, that both Alice's state, and Bob's, can be expressed as discrete sums of non-degenerate energy eigenstates, from Charlie's perspective,  
\begin{eqnarray} \label{energy_eigenstates_Bob}
\ket{\psi_{A}}_{A} = \sum_{n=1}^{\infty} \alpha_{n} \ket{\psi_{n}}_{A} \, ,  \quad \ket{\phi_{B}}_{B} = \sum_{n=1}^{\infty} \beta_{n} \ket{\phi_{n}}_{B} \, , 
\end{eqnarray}
where
\begin{eqnarray} \label{energy_eigenstates_Bob*}
\sum_{n=1}^{\infty} |\alpha_{n}|^2 = \sum_{n=1}^{\infty} |\beta_{n}|^2 = 1 \, , \quad \braket{\psi_{n}|\psi_{n'}}_{A} = \braket{\phi_{n}|\phi_{n'}}_{B} = \delta_{nn'} \, ,
\end{eqnarray}
and 
\begin{eqnarray} \label{energy_eigenstates_Bob**}
\hat{H}_{A}\ket{\psi_{n}}_{A} = E_{n}^{(A)}\ket{\psi_{n}}_{A} \, ,  \quad \hat{H}_{B}\ket{\phi_{n}}_{B} = E_{n}^{(B)}\ket{\phi_{n}}_{B} \, , 
\end{eqnarray}
where $\hat{H}_{A}$ and $\hat{H}_{B}$ denote Alice and Bob's local Hamiltonians, respectively. 
Clearly, each individual eigenstate, $\ket{\psi_{n}}_{A}$ and $\ket{\phi_{n}}_{B}$, can be expanded in terms of either a position or a momentum space wave function, 
\begin{eqnarray} \label{energy_eigenstates_Bob***}
\ket{\psi_{n}}_{A} = \int \psi_{n}({\bf q}_{A})\ket{{\bf q}_{A}}_{A}{\rm d}^{3}{\rm q}_{A} = \int \tilde{\psi}_{n}(\boldsymbol{\pi}_{A})\ket{\boldsymbol{\pi}_{A}}_{A}{\rm d}^{3}{\rm \pi}_{A} \, ,
\nonumber\\
\ket{\phi_{n}}_{B} = \int \phi_{n}({\bf q}_{B})\ket{{\bf q}_{B}}_{B}{\rm d}^{3}{\rm q}_{B} = \int \tilde{\phi}_{n}(\boldsymbol{\pi}_{B})\ket{\boldsymbol{\pi}_{B}}_{B}{\rm d}^{3}{\rm \pi}_{B} \, .
\end{eqnarray}

In this case, which operator should we apply to the state $\ket{\psi_{A}}_{A} \otimes \ket{\phi_{B}}_{B}$; $\hat{S}_x^{(C \rightarrow A)}$ or  $\hat{S}_p^{(C \rightarrow A)}$? 
Theoretically, one of these should tell us how the energy eigenspectrum of Bob and Charlie's state appears, to Alice, but there is no clear way to choose between them. 
Alternatively, one must define yet another unitary operator, to describe transformations between QRFs in the energy eigenstate basis, leading to at least a three-way ambiguity in the definition of such a transition.
\footnote{In fact, the situation is worse than this. In canonical QM, we are accustomed to speaking, somewhat lazily, about two representations of the wave function; the position space representation, $\psi({{\bf x}}) \equiv \psi(x^1,x^2,x^3)$, and the momentum space representation, $\tilde{\psi}({{\bf p}}) \equiv \tilde{\psi}(p_1,p_2,p_3)$. In reality, there are $2^3 = 8$ independent representations, comprised of the two aforementioned wave functions together with the six `hybrid' representations, $\tilde{\psi}(x^1,p_2,p_3)$, $\tilde{\psi}(p_1,x^2,p_3)$, etc. By the logic of the GCB formalism \cite{Giacomini:2017zju}, this leads to at least an eight-fold ambiguity in the definition of the transformed energy eigenspectra, discussed above. Furthermore, we could, in principle, choose to express each energy eigenstate in a different basis, $(x^1,x^2,x^3), \, (x^1,p_2,p_3), \dots (p_1,p_2,p_3)$. Experimentally, this corresponds to probing each energy level in the superposition via different set of measurements. This, in turn, requires us to associate $8^n$ inequivalent unitary operators, corresponding to $8^n$ different sets of coordinate transformations, with the single transition QRF $C$ $\rightarrow $ QRF $A$.} 

\subsection{Angular momentum and the spin degrees of freedom} \label{Sec.2.4}

Similar remarks also apply to the degrees of freedom associated with angular momentum, since, in general, we require three quantum numbers to specify a normalised eigenstate, $\ket{n \, l \, m}$. 
In fact, we note that even the `good' transformations produced by $\hat{S}_x^{(C \rightarrow A)}$ and $\hat{S}_p^{(C \rightarrow A)}$, i.e., those on the top line of Eqs. (\ref{S_x_coord_transf}) and those on the bottom line of Eqs. (\ref{S_p_coord_transf}), respectively, implicitly assume that the reference frames of $A$, $B$ and $C$ all share the same relative orientation. 
It is, therefore, by no means obvious how to describe transformations between QRFs in the energy-angular momentum basis, using the formalism developed in \cite{Giacomini:2017zju}. 
As a result, it is by no means clear how to describe transformations between quantum subsystems whose internal coordinates are rotated, or in a superposition of rotations, relative to one another. 

An additional complication also arises with respect to the spin degrees of freedom. 
Given the close connection between spin and orbital angular momentum, or, equivalently, between the groups $SU(2)$ and $SO(3)$, one would expect that, whatever the correct description of QRF-to-QRF transitions ought to be, in the angular momentum basis, the transformation of the spin degrees of freedom should share many of the same properties. 
In Sec. \ref{Sec.4}, we use the alternative formalism developed earlier in Sec. \ref{Sec.3} to try to resolve some of these outstanding issues.

\section{Proposal for an alternative formalism} \label{Sec.3}

\subsection{Gedanken experiment with a single particle system} \label{Sec.3.1}

Consider a canonical quantum single-particle state, labelled $A$, viewed from the perspective an arbitrary CRF, $O$. 
In keeping with accepted terminology we also refer to the state $A$ as `Alice', but we refer to $O$ as simply `$O$', in order to emphasise its different, classical-as-opposed-to-quantum nature. 
The CRF $O$ is equipped with coordinate systems, in both position and momentum space, whose origins are labelled by crosses in the diagrams presented in Fig. 3. 
The possible eigentates of Alice's wave function are labelled by dots and the shaded regions represent the volumes of physical space and momentum space, respectively, where $\psi_{A}$ and $\tilde{\psi}_{A}$ are nonzero. 
From $O$'s perspective, the state of Alice's system is,
\begin{eqnarray} \label{psi_A(O)}
\ket{\psi_{A}^{(O)}}_{A} = \int \psi_{A}({\bf r}_{A}) \ket{{\bf r}_{A}}_{A} {\rm d}^3{\rm r}_{A} = \int \tilde{\psi}_{A}(\boldsymbol{\kappa}_{A}) \ket{\boldsymbol{\kappa}_{A}}_{A} {\rm d}^3{\rm \kappa}_{A}  \, ,
\end{eqnarray}
where $\boldsymbol{\kappa}_{A}$ is the physical momentum conjugate to ${\bf r}_{A}$, 
\begin{eqnarray} \label{r_A_braket}
\braket{{\bf r}_{A} | \boldsymbol{\kappa}_{A}}_{A} = \left(\frac{1}{\sqrt{2\pi\hbar}}\right)^3e^{\frac{i}{\hbar} \boldsymbol{\kappa}_{A} . {\bf r}_{A}} \, ,
\end{eqnarray}
and $\tilde{\psi}_{A}(\boldsymbol{\kappa}_{A})$ is the Fourier transform of $\psi_{A}({\bf r}_{A})$.

Conversely, from Alice's perspective, the quantum state of $O$ is, 
\begin{eqnarray} \label{psi_O(A)}
\ket{\psi_{O}^{(A)}}_{O} &=& \int \psi_{O}({\bf r}_{A}) \ket{-{\bf r}_{A}}_{O} {\rm d}^3{\rm r}_{A} = \int \tilde{\psi}_{O}(\boldsymbol{\kappa}_{A}) \ket{-\boldsymbol{\kappa}_{A}}_{O} {\rm d}^3\kappa_{A}  
\nonumber\\
&=& \int \psi_{O}(-{\bf x}_{O}) \ket{{\bf x}_{O}}_{O} {\rm d}^3{\rm x}_{O} = \int \tilde{\psi}_{O}(-{\bf p}_{O}) \ket{{\bf p}_{O}}_{O} {\rm d}^3{\rm p}_{O}  \, ,
\end{eqnarray}
where ${\bf p}_{O}$ is the momentum conjugate to ${\bf x}_{O}$, 
\begin{eqnarray} \label{x_O_braket}
\braket{{\bf x}_{O} |{\bf p}_{O}}_{O} = \left(\frac{1}{\sqrt{2\pi\hbar}}\right)^3e^{\frac{i}{\hbar} {\bf p}_{O} . {\bf x}_{O}} \, , 
\end{eqnarray}
and $\tilde{\psi}_{A}({\bf p}_{O})$ is the Fourier transform of $\psi_{A}({\bf x}_{O})$, with
\begin{eqnarray} \label{single_particle_coord_transf}
{\bf x}_{O} = -{\bf r}_{A} \, , \quad {\bf p}_{O} = -\boldsymbol{\kappa}_{A} \, . 
\end{eqnarray}

In this case, it is clear that the operator affecting the switch from the CRF $O$ to the QRF $A$ is trivial, namely, it is simply the parity swap operator, $\hat{\mathcal{P}}_{O \leftrightarrow A}$, which acts on the basis vectors $\ket{{\bf r}_{A}}_{A}$ and $\ket{\boldsymbol{\kappa}_{A}}_{A}$ according to
\begin{eqnarray} \label{ParitySwap_OA}
\hat{\mathcal{P}}_{O \leftrightarrow A} \ket{{\bf r}_{A}}_{A} = \ket{-{\bf r}_{A}}_{O} \, , \quad  \hat{\mathcal{P}}_{O \leftrightarrow A} \ket{\boldsymbol{\kappa}_{A}}_{A} = \ket{-\boldsymbol{\kappa}_{A}}_{O} \, ,
\end{eqnarray}
giving
\begin{eqnarray} \label{psi_A(O)->psi_O(A)}
\ket{\psi_{O}^{(A)}}_{O} = \hat{\mathcal{P}}_{O \leftrightarrow A} \ket{\psi_{A}^{(O)}}_{A} \, .
\end{eqnarray}
The switch back from the QRF $A$ to the CRF $O$ is affected by the inverse operator, $\hat{\mathcal{P}}_{A \leftrightarrow O} \equiv \hat{\mathcal{P}}_{O \leftrightarrow A}$, since
\begin{eqnarray} \label{ParitySwap_OA*}
\hat{\mathcal{P}}_{O \leftrightarrow A} \ket{-{\bf r}_{A}}_{O} = \ket{{\bf r}_{A}}_{A} \, , \quad  \hat{\mathcal{P}}_{O \leftrightarrow A} \ket{-\boldsymbol{\kappa}_{A}}_{O} = \ket{\boldsymbol{\kappa}_{A}}_{A} \, .
\end{eqnarray}
In other words, a parity swap between two observers is a self-inverse operation. 

\begin{figure}[h] \label{Fig.3}
\begin{center}
\includegraphics[width=8cm]{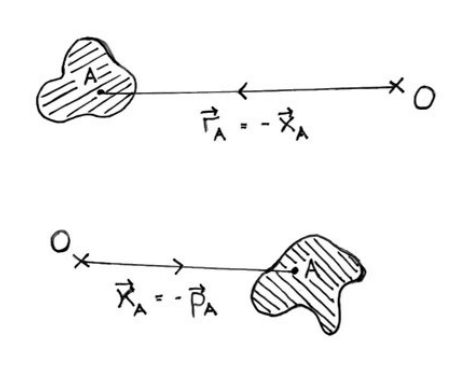}
\caption{A single-particle state, $A$, viewed from the perspective of an arbitrary classical observer, $O$, in both position and momentum space, is described the coordinates ${\bf r}_{A}$ and $\boldsymbol{\kappa}_{A}$, respectively. Conversely, from the perspective of $A$, the position and momentum space coordinates of $O$ are ${\bf x}_{O} = -{\bf r}_{A}$ and ${\bf p}_{O} = -\boldsymbol{\kappa}_{A}$.}
\end{center}
\end{figure}

Note that, here, $O$ is not genuinely associated with a quantum state. 
This may be easily verified by considering the perspective of another CRF, $O' \neq O$. 
However, since, by construction, a QRF should not be capable of obtaining any information about its own internal state, by means of an individual measurement, Alice perceives each point in the classical spatial background, and in its dual momentum space, as existing in a superposition of states.
\footnote{In Secs. \ref{Sec.3.2}-\ref{Sec.3.4}, we consider ways in which a QRF {\it can} obtain limited, statistical information about its own internal state. The implications of these findings for models of phenomenological quantum gravity are further explored in Sec. \ref{Sec.5}.}  
The position and momentum uncertainties of these states are
\begin{eqnarray} \label{uncertainties_OA}
\Delta_{\psi}x_{O}^{i} = \Delta_{\psi}r_{A}^{i} \, , \quad \Delta_{\psi}p_{Oj} = \Delta_{\psi}\kappa_{Aj} \, ,
\end{eqnarray}
where 
\begin{eqnarray} \label{operators_OA-1}
\hat{r}_{A}^{i} = \int r_{A}^{i} \ket{{\bf r}_{A}}\bra{{\bf r}_{A}}_{A} {\rm d}^{3} r_{A} \, , \quad \hat{\kappa}_{Aj} = \int \kappa_{Aj} \ket{\boldsymbol{\kappa}_{A}}\bra{\boldsymbol{\kappa}_{A}}_{A} {\rm d}^{3} \kappa_{A} \, , 
\end{eqnarray}
and
\begin{eqnarray} \label{operators_OA-2}
&&\hat{x}_{O}^{i} = \int x_{O}^{i} \ket{{\bf x}_{O}}\bra{{\bf x}_{O}}_{O} {\rm d}^{3} x_{O} = \int (-r_{A}^{i}) \ket{-{\bf r}_{A}}\bra{-{\bf r}_{A}}_{O} {\rm d}^{3} r_{A} \, , \quad 
\nonumber\\
&&\hat{p}_{Oj} = \int p_{Oj} \ket{{\bf p}_{O}}\bra{{\bf p}_{O}}_{O} {\rm d}^{3} p_{O} = \int (-\kappa_{Aj}) \ket{-\boldsymbol{\kappa}_{A}}\bra{-\boldsymbol{\kappa}_{A}}_{O} {\rm d}^{3} \kappa_{A} \, , 
\end{eqnarray}
and the analysis above remains valid, regardless of the fact that no physical observer can exist as a CRF. 

Physically, what happens when Alice's wave function is collapsed to a position eigenstate is an effective resetting of the origin of her internal coordinate system, ${\bf x}_{O}$. 
Unlike $O$'s coordinate system, this is not fixed on the classical background, but exists in a superposition of states, a different coordinate origin being associated with each point at which Alice's position space wave function, 
$\psi_{A}({\bf r}_{A})$, is nonzero. 
Similar remarks apply to her `coordinate system'  in momentum space - really, a superposition of coordinate systems - which is determined by $\tilde{\psi}_{A}(\boldsymbol{\kappa}_{A})$. 

These remarks conclude our analysis of the single-particle state, except for one final but absolutely crucial caveat. 
Not only is Alice forbidden, in her capacity as the `observing' rather than the `observed' quantum system, from obtaining non-statistical information about her own degrees of freedom, she is also restricted, as a material body, to performing measurement-type interactions only with other material bodies. 
Therefore, it is not strictly speaking true to say that `from Alice's perspective, the quantum state of $O$ is Eq. (\ref{psi_O(A)})', as stated above. 
More accurately, $\ket{\psi_{O}^{(A)}}_{O}$ is the quantum state that Alice would attribute to the spatial point $O$, {\it if} she were able to perform any physical measurements by which she could gauge her own superposition of states {\it relative} to the background geometry. 

In reality, she cannot, and we must account for this fact by tracing out the degrees of freedom that $O$ attributes to Alice, and that Alice would attribute to $O$, {\it if} she were able to physically interact with a CRF. 
This means tracing over $\mathcal{H}_{A}$ in Eq. (\ref{psi_A(O)}) and over $\mathcal{H}_{O}^{(A)} := \hat{\mathcal{P}}_{O \leftrightarrow A}\mathcal{H}_{A}$ - the {\it effective} Hilbert space that Alice nominally associates with $O$ - in Eq. (\ref{psi_O(A)}), which destroys all information about the system.
In this respect, the single-particle system is utterly trivial, as a QRF, although the arguments presented above are anything but. 
In the next subsection, we consider adding a second quantum particle $B$, or `Bob', to the single-particle system considered here. 
This allows us to compare, in the simplest nontrivial example, the information about the bipartite state that is available to the two QRFs, $A$ and $B$, and to an arbitrary CRF, $O$. 
By making use of, and extending, the arguments presented here, we show that the latter is complete, whereas the former is restricted, but nonzero.  

\subsection{Gedanken experiment with a bipartite system} \label{Sec.3.2}

Let us now consider a bipartite state, consisting of quantum subsystems $A$ and $B$, or Alice and Bob, viewed from the perspective an arbitrary CRF, $O$. 
We use the bold face letters $({\bf r}_{A/B},\boldsymbol{\kappa}_{A/B})$ to denote the real and momentum space coordinates of $A$ and $B$, as seen by $O$, the letters $({\bf x}_{O/B}, {\bf p}_{O/B})$ to denote the coordinates of $O$ and $B$, as seen by $A$, and $({\bf q}_{O/A},\boldsymbol{\pi}_{O/A})$ to denote the coordinates of $O$ and $A$, as seen by $B$. 
Let us assume that, from $O$'s perspective, the quantum state of the bipartite system $AB$ is separable,
\begin{eqnarray} \label{psi_AB(O)}
\ket{\Psi_{AB}^{(O)}}_{AB} &=& \int \int \psi_{A}({\bf r}_{A}) \phi_{B}({\bf r}_{B}) \ket{{\bf r}_{A}}_{A} \ket{{\bf r}_{B}}_{B} {\rm d}^3{\rm r}_{A}{\rm d}^3{\rm r}_{B} 
\nonumber\\
&=& \int \int \tilde{\psi}_{A}(\boldsymbol{\kappa}_{A}) \tilde{\phi}_{B}(\boldsymbol{\kappa}_{B}) \ket{\boldsymbol{\kappa}_{A}}_{A} \ket{\boldsymbol{\kappa}_{B}}_{B} {\rm d}^3{\rm \kappa}_{A} {\rm d}^3{\rm \kappa}_{B} \, .
\end{eqnarray}

\begin{figure}[h] \label{Fig.4}
\begin{center}
\includegraphics[width=16cm]{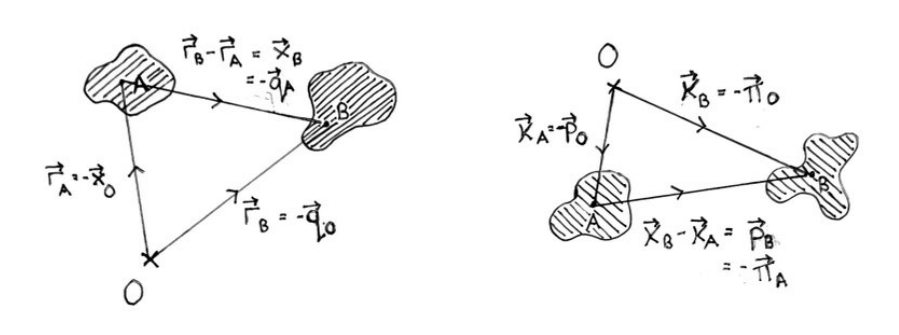}
\caption{A bipartite state, consisting of quantum subsystems $A$ and $B$, viewed from the perspective of an arbitrary classical observer, $O$, in both position and momentum space.}
\end{center}
\end{figure}

We may then ask: how does Alice perceive the `bipartite' state, consisting of $B$ and $O$? 
Again, we stress that the CRF $O$ is not genuinely associated with a quantum state, but we may apply the same logic as before to obtain the coordinate transform (\ref{single_particle_coord_transf}), which is associated with the parity swap operator, $\hat{\mathcal{P}}_{O \leftrightarrow A}$ (\ref{ParitySwap_OA}). 
The key difference between this example and the single particle state considered in Sec. \ref{Sec.3.1} is the existence of a second quantum state, and potential QRF, $B$. 
We must therefore ask: how does Alice perceive Bob? 

\subsubsection{Position measurements} \label{Sec.3.2.1}

To answer this question, let us first consider the position space representation. 
In order for Alice to resolve a concrete position relative to Bob, both Bob's wave function, and hers, must collapse to single-value eigenstates. 
From the `external' perspective of the CRF $O$ these states correspond to the positions ${\bf r}_{A}$ and ${\bf r}_{B}$ - that is, to the basis vector $\ket{{\bf r}_{A}}_{A} \ket{{\bf r}_{B}}_{B}$ of the bipartite system $AB$ - and the probability associated with the transition $\ket{\Psi_{AB}^{(O)}}_{AB} \mapsto \ket{{\bf r}_{A}}_{A} \ket{{\bf r}_{B}}_{B}$ is $|\psi_{A}({\bf r}_{A})|^2 |\phi_{B}({\bf r}_{B})|^2{\rm d}^{3}{\rm r}_{A}{\rm d}^{3}{\rm r}_{B}$. 

From Alice's perspective, this pair of states corresponds to a single measured value of her position relative to Bob, ${\bf x}_{B} = {\bf r}_{B} - {\bf r}_{A}$, or, equivalently, to a projection onto the basis vector $\ket{{\bf x}_{B}}_{B} = \ket{{\bf r}_{B} - {\bf r}_{A}}_{B}$. 
(We recall, again, that Alice has no physical means by which to measure the distance ${\bf x}_{O}=-{\bf r}_{A}$.)
Moreover, since there are infinitely many bipartite projections, $\ket{{\bf r}_{A}}_{A} \ket{{\bf r}_{B}}_{B}$, that give rise to single value of ${\bf x}_{B} = {\bf r}_{B} - {\bf r}_{A}$, the probability of obtaining a given value of ${\bf x}_{B}$ is not $|\psi_{A}({\bf r}_{A})|^2 |\phi_{B}({\bf r}_{B})|^2{\rm d}^{3}{\rm r}_{A}{\rm d}^{3}{\rm r}_{B} \equiv |\psi_{O}(-{\bf x}_{O})|^2 |\phi_{B}({\bf r}_{B})|^2{\rm d}^{3}{\rm x}_{O}{\rm d}^{3}{\rm r}_{B}$ but
\begin{eqnarray} \label{|Phi_B|^2}
{\rm d}^{3}P({\bf x}_{B}) := |\psi_{O}(-{\bf x}_{O})|^2 |\phi_{B}({\bf r}_{B})|^2{\rm d}^{3}{\rm x}_{O}{\rm d}^{3}{\rm r}_{B}\big|_{{\bf r}_{B} = {\bf x}_{B} - {\bf x}_{O}} \, , 
\end{eqnarray}
or, equivalently, 
\begin{eqnarray} \label{|Phi_B|^2*}
{\rm d}^{3}P({\bf x}_{B}) &:=& \left(\int |\psi_{O}(-{\bf x}_{O})|^2 |\phi_{B}({\bf x}_{B}-{\bf x}_{O})|^2{\rm d}^{3}{\rm x}_{O}\right) {\rm d}^{3}{\rm x}_{B} 
\nonumber\\
&=& |\psi_{O}|^2 * |\phi_{B}|^2({\bf x}_{B}) {\rm d}^{3}{\rm x}_{B}  \, .
\end{eqnarray}

Next, we note that it is possible to obtain (\ref{|Phi_B|^2*}) by performing a partial trace over the product of a density operator, 
\begin{eqnarray} \label{rho'_OB}
\hat{\rho}_{OB}'^{(A)} := \ket{\Psi_{OB}'^{(A)}}\bra{\Psi_{OB}'^{(A)}}_{OB} \, , 
\end{eqnarray}
where the state $\ket{\Psi_{OB}'^{(A)}}$ is defined as
\begin{eqnarray} \label{Psi'_OB(A)}
\ket{\Psi_{OB}'^{(A)}}_{OB} &:=& \int \int \psi_{O}(-{\bf x}_{O}) \phi_{B}({\bf x}_{B}-{\bf x}_{O}) \ket{{\bf x}_{O}}_{O} \ket{{\bf x}_{B}}_{B} {\rm d}^3{\rm x}_{O}{\rm d}^3{\rm x}_{B}
\nonumber\\
&=& \int \int \psi_{O}({\bf r}_{A}) \phi_{B}({\bf r}_{B}) \ket{-{\bf r}_{A}}_{O} \ket{{\bf r}_{B}-{\bf r}_{A}}_{B} {\rm d}^3{\rm r}_{A}{\rm d}^3{\rm r}_{B} 
\nonumber\\
&\equiv&  \int \int \tilde{\psi}_{O}(-({\bf p}_{B}+{\bf p}_{O})) \tilde{\phi}_{B}({\bf p}_{B}) \ket{{\bf p}_{O}}_{O} \ket{{\bf p}_{B}}_{B} {\rm d}^3{\rm p}_{O}{\rm d}^3{\rm p}_{B}
\nonumber\\
&=& \int \int \tilde{\psi}_{O}(\boldsymbol{\kappa}_{A}) \tilde{\phi}_{B}(\boldsymbol{\kappa}_{B}) \ket{-(\boldsymbol{\kappa}_{B}+\boldsymbol{\kappa}_{A})}_{O}\ket{\boldsymbol{\kappa}_{B}}_{B} {\rm d}^3{\rm \kappa}_{A}{\rm d}^3{\rm \kappa}_{B} \, ,
\end{eqnarray}
and an appropriate projection operator,
\begin{eqnarray} \label{Pi(x0,xB)}
\hat{\Pi}_{\ket{{\bf x}_{O}}_{O}\ket{{\bf x}_{B}}_{B}} := \ket{{\bf x}_{O}}\bra{{\bf x}_{O}}_{O} \otimes \ket{{\bf x}_{B}}\bra{{\bf x}_{B}}_{B} \, . 
\end{eqnarray}
Thus, we have 
\begin{eqnarray} \label{Tr_rho'_OB}
{\rm d}^{3}P({\bf x}_{B}) = {\rm tr}_{O}\left[\hat{\Pi}_{\ket{{\bf x}_{O}}_{O}\ket{{\bf x}_{B}}_{B}}  \hat{\rho}'_{OB}\right] {\rm d}^{3}{\rm x}_{B} = |\psi_{O}|^2 * |\phi_{B}|^2({\bf x}_{B}) {\rm d}^{3}{\rm x}_{B} \, , 
\end{eqnarray}
where ${\rm tr}_{O}$ denotes the partial trace over the $\mathcal{H}_{O}^{(A)}$ subspace of $\mathcal{H}_{OB}^{(A)} = \mathcal{H}_{O}^{(A)} \otimes \mathcal{H}_{B}$. 
Here, we again use the notation $\mathcal{H}_{O}^{(A)} := \hat{\mathcal{P}}_{O \leftrightarrow A}\mathcal{H}_{A}$ to refer to the {\it effective} Hilbert space that Alice nominally associates with $O$. 

This seems to suggest the following interpretation, by analogy with the state $\ket{\psi_{O}^{(A)}}_{O}$ (\ref{psi_A(O)}) considered in Sec. \ref{Sec.3.1}: `$\ket{\Psi_{OB}'^{(A)}}_{OB}$ is the state that of the `bipartite' system, $OB$, that Alice would see, {\it if} she were able to physically interact with the CRF $O$'. 
Indeed, the state (\ref{Psi'_OB(A)}) was considered in \cite{Giacomini:2017zju} (using different notation) and was interpreted by the authors as representing the state of a tripartite system, consisting of two quantum particles viewed from the perspective of a third. 
Therefore, if the interpretation given above is valid, the QRF formalism presented here is equivalent to the GCB formalism, except for the fact that their three-particle state is equivalent to our two-particle state, their four-particle state is equivalent to our three-particle state, and so on. 
However, as we will now show, this is certainly not the case. 
We chose to denote the state (\ref{Psi'_OB(A)}) with a prime, as $\ket{\Psi_{OB}'^{(A)}}_{OB}$, in order to distinguish it from the state that can be interpreted, legitimately, in the way described above.
The latter is denoted without a prime, as $\ket{\Psi_{OB}^{(A)}}_{OB}$, and its explicit form is derived later in this section. 
It is the density matrix constructed from {\it this} state that we must trace over, in the $\mathcal{H}_{O}^{(A)}$ subspace, in order to extract the predictions of our model, for the QRF $A$.

In order to see why $\ket{\Psi_{OB}'^{(A)}}_{OB}$ is, in fact, not a good choice for `the state of the `bipartite' system $OB$, as seen by $A$', we must consider the momentum-basis expansion, given in the last two lines of Eq. (\ref{Psi'_OB(A)}). 
First, we note that this follows directly from the definition of the position-basis expansion, given in the top two lines, plus what GCB refer to as the `canonical completion' of the coordinate transform \cite{Giacomini:2017zju}. 
By this, they mean simply that one must obtain the same expression for the wave function, $\Psi_{OB}'^{(A)}({\bf x}_{O},{\bf x}_{B}) = \psi_{O}(-{\bf x}_{O}) \phi_{B}({\bf x}_{B}-{\bf x}_{O})$, when acting with the bras $\bra{{\bf x}_{O}}_{O}\bra{{\bf x}_{B}}_{B}$ on the ket $\ket{\Psi_{OB}'^{(A)}}_{OB}$, regardless of whether one works with the position-basis expansion or the momentum-basis expansion of the latter. 
The dual-basis completion then holds automatically, i.e., acting with $\bra{{\bf p}_{O}}_{O}\bra{{\bf p}_{B}}_{B}$ on $\ket{\Psi_{OB}'^{(A)}}_{OB}$ gives the same wave function, $\tilde{\Psi}_{OB}'^{(A)}({\bf p}_{O},{\bf p}_{B}) =  \tilde{\psi}_{O}(-({\bf p}_{B}+{\bf p}_{A})) \tilde{\phi}_{B}({\bf p}_{B})$, regardless of which basis one chooses for the expansion of $\ket{\Psi_{OB}'^{(A)}}_{OB}$. 

From this simple requirement it follows immediately that, {\it if} the unitary transformation $\ket{\Psi_{AB}^{(O)}}_{AB} \mapsto \ket{\Psi_{OB}'^{(A)}}_{OB}$ represents a physical transformation from the CRF $O$ to the QRF $A$, this `jump' is characterised by the coordinate transformations
\begin{eqnarray} \label{GCB_coord_transfs'}
{\bf r}_{A} \mapsto {\bf x}_{O} = -{\bf r}_{A} \, , \quad {\bf r}_{B} \mapsto {\bf x}_{B} = {\bf r}_{B} - {\bf r}_{A} \, , 
\nonumber\\
\boldsymbol{\kappa}_{A} \mapsto {\bf p}_{O} = -(\boldsymbol{\kappa}_{B}+\boldsymbol{\kappa}_{A}) \, , \quad \boldsymbol{\kappa}_{B} \mapsto {\bf p}_{B} = \boldsymbol{\kappa}_{B} \, ,
\end{eqnarray}
as in the position space definition of a QRF-to-QRF transition, in the GCB model \cite{Giacomini:2017zju}. 

Unfortunately, only the position space transformations, given in Eqs. (\ref{GCB_coord_transfs'}), are consistent with the intuitive diagram Fig. 4. 
The momentum space transformations, by contrast, are extremely difficult to explain. 
So, if $\ket{\Psi_{OB}'^{(A)}}_{OB}$ is not the correct description of the state of the $OB$ system, as viewed by Alice, then what is? 

\subsubsection{Momentum measurements} \label{Sec.3.2.2}

To answer this question, we must also consider measurements of momentum, as viewed by both $O$ and $A$. 
We proceed by analogy with our previous arguments, regarding position measurements, since the same basic logic holds, with respect to the appropriate representation, in both cases. 

We begin by noting that, in order for Alice to resolve a concrete momentum relative to Bob, both her wave function and his must collapse to momentum eigenstates. 
From $O$'s perspective, these correspond to the values $\boldsymbol{\kappa}_{A}$ and $\boldsymbol{\kappa}_{B}$, or, equivalently, to a projection onto the basis vectors $\ket{\Psi_{AB}^{(O)}}_{AB} \mapsto \ket{\boldsymbol{\kappa}_{A}}_{A}\ket{\boldsymbol{\kappa}_{A}}_{B}$, whose probability is $|\tilde{\psi}_{A}(\boldsymbol{\kappa}_{A})|^2|\tilde{\phi}_{B}(\boldsymbol{\kappa}_{B})|^2 {\rm d}^{3}{\rm \kappa}_{A}{\rm d}^{3}{\rm \kappa}_{B}$. 
From Alice's perspective, this transition corresponds to a single measured value, ${\bf p}_{B} = \boldsymbol{\kappa}_{B}-\boldsymbol{\kappa}_{A}$, or, equivalently, a projection onto the basis vector $\ket{{\bf p}_{B}}_{B} = \ket{\boldsymbol{\kappa}_{B}-\boldsymbol{\kappa}_{A}}_{B}$. 
Again, as there are an infinite number of bipartite projections, $\ket{\boldsymbol{\kappa}_{A}}_{A}\ket{\boldsymbol{\kappa}_{A}}_{B}$, that give rise to single measured value of ${\bf p}_{B} = \boldsymbol{\kappa}_{B}-\boldsymbol{\kappa}_{A}$, the probability of obtaining any given value of ${\bf p}_{B}$ is not $|\tilde{\psi}_{A}(\boldsymbol{\kappa}_{A})|^2|\tilde{\phi}_{B}(\boldsymbol{\kappa}_{B})|^2 {\rm d}^{3}{\rm \kappa}_{A}{\rm d}^{3}{\rm \kappa}_{B} \equiv |\tilde{\psi}_{O}(-{\bf p}_{O})|^2|\tilde{\phi}_{B}(\boldsymbol{\kappa}_{B})|^2 {\rm d}^{3}{\rm p}_{O}{\rm d}^{3}{\rm \kappa}_{B}$ but 
\begin{eqnarray} \label{|tildePhi_B|^2}
{\rm d}^{3}P({\rm p}_{B}) := |\tilde{\psi}_{O}(-{\bf p}_{O})|^2 |\phi_{B}(\boldsymbol{\kappa}_{B})|^2{\rm d}^{3}{\rm p}_{O}{\rm d}^{3}{\rm \kappa}_{B}\big|_{\boldsymbol{\kappa}_{B} = {\bf p}_{B} - {\bf p}_{O}} \, , 
\end{eqnarray}
or, equivalently, 
\begin{eqnarray} \label{|tildePhi_B|^2*}
{\rm d}^{3}P({\bf p}_{B})  &:=& \left(\int |\tilde{\psi}_{O}(-{\bf p}_{O})|^2 |\tilde{\phi}_{B}({\bf p}_{B}-{\bf p}_{O})|^2{\rm d}^{3}{\rm p}_{O}\right) {\rm d}^{3}{\rm p}_{B} 
\nonumber\\
&=& |\tilde{\psi}_{O}|^2 * |\tilde{\phi}_{B}|^2({\bf p}_{B}) {\rm d}^{3}{\rm p}_{B}  \, .
\end{eqnarray}

Next, we note that (\ref{|tildePhi_B|^2*}) can be obtained by performing a partial trace over the product of the density operator 
\begin{eqnarray} \label{rho''_OB}
\hat{\rho}_{OB}''^{(A)} := \ket{\Psi_{OB}''^{(A)}}\bra{\Psi_{OB}''^{(A)}}_{OB} \, ,
\end{eqnarray}
where the state $\ket{\Psi_{OB}''^{(A)}}$ is defined as
\begin{eqnarray} \label{Psi''_OB(A)}
\ket{\Psi_{OB}''^{(A)}}_{OB} &:=& \int \int \tilde{\psi}_{O}(-{\bf p}_{O}) \tilde{\phi}_{B}({\bf p}_{B}-{\bf p}_{O}) \ket{{\bf p}_{O}}_{O} \ket{{\bf p}_{B}}_{B} {\rm d}^3{\rm p}_{O}{\rm d}^3{\rm p}_{B}
\nonumber\\
&=& \int \int \tilde{\psi}_{O}(\boldsymbol{\kappa}_{A}) \tilde{\phi}_{B}(\boldsymbol{\kappa}_{B}) \ket{-\boldsymbol{\kappa}_{A}}_{O} \ket{\boldsymbol{\kappa}_{B}-\boldsymbol{\kappa}_{A}}_{B} {\rm d}^3{\rm \kappa}_{A}{\rm d}^3{\rm \kappa}_{B} 
\nonumber\\
&\equiv&  \int \int \psi_{O}(-({\bf x}_{B}+{\bf x}_{O})) \phi_{B}({\bf x}_{B}) \ket{{\bf x}_{O}}_{O} \ket{{\bf x}_{B}}_{B} {\rm d}^3{\rm x}_{O}{\rm d}^3{\rm x}_{B}
\nonumber\\
&=& \int \int \psi_{O}({\bf r}_{A}) \phi_{B}({\bf r}_{B}) \ket{-({\bf r}_{B}+{\bf r}_{A})}_{O}\ket{{\bf r}_{B}}_{B} {\rm d}^3{\rm r}_{A}{\rm d}^3{\rm r}_{B} \, ,
\end{eqnarray}
and the projection operator 
\begin{eqnarray} \label{Pi(p0,pB)}
\hat{\Pi}_{\ket{{\bf p}_{O}}_{O}\ket{{\bf p}_{B}}_{B}} := \ket{{\bf p}_{O}}\bra{{\bf p}_{O}} \otimes \ket{{\bf p}_{B}}\bra{{\bf p}_{B}} \, . 
\end{eqnarray}
Thus, we have
\begin{eqnarray} \label{Tr_rho''_OB}
{\rm d}^{3}P({\rm p}_{B}) = {\rm tr}_{O}\left[\hat{\Pi}_{_{\ket{{\bf p}_{O}}_{O}\ket{{\bf p}_{B}}_{B}}}\hat{\rho}''_{OB}\right] {\rm d}^{3}{\rm p}_{B} = |\tilde{\psi}_{O}|^2 * |\tilde{\phi}_{B}|^2({\bf p}_{B}) {\rm d}^{3}{\rm p}_{B} \, .
\end{eqnarray}

The position-basis expansion of  $\ket{\Psi_{OB}''^{(A)}}_{OB}$, given on the last two lines of (\ref{Psi''_OB(A)}), follows directly from its definition in terms of the momentum-basis expansion, plus the canonical completion $\bra{{\bf p}_{O}}_{O}\bra{{\bf p}_{B}}_{B}\ket{\Psi_{OB}''^{(A)}}_{OB} = \tilde{\Psi}_{OB}'^{(A)}({\bf p}_{O},{\bf p}_{B}) = \tilde{\psi}_{O}(-{\bf p}_{O}) \tilde{\phi}_{B}({\bf p}_{B}-{\bf p}_{O})$. 
The dual-basis completion, $\bra{{\bf x}_{O}}_{O}\bra{{\bf x}_{B}}_{B}\ket{\Psi_{OB}''^{(A)}}_{OB} = \Psi_{OB}'^{(A)}({\bf x}_{O},{\bf x}_{B}) = \psi_{O}(-({\bf x}_{B}+{\bf x}_{O})) \phi_{B}({\bf x}_{B})$, then holds automatically, and, taken together, these imply the coordinate transformations
\begin{eqnarray} \label{GCB_coord_transfs''}
{\bf r}_{A} \mapsto {\bf p}_{O} = -({\bf r}_{B}+{\bf r}_{A}) \, , \quad {\bf r}_{B} \mapsto {\bf x}_{B} = {\bf r}_{B} \, , 
\nonumber\\
\boldsymbol{\kappa}_{A} \mapsto {\bf p}_{O} = -\boldsymbol{\kappa}_{A} \, , \quad \boldsymbol{\kappa}_{B} \mapsto {\bf p}_{B} = \boldsymbol{\kappa}_{B} - \boldsymbol{\kappa}_{A} \, ,
\end{eqnarray}
as in the momentum space definition of a QRF-to-QRF transition, in the GCB model \cite{Giacomini:2017zju}. 

In this case, only the momentum space transformations, given in Eqs. (\ref{GCB_coord_transfs''}), are consistent with the intuitive diagram Fig. 4 and it is the position space transformations that are difficult to explain intuitively. 
In fact, both the transition $\ket{\Psi_{AB}^{(O)}}_{AB} \mapsto \ket{\Psi_{OB}'^{(A)}}_{OB}$ and the transition $\ket{\Psi_{AB}^{(O)}}_{AB} \mapsto \ket{\Psi_{OB}''^{(A)}}_{OB}$ were considered in \cite{Giacomini:2017zju} (using different notation) and were attributed, by the authors, to the {\it same} physical transition between two QRFs in a tripartite system, despite that fact that they imply different coordinate transformations. 
Thus, GCB state that ``when we promote a physical system to a reference frame, the question what the description of the rest of the world is relative to the reference frame is ill-posed unless a choice of relative coordinates is met. An equivalent statement is that, when the reference frame is considered as a physical system, there is no unambiguous notion of ‘jumping’ to a reference frame.''. 
This ambiguity is justified, in \cite{Giacomini:2017zju}, on the grounds that ``this feature arises both in classical and quantum mechanics from the requirement of canonicity of the reference-frame transformation when the reference frames are considered as physical degrees of freedom, and therefore attributed a phase space.'' 

\subsubsection{The non-unitary nature of CRF to QRF transitions} \label{Sec.3.2.3}

We find the statements in \cite{Giacomini:2017zju} paradoxical. 
The coordinates $({\bf r}_{A/B},\boldsymbol{\kappa}_{A/B})$ are not arbitrary generalised coordinates, but global Cartesians, each spanning one of the three-dimensional Euclidean subspaces of the canonical phase space geometry \cite{Frankel:1997ec,Nakahara:2003nw}. 
It is straightforward to verify that, for such coordinates, the transformations 
\begin{eqnarray} \label{canonical_coord_transfs}
{\bf r}_{A} \mapsto {\bf x}_{O} = -{\bf r}_{A} \, , \quad {\bf r}_{B} \mapsto {\bf x}_{B} = {\bf r}_{B} - {\bf r}_{A} \, ,
\nonumber\\
\boldsymbol{\kappa}_{A} \mapsto {\bf p}_{O} = -{\bf p}_{A} \, , \quad \boldsymbol{\kappa}_{B} \mapsto {\bf p}_{B} = \boldsymbol{\kappa}_{B} - \boldsymbol{\kappa}_{A} \, ,
\end{eqnarray}
which correspond to the intuitive diagram Fig. 4, satisfy the requirements of a canonical transformation \cite{Classical_Mechanics_Kibble}. 
We contend, therefore, that it is possible to implement the transformations given on the top and bottom lines of (\ref{canonical_coord_transfs}), self-consistently, where the former is applied in the position space representation of the composite quantum state and the latter is applied in the momentum space representation. 

The key point is that, in our construction of the QRF $A$, the coordinates $({\bf r}_{A/B},\boldsymbol{\kappa}_{A/B})$ are {\it not} assigned to a system with its own, additional, internal degrees of freedom. 
Instead, they refer to the position and momentum space coordinates of two quantum subsystems, Alice and Bob, viewed from the perspective of an external classical observer, $O$. 
In this scenario, performing the transformation $({\bf r}_{A/B},\boldsymbol{\kappa}_{A/B}) \mapsto ({\bf x}_{O/B},{\bf p}_{O/B})$, according to Eqs. (\ref{canonical_coord_transfs}), is unproblematic, since we are able to apply the same canonical transformation that we would apply to the position (momentum) coordinates of a {\it classical} point-particle to each and every point in the superposition of states that make up Alice's wave function, $\psi_{A}({\bf r}_{A})$ ($\tilde{\psi}_{A}(\boldsymbol{\kappa}_{A})$). 
Our ability to do this is an important physical consequence of the fact that both $\psi_{A}({\bf r}_{A})$ and $\tilde{\psi}_{A}(\boldsymbol{\kappa}_{A})$ are defined as complex scalars in classical background spaces, or, in other words, of the fact that classical frames of reference {\it do} exist in canonical quantum mechanics \cite{Jones:1998,Giachetta:2011,Stoica:2021owy}. 
If, on the other hand, one does not admit the validity of such frames, then it is impossible to implement both the top and bottom lines of (\ref{canonical_coord_transfs}), in a self-consistent way, as concluded in \cite{Giacomini:2017zju}. 

These arguments suggest that it {\it is} possible to define an unambiguous `jump' from an arbitrary CRF, $O$, to a given QRF, $A$ or $B$, in the context of canonical quantum theory. 
Furthermore, since the inverse transformation must also be well defined, we may perform the sequence of `jumps', QRF $A$ $\rightarrow$ CRF $O$ $\rightarrow$ QRF $B$, giving rise to an unambiguous notion of `jumping' from one QRF to another. 

More specifically, the arguments above imply that there must exist a `bipartite' quantum state, $\ket{\Psi_{OB}^{(A)}}_{OB}$, such that the map $\ket{\Psi_{AB}^{(O)}}_{AB} \mapsto \ket{\Psi_{OB}^{(A)}}_{OB}$ corresponds to the transformations (\ref{canonical_coord_transfs}). 
The operator that implements this map is then the first step in the transition from the CRF $O$ to the QRF $A$. 
The former has access to complete information about the state of the bipartite quantum system $AB$, while Alice's description depends on the state of the `bipartite' system $OB$, in which $O$ is not genuinely quantum mechanical. 
To account for this, $O$'s {\it effective} degrees of freedom must be traced out, by performing a partial trace over the product of the density operator $\hat{\rho}_{OB}^{(A)} := \ket{\Psi_{OB}^{(A)}}_{OB}\bra{\Psi_{OB}^{(A)}}_{OB}$ and an appropriate projection operator, with respect to the subspace $\mathcal{H}_{O}^{(A)} := \mathcal{P}_{A \leftrightarrow O}\mathcal{H}_{A}$. 
The choice of projection operator is important because this choice must be consistent with the transformations (\ref{canonical_coord_transfs}).
\footnote{We recall, again, that the basis vectors corresponding to the `canonical completions' of the states $\ket{\Psi_{OB}'^{(A)}}_{OB}$ (\ref{Psi'_OB(A)}) and $\ket{\Psi_{OB}''^{(A)}}_{OB}$ (\ref{Psi''_OB(A)}), which form the foundation of the GCB model \cite{Giacomini:2017zju}, are {\it not} consistent with the transformations (\ref{canonical_coord_transfs}). This leads to ambiguities in the description of the post-transformation state, and, hence, in the definition of the transformation itself, which we now attempt to resolve.} 

In light of our previous considerations, two natural candidates for the definition of the state $\ket{\Psi_{OB}^{(A)}}_{OB}$ are immediately apparent. 
The first of these, which we denote $\ket{{}^{({\rm x})}\Psi_{OB}^{(A)}}_{OB}$, is defined in the position-basis by performing the coordinate transformation ${\bf r}_{A} \mapsto {\bf x}_{O} = -{\bf r}_{A}$, ${\bf r}_{B} \mapsto {\bf x}_{B} = {\bf r}_{B} - {\bf r}_{A}$ on the state $\ket{\Psi_{AB}^{(O)}}_{AB}$ (\ref{psi_AB(O)}). 
The momentum-basis expansion is then defined, with respect to this, so as to implement the transformation $\boldsymbol{\kappa}_{A} \mapsto {\bf p}_{O} = -{\bf p}_{A}$, $\boldsymbol{\kappa}_{B} \mapsto {\bf p}_{B} = \boldsymbol{\kappa}_{B} - \boldsymbol{\kappa}_{A}$, while maintaining the basis-independence of the normalisation condition, $\braket{{}^{({\rm x})}\Psi_{OB}^{(A)}|{}^{({\rm x})}\Psi_{OB}^{(A)}}_{OB} = 1$. 
Thus, we have
\begin{eqnarray} \label{(x)_Psi_OB(A)}
\ket{{}^{({\rm x})}\Psi_{OB}^{(A)}}_{OB} &:=& \int \int \psi_{O}(-{\bf x}_{O}) \phi_{B}({\bf x}_{B}-{\bf x}_{O}) \ket{{\bf x}_{O}}_{O} \ket{{\bf x}_{B}}_{B} {\rm d}^3{\rm x}_{O}{\rm d}^3{\rm x}_{B}
\nonumber\\
&=& \int \int \psi_{O}({\bf r}_{A}) \phi_{B}({\bf r}_{B}) \ket{-{\bf r}_{A}}_{O} \ket{{\bf r}_{B}-{\bf r}_{A}}_{B} {\rm d}^3{\rm r}_{A}{\rm d}^3{\rm r}_{B} 
\nonumber\\
&:=&  \int \int \tilde{\psi}_{O}(-{\bf p}_{A}) \tilde{\phi}_{B}({\bf p}_{B}-{\bf p}_{O}) \ket{{\bf p}_{O} \, {\bf p}_{B}}_{OB} {\rm d}^3{\rm p}_{O}{\rm d}^3{\rm p}_{B}
\nonumber\\
&=& \int \int \tilde{\psi}_{O}(\boldsymbol{\kappa}_{A}) \tilde{\phi}_{B}(\boldsymbol{\kappa}_{B}) \ket{\boldsymbol{\kappa}_{A} \, \boldsymbol{\kappa}_{B}}_{OB} {\rm d}^3{\rm \kappa}_{A}{\rm d}^3{\rm \kappa}_{B} \, ,
\end{eqnarray}
where
\begin{eqnarray} \label{entangled_bases_bipartite_system_P-1}
\bra{-{\bf r}_{A}}_{O}\bra{{\bf r}_{B}-{\bf r}_{A}}_{B} \ket{\boldsymbol{\kappa}_{A} \, \boldsymbol{\kappa}_{B}}_{OB} 
:= \left(\frac{1}{\sqrt{2\pi\hbar}}\right)^{6} \exp\left[\frac{i}{\hbar}(\boldsymbol{\kappa}_{A}.{\bf r}_{A} + \boldsymbol{\kappa}_{B}.{\bf r}_{B})\right] \, ,
\end{eqnarray}
\begin{eqnarray} \label{entangled_bases_bipartite_system_P-2}
\bra{{\bf x}_{O}}_{O}\bra{{\bf x}_{B}}_{B}\ket{{\bf p}_{O} \, {\bf p}_{B}}_{OB} := \left(\frac{1}{\sqrt{2\pi\hbar}}\right)^{6} \exp\left[\frac{i}{\hbar}\left\{{\bf p}_{O}.{\bf x}_{O} + ({\bf p}_{B}-{\bf p}_{O}).({\bf x}_{B}-{\bf x}_{O})\right\} \right] \, ,
\end{eqnarray}
and
\begin{eqnarray} \label{entangled_bases_bipartite_system_X-1*}
\braket{{\bf x}_O | -{\bf r}_A}_{O} = \delta^{3}({\bf x}_O + {\bf r}_A) \, , \quad \braket{{\bf x}_B | {\bf r}_B-{\bf r}_A}_{B} = \delta^{3}({\bf x}_B - ({\bf r}_B-{\bf r}_A)) \, , 
\end{eqnarray}
\begin{eqnarray} \label{entangled_bases_bipartite_system_X-2*}
\braket{{\bf p}_{O} \, {\bf p}_{B} | \boldsymbol{\kappa}_{A} \, \boldsymbol{\kappa}_{B}}_{OB} := \delta^{3}({\bf p}_{O} +  \boldsymbol{\kappa}_{A}) \, \delta^{3}({\bf p}_{B} - (\boldsymbol{\kappa}_{B}-\boldsymbol{\kappa}_{A})) \, .
\end{eqnarray}

The second option, which we denote $\ket{{}^{({\rm p})}\Psi_{OB}^{(A)}}_{OB}$, is defined in the momentum-basis by performing the coordinate transformation $\boldsymbol{\kappa}_{A} \mapsto {\bf p}_{O} = -{\bf p}_{A}$, $\boldsymbol{\kappa}_{B} \mapsto {\bf p}_{B} = \boldsymbol{\kappa}_{B} - \boldsymbol{\kappa}_{A}$ on the state $\ket{\Psi_{AB}^{(O)}}_{AB}$ (\ref{psi_AB(O)}). 
The position-basis expansion is then defined, with respect to this, so as to implement the transformation ${\bf r}_{A} \mapsto {\bf x}_{O} = -{\bf r}_{A}$, ${\bf r}_{B} \mapsto {\bf x}_{B} = {\bf r}_{B} - {\bf r}_{A}$, while maintaining the basis-independence of the condition $\braket{{}^{({\rm x})}\Psi_{OB}^{(A)}|{}^{({\rm x})}\Psi_{OB}^{(A)}}_{OB} = 1$. 
This gives
\begin{eqnarray} \label{(p)_Psi_OB(A)}
\ket{{}^{(p)}\Psi_{OB}^{(A)}}_{OB} &:=& \int \int \tilde{\psi}_{O}(-{\bf p}_{O}) \tilde{\phi}_{B}({\bf p}_{B}-{\bf p}_{O}) \ket{{\bf p}_{O}}_{O} \ket{{\bf p}_{B}}_{B} {\rm d}^3{\rm p}_{O}{\rm d}^3{\rm p}_{B}
\nonumber\\
&=& \int \int \tilde{\psi}_{O}(\boldsymbol{\kappa}_{A}) \tilde{\phi}_{B}(\boldsymbol{\kappa}_{B}) \ket{-\boldsymbol{\kappa}_{A}}_{O} \ket{\boldsymbol{\kappa}_{B}-\boldsymbol{\kappa}_{A}}_{B} {\rm d}^3{\rm \kappa}_{A}{\rm d}^3{\rm \kappa}_{B} 
\nonumber\\
&:=&  \int \int \psi_{O}(-{\bf x}_{O}) \phi_{B}({\bf x}_{B}-{\bf x}_{O}) \ket{{\bf x}_{O} \, {\bf x}_{B}}_{OB} {\rm d}^3{\rm x}_{O}{\rm d}^3{\rm x}_{B}
\nonumber\\
&=& \int \int \psi_{O}({\bf r}_{A}) \phi_{B}({\bf r}_{B}) \ket{-{\bf r}_{A}}_{O}\ket{{\bf r}_{A} \, {\bf r}_{B}}_{OB} {\rm d}^3{\rm r}_{A}{\rm d}^3{\rm r}_{B} \, ,
\end{eqnarray}
where 
\begin{eqnarray} \label{entangled_bases_bipartite_system_X-1}
\bra{{\bf r}_{A} \, {\bf r}_{B}}_{OB} \ket{-\boldsymbol{\kappa}_{A}}_{O}\ket{\boldsymbol{\kappa}_{B}-\boldsymbol{\kappa}_{A}}_{B}
:= \left(\frac{1}{\sqrt{2\pi\hbar}}\right)^{6} \exp\left[\frac{i}{\hbar}(\boldsymbol{\kappa}_{A}.{\bf r}_{A} + \boldsymbol{\kappa}_{B}.{\bf r}_{B})\right] \, ,
\end{eqnarray}
\begin{eqnarray} \label{entangled_bases_bipartite_system_X-2}
\bra{{\bf x}_{O} \, {\bf x}_{B}}_{OB}\ket{{\bf p}_{O}}_{O}\ket{{\bf p}_{B}}_{B} := \left(\frac{1}{\sqrt{2\pi\hbar}}\right)^{6} \exp\left[\frac{i}{\hbar}\left\{{\bf p}_{O}.{\bf x}_{O} + ({\bf p}_{B}-{\bf p}_{O}).({\bf x}_{B}-{\bf x}_{O})\right\} \right] \, . 
\end{eqnarray}
and 
\begin{eqnarray} \label{entangled_bases_bipartite_system_P-1*}
\braket{{\bf x}_{O} \, {\bf x}_{B} | {\bf r}_{A} \, {\bf r}_{B}}_{OB} := \delta^{3}({\bf x}_O + {\bf r}_A) \,  \delta^{3}({\bf x}_B - ({\bf r}_B-{\bf r}_A)) \, , 
\end{eqnarray}
\begin{eqnarray} \label{entangled_bases_bipartite_system_P-2*}
\braket{{\bf p}_O | -\boldsymbol{\kappa}_A}_{O} = \delta^{3}({\bf p}_O + \boldsymbol{\kappa}_A) \, , \quad \braket{{\bf p}_B | \boldsymbol{\kappa}_B-\boldsymbol{\kappa}_A}_{B} = \delta^{3}({\bf p}_B - (\boldsymbol{\kappa}_B-\boldsymbol{\kappa}_A)) \, .
\end{eqnarray}

Here, we have used the cumbersome notation $\ket{{}^{({\rm x})}\Psi_{OB}^{(A)}}_{OB}$ to indicate that the position-basis utilised in the expansion of the state (\ref{(x)_Psi_OB(A)}) is unentangled, whereas the corresponding momentum-basis is entangled according to Eqs. (\ref{entangled_bases_bipartite_system_P-1})-(\ref{entangled_bases_bipartite_system_X-2*}). 
Likewise, $\ket{{}^{({\rm p})}\Psi_{OB}^{(A)}}_{OB}$ indicates that the momentum-basis utilised in the expansion of the state (\ref{(p)_Psi_OB(A)}) is unentangled, whereas the corresponding position-basis expansion is entangled according to Eqs. (\ref{entangled_bases_bipartite_system_X-1})-(\ref{entangled_bases_bipartite_system_P-2*}). 
Notationally, we indicate the non-separability of the bases $\ket{{\bf r}_{A} \, {\bf r}_{B}}_{OB}$, $\ket{{\bf x}_{O} \, {\bf x}_{B}}_{OB}$ and $\ket{\boldsymbol{\kappa}_{A} \, \boldsymbol{\kappa}_{B}}_{OB}$, $\ket{{\bf p}_{O} \, {\bf p}_{B}}_{OB}$ by not separating the variables enclosed within the ket by a comma, $\ket{{\bf x}_{O} \, {\bf x}_{B}}_{OB} \neq \ket{{\bf x}_{O}, {\bf x}_{B}}_{OB} := \ket{{\bf x}_{O}}_{O} \otimes \ket{{\bf x}_{B}}_{B}$, etc. 

If we choose $\ket{{}^{({\rm x})}\Psi_{OB}^{(A)}}_{OB}$ to represent the state of the `bipartite' system $OB$, the associated density operator is
\begin{eqnarray} \label{(x)_rho_OB}
{}^{({\rm x})}\hat{\rho}_{OB}^{(A)} := \ket{{}^{({\rm x})}\Psi_{OB}^{(A)}}\bra{{}^{({\rm x})}\Psi_{OB}^{(A)}}_{OB} \, .
\end{eqnarray}
We may obtain the desired probability densities, for both position and momentum measurements, by acting on this matrix with the projection operators
\begin{eqnarray} \label{Pi(x0,xB;p0pB)}
\hat{\Pi}_{\ket{{\bf x}_{O}}_{O}\ket{{\bf x}_{B}}_{B}} := \ket{{\bf x}_{O}}\bra{{\bf x}_{O}}_{O} \otimes \ket{{\bf x}_{B}}\bra{{\bf x}_{B}}_{B} \, , \quad
\hat{\Pi}_{\ket{{\bf p}_{O} \, {\bf p}_{B}}_{OB}} := \ket{{\bf p}_{O} \, {\bf p}_{B}}\bra{{\bf p}_{O} \, {\bf p}_{B}}_{OB} \, ,
\end{eqnarray}
respectively, before tracing over the $\mathcal{H}_{O}^{(A)}$ subspace of the full Hilbert space, $\mathcal{H}_{OB}^{(A)} = \mathcal{H}_{O}^{(A)} \otimes \mathcal{H}_{B}$. 
Thus, we have
\begin{eqnarray} \label{1}
{\rm d}^{3}P({\bf x}_{B}) = {\rm tr}_{O}\left[\hat{\Pi}_{\ket{{\bf x}_{O}}_{O}\ket{{\bf x}_{B}}_{B}}{}^{({\rm x})}\hat{\rho}_{OB}^{(A)}\right] {\rm d}^{3}{\rm x}_{B} = |\psi_{O}|^2 * |\phi_{B}|^2({\bf x}_{B}) {\rm d}^{3}{\rm x}_{B}  \, ,
\end{eqnarray}
and 
\begin{eqnarray} \label{1*}
{\rm d}^{3}P({\bf p}_{B}) = {\rm tr}_{O}\left[\hat{\Pi}_{\ket{{\bf p}_{O} \, {\bf p}_{B}}_{OB}}{}^{({\rm x})}\hat{\rho}_{OB}^{(A)}\right] {\rm d}^{3}{\rm p}_{B} = |\tilde{\psi}_{O}|^2 * |\tilde{\phi}_{B}|^2({\bf p}_{B}) {\rm d}^{3}{\rm p}_{B}  \, ,
\end{eqnarray}
as required. 

Similarly, if we choose $\ket{{}^{({\rm p})}\Psi_{OB}^{(A)}}_{OB}$ to represent the state of $OB$, as seen by $A$, we must instead define the density operator 
\begin{eqnarray} \label{(p)_rho_OB}
{}^{({\rm p})}\hat{\rho}_{OB}^{(A)} := \ket{{}^{(p)}\Psi_{OB}^{(A)}}_{OB}\bra{{}^{(p)}\Psi_{OB}^{(A)}}_{OB} \, . 
\end{eqnarray}
To obtain the required position and momentum probability densities, we must act on this with the projection operators 
\begin{eqnarray} \label{Pi(x0xB;p0,pB)}
\hat{\Pi}_{\ket{{\bf x}_{O} \, {\bf x}_{B}}_{OB}} := \ket{{\bf x}_{O} \, {\bf x}_{B}}\bra{{\bf x}_{O} \, {\bf x}_{B}}_{OB} \, , \quad
\hat{\Pi}_{\ket{{\bf p}_{O}}_{O}\ket{{\bf p}_{B}}_{B}} := \ket{{\bf p}_{O}}\bra{{\bf p}_{O}}_{O} \otimes \ket{{\bf p}_{B}}\bra{{\bf p}_{B}}_{B} \, , 
\end{eqnarray}
respectively, before performing the partial trace over $\mathcal{H}_{O}^{(A)}$, giving
\begin{eqnarray} \label{2}
{\rm d}^{3}P({\bf x}_{B}) = {\rm tr}_{O}\left[\hat{\Pi}_{\ket{{\bf x}_{O} \, {\bf x}_{B}}_{OB}}{}^{({\rm p})}\hat{\rho}_{OB}^{(A)}\right] {\rm d}^{3}{\rm x}_{B} = |\psi_{O}|^2 * |\phi_{B}|^2({\bf x}_{B}) {\rm d}^{3}{\rm x}_{B}  \, ,
\end{eqnarray}
and 
\begin{eqnarray} \label{2*}
{\rm d}^{3}P({\bf p}_{B}) = {\rm tr}_{O}\left[\hat{\Pi}_{\ket{{\bf p}_{O}}_{O}\ket{{\bf p}_{B}}_{B}}{}^{({\rm p})}\hat{\rho}_{OB}^{(A)}\right] {\rm d}^{3}{\rm p}_{B} = |\tilde{\psi}_{O}|^2 * |\tilde{\phi}_{B}|^2({\bf p}_{B}) {\rm d}^{3}{\rm p}_{B}  \, .
\end{eqnarray}

However, it is most convenient to express the state of $O$ and Bob, from Alice's perspective, in terms of the symmetric bases $\left\{\ket{{\bf x}_{O}}_{O}\ket{{\bf x}_{B}-{\bf x}_{O}}_{B}\right\}$ and $\left\{\ket{{\bf p}_{O}}_{O}\ket{{\bf p}_{B}-{\bf p}_{O}}_{B}\right\}$, as
\begin{eqnarray} \label{Psi_OB(A)}
\ket{\Psi_{OB}^{(A)}}_{OB} &:=&  \int \int \psi_{O}(-{\bf x}_{O}) \phi_{B}({\bf x}_{B}-{\bf x}_{O}) \ket{{\bf x}_{O}}_{O} \ket{{\bf x}_{B}-{\bf x}_{O}}_{B} {\rm d}^3{\rm x}_{O}{\rm d}^3{\rm x}_{B}
\nonumber\\
&=& \int \int \psi_{O}({\bf r}_{A}) \phi_{B}({\bf r}_{B}) \ket{-{\bf r}_{A}}_{O}\ket{{\bf r}_{B}}_{B} {\rm d}^3{\rm r}_{A}{\rm d}^3{\rm r}_{B} 
\nonumber\\
&:=& \int \int \tilde{\psi}_{O}(-{\bf p}_{O}) \tilde{\phi}_{B}({\bf p}_{B}-{\bf p}_{O}) \ket{{\bf p}_{O}}_{O} \ket{{\bf p}_{B}-{\bf p}_{O}}_{B} {\rm d}^3{\rm p}_{O}{\rm d}^3{\rm p}_{B}
\nonumber\\
&=& \int \int \tilde{\psi}_{O}(\boldsymbol{\kappa}_{A}) \tilde{\phi}_{B}(\boldsymbol{\kappa}_{B}) \ket{-\boldsymbol{\kappa}_{A}}_{O} \ket{\boldsymbol{\kappa}_{B}}_{B} {\rm d}^3{\rm \kappa}_{A}{\rm d}^3{\rm \kappa}_{B} \, .
\end{eqnarray}
The corresponding density matrix is
\begin{eqnarray} \label{rho_OB(A)}
\hat{\rho}_{OB}^{(A)} := \ket{\Psi_{OB}^{(A)}}_{OB}\bra{\Psi_{OB}^{(A)}}_{OB} \, , 
\end{eqnarray}
and we must act on this with the projection operators 
\begin{eqnarray} \label{Pi(x0,xB-x0;p0,pB-p0)}
&&\hat{\Pi}_{\ket{{\bf x}_{O}}_{O}\ket{{\bf x}_{B}-{\bf x}_{O}}_{B}} := \ket{{\bf x}_{O}}_{O}\bra{{\bf x}_{O}}_{O} \otimes \ket{{\bf x}_{B}-{\bf x}_{O}}_{B}\bra{{\bf x}_{B}-{\bf x}_{O}}_{B} \, , 
\nonumber\\
&&\hat{\Pi}_{\ket{{\bf p}_{O}}_{O}\ket{{\bf p}_{B}-{\bf p}_{O}}_{B}} := \ket{{\bf p}_{O}}_{O}\bra{{\bf p}_{O}}_{O} \otimes \ket{{\bf p}_{B}-{\bf p}_{O}}_{B}\bra{{\bf p}_{B}-{\bf p}_{O}}_{B} \, , 
\end{eqnarray}
before performing the partial trace over $\mathcal{H}_{O}^{(A)}$. 
These operations give
\begin{eqnarray} \label{3}
{\rm d}^{3}P({\bf x}_{B}) = {\rm tr}_{O}\left[\hat{\Pi}_{\ket{{\bf x}_{O}}_{O}\ket{{\bf x}_{B}-{\bf x}_{O}}_{B}}\hat{\rho}_{OB}^{(A)}\right] {\rm d}^{3}{\rm x}_{B} = |\psi_{O}|^2 * |\phi_{B}|^2({\bf x}_{B}) {\rm d}^{3}{\rm x}_{B}  \, ,
\end{eqnarray}
and
\begin{eqnarray} \label{3*}
{\rm d}^{3}P({\bf p}_{B}) = {\rm tr}_{O}\left[\hat{\Pi}_{\ket{{\bf p}_{O}}_{O}\ket{{\bf p}_{B}-{\bf p}_{O}}_{B}}\hat{\rho}_{OB}^{(A)}\right] {\rm d}^{3}{\rm p}_{B} = |\tilde{\psi}_{O}|^2 * |\tilde{\phi}_{B}|^2({\bf p}_{B}) {\rm d}^{3}{\rm p}_{B}  \, ,
\end{eqnarray}
as required, since the symmetrised position and momentum basis vectors form canonical-type brakets, according to 
\begin{eqnarray} \label{canonical-type_brakets-1}
\braket{{\bf r}_{A}|\boldsymbol{\kappa}_{A}}_{O} = \left(\frac{1}{\sqrt{2\pi\hbar}}\right)^{3} \exp\left[\frac{i}{\hbar}\boldsymbol{\kappa}_{A} \, . \, {\bf r}_{A}\right] \, , \quad 
\braket{{\bf r}_{B}|\boldsymbol{\kappa}_{B}}_{B} = \left(\frac{1}{\sqrt{2\pi\hbar}}\right)^{3} \exp\left[\frac{i}{\hbar}\boldsymbol{\kappa}_{B} \, . \, {\bf r}_{B}\right] \, ,
\end{eqnarray}
\begin{eqnarray} \label{canonical-type_brakets-2}
\braket{{\bf x}_{O}|{\bf p}_{O}}_{O} = \left(\frac{1}{\sqrt{2\pi\hbar}}\right)^{3} \exp\left[\frac{i}{\hbar}{\bf p}_{O} \, . \, {\bf x}_{O}\right] \, , 
\nonumber\\
\braket{{\bf x}_{B}-{\bf x}_{O}|{\bf p}_{B}-{\bf p}_{O}}_{B} = \left(\frac{1}{\sqrt{2\pi\hbar}}\right)^{3} \exp\left[\frac{i}{\hbar}({\bf p}_{B}-{\bf p}_{O}) \, . \, ({\bf x}_{B}-{\bf x}_{O})\right] \, ,
\end{eqnarray}
and
\begin{eqnarray} \label{canonical-type_brakets-3}
\braket{{\bf x}_{O} | -{\bf r}_{A}}_{O} = \delta^{3}({\bf x}_{O}+{\bf r}_{A}) \, , \quad \braket{{\bf x}_{B}-{\bf x}_{O} | {\bf r}_{B}}_{B} = \delta^{3}(({\bf x}_{B}-{\bf x}_{O}) - {\bf r}_{B}) \, , 
\end{eqnarray}
\begin{eqnarray} \label{canonical-type_brakets-4}
\braket{{\bf p}_{O} | -\boldsymbol{\kappa}_{A}}_{O} = \delta^{3}({\bf p}_{O}+\boldsymbol{\kappa}_{A}) \, , \quad \braket{{\bf p}_{B}-{\bf p}_{O} | \boldsymbol{\kappa}_{B}}_{B} = \delta^{3}(({\bf p}_{B}-{\bf p}_{O}) - \boldsymbol{\kappa}_{B}) \, .
\end{eqnarray}

After all this work, we have finally obtained our desired result; a concrete model of the transition from a classical frame of reference, $O$, to the quantum frame of reference corresponding to Alice's system. 
For the bipartite quantum system $AB$, the probability that Alice perceives Bob's state as the state $\ket{\varphi_{B}^{(A)}}_{B}$ is given by the map $S^{(O \rightarrow A)}_{OAB}(\varphi_{B}^{(A)}): \ket{\Psi_{AB}^{(O)}}_{AB} \mapsto {\rm d}^{3}P(\varphi_{B}^{(A)}|\Psi_{AB}^{(O)})$, which is defined by the relation
\begin{eqnarray} \label{S^(O->A)_OAB}
&&{\rm d}^{3}P(\varphi_{B}^{(A)}|\Psi_{AB}^{(O)}) := S^{(O \rightarrow A)}_{OAB}(\varphi_{B}^{(A)}) \ket{\Psi_{AB}^{(O)}}_{AB} 
\nonumber\\
&:=& {\rm tr}_{O}\left[\hat{\Pi}_{\ket{{\rm Bob}=\varphi_{B}^{(A)}}_{OB}} \, \hat{P}_{O \leftrightarrow A}\ket{\Psi_{AB}^{(O)}}_{AB}\bra{\Psi_{AB}^{(O)}}_{AB}\right] {\rm d}^{3}{\rm V}_{\varphi}
\nonumber\\
&:=& {\rm tr}_{O}\left[\hat{\Pi}_{\ket{{\rm Bob}=\varphi_{B}^{(A)}}_{OB}} \hat{\rho}_{OB}^{(A)}\right] {\rm d}^{3}{\rm V}_{\varphi} \, , 
\end{eqnarray}
where $\hat{\Pi}_{\ket{{\rm Bob}=\varphi_{B}^{(A)}}_{OB}}$ is an appropriate projection operator, that acts on the total Hilbert space of the composite state, and ${\rm d}^{3}{\rm V}_{\varphi}$ is the corresponding volume element for Bob's degrees of freedom (e.g., ${\rm d}^{3}{\rm V}_{\varphi} = {\rm d}^{3}{\rm x}_{B}$  (\ref{3}) or ${\rm d}^{3}{\rm V}_{\varphi} = {\rm d}^{3}{\rm p}_{B}$ (\ref{3*}), as in the examples above). 
In general, this may be a projection onto an entangled basis of the $OB$ system, $\mathcal{H}_{O}^{(A)} \otimes \mathcal{H}_{B} := \hat{P}_{O \leftrightarrow A}\mathcal{H}_{A} \otimes \mathcal{H}_{B}$. 

Clearly, (\ref{S^(O->A)_OAB}) does not define a single map, but a family of projection-dependent maps. 
To our horror, we seem to have encountered a similar problem to that faced in \cite{Giacomini:2017zju}, namely, a basis-dependent ambiguity in the definition of the post-transformation state. 
But this is not really the case. 
The predictions of our QRF model depend on the chosen projector, $\hat{\Pi}_{\ket{{\rm Bob}=\varphi_{B}^{(A)}}_{OB}}$, but not on the basis used to express the `objective' state, $\ket{\Psi_{AB}^{(O)}}_{AB}$. 
This is as it should be, since the former is determined by the the kind of physical measurements we choose to make on the system, whereas the latter is not. 
In order to extract predictions for the results of position measurements, we must project onto the basis $\left\{\ket{{\bf x}_{O}}_{O}\ket{{\bf x}_{B}-{\bf x}_{O}}_{B}\right\}$, which implements the coordinate transformation ${\bf r}_{A} \mapsto {\bf x}_{O} = -{\bf r}_{A}$, ${\bf r}_{B} \mapsto {\bf x}_{B} = {\bf r}_{B}-{\bf r}_{A}$, whereas, in order to extract predictions for the results of momentum measurements, we must project onto the basis $\left\{\ket{{\bf p}_{O}}_{O}\ket{{\bf p}_{B}-{\bf p}_{O}}_{B}\right\}$, which is equivalent to implementing the coordinate transformation $\boldsymbol{\kappa}_{A} \mapsto {\bf p}_{A} = -\boldsymbol{\kappa}_{A}$, $\boldsymbol{\kappa}_{B} \mapsto {\bf p}_{B} = \boldsymbol{\kappa}_{B}-\boldsymbol{\kappa}_{A}$. 
Crucially, we see that these position and momentum space transformations are not mutually exclusive. 
There exists a single bipartite state, $\ket{\Psi_{OB}^{(A)}}_{OB} := \hat{P}_{O \leftrightarrow A}\ket{\Psi_{AB}^{(O)}}_{AB}$ (\ref{Psi_OB(A)}), that is compatible with {\it both} transformations, (\ref{canonical_coord_transfs}).

In a completely analogous way, we can define a map from the CRF $O$ to the QRF $B$, for the bipartite $AB$ system, as
\begin{eqnarray} \label{S^(O->B)_OAB}
&&{\rm d}^{3}P(\varphi_{A}^{(B)}|\Psi_{AB}^{(O)}) := S^{(O \rightarrow B)}_{OAB}(\varphi_{A}^{(B)}) \ket{\Psi_{AB}^{(O)}}_{AB} 
\nonumber\\
&:=& {\rm tr}_{O}\left[\hat{\Pi}_{\ket{{\rm Alice}=\varphi_{A}^{(B)}}_{AO}} \, \hat{P}_{O \leftrightarrow B}\ket{\Psi_{AB}^{(O)}}_{AB}\bra{\Psi_{AB}^{(O)}}_{AB}\right] {\rm d}^{3}{\rm V}_{\varphi}
\nonumber\\
&:=& {\rm tr}_{O}\left[\hat{\Pi}_{\ket{{\rm Alice}=\varphi_{A}^{(B)}}_{AO}} \hat{\rho}_{AO}^{(B)}\right] {\rm d}^{3}{\rm V}_{\varphi} \, , 
\end{eqnarray}
where $\hat{\Pi}_{\ket{{\rm Alice}=\varphi_{A}^{(B)}}_{AO}}$ is an appropriate projection operator, that acts on the total Hilbert space of the composite state, $\mathcal{H}_{A} \otimes \mathcal{H}_{O}^{(B)} := \mathcal{H}_{A} \otimes \hat{\mathcal{P}}_{B \leftrightarrow O}\mathcal{H}_{B}$, and ${\rm d}^{3}\varphi_{A}^{(B)}$ denotes the corresponding volume element for Alice's degrees of freedom. 
By projecting onto the basis $\left\{\ket{{\bf q}_{A}-{\bf q}_{O}}_{A}\ket{{\bf q}_{O}}_{O}\right\}$, we obtain the predictions of our model for the outcomes of position measurements, viewed from $B$'s perspective. 
In like manner, projecting onto the basis $\left\{\ket{\boldsymbol{\pi}_{A}-\boldsymbol{\pi}_{O}}_{A}\ket{\boldsymbol{\pi}_{O}}_{O}\right\}$ yields predictions for the outcomes of momentum measurements, as seen by $B$'s QRF. 
However, in general, we may determine the probability that Alice's effective state, as seen by Bob, is an arbitrary wave function, denoted $\varphi_{A}^{(B)}$.

From these results, we see immediately that the CRF-to-QRF transition is {\it non-unitary}. 
This makes intuitive sense, since, although the CRF $O$ has access to complete information about the bipartite quantum system $AB$, the information that can be obtained by either subspace, $A$ or $B$, is limited to measurements of their {\it relative} position and momentum. 
By `jumping' from a CRF to a QRF, therefore, information is effectively destroyed, in the sense that it become unavailable to the designated `observer'. 

\subsubsection{The non-unitary nature of QRF to QRF transitions} \label{Sec.3.2.4}

Moreover, since the trace over the effective subspace attributed to $O$ is performed {\it after} the parity swap operation, the transformation from Alice's QRF to Bob's QRF must also be represented by a non-unitary operation. 
When performing the transition CRF $O \rightarrow$ QRF $A$, we trace over the subspace $\mathcal{H}_{O}^{(A)} := \hat{P}_{O \leftrightarrow A}\mathcal{H}_{A} \equiv \mathcal{H}_{A}$ of the total $AB$ Hilbert space, 
$\mathcal{H}_{AB} = \mathcal{H}_{A} \otimes \mathcal{H}_{B}$, whereas, when performing the transition CRF $O \rightarrow$ QRF $B$, we trace over $\mathcal{H}_{O}^{(B)} := \hat{P}_{O \leftrightarrow B}\mathcal{H}_{B} \equiv \mathcal{H}_{B}$. 
Each operation renders certain information about the total state physically inaccessible, but the information that is effectively destroyed in each case is different, since there is no a priori relation between the wave functions of $A$ and $B$. 
We discuss the physical implications of this, in greater detail, in Sec. \ref{Sec.3.2.5}.

\subsubsection{Generalised uncertainty relations} \label{Sec.3.2.5}

To illustrate the physical consequences of the conclusions drawn in Sec. \ref{Sec.3.2.3}, regarding the non-unitary nature of the CRF-to-QRF transition, we now consider the uncertainty relations corresponding to Alice's measurements of Bob's relative position and momentum. 
To proceed, it is convenient to define the generalised position and momentum operators, $\hat{{\bf X}}_{B}$ and  $\hat{{\bf P}}_{B}$, whose eigenvalues are ${\bf x}_{B}$ and ${\bf p}_{B}$, respectively. 
Here, we use capital letters to emphasise the fact that these operators act on the tensor product Hilbert space, $\mathcal{H}_{OB}^{(A)} = \mathcal{H}_{O}^{(A)} \otimes \mathcal{H}_{B}$, and not on $\mathcal{H}_{B}$ alone. 
The components of these operators are
\begin{eqnarray} \label{X_{B}^{i}_bipartite}
\hat{X}_{B}^{i} &:=& \hat{\mathcal{X}}_{B}^{i} + \hat{\mathcal{X}}_{B}'^{i}
\nonumber\\
&:=& \hat{x}_{O}^{i} \otimes \hat{\mathbb{I}}_{B} + \hat{\mathbb{I}}_{O} \otimes \widehat{x_{B}^{i}-x_{O}^{i}}
\nonumber\\
&=& \int\int x_{O}^{i} \ket{{\bf x}_{O}}\bra{{\bf x}_{O}}_{O} \otimes \ket{{\bf x}_{B}-{\bf x}_{O}}\bra{{\bf x}_{B}-{\bf x}_{O}}_{B} {\rm d}^3{\rm x}_{O}{\rm d}^3{\rm x}_{B}
\nonumber\\
&+& \int\int (x_{B}^{i}-x_{O}^{i}) \ket{{\bf x}_{O}}\bra{{\bf x}_{O}}_{O} \otimes \ket{{\bf x}_{B}-{\bf x}_{O}}\bra{{\bf x}_{B}-{\bf x}_{O}}_{B} {\rm d}^3{\rm x}_{O}{\rm d}^3{\rm x}_{B}
\nonumber\\
&\equiv& (-\hat{r}_{A}^{i})_{O} \otimes \hat{\mathbb{I}}_{B} + \hat{\mathbb{I}}_{O} \otimes \hat{r}_{B}^{i}
\nonumber\\
&=& \int\int (-r_{A}^{i}) \ket{-{\bf r}_{A}}\bra{-{\bf r}_{A}}_{O} \otimes \ket{{\bf r}_{B}}\bra{{\bf r}_{B}}_{B}  {\rm d}^3{\rm r}_{A}{\rm d}^3{\rm r}_{B}
\nonumber\\
&+& \int\int r_{B}^{i} \ket{-{\bf r}_{A}}\bra{-{\bf r}_{A}}_{O} \otimes \ket{{\bf r}_{B}}\bra{{\bf r}_{B}}_{B}  {\rm d}^3{\rm r}_{A}{\rm d}^3{\rm r}_{B} \, ,
\end{eqnarray}
and
\begin{eqnarray} \label{P_{Bj}_bipartite} 
\hat{P}_{Bj} &:=& \hat{\mathcal{P}}_{Bj} +  \hat{\mathcal{P}}_{Bj}'
\nonumber\\
&:=& \hat{p}_{Oj} \otimes \hat{\mathbb{I}}_{B} + \hat{\mathbb{I}}_{O} \otimes \widehat{p_{Bj}-p_{Oj}}
\nonumber\\
&=& \int\int p_{Oj} \ket{{\bf p}_{O}}\bra{{\bf p}_{O}}_{O} \otimes \ket{{\bf p}_{B}-{\bf p}_{O}}\bra{{\bf p}_{B}-{\bf p}_{O}}_{B} {\rm d}^3{\rm p}_{O}{\rm d}^3{\rm p}_{B}
\nonumber\\
&+& \int\int (p_{Bj}-p_{Oj}) \ket{{\bf p}_{O}}\bra{{\bf p}_{O}}_{O} \otimes \ket{{\bf p}_{B}-{\bf p}_{O}}\bra{{\bf p}_{B}-{\bf p}_{O}}_{B} {\rm d}^3{\rm p}_{O}{\rm d}^3{\rm p}_{B}
\nonumber\\
&\equiv& (-\hat{\kappa}_{Aj})_{O} \otimes \hat{\mathbb{I}}_{B} + \hat{\mathbb{I}}_{O} \otimes \hat{\kappa}_{Bj}
\nonumber\\
&=& \int\int (-\kappa_{Aj}) \ket{-\boldsymbol{\kappa}_{A}}\bra{-\boldsymbol{\kappa}_{A}}_{O} \otimes \ket{\boldsymbol{\kappa}_{B}}\bra{\boldsymbol{\kappa}_{B}}_{B}  {\rm d}^3{\rm \kappa}_{A}{\rm d}^3{\rm \kappa}_{B}
\nonumber\\
&+& \int\int \kappa_{Bj} \ket{-\boldsymbol{\kappa}_{A}}\bra{-\boldsymbol{\kappa}_{A}}_{O} \otimes \ket{\boldsymbol{\kappa}_{B}}\bra{\boldsymbol{\kappa}_{B}}_{B}  {\rm d}^3{\rm \kappa}_{A}{\rm d}^3{\rm \kappa}_{B} \, .
\end{eqnarray}

Using these definitions, plus the definition of the state $\ket{\Psi_{OB}^{(A)}}_{OB}$ (\ref{Psi_OB(A)}), it follows immediately that 
\begin{eqnarray} \label{indep_uncertainties_O_AB_X}
\Delta_{\Psi}X_{B}^{i} = \sqrt{(\Delta_{\Psi}\mathcal{X}_{B}^{i})^2 + (\Delta_{\Psi}\mathcal{X}_{B}'^{i})^2} \equiv \Delta_{\rho}x_{B}^{i} = \sqrt{(\Delta_{\psi}r_{A}^{i})^2 + (\Delta_{\phi}r_{B}^{i})^2} \, , 
\end{eqnarray}
and
\begin{eqnarray} \label{indep_uncertainties_O_AB_P}
\Delta_{\Psi}P_{Bj} = \sqrt{(\Delta_{\Psi}\mathcal{P}_{Bj})^2 + (\Delta_{\Psi}\mathcal{P}_{Bj}')^2} \equiv \Delta_{\rho}p_{Bj} = \sqrt{(\Delta_{\psi}\kappa_{Aj})^2 + (\Delta_{\phi}\kappa_{Bj})^2} \, ,
\end{eqnarray}
where
\begin{eqnarray} \label{indep_uncertainty_relations_O_AB_psi_phi}
\Delta_{\psi}r_{A}^{i} \, \Delta_{\psi}\kappa_{Aj} \geq \frac{\hbar}{2} \, \delta^{i}{}_{j} \, , \quad \Delta_{\phi}r_{B}^{i} \, \Delta_{\phi}\kappa_{Bj} \geq \frac{\hbar}{2} \, \delta^{i}{}_{j} 
\end{eqnarray}
are the fundamental uncertainty relations for the wave functions $\psi_{A}$ and $\phi_{B}$. 
Here, the subscript $\rho$ denotes the effective state of Bob's system, as seen by Alice, which may be mixed rather than pure. 
(We consider this scenario, in greater detail, in Sec. \ref{Sec.3.2.6}.)

The important point is that, {\it if} a real physical observer could be embodied as a classical object - that is, as a material body that is perfectly well localised in both position and momentum space - they would be able to measure not only the relative position and momentum of Alice and Bob, but also their `objective' positions and momenta, relative to a classical frame of reference defined with respect to the background geometry. 
In other words, they would be able to measure the quantities ${\bf r}_{A}$ and ${\bf r}_{B}$, or $\boldsymbol{\kappa}_{A}$ and $\boldsymbol{\kappa}_{B}$, independently, and, thus, to determine the individual uncertainties
\begin{eqnarray} \label{indep_uncertainties_O_AB_X*}
\Delta_{\psi}r_{A}^{i} \, , \quad \Delta_{\phi}r_{B}^{i} \, ,
\end{eqnarray}
and
\begin{eqnarray} \label{indep_uncertainties_O_AB_P*}
\Delta_{\psi}\kappa_{Aj} \, , \quad \Delta_{\phi}\kappa_{Bj} \, ,
\end{eqnarray}
by direct measurement. 
By contrast, in the bipartite $AB$ system, Alice can only perform physical measurements to determine the quantities ${\bf x}_{B} = {\bf r}_{B}-{\bf r}_{A}$ and ${\bf p}_{B} = \boldsymbol{\kappa}_{B}-\boldsymbol{\kappa}_{A}$. 
She is able to calculate the generalised uncertainties $\Delta_{\Psi}X_{B}^{i} \equiv \Delta_{\rho}x_{B}^{i}$ (\ref{indep_uncertainties_O_AB_X}) by performing repeated measurements of Bob's position on an ensemble of identically prepared experiments - taking care not to alter the preparation of her own state, too - but she cannot determine either $\Delta_{\psi}r_{A}^{i}$ or $\Delta_{\phi}r_{B}^{i}$ in this way.
Similarly, she is able to calculate the generalised uncertainties $\Delta_{\Psi}P_{Bj} \equiv \Delta_{\rho}p_{Bj}$ (\ref{indep_uncertainties_O_AB_P}) by performing repeated measurements of his relative momentum, in like manner, but she cannot determine either $\Delta_{\psi}\kappa_{Aj}$ or $\Delta_{\phi}\kappa_{Bj}$ using this procedure. 

Thus, she is unable to detect her own `quantumness', from the result of any individual measurement, or from the statistical spreads of any one set of measurements, of either ${\bf x}_{B}$ or ${\bf p}_{B}$. 
Nonetheless, as we will now show, Alice {\it is} able to infer limited information about her own quantum state, using the generalised uncertainty relations implied by Eqs. (\ref{indep_uncertainties_O_AB_X})-(\ref{indep_uncertainties_O_AB_P}). 
To construct these, we combine both equations to give two alternative expressions, which together contain the same information as (\ref{indep_uncertainties_O_AB_X})-(\ref{indep_uncertainties_O_AB_P}), namely
\begin{eqnarray} \label{indep_uncertainties_O_AB**}
(\Delta_{\rho}x_{B}^{i})^2 \, (\Delta_{\rho}p_{Bj})^2 &=& (\sigma_{A}^{i})^2 \, (\Delta_{\rho}p_{Bj})^2 + (\Delta_{\rho}x_{B}^{i})^2 \, (\tilde{\sigma}_{Aj})^2 
\nonumber\\ 
&+ & (\sigma_{A}^{i})^2 \, (\tilde{\sigma}_{Aj})^2 - (\sigma_{B}^{i})^2 \, (\tilde{\sigma}_{Bj})^2 \, , 
\nonumber\\ 
(\Delta_{\rho}x_{B}^{i})^2 \, (\Delta_{\rho}p_{Bj})^2 &=& (\sigma_{B}^{i})^2 \, (\Delta_{\rho}p_{Bj})^2 + (\Delta_{\rho}x_{B}^{i})^2 \, (\tilde{\sigma}_{Bj})^2 
\nonumber\\ 
&+ & (\sigma_{B}^{i})^2 \, (\tilde{\sigma}_{Bj})^2 - (\sigma_{A}^{i})^2 \, (\tilde{\sigma}_{Aj})^2 \, , 
\end{eqnarray}
where we have rewritten 
\begin{eqnarray} \label{sigmas_OAB}
\Delta_{\psi}r_{A}^{i} = \sigma_{A}^{i} \, , \quad \Delta_{\psi}\kappa_{Aj} = \tilde{\sigma}_{Aj} \, , 
\nonumber\\ 
\Delta_{\phi}r_{B}^{i} = \sigma_{B}^{i} \, , \quad \Delta_{\phi}\kappa_{Bj} = \tilde{\sigma}_{Bj} \, . 
\end{eqnarray}

This new notation is intended to emphasise the fact that it is not possible to determine the uncertainties $\Delta_{\psi}r_{A}^{i}$ and $\Delta_{\phi}r_{B}^{i}$, individually, from the statistical spread of position measurements performed by any {\it physical} observer. 
Likewise, it is not possible to determine either $\Delta_{\psi}\kappa_{Aj}$ or $\Delta_{\phi}\kappa_{Bj}$, individually, from the statistical spread of physical momentum measurements. 
In the GURs (\ref{indep_uncertainties_O_AB**}), these quantities therefore play the role of numerical coefficients, multiplying the empirically measurable quantities $\Delta_{\rho}x_{B}^{i}$ and $\Delta_{\rho}p_{Bj}$. 
By combining her data on both position and momentum measurements, and using these relations, Alice is, at least in principle, able to determine the values of $\sigma_{A}^{i}$, $\tilde{\sigma}_{Aj}$, $\sigma_{B}^{i}$ and $\tilde{\sigma}_{Bj}$, for a given ensemble of identically prepared systems. 
She is therefore able to determine her own position and momentum uncertainties, albeit {\it indirectly}, via their effect on the uncertainty relations she attributes to Bob. 
If the latter are found to deviate from the canonical Heisenberg form, according to Eqs. (\ref{indep_uncertainties_O_AB**}), then Alice is able to determine that both she and Bob are quantum mechanical in nature. 
The product of uncertainties $\Delta_{\rho}x_{B}^{i} \, \Delta_{\rho}p_{Bj}$ remains bounded by the Schr{\" o}dinger-Robertson limit, 
\begin{eqnarray} \label{bipartite_SR_limit}
\Delta_{\rho}x_{B}^{i} \, \Delta_{\rho}p_{Bj} \geq \frac{(\hbar + \hbar)}{2} \, \delta^{i}{}_{j} = \hbar \, \delta^{i}{}_{j} \, , 
\end{eqnarray}
and this inequality is saturated when either
\begin{eqnarray} \label{bipartite_SR_limit_saturation_cond-1}
\sigma_{A}^{i}\tilde{\sigma}_{Ai} = \frac{\hbar}{2} \, ; \quad 
\sigma_{B}^{i} = \sqrt{\frac{\hbar}{2}\frac{\sigma_{A}^{i}}{\tilde{\sigma}_{Ai}}} \, , \ \tilde{\sigma}_{Bi} = \sqrt{\frac{\hbar}{2}\frac{\tilde{\sigma}_{Ai}}{\sigma_{A}^{i}}} \, ,
\end{eqnarray}
or
\begin{eqnarray} \label{bipartite_SR_limit_saturation_cond-2}
\sigma_{A}^{i} = \sqrt{\frac{\hbar}{2}\frac{\sigma_{B}^{i}}{\tilde{\sigma}_{Bi}}} \, , \ \tilde{\sigma}_{Ai} = \sqrt{\frac{\hbar}{2}\frac{\tilde{\sigma}_{Bi}}{\sigma_{B}^{i}}} \, ; \quad 
\sigma_{B}^{i}\tilde{\sigma}_{Bi} = \frac{\hbar}{2} \, .
\end{eqnarray}

This situation is analogous to that proposed in phenomenological quantum gravity, in which the quantised spatial background is expected to act, effectively, as a non-material QRF, giving rise to GURs like the GUP, EUP or EGUP, which have been extensively studied in the literature \cite{Maggiore:1993rv,Adler:1999bu,Scardigli:1999jh,Bolen:2004sq,Park:2007az,Bambi:2007ty,Kempf:1996ss,Hossenfelder:2012jw,Tawfik:2015rva,Tawfik:2014zca,Bosso:2023aht}. 
In this scenario, it is impossible to detect the quantum nature of spacetime, directly, through its effects on any individual measurement of a material system. 
Nonetheless, it is still possible to infer the existence of quantum gravitational effects - that is, effects due to the quantum nature of the background geometry - {\it indirectly}, via their induced modifications of the Heisenberg uncertainty principle \cite{Lake:2018zeg,Lake:2019nmn,Lake:2019oaz,Lake:2020rwc,Lake:2021beh,Lake:2020chb,Lake:2021gbu,Lake:2022hzr,Lake:2023lvh,Lake:2023uoi}. 
We discuss this point, in greater detail, in Sec. \ref{Sec.5}.

Finally, before concluding this section, we note that using the `sub-operators' $\hat{\mathcal{X}}_{B}^{i}$, $\hat{\mathcal{X}}_{B}'^{i}$ and $\hat{\mathcal{P}}_{Bj}$, $\hat{\mathcal{P}}'_{Bj}$, defined in in Eqs. (\ref{X_{B}^{i}_bipartite})-(\ref{P_{Bj}_bipartite}), we can construct the useful unitary operators
\begin{eqnarray} \label{U'(A->O)OAB}
\hat{U}_{OAB}'^{(A \rightarrow O)} := \hat{\mathcal{P}}_{A \leftrightarrow O} \, \hat{U}_{OB}'^{(A)} \, ,  \quad 
\hat{U}_{OB}'^{(A)} := \exp\left[\frac{i}{\hbar} \hat{{\bf \mathcal{P}}}'_{B} \, . \, \hat{{\bf \mathcal{X}}}_{B}\right] \, ,
\end{eqnarray}
and 
\begin{eqnarray} \label{U''(O->A)OAB}
\hat{U}_{OAB}''^{(A \rightarrow O)} := \hat{\mathcal{P}}_{A \leftrightarrow O} \, \hat{U}_{OB}''^{(A)} \, ,  \quad 
\hat{U}_{OB}''^{(A)} := \exp\left[-\frac{i}{\hbar} \hat{{\bf \mathcal{P}}}_{B} \, . \, \hat{{\bf \mathcal{X}}}_{B}'\right] \, .
\end{eqnarray}
Although we use very different notation, the operators $\hat{U}_{OAB}'^{(A \rightarrow O)}$ and $\hat{U}_{OAB}''^{(A \rightarrow O)}$ are completely equivalent to the operators that would be denoted as $\hat{S}_{x}^{(A \rightarrow O)}$ and $\hat{S}_{p}^{(A \rightarrow O)}$, respectively, in the GCB formalism. 
(In which $O$ would be regarded as a {\it genuine} quantum system.)
In our notation, the corresponding transitions between states are expressed as
\begin{eqnarray} \label{U'_U''_action_GCB}
\ket{\Psi_{OB}'^{(A)}}_{OB} \stackrel{\hat{U}_{OAB}'^{(A \rightarrow O)\dagger}}{\longrightarrow} \ket{\Psi_{AB}^{(O)}}_{AB} \stackrel{\hat{U}_{OAB}''^{(A \rightarrow O)\dagger}}{\longleftarrow} \ket{\Psi_{OB}''^{(A)}}_{OB} \, , 
\end{eqnarray}
where $\ket{\Psi_{AB}^{(O)}}_{AB}$ is the state of the bipartite quantum system $AB$, as seen by the classical observer $O$ (\ref{psi_AB(O)}), and the states $\ket{\Psi_{OB}'^{(A)}}_{OB}$ and $\ket{\Psi_{OB}''^{(A)}}_{OB}$ are defined in Eqs. (\ref{Psi'_OB(A)}) and (\ref{Psi''_OB(A)}). 
We recall that these states are our failed candidates, for the state of the $OB$ system, as nominally seen by Alice, and that each was rejected on the grounds that it is not compatible with the required coordinate transformations. 

In the GCB model, both operators, $\hat{U}_{OAB}'^{(A \rightarrow O)} \equiv \hat{S}_{x}^{(A \rightarrow O)}$ and $\hat{U}_{OAB}''^{(A \rightarrow O)} \equiv \hat{S}_{p}^{(A \rightarrow O)}$, are associated with a single transition between two QRFs and it is stated that ``whenever we perform a transformation to a quantum reference frame, we have to choose the relative variables we are interested in, and then complete the transformation of the conjugated variable by canonicity'' \cite{Giacomini:2017zju}. 
As we have shown above, this is not the only possibility, and we can remove the ambiguity in the definitions of both CRF-to-QRF and QRF-to-QRF transitions by choosing an appropriate set of basis states, which are compatible with the canonical coordinate transformations we require, Eqs. (\ref{canonical_coord_transfs}).   

Nonetheless, it turns out that much of the mathematical machinery, originally developed in \cite{Giacomini:2017zju}, is still useful in our model. 
In particular, the operators $\hat{U}_{OB}'^{(A)}$ and $\hat{U}_{OB}''^{(A)}$, which act on the `bipartite' state $OB$ and are simply $\hat{U}_{OAB}'^{(A \rightarrow O)} \equiv \hat{S}_{x}^{(A \rightarrow O)}$ and $\hat{U}_{OAB}''^{(A \rightarrow O)} \equiv \hat{S}_{p}^{(A \rightarrow O)}$, without the parity swap $\hat{\mathcal{P}}_{A \leftrightarrow O}$, transform between the entangled and separable bases, introduced in Eqs. (\ref{(x)_Psi_OB(A)}), (\ref{(p)_Psi_OB(A)}) and (\ref{Psi_OB(A)}), according to
\begin{eqnarray} \label{U'_U''_action_bases}
\ket{{\bf x}_{O}}_{O}\ket{{\bf x}_{B}}_{B} \, , \, \ket{{\bf p}_{O} \, {\bf p}_{B}}_{OB} \stackrel{\hat{U}_{OB}'^{(A)\dagger}}{\longrightarrow} \ket{{\bf x}_{O}}_{O}\ket{{\bf x}_{B}-{\bf x}_{O}}_{B} \, , \, \ket{{\bf p}_{O}}_{O}
\ket{{\bf p}_{B}-{\bf p}_{O}}_{B} 
\nonumber\\
\ket{{\bf x}_{O}}_{O}\ket{{\bf x}_{B}-{\bf x}_{O}}_{B} \, , \, \ket{{\bf p}_{O}}_{O}
\ket{{\bf p}_{B}-{\bf p}_{O}}_{B} \stackrel{\hat{U}_{OB}''^{(A)\dagger}}{\longleftarrow} 
\ket{{\bf x}_{O} \, {\bf x}_{B}}_{OB} \, , \, \ket{{\bf p}_{O}}_{O}\ket{{\bf p}_{B}}_{B} \, . 
\end{eqnarray}
They therefore transform between our alternative `candidates' for the state of the $OB$ system, as seen by Alice, according to
\begin{eqnarray} \label{U'_U''_action_states}
\ket{{}^{(x)}\Psi_{OB}^{(A)}}_{OB} \stackrel{\hat{U}_{OB}'^{(A)\dagger}}{\longrightarrow} \ket{\Psi_{OB}^{(A)}}_{OB} \stackrel{\hat{U}_{OB}''^{(A)\dagger}}{\longleftarrow} \ket{{}^{(p)}\Psi_{OB}^{(A)}}_{OB} \, . 
\end{eqnarray}

However, here, the states $\ket{{}^{(x)}\Psi_{OB}^{(A)}}_{OB}$ (\ref{(x)_Psi_OB(A)}), $\ket{{}^{(p)}\Psi_{OB}^{(A)}}_{OB}$ (\ref{Psi_OB(A)}) and $\ket{\Psi_{OB}^{(A)}}_{OB}$ (\ref{(p)_Psi_OB(A)}), are not really alternative candidates for the state of $OB$. 
We have shown that the map between a CRF and a QRF, as well as the map between two QRFs, is necessarily non-unitary. 
In our formalism, the unitary equivalence between these states indicates their {\it physical equivalence}, as descriptions of the $OB$ system, viewed from Alice's perspective. 
As shown, explicitly, in Sec. \ref{Sec.3.2.3}, all three states are compatible with the same set of canonical coordinate transformations, Eqs. (\ref{canonical_coord_transfs}), and give rise to the same predictions for the probability densities associated with measurements of ${\bf x}_{B}$ or ${\bf p}_{B}$, following the partial trace over $O$'s {\it effective} degrees of freedom. 

\subsubsection{Phase information} \label{Sec.3.2.6}

So far, we have considered only the squared absolute value of Bob's effective state, as seen by Alice. 
This was all that we required, in order to determine the probabilities associated with measurements of their relative position or momentum, ${\rm d}^{3}P({\bf x}_{B}) = |\psi_{O}|^2*|\phi_{B}|^2({\bf x}_{B}) {\rm d}^{3}{\rm x}_{B}$ and ${\rm d}^{3}P({\bf p}_{B}) = |\tilde{\psi}_{O}|^2*|\tilde{\phi}_{B}|^2({\bf p}_{B}){\rm d}^{3}{\rm p}_{B}$. 
Now, in order to complete our description of Alice's QRF, we must consider what information she is able to obtain about the phase distribution of Bob's state. 

To proceed, let us first consider the position space representation. 
Again, the key point is that a single value of the relative displacement, ${\bf x}_{B} = {\bf r}_{B}-{\bf r}_{A}$, corresponds to multiple pairings $({\bf r}_{A},{\bf r}_{B})$, and hence to multiple points in Bob's wave function $\phi_{B}({\bf r}_{B}) = |\phi_{B}({\bf r}_{B})|e^{i\theta_{B}({\bf r}_{B})} \equiv \phi_{B}({\bf x}_{B}-{\bf x}_{O}) = |\phi_{B}({\bf x}_{B}-{\bf x}_{O})|e^{i\theta_{B}({\bf x}_{B}-{\bf x}_{O})}$. 
Generically, these points will have both different phases and different absolute values of the amplitude density, 
\begin{eqnarray} \label{relative_phase_AB_x}
\theta_{B}({\bf r}_{B}) = \theta_{B}({\bf x}_{B}-{\bf x}_{O}) \, ,  \quad |\phi_{B}({\bf r}_{B})| = |\phi_{B}({\bf x}_{B}-{\bf x}_{O})| \, . 
\end{eqnarray}
Bob's pure state therefore appears, to Alice, as a mixed state. 

To determine it's form, let us consider a measurement that collapses Alice's wave function to a single position eigenstate, ${\bf r}_{A}$, but leaves Bob's wave function unchanged. 
Experimentally, we may imagine interfering Bob's state with a given reference state, $R$, with known phase and absolute modulus distributions, i.e., a known wave function $\psi_{R} = |\psi_{R}|e^{i\theta_{R}}$.
\footnote{This is clearly an idealised scenario, that is not, strictly speaking, ever reachable in practice. However, we may consider a state $\psi_{R}$ that has been pre-prepared by a macroscopic, {\it effectively} classical observer (see Sec. \ref{Sec.1.1.3}), whose knowledge has then been imparted to Alice.} 
This gives rise to a fixed interference pattern, and Alice may then ask the question, `how far away am I from a given point in this pattern?' 
Once her displacement from one point in the pattern is fixed, her displacement from every other point is also fixed, and this fixes her relative displacement from each and every point in Bob's wave function, $\phi_{B}$. 
In this sense, she `sees' Bob's wave function, as whole, including all relevant phase and modulus information, that is also available to the CRF $O$. 
In fact, {\it if}, by chance, Alice's wave function collapses to the point ${\bf r}_{A}=0$, her perception of Bob's state is identical $O$'s, $\ket{\phi_{B}}_{B}$. 
For an arbitrary displacement, ${\bf r}_{A} \neq 0$, she instead sees Bob's state translated by the vector ${\bf x}_{O} = - {\bf r}_{A}$, which we denote as $\ket{\phi_{B}({\bf x}_{O})}_{B}$. 
In this notation, $\ket{\phi_{B}}_{B} \equiv \ket{\phi_{B}(0)}_{B}$, and the general displaced state is defined as
\begin{eqnarray} \label{Bob_displaced_state}
\ket{\phi_{B}({\bf x}_{O})}_{B} := \hat{U}_{B}({\bf x}_{O})\ket{\phi_{B}}_{B} := \int \phi_{B}({\bf x}_{B}-{\bf x}_{O}) \ket{{\bf x}_{B}}_{B}{\rm d}^3{\rm x}_{B} \, ,
\end{eqnarray}
where 
\begin{eqnarray} \label{d_B(x_O)}
\hat{U}_{B}({\bf x}_{O}) := \exp\left(\frac{i}{\hbar}\hat{\bold{p}}_{B} \, . \, \bold{x}_{O}\right)
\end{eqnarray}
is the canonical displacement operator. 
Bob's mixed state, as seen by Alice, is therefore described by the density operator
\begin{eqnarray} \label{Bob_effective_mixed_state_x}
\hat{\rho}_{B}^{(A)} := \int \frac{{\rm d}^{3}P({\bf x}_{O})}{{\rm d}^{3}{\rm x}_{O}} \ket{\phi_{B}({\bf x}_{O})}\bra{\phi_{B}({\bf x}_{O})}_{B} {\rm d}^{3}{\rm x}_{O} \, ,
\end{eqnarray}
where we have used the shorthand notation
\begin{eqnarray} \label{P(theta_O|theta_B)}
{\rm d}^{3}P({\bf x}_{O}) &:=& {\rm d}^{3}P({\bf r}_{A}|\Psi_{AB}^{(O)}) = {\rm d}^{3}P(-{\bf x}_{O}|\Psi_{OB}^{(A)})
\nonumber\\
&=& {\rm d}^{3}P({\bf r}_{A}|\psi_{A}) = {\rm d}^{3}P(-{\bf x}_{O}|\psi_{O})
\nonumber\\
&=& |\psi_{A}({\bf r}_{A})|^2 {\rm d}^{3}{\rm r}_{A} = |\psi_{O}(-{\bf x}_{O})|^2 {\rm d}^{3}{\rm x}_{O} \, .
\end{eqnarray}

To recap; in order to see that Eq. (\ref{Bob_effective_mixed_state_x}) is the correct description of Bob's state, as seen by Alice, we need only remark that the same value of relative position, ${\bf x}_{B}$, can correspond to {\it any} individual values of ${\bf r}_{A}$ and ${\bf r}_{B}$ for which the relation ${\bf x}_{B} = {\bf r}_{B}-{\bf r}_{A}$ holds. 
In general, there will be an infinite number of pairs $({\bf r}_{A},{\bf r}_{B}) = (-{\bf x}_{O},{\bf x}_{B}-{\bf x}_{O})$ for which this relation is satisfied, and each individual pair can, in principle, correspond to both a different value of Bob's phase, and a different value of his wave function's absolute magnitude (\ref{relative_phase_AB_x}). 
From Alice's perspective, it appears that a different amplitude, $\phi_{B}({\bf x}_{B})$, is associated with the same point in physical space, in each individual state of an ensemble of identically prepared systems; or, in other words, that Bob's state is mixed. 
This scenario is depicted, heuristically, in Fig. 5.

Using the definitions (\ref{Bob_effective_mixed_state_x}) and (\ref{P(theta_O|theta_B)}), we then obtain
\begin{eqnarray} \label{|Phi_B|^2_final}
{\rm d}^{3}P({\bf x}_{B}) = {\rm tr}\left[\ket{{\bf x}_{B}}\bra{{\bf x}_{B}}_{B}\hat{\rho}_{B}^{(A)}\right] {\rm d}^{3}{\rm x}_{B} = |\psi_{O}|^2*|\phi_{B}|^2({\bf x}_{B}){\rm d}^{3}{\rm x}_{B} \, , 
\end{eqnarray}
and
\begin{eqnarray} \label{|Phi_B|^2_final*}
{\rm d}^{3}P({\bf x}_{O}) = {\rm tr}\left[\ket{\phi_{B}({\bf x}_{O})}\bra{\phi_{B}({\bf x}_{O})}_{B}\hat{\rho}_{B}^{(A)}\right] {\rm d}^{3}{\rm p}_{B} = |\psi_{O}(-{\bf x}_{O})|^2{\rm d}^{3}{\rm x}_{O} \, , 
\end{eqnarray}
as required. 
Finally, we are able to obtain the correct GUR, 
\begin{eqnarray} \label{GUR_X_bipartite}
\Delta_{\rho}x_{B}^{i} = \sqrt{(\Delta_{\psi}x_{B}^{i})^2 + (\Delta_{\phi}x_{B}^{i})^2} \, , 
\end{eqnarray}
using the trace 
\begin{eqnarray} \label{GUR_X_op_bipartite}
\langle (\hat{x}_{B}^{i})^{n} \rangle_{\rho} = {\rm tr}\left[(\hat{x}_{B}^{i})^{n}\hat{\rho}_{B}^{(A)}\right] 
\end{eqnarray}
where 
\begin{eqnarray} \label{canonical_x_B^i}
\hat{x}_{B}^{i} = \int x_{B}^{i} \ket{{\bf x}_{B}}\bra{{\bf x}_{B}}_{B} {\rm d}^{3}{\rm x}_{B} 
\end{eqnarray}
is the canonical one-particle position operator, without referring at all to the fictitious quantum subsystem, $\mathcal{H}_{O}^{(A)} := \hat{\mathcal{P}}_{O \leftrightarrow A}\mathcal{H}_{A} \equiv \mathcal{H}_{A}$. 
Unlike the operator $\hat{X}_{B}^{i}$ (\ref{X_{B}^{i}_bipartite}), which acts on the state $\ket{\Psi_{OB}^{(A)}} \in \mathcal{H}_{OB}^{(A)} := \mathcal{H}_{O}^{(A)} \otimes \mathcal{H}_{B}$ (\ref{Psi_OB(A)}), or, equivalently, $\hat{\rho}_{OB}^{(A)}$ (\ref{rho_OB(A)}), the canonical position operator $\hat{x}_{B}^{i}$ (\ref{canonical_x_B^i}) acts directly on Bob's effective mixed state, $\hat{\rho}_{B}^{(A)}$ (\ref{Bob_effective_mixed_state_x}), where $\ket{\phi_{B}({\bf x}_{O})}_{B} \in \mathcal{H}_{B}$ (\ref{Bob_displaced_state}). 
The exact relation between $\hat{\rho}_{OB}^{(A)}$ (\ref{rho_OB(A)}) and $\rho_{B}^{(A)}$ (\ref{Bob_effective_mixed_state_x}) is 
\begin{eqnarray} \label{rho_B_from_rho_OB}
\hat{\rho}_{B}^{(A)} := \int {\rm tr}_{O}\left[\ket{\bold{x}_{O}}\bra{\bold{x}_{O}}_{O} \otimes \hat{U}_{B}({\bf x}_{O}) \, \hat{\rho}_{OB}^{(A)}\right]  {\rm d}^{3}{\rm x}_{O} \, .
\end{eqnarray}

\begin{figure}[h] \label{Fig.5}
\begin{center}
\includegraphics[width=10cm]{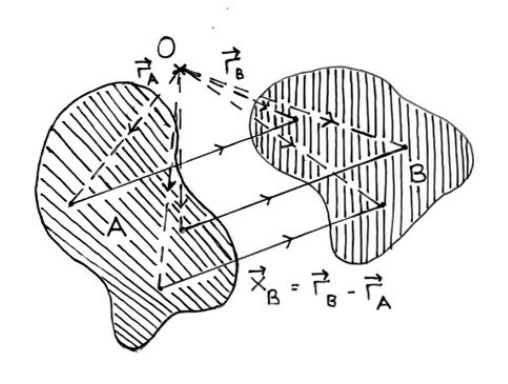}
\caption{The same relative displacement between Alice and Bob, ${\bf x}_{B} = {\bf r}_{B}-{\bf r}_{A}$, can be associated with different values of Bob's amplitude, $\phi_{B}({\bf r}_{B}) = \phi_{B}({\bf x}_{B}-{\bf x}_{O})$, depending on the position eigenstate, ${\bf r}_{A} = -{\bf x}_{O}$, that Alice's wave function collapses to. Bob's pure state therefore appears, to Alice, as a mixed state. In this diagram, the three solid black lines represent the same measured value of 
${\bf x}_{B}$, but each corresponds to a different pair of values $({\bf r}_{A},{\bf r}_{B})$ from the perspective of a classical observer $O$. An analogous diagram can also be constructed for the momentum space representation.}
\end{center}
\end{figure}

Analogous reasoning can also be applied to the momentum space picture, where $\tilde{\phi}_{B}(\boldsymbol{\kappa}_{B}) = |\tilde{\phi}_{B}(\boldsymbol{\kappa}_{B})|e^{i\tilde{\theta}_{B}(\boldsymbol{\kappa}_{B})} \equiv \tilde{\phi}_{B}({\bf p}_{B}-{\bf p}_{O}) = |\tilde{\phi}_{B}({\bf p}_{B}-{\bf p}_{O})|e^{i\theta_{B}({\bf p}_{B}-{\bf p}_{O})}$, and the same value of the relative momentum, ${\bf p}_{B} = \boldsymbol{\kappa}_{B}-\boldsymbol{\kappa}_{A}$, can be associated with different values of the phase and absolute modulus of Bob's momentum space wave function, 
\begin{eqnarray} \label{relative_phase_AB_p}
\tilde{\theta}_{B}(\boldsymbol{\kappa}_{B}) = \tilde{\theta}_{B}({\bf p}_{B}-{\bf p}_{O}) \, , \quad  |\tilde{\phi}_{B}(\boldsymbol{\kappa}_{B})| = |\tilde{\phi}_{B}({\bf p}_{B}-{\bf p}_{O})| \, .  
\end{eqnarray}
In this case, Bob's state appears, to Alice, as the mixed state given by the density matrix
\begin{eqnarray} \label{Bob_effective_mixed_state_p}
\hat{\tilde{\rho}}_{B}^{(A)} := \int \frac{{\rm d}^{3}P({\bf p}_{O})}{{\rm d}^{3}{\rm p}_{O}} \ket{\phi_{B}({\bf p}_{O})}\bra{\phi_{B}({\bf p}_{O})}_{B}  {\rm d}^{3}{\rm p}_{O} \, ,
\end{eqnarray}
where the translated state is defined as
\begin{eqnarray} \label{Bob_effective_mixed_state_p*}
\ket{\phi_{B}({\bf p}_{O})}_{B} := \hat{\tilde{U}}_{B}({\bf p}_{O})\ket{\phi_{B}}_{B} := \int \tilde{\phi}_{B}({\bf p}_{B}-{\bf p}_{O}) \ket{{\bf p}_{B}}_{B}{\rm d}^3{\rm p}_{B} \, ,
\end{eqnarray}
$\hat{\tilde{U}}_{B}({\bf p}_{O})$ is the momentum space translation operator, 
\begin{eqnarray} \label{d_B(x_O)}
\hat{\tilde{U}}_{B}({\bf p}_{O}) := \exp\left(-\frac{i}{\hbar}\bold{p}_{O} \, . \, \hat{\bold{x}}_{B}\right) \, , 
\end{eqnarray}
and we have introduced the shorthand notation
\begin{eqnarray} \label{tildeP(theta_O|theta_B)}
{\rm d}^{3}P({\bf p}_{O}) &:=& {\rm d}^{3}P(\boldsymbol{\kappa}_{A}|\tilde{\Psi}_{AB}^{(O)}) = {\rm d}^{3}P(-{\bf p}_{O}|\tilde{\Psi}_{OB}^{(A)})
\nonumber\\
&=& {\rm d}^{3}P(\boldsymbol{\kappa}_{A}|\tilde{\psi}_{A}) = {\rm d}^{3}P(-{\bf p}_{O}|\tilde{\psi}_{O})
\nonumber\\
&=& |\tilde{\psi}_{A}(\boldsymbol{\kappa}_{A})|^2 {\rm d}^{3}{\rm \kappa}_{A}  = |\tilde{\psi}_{O}(-{\bf p}_{O})|^2 {\rm d}^{3}{\rm p}_{O} \, .
\end{eqnarray}

Experimentally, we may imagine performing a measurement that collapses Alice's momentum, relative to $O$, to a particular eigenstate $\boldsymbol{\kappa}_{A} = -{\bf p}_{O}$, but leaves Bob's wave function unchanged. 
She may then employ the reference state $R$, $\tilde{\psi}_{R} = |\tilde{\psi}_{R}| e^{i\tilde{\theta}_{R}} = \mathcal{F}[\psi_{R}e^{i\theta_{R}}]$, where $\mathcal{F}$ denotes the Fourier transform, to create a momentum space interference pattern. 
This procedure fixes her distance, in momentum space, from each and every point in the momentum space representation of Bob's wave function, $\tilde{\psi}_{B}({\bf p}_{B}-{\bf p}_{O})$. 

Using the definitions (\ref{Bob_effective_mixed_state_p})-(\ref{tildeP(theta_O|theta_B)}), we then obtain
\begin{eqnarray} \label{|tildePhi_B|^2_final}
{\rm d}^{3}P({\bf p}_{B}) = {\rm tr}\left[\ket{{\bf p}_{B}}\bra{{\bf p}_{B}}_{B}\hat{\tilde{\rho}}_{B}^{(A)}\right] {\rm d}^{3}{\rm x}_{B} = |\tilde{\psi}_{O}|^2*|\tilde{\phi}_{B}|^2({\bf p}_{B}){\rm d}^{3}{\rm p}_{B} \, ,
\end{eqnarray}
and
\begin{eqnarray} \label{|tildePhi_B|^2_final*}
{\rm d}^{3}P({\bf p}_{O}) = {\rm tr}\left[\ket{\phi_{B}({\bf p}_{B})}\bra{\phi_{B}({\bf p}_{B})}_{B}\hat{\tilde{\rho}}_{B}^{(A)}\right] {\rm d}^{3}{\rm p}_{B} = |\tilde{\psi}_{O}(-{\bf p}_{B})|^2{\rm d}^{3}{\rm p}_{O} \, ,
\end{eqnarray}
as required. 
The required GUR for momentum measurements,
\begin{eqnarray} \label{GUR_P_bipartite}
\Delta_{\rho}p_{Bj} = \sqrt{(\Delta_{\psi}p_{Aj})^2 + (\Delta_{\phi}p_{Bj})^2} \, ,
\end{eqnarray}
is obtained from the trace 
\begin{eqnarray} \label{GUR_P_op_bipartite}
\langle (\hat{p}_{Bj})^{n} \rangle_{\rho} = {\rm tr}\left[(\hat{p}_{Bj})^{n}\hat{\rho}_{B}^{(A)}\right] \, ,
\end{eqnarray}
where 
\begin{eqnarray} \label{canonical_p_Bj}
\hat{p}_{Bj} = \int p_{Bj} \ket{{\bf p}_{B}}\bra{{\bf p}_{B}}_{B} {\rm d}^{3}{\rm p}_{B} 
\end{eqnarray}
is the canonical one-particle momentum operator. 
Again, in this formulation of our QRF model, we have no need to to refer to the fictitious quantum subsystem, $\mathcal{H}_{O}^{(A)} := \hat{\mathcal{P}}_{O \leftrightarrow A}\mathcal{H}_{A} \equiv \mathcal{H}_{A}$. 
Unlike the operator $\hat{P}_{Bj}$ (\ref{P_{Bj}_bipartite}), which acts on the state $\ket{\Psi_{OB}^{(A)}} \in \mathcal{H}_{OB}^{(A)} := \mathcal{H}_{O}^{(A)} \otimes \mathcal{H}_{B}$ (\ref{Psi_OB(A)}), or, equivalently, $\hat{\rho}_{OB}^{(A)}$ (\ref{rho_OB(A)}), the canonical momentum operator $\hat{p}_{Bj}$ (\ref{canonical_p_Bj}) acts directly on Bob's effective mixed state in momentum space, $\hat{\tilde{\rho}}_{B}^{(A)}$ (\ref{Bob_effective_mixed_state_p}), where $\ket{\tilde{\phi}_{B}({\bf p}_{O})}_{B} \in \mathcal{H}_{B}$ (\ref{Bob_displaced_state}). 
The exact relation between $\hat{\tilde{\rho}}_{OB}^{(A)}$ (\ref{rho_OB(A)}) and $\hat{\tilde{\rho}}_{B}^{(A)}$ (\ref{Bob_effective_mixed_state_p}) is 
\begin{eqnarray} \label{tilde_rho_B_from_rho_OB}
\hat{\tilde{\rho}}_{B}^{(A)} := \int {\rm tr}_{O}\left[\ket{\bold{p}_{O}}\bra{\bold{p}_{O}}_{O} \otimes \hat{\tilde{U}}_{B}({\bf p}_{O}) \, \hat{\rho}_{OB}^{(A)}\right]  {\rm d}^{3}{\rm p}_{O} \, .
\end{eqnarray}

Yet again, we appear to have a similar problem to that encountered in the GCB formalism. 
We have associated {\it two inequivalent mixed states}, $\hat{\rho}_{B}^{(A)}$ (\ref{Bob_effective_mixed_state_x}) and $\hat{\tilde{\rho}}_{B}^{(A)}$ (\ref{Bob_effective_mixed_state_p}), which are related by {\it non-unitary} transformations, with the same QRF $A$, as opposed to associating this with {\it two inequivalent pure states}, which are related by {\it unitary} transformations, as in \cite{Giacomini:2017zju}. 
However, in reality, there is no ambiguity. 
As we have already shown, the mixed states $\hat{\rho}_{B}^{(A)}$ (\ref{Bob_effective_mixed_state_x}) and $\hat{\tilde{\rho}}_{B}^{(A)}$ (\ref{Bob_effective_mixed_state_p}) yield the GURs (\ref{GUR_X_bipartite}) and (\ref{GUR_P_bipartite}), respectively. 
In Sec. \ref{Sec.3.2.5}, we showed how, by combining these relations, Alice is in principle able to determine that she, like Bob, is quantum mechanical in nature. 
To do this, she must combine the statistical results obtained by performing repeated position measurements, on an ensemble of identically prepared systems, with the statistics obtained from repeated momentum measurements, performed on a second ensemble of states. 

Performing only position measurements leads Alice to conclude that Bob's state is the mixed state $\hat{\rho}_{B}^{(A)}$ (\ref{Bob_effective_mixed_state_x}), whereas performing only momentum measurements leads her to identify his state as $\hat{\tilde{\rho}}_{B}^{(A)}$ (\ref{Bob_effective_mixed_state_p}). 
Only when she combines these two sets of complimentary results does she notice an {\it apparent} contradiction; the inequivalence of these two viewpoints alerts her to the fact that, in order to explain the observational facts, she must account, also, for the quantum mechanical nature of her own system. 
This suggests an observational protocol, capable of distinguishing between competing definitions of the term `quantum reference frame'. 

In summary, if the states $\hat{\rho}_{B}^{(A)}$ (\ref{Bob_effective_mixed_state_x}) and $\hat{\tilde{\rho}}_{B}^{(A)}$ (\ref{Bob_effective_mixed_state_p}) are dissimilar enough for Alice to be able to detect a difference between the two, within the limits of her experimental precision, then she is able to discern the fact that she, like Bob, is a quantum system. 
At the level of the uncertainty relations, this is equivalent to observing the non-Heisenberg statistics implied by Eqs. (\ref{indep_uncertainties_O_AB**}). 
Such deviations from the canonical HUP do not appear to be possible within the standard QRF formalism \cite{Giacomini:2017zju}, so that designing an experiment to detect them should enable us to determine, once and for all, whether wave functions objectively exist in the physical background space \cite{Stoica:2021owy}, and, hence, what the correct definition of a QRF should be.

These remarks complete our picture of Alice's QRF, but the findings above beg an important question: If all physical observers are QRFs, and if all QRFs, like Alice, perceive pure states as mixed, then how can we reconcile this with the observation of pure states in physical experiments? 
The answer, of course, is that the mixed nature of the states $\hat{\rho}_{B}^{(A)}$ (\ref{Bob_effective_mixed_state_x}) and $\hat{\tilde{\rho}}_{B}^{(A)}$ (\ref{Bob_effective_mixed_state_p}) is obscured for observers with macroscopic proportions. 
To see this, it is first easier to rewrite (\ref{Bob_effective_mixed_state_p}) in terms of velocities rather than momenta, using the relations
\begin{eqnarray} \label{velocities}
{\bf v}_{O} = {\bf p}_{O}/M_O \, , \quad {\bf v}_{B} = {\bf p}_{B}/m_B \, , 
\end{eqnarray}
which are valid in the nonrelativistic approximation. 
The corresponding uncertainty relations are then
\begin{eqnarray} \label{velocities_UR}
\Delta_{\psi}x_{O}^{i} \, \Delta_{\psi}v_{Oj} \geq \frac{\hbar}{2M_{O}} \, , \quad \Delta_{\psi}x_{B}^{i} \, \Delta_{\psi}v_{Bj} \geq \frac{\hbar}{2m_{B}} \, .
\end{eqnarray}
Here, we assume that the mass of the observed system, Bob, is very small compared to that of the observer, Alice, $m_{B} \ll M_{O} \equiv M_{A}$, and note that it is possible to obtain the limit 
\begin{eqnarray} \label{M_O->Infinity}
\Delta_{\psi}x_{O}^{i} \, \Delta_{\psi}v_{Oj} \rightarrow 0 \, , \quad {\rm as} \quad M_{O} \rightarrow \infty \, . 
\end{eqnarray}
This is compatible with the approximations 
\begin{eqnarray} \label{M_O->Infinity_approximation}
|\psi_{O}(-{\bf x}_{O})|^2 \simeq \delta^{3}(-{\bf x}_{O}) \quad {\rm and} \quad |\tilde{\psi}_{O}(-{\bf v}_{O})|^2 \simeq \delta^{3}(-{\bf v}_{O}) \, , 
\end{eqnarray}
giving
\begin{eqnarray} \label{M_O->Infinity_approximation*}
\Delta_{\psi}x_{O}^{i} \simeq 0 \quad {\rm and} \quad \Delta_{\psi}v_{Oj} \simeq 0 \, , 
\end{eqnarray}
individually, for large enough $M_{O}$.

Next consider, for example, the scenario in which the system corresponding to Alice's QRF is not a microscopic particle, but a massive composite body consisting of many individual particles. 
In this case, her position coordinate ${\bf r}_{A} = -{\bf x}_{O}$ does not represent the position of a single particle, but the position of the centre-of-mass of the composite body, and $\boldsymbol{\kappa}_{A} = -{\bf p}_{O} = -M_{O}{\bf v}_{O}$ is its conjugate momentum. 
Now let us assume, for the sake of argument, that Alice's wave function has a positional uncertainty of the order of an angstrom and that her mass is of the order of $\sim100 {\rm kg}$, which would not be unreasonable for a realistic piece of laboratory equipment, or for the human observer operating it. 
Her corresponding velocity uncertainty is of the order of $\sim 10^{-26}$ ${\rm ms^{-1}}$.
\footnote{For context, we note that, if this were a true velocity, rather than the standard deviation of a statistical distribution, it would take an object travelling at this speed roughly the age of the Universe, $t_{\rm U} \simeq 10^{17}$ s, to traverse the diameter of a single atom.} 
In short, her wave function is extremely narrow, in both position space and velocity space, due to the large value of her mass, so that both $|\psi_{O}({\bf x}_{O})|^2$ and $|\tilde{\psi}_{O}({\bf v}_{O})|^2$ have extremely narrow and effectively compact support. 
In this limit, both the mixed states $\hat{\rho}_{B}^{(A)}$ (\ref{Bob_effective_mixed_state_x}) and $\hat{\tilde{\rho}}_{B}^{(A)}$ (\ref{Bob_effective_mixed_state_p}) reduce, approximately, to the {\it same} density matrix, which is simply the density matrix of the canonical pure state of Bob's system, as seen by the CRF $O$, $\ket{\phi_{B}}\bra{\phi_{B}}_{B}$, though just how good this approximation is will depend on how tight Alice's positional uncertainty is, and whether or not she is heavy enough to compensate for the corresponding spread in momentum.
\footnote{Specifically, we obtain $\hat{\rho}_{B}^{(A)} \, , \, \hat{\tilde{\rho}}_{B}^{(A)} \rightarrow \ket{\phi_{B}}\bra{\phi_{B}}_{B}$ as $|\psi_{O}({\bf x}_{O})|^2 \rightarrow \delta^{3}({\bf x}_{O})$ and $|\tilde{\psi}_{O}({\bf v}_{O})|^2 \rightarrow \delta^{3}({\bf v}_{O})$ ($M_{O} \rightarrow \infty$), but the correct pure state is obtained from the limits of the corresponding state vectors, $\ket{\phi_{B}({\bf x}_{O})}_{B} \, , \, \ket{\phi_{B}({\bf v}_{O})}_{B} \rightarrow \ket{\phi_{B}}_{B}$, as integration over the aforementioned probability distributions, in Eqs. (\ref{Bob_effective_mixed_state_x}) and (\ref{Bob_effective_mixed_state_p}), sets ${\bf x}_{O} = 0$ and ${\bf v}_{O} = 0$ in Eqs. (\ref{Bob_displaced_state}) and (\ref{Bob_effective_mixed_state_p*}), respectively.}

In our QRF model, pure quantum states exist {\it objectively} in the classical background geometry but these can only be perceived as pure, with complete accuracy, by classical observers (i.e., by CRFs defined by the background itself \cite{Jones:1998,Stoica:2021owy}). 
By contrast, all physical observers perceive pure states as mixed. 
However, as the quantum subsystem that acts as a QRF becomes `more classical' (i.e., larger and heavier), it more accurately perceives the nature of its fellow subsystems. 
In this sense, taking the $M_O \rightarrow \infty$ limit of Eqs. (\ref{Bob_effective_mixed_state_x}) and (\ref{Bob_effective_mixed_state_p}) provides a model of the effective quantum-to-classical transition, for a designated `observer', and shows how this observer's perception of the `observed' quantum subsystem changes, accordingly. 

\subsection{Gedanken experiment with a tripartite system} \label{Sec.3.3}

We may construct a model in which a single subsystem of a tripartite state acts as a QRF by complete analogy with the bipartite state analysis, given in Sec. \ref{Sec.3.2}. 
Here, we use the bold face letters $({\bf r}_{A/B/C},\boldsymbol{\kappa}_{A/B/C})$ to denote the phase space coordinates of Alice, Bob and Charlie, as seen by an arbitrary CRF $O$, $({\bf x}_{O/B/C},{\bf p}_{O/B/C})$ to denote the coordinates of $O$, Bob and Charlie, as seen by Alice, $({\bf q}_{O/A/C},\boldsymbol{\pi}_{O/A/C})$ to denote the coordinates of $O$, Alice and Charlie, as seen by Bob, and $({\bf a}_{O/A/B},\boldsymbol{\alpha}_{O/A/B})$ to denote the coordinates of $O$, Alice and Bob, as seen by Charlie, as shown in Fig. 6. 
The transition CRF $O$ $\rightarrow$ QRF $A$ corresponds to the coordinate transformation 
\begin{eqnarray} \label{CRF-O->QRF-A}
{\bf r}_{A} \mapsto {\bf x}_{O} = -{\bf r}_{A} \, , \quad {\bf r}_{B} \mapsto {\bf x}_{B} = {\bf r}_{B} - {\bf r}_{A} \, , \quad {\bf r}_{C} \mapsto {\bf x}_{C} = {\bf r}_{C} - {\bf r}_{A} \, , 
\nonumber\\
\boldsymbol{\kappa}_{A} \mapsto {\bf p}_{O} = -\boldsymbol{\kappa}_{A} \, , \quad \boldsymbol{\kappa}_{B} \mapsto {\bf p}_{B} = \boldsymbol{\kappa}_{B} - \boldsymbol{\kappa}_{A} \, , \quad \boldsymbol{\kappa}_{C} \mapsto {\bf p}_{C} = \boldsymbol{\kappa}_{C} - \boldsymbol{\kappa}_{A} \, , 
\end{eqnarray}
the transition CRF $O$ $\rightarrow$ QRF $B$ corresponds to 
\begin{eqnarray} \label{CRF-O->QRF-B}
{\bf r}_{A} \mapsto {\bf q}_{A} = -({\bf r}_{B}-{\bf r}_{A}) \, , \quad {\bf r}_{B} \mapsto {\bf q}_{O} = -{\bf r}_{B} \, , \quad {\bf r}_{C} \mapsto {\bf q}_{C} = {\bf r}_{C}-{\bf r}_{AB} \, ,
\nonumber\\
\boldsymbol{\kappa}_{A} \mapsto \boldsymbol{\pi}_{A} = -(\boldsymbol{\kappa}_{B}-\boldsymbol{\kappa}_{A}) \, , \quad \boldsymbol{\kappa}_{B} \mapsto \boldsymbol{\pi}_{O} = -\boldsymbol{\kappa}_{B} \, , \quad \boldsymbol{\kappa}_{C} \mapsto \boldsymbol{\pi}_{C} = \boldsymbol{\kappa}_{C} - \boldsymbol{\kappa}_{B} \, ,
\end{eqnarray}
and the transition CRF $O$ $\rightarrow$ QRF $C$ corresponds to 
\begin{eqnarray} \label{CRF-O->QRF-C}
{\bf r}_{A} \mapsto {\bf a}_{A} = -({\bf r}_{C}-{\bf r}_{A}) \, , \quad {\bf r}_{B} \mapsto {\bf a}_{B} = -({\bf r}_{C}-{\bf r}_{B}) \, , \quad {\bf r}_{C} \mapsto {\bf a}_{O} = -{\bf r}_{C} \, , 
\nonumber\\
\boldsymbol{\kappa}_{A} \mapsto \boldsymbol{\alpha}_{A} = -(\boldsymbol{\kappa}_{C}-\boldsymbol{\kappa}_{A}) \, , \quad \boldsymbol{\kappa}_{B} \mapsto \boldsymbol{\alpha}_{B} = -(\boldsymbol{\kappa}_{C}-\boldsymbol{\kappa}_{B}) \, , \quad \boldsymbol{\kappa}_{C} \mapsto \boldsymbol{\alpha}_{O} = -\boldsymbol{\kappa}_{C} \, .
\end{eqnarray}

\begin{figure}[h] \label{Fig.6}
\begin{center}
\includegraphics[width=16cm]{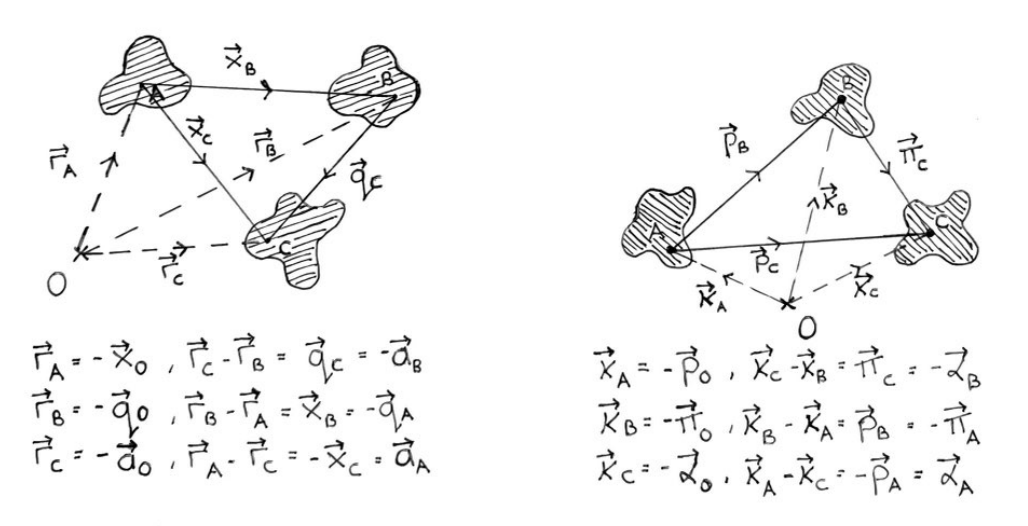}
\caption{A tripartite state, consisting of quantum subsystems $A$, $B$ and $C$, viewed from the perspective of an arbitrary classical observer, $O$, in both position and momentum space.}
\end{center}
\end{figure}

The composite state of the ABC system, from $O$'s perspective, is
\begin{eqnarray} \label{|ABC>}
\ket{\Psi_{ABC}^{(O)}}_{ABC} = \int\int\int \psi_{A}({\bf r}_{A}) \phi_{B}({\bf r}_{B}) \xi_{C}({\bf r}_{C}) \ket{{\bf r}_{A}}_{A} \ket{{\bf r}_{B}}_{B} \ket{{\bf r}_{C}}_{C} {\rm d}^{3}{\rm r}_{A} {\rm d}^{3}{\rm r}_{B} {\rm d}^{3}{\rm r}_{C}
\nonumber\\
= \int\int\int \tilde{\psi}_{A}(\boldsymbol{\kappa}_{A}) \tilde{\phi}_{B}(\boldsymbol{\kappa}_{B}) \tilde{\xi}_{C}(\boldsymbol{\kappa}_{C}) \ket{\boldsymbol{\kappa}_{A}}_{A} \ket{\boldsymbol{\kappa}_{B}}_{B} \ket{\boldsymbol{\kappa}_{C}}_{C} {\rm d}^{3}{\rm \kappa}_{A} {\rm d}^{3}{\rm \kappa}_{B} {\rm d}^{3}{\rm \kappa}_{C} \, , 
\end{eqnarray}
which is equivalent to the density operator
\begin{eqnarray} \label{|ABC>*}
\hat{\rho}_{ABC}^{(O)} := \ket{\Psi_{ABC}^{(O)}}\bra{\Psi_{ABC}^{(O)}}_{ABC} \, , 
\end{eqnarray}
where we again assume separability, for the sake of simplicity, and the states $\ket{\Psi_{OBC}^{(A)}}_{OBC}$, $\ket{\Psi_{AOC}^{(B)}}_{AOC}$ and $\ket{\Psi_{ABO}^{(C)}}_{ABO}$ are obtained by acting on (\ref{|ABC>}) with the relevant parity swap operators. 
The probabilities associated with the outcomes of physical measurements, from the perspective of the QRFs $A$, $B$ or $C$, are then determined by performing an appropriate projection, followed by the relevant partial trace,
\begin{eqnarray} 
{\rm d}^{6}P(\Phi_{BC}^{(A)}|\Psi_{ABC}^{(O)}) := S_{OABC}^{O \rightarrow A}(\Phi_{BC}^{(A)})\ket{\Psi_{ABC}^{(O)}}_{ABC}
:= {\rm tr}_{O}\left[\hat{\Pi}_{\ket{{\rm Bob-Charlie}=\Phi_{BC}^{(A)}}_{OBC}}\hat{\rho}_{OBC}^{(A)}\right] {\rm d}^{6}{\rm V}_{\Phi} \, , 
\nonumber\\ 
{\rm d}^{6}P(\Phi_{AC}^{(B)}|\Psi_{ABC}^{(O)}) := S_{OABC}^{O \rightarrow B}(\Phi_{AC}^{(B)})\ket{\Psi_{ABC}^{(O)}}_{ABC} 
:= {\rm tr}_{O}\left[\hat{\Pi}_{\ket{{\rm Bob-Charlie}=\Phi_{AC}^{(B)}}_{AOC}}\hat{\rho}_{AOC}^{(B)}\right] {\rm d}^{6}{\rm V}_{\Phi} \, , 
\nonumber\\ 
{\rm d}^{6}P(\Phi_{AB}^{(C)}|\Psi_{ABC}^{(O)}) := S_{OABC}^{O \rightarrow C}(\Phi_{AB}^{(C)})\ket{\Psi_{ABC}^{(O)}}_{ABC} 
:= {\rm tr}_{O}\left[\hat{\Pi}_{\ket{{\rm Bob-Charlie}=\Phi_{AB}^{(C)}}_{ABO}}\hat{\rho}_{ABO}^{(C)}\right] {\rm d}^{6}{\rm V}_{\Phi} \, .
\nonumber
\end{eqnarray}
\begin{eqnarray} \label{S_operators_tripartite}
{}
\end{eqnarray}
\indent We may determine the probabilities associated with the results of simultaneous position (momentum) measurements, on the composite state of Bob and Charlie, from Alice's perspective, by projecting onto the bases 
\begin{eqnarray} \label{bases_Alice}
\left\{\ket{{\bf x}_{O}}_{O}\ket{{\bf x}_{B}-{\bf x}_{O}}_{B}\ket{{\bf x}_{C}-{\bf x}_{O}}_{C}\right\} \quad (\left\{\ket{{\bf p}_{O}}_{O}\ket{{\bf p}_{B}-{\bf p}_{O}}_{B}\ket{{\bf p}_{C}-{\bf p}_{O}}_{C}\right\}) \, .
\end{eqnarray}
Likewise, we may determine the probabilities associated with simultaneous position (momentum) measurements, on the composite state of Alice and Charlie, from Bob's perspective, by projecting onto
\begin{eqnarray} \label{bases_Bob}
\left\{\ket{{\bf q}_{A}-{\bf q}_{O}}_{A}\ket{{\bf q}_{O}}_{O}\ket{{\bf q}_{C}-{\bf q}_{O}}_{C}\right\} \quad (\left\{\ket{\boldsymbol{\pi}_{A}-\boldsymbol{\pi}_{O}}_{A}\ket{\boldsymbol{\pi}_{O}}_{O}\ket{\boldsymbol{\pi}_{C}-\boldsymbol{\pi}_{O}}_{C}\right\}) \, .
\end{eqnarray}
Lastly, we may determine the probabilities associated with simultaneous position (momentum) measurements, on the composite state of Alice and Bob, from Charlie's perspective, by projecting onto
\begin{eqnarray} \label{bases_Charlie}
\left\{\ket{{\bf a}_{A}-{\bf a}_{O}}_{A}\ket{{\bf a}_{B}-{\bf a}_{O}}_{B}\ket{{\bf a}_{O}}_{O}\right\} \quad (\left\{\ket{\boldsymbol{\alpha}_{A}-\boldsymbol{\alpha}_{O}}_{A}\ket{\boldsymbol{\alpha}_{B}-\boldsymbol{\alpha}_{O}}_{B}\ket{\boldsymbol{\alpha}_{O}}_{O}\right\}) \, . 
\end{eqnarray}

The explicit calculation for the determination of ${\rm d}^{6}P({\bf x}_{B} , {\bf x}_{C}|\Psi_{OBC}^{(A)})$, for which we introduce the shorthand notation ${\rm d}^{6}P({\bf x}_{B},{\bf x}_{C})$, for simplicity, is
\begin{eqnarray} \label{partial_trace_tripartite_explicit-A-X}
&&{\rm d}^{6}P({\bf x}_{B},{\bf x}_{C})  
\nonumber\\
&:=& {\rm tr}_{O}\left[\ket{{\bf x}_{O}}\bra{{\bf x}_{O}}_{O} \otimes \ket{{\bf x}_{B}-{\bf x}_{O}}\bra{{\bf x}_{B}-{\bf x}_{O}}_{B}
\otimes \ket{{\bf x}_{C}-{\bf x}_{O}}\bra{{\bf x}_{C}-{\bf x}_{O}}_{C} \hat{\rho}_{OBC}^{(A)}\right] {\rm d}^{3}{\rm x}_{B}{\rm d}^{3}{\rm x}_{C}
\nonumber\\
&=& \left(\int |\psi_{O}(-{\bf x}_{O})|^2 |\phi_{B}({\bf x}_{B}-{\bf x}_{O})|^2 |\xi_{C}({\bf x}_{C}-{\bf x}_{O})|^2 {\rm d}^{3}{\rm x}_{O}\right) {\rm d}^{3}{\rm x}_{B}{\rm d}^{3}{\rm x}_{C}
\nonumber\\
&=& |\psi_{O}|^2 * (|\phi_{B}({\bf x}_{B})|^2 |\xi_{C}({\bf x}_{C})|^2) \, {\rm d}^{3}{\rm x}_{B}{\rm d}^{3}{\rm x}_{C} \, ,
\end{eqnarray}
where the notation $|\psi_{O}|^2 *(|\phi_{B}({\bf x}_{B})|^2 |\xi_{C}({\bf x}_{C})|^2)$ indicates that $|\psi_{O}|^2$ is convolved with the joint distribution $|\phi_{B}({\bf x}_{B})|^2|\xi_{C}({\bf x}_{C})|^2$, through integration over a single dummy variable, but that $|\phi_{B}({\bf x}_{B})|^2$ and $|\xi_{C}({\bf x}_{C})|^2$ are not convolved with each other, since each is translated by the same parameter in the convolution, in this case ${\bf x}_{O}$.
\footnote{This should not be confused with either $|\psi_{O}|^2 * |\phi_{B}|^2 * |\xi_{C}|^2({\bf x}_{B})$ or $|\psi_{O}|^2 * |\phi_{B}|^2 * |\xi_{C}|^2({\bf x}_{C})$. The former requires integration over both ${\bf x}_{O}$ and ${\bf x}_{C}$, whereas the latter requires integration over both ${\bf x}_{O}$ and ${\bf x}_{B}$.}

The same probability density can also be obtained, by complete analogy with our previous analysis of the bipartite state, by acting with the canonical projection $\ket{{\bf x}_{B},{\bf x}_{C}}\bra{{\bf x}_{B},{\bf x}_{C}}_{BC}$ on the density operator of the effective $BC$ state, as seen by $A$, which is defined as
\begin{eqnarray} \label{Bob-Charlie_effective_mixed_state_x}
\hat{\rho}_{BC}^{(A)} := \int \frac{{\rm d}^{3}P({\bf x}_{O})}{{\rm d}^{3}{\rm x}_{O}} \ket{\phi_{B}({\bf x}_{O})}\bra{\phi_{B}({\bf x}_{O})}_{B} \otimes \ket{\xi_{C}({\bf x}_{O})}\bra{\xi_{C}({\bf x}_{O})}_{C}  {\rm d}^{3}{\rm x}_{O} \, ,
\end{eqnarray}
where ${\rm d}^{3}P({\bf x}_{O}) = |\psi_{O}(-{\bf x}_{O})|^2$ (\ref{P(theta_O|theta_B)}), $\ket{\phi_{B}({\bf x}_{O})}_{B}$ is given by (\ref{Bob_displaced_state}), and
\begin{eqnarray} \label{Bob-Charlie_effective_mixed_state_x*}
\ket{\xi_{C}({\bf x}_{O})}_{C} := \hat{U}_{C}({\bf x}_{O})\ket{\xi_{C}}_{C} := \int \xi_{C}({\bf x}_{C}-{\bf x}_{O}) \ket{{\bf x}_{C}}_{C}{\rm d}^3{\rm x}_{C} \, .
\end{eqnarray}

However, here, an additional subtlety arises that was not present in the case of the bipartite quantum system. 
If Alice performs separate measurements of ${\bf x}_{B}$ and ${\bf x}_{C}$, on different copies of identically prepared states in an ensemble, then the tripartite system reduces, in effect, to two distinct bipartite systems, $AB$ and $AC$. 
The former is analysed in Sec. \ref{Sec.3.2}, above, and gives rise to the mixed state $\hat{\rho}_{B}^{(A)}$ (\ref{Bob_effective_mixed_state_x}), together with the corresponding GUR (\ref{GUR_X_bipartite}). 
The $AC$ system then gives rise to completely analogous expressions, namely
\begin{eqnarray} \label{Charlie_effective_mixed_state_x}
\hat{\rho}_{C}^{(A)} := \int \frac{{\rm d}^{3}P({\bf x}_{O})}{{\rm d}^{3}{\rm x}_{O}} \ket{\xi_{C}^{(A)}({\bf x}_{O})}\bra{\xi_{C}^{(A)}({\bf x}_{O})}_{C} {\rm d}^{3}{\rm x}_{O} \, ,
\end{eqnarray}
and
\begin{eqnarray} \label{GUR_X_bipartite_AC}
\Delta_{\rho}x_{C}^{i} = \sqrt{(\Delta_{\psi}x_{C}^{i})^2 + (\Delta_{\xi}x_{C}^{i})^2} \, .
\end{eqnarray}
In this case, the probabilities of obtaining the relative displacements ${\bf x}_{B}$ and ${\bf x}_{C}$ are independent,
\begin{eqnarray} \label{Bob+Charlie_effective_mixed_state_independent_measurements}
{\rm d}^{3}P({\bf x}_{B}) &=& |\psi_{O}|^2 * |\phi_{B}|^2({\bf x}_{B}) \, {\rm d}^{3}{\rm x}_{B} \, , 
\nonumber\\
{\rm d}^{3}P({\bf x}_{C}) &=& |\psi_{O}|^2 * |\xi_{C}|^2({\bf x}_{C}) \, {\rm d}^{3}{\rm x}_{C} \, .
\end{eqnarray}
We then see, immediately, that
\begin{eqnarray} \label{experimental_uncertainties-1}
{\rm d}^{6}P({\bf x}_{B},{\bf x}_{C}) \neq {\rm d}^{3}P({\bf x}_{B}){\rm d}^{3}P({\bf x}_{C}) \, , 
\end{eqnarray}
where ${\rm d}^{6}P({\bf x}_{B},{\bf x}_{C})$ is given by Eq. (\ref{partial_trace_tripartite_explicit-A-X}).

This result can be compared with the analogous results for both a classical observer in canonical QM and for a QRF in the GCB model \cite{Giacomini:2017zju}. 
First, we note that in the canonical theory it is {\it always} possible to prepare a bipartite pure state that is separable, from the perspective of a given CRF $O$. 
In this case, the joint-probability distribution ${\rm d}^{6}P({\bf x}_{B},{\bf x}_{C})$, which represents the probability of obtaining the measured values ${\bf x}_{B}$ {\it and} ${\bf x}_{C}$, is related to the individual (marginal) probability distributions via 
\begin{eqnarray} \label{experimental_uncertainties-1*}
{\rm d}^{6}P({\bf x}_{B},{\bf x}_{C}) = {\rm d}^{3}P({\bf x}_{B}){\rm d}^{3}P({\bf x}_{C}) \, .
\end{eqnarray}
In particular, we note that this relation holds true {\it regardless} of whether the measurements are performed simultaneously, on a single state in an ensemble, or separately, on different states therein. 
Second, we note that in the GCB model a tripartite quantum state is described by only two wave functions. 
Hence, every tripartite state in the GCB formalism has the same mathematical form as {\it some} bipartite state in canonical QM, and vice versa. 
It is, therefore, {\it always} possible to prepare a state $BC$ that is separable, from the perspective of a given QRF $A$. 
For such states, Eq. (\ref{experimental_uncertainties-1*}) also holds, regardless of whether the measurements are performed simultaneously, on a single state in an ensemble, or separately on different states. 

These two inequivalent measurement procedures are illustrated in Figs. 7(a) and 7(b), respectively, and it is clear that they suggest a simple experimental procedure to discriminate between competing QRF models. 
In the GCB formalism, if the composite state of the $BC$ system, as seen by Alice, is separable, then the relevant probabilities are ${\rm d}^{6}P({\bf x}_{B},{\bf x}_{C}) = |\psi({\bf x}_{C})|^2|\phi({\bf x}_{B})|^2{\rm d}^{3}{\rm x}_{B}{\rm d}^{3}{\rm x}_{C}$ and ${\rm d}^{3}P({\bf x}_{B}) = |\phi({\bf x}_{B})|^2{\rm d}^{3}{\rm x}_{B}$, ${\rm d}^{3}P({\bf x}_{C}) = |\psi({\bf x}_{C})|^2{\rm d}^{3}{\rm x}_{C}$, which clearly obey the relation (\ref{experimental_uncertainties-1*}). 
By contrast, in our model, Eq. (\ref{experimental_uncertainties-1*}) holds only for separate measurements, on different states in an ensemble, and {\it cannot} hold for simultaneous measurements on a single tripartite state, regardless of how we prepare the initial states of the subsystems $A$, $B$ and $C$. 

If the wave functions of Alice, Bob and Charlie exist objectively in the classical background space \cite{Jones:1998,Stoica:2021owy}, as we propose here, these different experimental procedures give rise to different joint-probability densities because the former projects pairs of wave functions onto pairs of position eigenstates, whereas the latter projects three wave functions onto three position eigenstates, at the same time.
The relation (\ref{experimental_uncertainties-1*}) is then satisfied, for simultaneous measurements, only in the limit $M_{O} \equiv M_{A} \rightarrow \infty$. 
In this limit, there is again no distinction between separate and simultaneous measurements, since Alice's positional uncertainty tends to zero ($|\psi_{O}|^2 \equiv |\psi_{A}|^2 \rightarrow \delta^{3}(-{\bf x}_{O})$) and the effective mixed-state density matrix $\hat{\rho}_{BC}^{(A)}$ (\ref{Bob-Charlie_effective_mixed_state_x}) reduces to canonical pure-state form, $\hat{\rho}_{BC}^{(A)} \rightarrow \hat{\rho}_{BC} = \ket{\phi_{B}}\bra{\phi_{B}}_{B} \otimes \ket{\xi_{C}}\bra{\xi_{C}}_{C}$, which is valid for a classical observer.

In the momentum space representation, the explicit calculation for the determination of the probability density ${\rm d}^{6}P({\bf p}_{B} , {\bf p}_{C}|\Psi_{OBC}^{(A)})$, for which we introduce the shorthand notation ${\rm d}^{6}P({\bf p}_{B},{\bf p}_{C})$, proceeds as
\begin{eqnarray} \label{partial_trace_tripartite_explicit-A-P}
&&{\rm d}^{6}P({\bf p}_{B},{\bf p}_{C}) 
\nonumber\\
&:=& {\rm tr}_{O}\left[\ket{{\bf p}_{O}}\bra{{\bf p}_{O}}_{O} \otimes \ket{{\bf p}_{B}-{\bf p}_{O}}\bra{{\bf p}_{B}-{\bf p}_{O}}_{B} \otimes \ket{{\bf p}_{C}-{\bf p}_{O}}\bra{{\bf p}_{C}-{\bf p}_{O}}_{C} \hat{\rho}_{OBC}^{(A)}\right]
{\rm d}^{3}{\rm p}_{B}{\rm d}^{3}{\rm p}_{C}
\nonumber\\
&=& \left(\int |\tilde{\psi}_{O}(-{\bf p}_{O})|^2 |\tilde{\phi}_{B}({\bf p}_{B}-{\bf p}_{O})|^2 |\tilde{\xi}_{C}({\bf p}_{C}-{\bf p}_{O})|^2 {\rm d}^{3}{\rm p}_{O}\right) {\rm d}^{3}{\rm p}_{B}{\rm d}^{3}{\rm p}_{C}
\nonumber\\
&=& |\tilde{\psi}_{O}|^2 * (|\tilde{\phi}_{B}({\bf p}_{B})|^2 |\tilde{\xi}_{C}({\bf p}_{C})|^2) \, {\rm d}^{3}{\rm p}_{B}{\rm d}^{3}{\rm p}_{C} \, .
\end{eqnarray}
The same probability density can be obtained by acting with the canonical projection operator $\ket{{\bf p}_{B},{\bf p}_{C}}\bra{{\bf p}_{B},{\bf p}_{C}}_{BC}$ on the density matrix 
\begin{eqnarray} \label{Bob-Charlie_effective_mixed_state_p}
\hat{\tilde{\rho}}_{BC}^{(A)} := \int \frac{{\rm d}^{3}P({\bf p}_{O})}{{\rm d}^{3}{\rm p}_{O}} \ket{\phi_{B}({\bf p}_{O})}\bra{\phi_{B}({\bf p}_{O})}_{B} \otimes \ket{\xi_{C}({\bf p}_{O})}\bra{\xi_{C}({\bf p}_{O})}_{C}  {\rm d}^{3}{\rm p}_{O} \, ,
\end{eqnarray}
where ${\rm d}^{3}P({\bf p}_{O}) = |\tilde{\psi}_{O}(-{\bf p}_{O})|^2$ (\ref{tildeP(theta_O|theta_B)}), $\ket{\phi_{B}({\bf p}_{O})}_{B}$ is given by (\ref{Bob_effective_mixed_state_p*}), and
\begin{eqnarray} \label{Bob-Charlie_effective_mixed_state_x*}
\ket{\xi_{C}({\bf p}_{O})}_{C} := \hat{\tilde{U}}_{C}({\bf p}_{O})\ket{\xi_{C}}_{C} := \int \xi_{C}({\bf p}_{C}-{\bf p}_{O}) \ket{{\bf p}_{C}}_{C}{\rm d}^3{\rm p}_{C} \, ,
\end{eqnarray}
by complete analogy with our previous analysis for position measurements.

If, instead, Alice performs separate measurements of ${\bf p}_{B}$ and ${\bf p}_{C}$, on different systems in an ensemble of identically prepared states, the tripartite system again reduces, in effect, to two bipartite systems, $AB$ and $AC$. 
Performing measurements of ${\bf p}_{B}$ on the former gives rise to the mixed state $\hat{\tilde{\rho}}_{B}^{(A)}$ (\ref{Bob_effective_mixed_state_p}), together with the corresponding GUR (\ref{GUR_P_bipartite}), and perfomring measurements of ${\bf p}_{C}$ on the latter generates completely analogous expressions, 
\begin{eqnarray} \label{Charlie_effective_mixed_state_p}
\hat{\tilde{\rho}}_{C}^{(A)} := \int \frac{{\rm d}^{3}P({\bf p}_{O})}{{\rm d}^{3}{\rm p}_{O}} \ket{\xi_{C}^{(A)}({\bf p}_{O})}\bra{\xi_{C}^{(A)}({\bf p}_{O})}_{C} {\rm d}^{3}{\rm p}_{O} \, ,
\end{eqnarray}
and
\begin{eqnarray} \label{GUR_P_bipartite_AC}
\Delta_{\rho}p_{C}^{i} = \sqrt{(\Delta_{\psi}p_{C}^{i})^2 + (\Delta_{\xi}p_{C}^{i})^2} \, .
\end{eqnarray}
In the is case, the probabilities of obtaining the relative momenta ${\bf p}_{B}$ and ${\bf p}_{C}$ are independent, 
\begin{eqnarray} \label{Bob+Charlie_effective_mixed_state_independent_measurements*}
{\rm d}^{3}P({\bf p}_{B}) &=& |\tilde{\psi}_{O}|^2 * |\tilde{\phi}_{B}|^2({\bf p}_{B}) \, {\rm d}^{3}{\rm p}_{B} \, , 
\nonumber\\
{\rm d}^{3}P({\bf p}_{C}) &=& |\tilde{\psi}_{O}|^2 * |\tilde{\xi}_{C}|^2({\bf p}_{C}) \, {\rm d}^{3}{\rm p}_{C} \, .
\end{eqnarray}
It follows immediately that 
\begin{eqnarray} \label{experimental_uncertainties-2}
{\rm d}^{6}P({\bf p}_{B},{\bf p}_{C}) \neq {\rm d}^{3}P({\bf p}_{B}){\rm d}^{3}P({\bf p}_{C}) \, , 
\end{eqnarray}
where ${\rm d}^{6}P({\bf p}_{B},{\bf p}_{C})$ is given by Eq. (\ref{partial_trace_tripartite_explicit-A-P}).

This result should also be compared with the analogous results for a classical observer in canonical QM and for a QRF in the GCB model \cite{Giacomini:2017zju}. 
In the canonical theory the relevant probability distributions for a bipartite, separable, pure quantum state $BC$ obey the relation
\begin{eqnarray} \label{experimental_uncertainties-2*}
{\rm d}^{6}P({\bf p}_{B},{\bf p}_{C}) = {\rm d}^{3}P({\bf p}_{B}){\rm d}^{3}P({\bf p}_{C}) \, ,
\end{eqnarray}
regardless of whether the measurements are performed separately, on different states in an ensemble, or simultaneously on a single state.
A completely analogous relation therefore holds, in the GCB formalism, for all $BC$ states that appear separable from the perspective of the QRF $A$.

In our QRF model, the relation (\ref{experimental_uncertainties-2*}) can only satisfied be satisfied in the limit $M_{O} \equiv M_{A} \rightarrow \infty$, when the mixed state density matrix for momentum-measurements, $\hat{\tilde{\rho}}_{BC}^{(A)}$ (\ref{Bob-Charlie_effective_mixed_state_p}), reduces to the density matrix for the objective pure state, $\hat{\tilde{\rho}}_{BC}^{(A)} \rightarrow \hat{\rho}_{BC} = \ket{\phi_{B}}\bra{\phi_{B}}_{B} \otimes \ket{\xi_{B}}\bra{\xi_{B}}_{B}$. 
To see this, we need only rewrite (\ref{Bob-Charlie_effective_mixed_state_p}) as an integral over the velocity-space volume element, ${\rm d}^{3}{\rm v}_{O}$, and note that ${\rm d}^{3}P({\bf v}_{O})/{\rm d}^{3}{\rm v}_{O} = |\tilde{\psi}_{O}(-{\bf v}_{O})|^2 \rightarrow \delta^{3}(-{\bf v}_{O})$, for $M_{O} \rightarrow \infty$.
Hence, in this limit, there is no distinction between separate and simultaneous measurements, of either position or momentum, since Alice's position and velocity uncertainties both tend to zero, and she becomes `fully classical', as expected.

Finally, we note that the generalised position and momentum operators, representing Alice's measurements on the $BC$ subsystem of the objective tripartite state $\ket{\Psi_{ABC}^{(O)}}_{ABC}$ (\ref{|ABC>}), can be denoted as $\hat{X}_{B/C}^{i} := \hat{\mathcal{X}}_{B/C}^{i} + \hat{\mathcal{X}}_{B/C}'^{i}$ and $\hat{P}_{B/Cj} := \hat{\mathcal{P}}_{B/Cj} + \hat{\mathcal{P}}_{B/Cj}'$, respectively. 
These are constructed, explicitly, via a straightforward extension of the definitions (\ref{X_{B}^{i}_bipartite}) and (\ref{P_{Bj}_bipartite}), given in Sec. \ref{Sec.3.2}. 
By exponentiating commuting pairs of the sub-operators $\left\{\hat{{\bf \mathcal{X}}}_{B/C},\hat{{\bf \mathcal{X}}}_{B/C}',\hat{{\bf \mathcal{P}}}_{B/C},\hat{{\bf \mathcal{P}}}_{B/C}'\right\}$ we can again construct useful unitary operators that shuffle between entangled and unentangled bases of $\ket{\Psi_{ABC}^{(O)}}_{ABC}$ (\ref{|ABC>}), in whatever coordinate system we wish to employ, $({\bf r}_{A/B/C},\boldsymbol{\kappa}_{A/B/C})$, $({\bf x}_{O/B/C},{\bf p}_{O/B/C})$, $({\bf q}_{O/A/C},\boldsymbol{\pi}_{O/A/C})$ or $({\bf a}_{O/A/B},\boldsymbol{\alpha}_{O/A/B})$. 

We stress, however, that such transformations do not change the fundamental relations between these coordinate systems. 
All such unitarily equivalent bases are consistent with the perspective of the CRF $O$, and with the relevant coordinate transformation, (\ref{CRF-O->QRF-A}), (\ref{CRF-O->QRF-B}) or (\ref{CRF-O->QRF-C}),  for whichever coordinates we choose to use. 
Generalised operators representing Bob's measurements on the $AC$ subsystem and Charlie's measurements on the $AB$ subsystem can also be constructed in like manner, although the corresponding proliferation of notation becomes rather cumbersome. 
We address this issue in our treatment of the general $N$-partite system, given in Sec. \ref{Sec.3.4}.

\begin{figure}[h] \label{Fig.7}
\centering 
	\includegraphics[width=7cm]{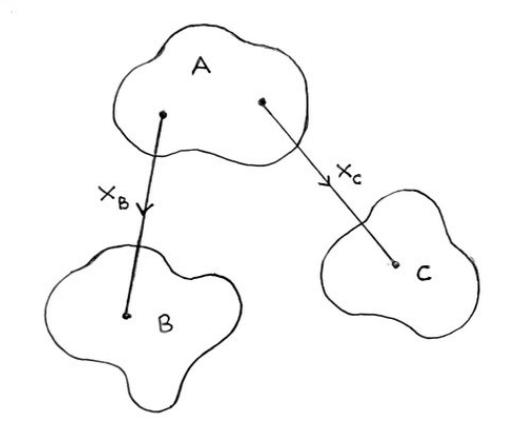}
	\includegraphics[width=7cm]{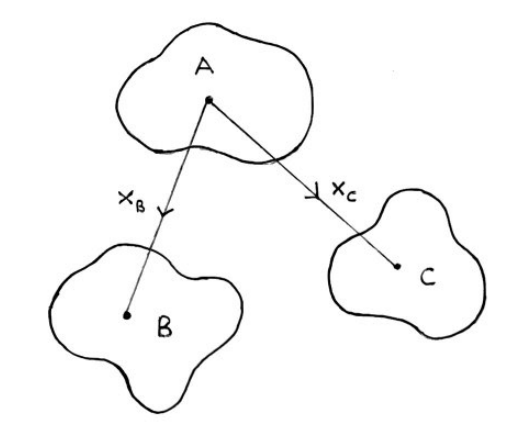}
\caption{Fig. 7(a) (left) illustrates an experimental procedure in which separate measurements of ${\bf x}_{B}$ and ${\bf x}_{C}$ are performed on different states in an ensemble, causing Alice's wave function to collapse, in general, to different position eigenstates. Fig. 7(b) (right) illustrates the simultaneous measurement of ${\bf x}_{B}$ and ${\bf x}_{C}$ on a single state of the tripartite system. These procedures generate different joint-probability distribtutions, which can be used to differentiate between competing definitions of the term `quantum reference frame'. An analogous diagram can also be constructed for measurements of the relative momenta, ${\bf p}_{B}$ and ${\bf p}_{C}$. In these figures, we indicate the regions in which the wave functions of $A$, $B$ and $C$ are nonzero with three closed curves, but neglect to shade their interiors, so that the different configurations to which the subsystems collapse can be seen more clearly.}
\end{figure}

\subsection{Generalisation to $N$-particle states} \label{Sec.3.4}

In generalising our treatment of the tripartite state to a general $N$-partite system, for arbitrarily large $N$, the biggest challenge we face is that of choosing an appropriate notation. 
For $N \gtrsim 10^2$, it is not only cumbersome, but also impracticable, to use different Latin or Greek letters (and variants thereof) to denote the different coordinates employed by different subsystems. 

We therefore choose, instead, to label the coordinates used by an arbitrary CRF with a zero, rather than an $O$, and to label those employed by quantum subsystems numerically as $1,2, \dots N$. 
The corresponding phase space coordinates are written explicitly as $({\bf x}_{(0)}^{[m]},{\bf p}_{(0)}^{[m]})$ and $({\bf x}_{(m)}^{[n]},{\bf p}_{(m)}^{[n]})$, respectively, for $m \in \left\{1,2, \dots N\right\}$ and $n \in \left\{0,1,2, \dots N\right\}$, with $n \neq m$. 
In other words, we use round brackets to indicate the subsystem that is acting as the QRF and square brackets to label all other subsystems. 
The individual components of these displacements and momenta can then be labelled using unbracketed suffixes, as usual, i.e., as $({\bf x}_{(0)}^{[m]i},{\bf p}_{(0)j}^{[m]})$ and $({\bf x}_{(m)}^{[n]i},{\bf p}_{(m)}^{[n]j})$, where $i,j \in \left\{1,2,3\right\}$. 
In this notation, the objective state of the $N$-partite system, from the perspective of CRF-0, is written as
\begin{eqnarray} \label{Psi_N-partite}
\ket{\Psi_{1,2, \dots N}^{(0)}}_{1,2, \dots N} = \int\int \dots \int \bigotimes_{n=1}^{N} \psi_{n}({\bf x}_{(0)}^{[n]}) \ket{{\bf x}_{(0)}^{[n]}}_{n}{\rm d}^{3}{\rm x}_{(0)}^{[n]} \, .
\end{eqnarray}
Again, we assume the state is separable, from the perspective of the classical observer, for simplicity. 

The probabilities associated with the outcomes of physical measurements, from the perspective of the QRF-$m$, for an arbitrary value of $m$ in the range $2 < m < N$, are then given by
\begin{eqnarray} \label{CRF-0->QRF-m}
&&{\rm d}^{3(N-1)}P(\Phi_{1,2,\dots m-1,m+1,\dots N}^{(m)}|\Psi_{1,2, \dots N}^{(0)}) 
\nonumber\\
&=& S_{0,1,2, \dots N}^{(0 \rightarrow m)}(\Phi_{1,2,\dots m-1,m+1,\dots N}^{(m)}) \, \ket{\Psi_{1,2, \dots N}^{(0)}}_{1,2, \dots N} 
\nonumber\\
&:=& {\rm tr}_{0}\left[\hat{\Pi}_{\ket{\Phi_{1,2,\dots m-1,m+1,\dots N}^{(m)}}_{1,2,\dots N}} \hat{\rho}_{1,2,\dots m-1,0,m+1,\dots N}^{(m)}\right] {\rm d}^{3(N-1)}{\rm V}_{\Phi} \, , 
\end{eqnarray}
where the partial trace is taken over the subspace $\mathcal{H}_{0}^{(m)} := \hat{\mathcal{P}}_{0 \leftrightarrow m} \, \mathcal{H}_{m}^{(0)} \equiv \mathcal{H}_{m}$ of the total Hilbert space of the $N$-partite system. 
Those for $m=1,2$ and $N$ are defined in like manner, with index addition taken modulo $N$, but we select $m$ in the range $2 < m < N$, to illustrate our procedure. 
Hence, in order to determine probability density associated with simultaneous measurements of the $N-1$ relative displacements, ${\bf x}_{(m)}^{[1]}$, ${\bf x}_{(m)}^{[2]}$, . . , we must project onto the bases 
\begin{eqnarray} \label{N-partite_X_basis}
\left\{\ket{{\bf x}_{(m)}^{[1]}-{\bf x}_{(m)}^{[0]}}_{1} \dots \otimes \ket{{\bf x}_{(m)}^{[m-1]}-{\bf x}_{(m)}^{[0]}}_{m-1} \otimes \ket{{\bf x}_{(m)}^{[0]}}_{0} \otimes \ket{{\bf x}_{(m)}^{[m+1]}-{\bf x}_{(m)}^{[0]}}_{m+1} \dots \right\} \, ,
\end{eqnarray}
whereas, to determine the probability density associated with simultaneous measurements of the $N-1$ relative momenta,  ${\bf p}_{(m)}^{[1]}$, ${\bf p}_{(m)}^{[2]}$, . . , we must project onto the bases
\begin{eqnarray} \label{N-partite_P_basis}
\left\{\ket{{\bf p}_{(m)}^{[1]}-{\bf p}_{(m)}^{[0]}}_{1} \dots \otimes \ket{{\bf p}_{(m)}^{[m-1]}-{\bf p}_{(m)}^{[0]}}_{m-1} \otimes \ket{{\bf p}_{(m)}^{[0]}}_{0} \otimes \ket{{\bf p}_{(m)}^{[m+1]}-{\bf p}_{(m)}^{[0]}}_{m+1} \dots \right\} \, .
\end{eqnarray}

The generalised operators, corresponding to measurements of relative position and momentum, performed by the QRF-$m$, are denoted as
\begin{eqnarray} \label{N-partite_operators}
\hat{X}_{(m)}^{[n]i} := \hat{\mathcal{X}}_{(m)}^{[n]i} + \hat{\mathcal{X}}_{(m)}'^{[n]i} \, , \quad \hat{P}_{(m)j}^{[n]} := \hat{\mathcal{P}}_{(m)j}^{[n]} + \hat{\mathcal{P}}_{(m)j}'^{[n]} \, , 
\end{eqnarray}
and the definitions of the sub-operators $\left\{\hat{\mathcal{X}}_{(m)}^{[n]i},\hat{\mathcal{X}}_{(m)}'^{[n]i},\hat{\mathcal{P}}_{(m)j}^{[n]},\hat{\mathcal{P}}_{(m)j}'^{[n]} \right\}$ should be obvious to the reader who has patiently followed our arguments thus far. 
Using these basic ingredients, the density operators of the effective mixed states seen by an arbitrary QRF, labelled QRF-$m$, in both position and momentum space, can be constructed, and it is straightforward to show that these reduce to the density operator of the objective pure state seen by CRF-$0$ - absent the $\mathcal{H}_{m}$ subsystem, which has been traced out - in the limit $M_{0} \equiv M_{m} \rightarrow \infty$.

\section{Advantages of the alternative formalism} \label{Sec.4}

\subsection{The spatial background and CRFs} \label{Sec.4.1}

One can view the QRF formalism developed in \cite{Giacomini:2017zju}, and in subsequent follow-up works \cite{Vanrietvelde:2018pgb,Vanrietvelde:2018dit,Hohn:2018toe,Hohn:2018iwn,Krumm:2020fws,Ballesteros:2020lgl,QRF_Bell_Test:2021,Giacomini:2021gei,delaHamette:2021iwx,Cepollaro:2021ccc,Castro-Ruiz:2021vnq,AliAhmad:2021adn,Hoehn:2021flk,Carrozza:2021gju,delaHamette:2021oex,delaHamette:2021piz,Giacomini:2022hco,Overstreet:2022zgq,Kabel:2022cje,Apadula:2022pxk,Amelino-Camelia:2022dsj,Kabel:2023jve,Hoehn:2023ehz,Hoehn:2023axh,Wang:2023koz}, as encoding the idea that even classical space is {\it relational}, and that there is no `objective' background spacetime.
\footnote{We thank Flaminia Giacomini for personal communications, clarifying this point.} 
In our view, this is at least {\it potentially} inconsistent, since even QRF formalisms must rely on some version of the Born rule, as discussed in Sec. \ref{Sec.2.1}.

By contrast, our QRF formalism pre-supposes the existence of a classical spacetime background, in which material quantum systems `live', as superpositions of position (momentum) eigenstates, relative to an arbitrary CRF. 
Our justification for this assumption is the formal equivalence between geometries and the groups that define their symmetries, according to the Erlangen program \cite{ErlangenProgram_Klein_1872,ErlangenProgram_EMS_2015,Zuber:2013rha,Kisil:2010,Horwood:2004uh,ErlangenProgram_Encycolpedia_Srpinger,Lev:2020igj,Goenner:2015}. 
These symmetries also define the relationships between arbitrary CRFs, so that a one-to-one correspondence exists between the set of {\it all} possible classical frames of reference, in a given geometry, and the geometry itself. 
For three-dimensional Euclidean space, the corresponding momentum space is also Euclidean, and the relevant symmetry group is the three-dimensional Galilean group, ${\rm G}^3$ \cite{Ibragimov:2015, Jones:1998}. 

We regard this as a distinct advantage of our model, since the relationship between the wave functions of material quantum systems and the classical geometric space, on which they are defined as complex-valued scalars, is also clearly defined \cite{Stoica:2021owy}. 
The correct definition of `relational' space, on the other hand, appears to us to be fraught with metaphysical difficulties, and the debate regarding its validity goes all the way back to the debate between Newton and Leibniz \cite{Newton_vs_Leibniz_Stanford_Encyclopedia_of_Philosophy-1,Newton_vs_Leibniz_Stanford_Encyclopedia_of_Philosophy-2,Newton_vs_Leibniz_Stanford_Encyclopedia_of_Philosophy-3,Newton_vs_Leibniz_Stanford_Encyclopedia_of_Philosophy-4,IEP,Ballard:1960,Barbour:1982,Pooley:2001,Khamara:2006,Anderson:2012az,Belkind:2013,Verelst:2014,Anderson_in_Book:2017,Evangelidis:2018,Jorati:2020}. 
The formalism developed here places us very much within the Newtonian camp, for better or worse.     

In the remainder of this subsection, we investigate the consequences of our metaphysical position for the time evolution of quantum systems, from the perspective of an arbitrary QRF. 
For concreteness, we consider the tripartite state analysed in Sec. \ref{Sec.3.3}, though our results generalise naturally to higher-order systems. 

From the perspective of an arbitrary CRF $O$, the Hamiltonian of the tripartite system $ABC$ is 
\begin{eqnarray} \label{tripartite_Hamiltonian_O}
\hat{H}_{ABC}^{(O)} = \frac{\hat{\kappa}_{A}^2}{2m_{A}} + \frac{\hat{\kappa}_{B}^2}{2m_{B}} + \frac{\hat{\kappa}_{C}^2}{2m_{C}} + \hat{V}_{ABC}(\hat{{\bf r}}_{A},\hat{{\bf r}}_{B},\hat{{\bf r}}_{C}) \, , 
\end{eqnarray}
where 
\begin{eqnarray} \label{tripartite_Hamiltonian_potential_O}
\hat{V}_{ABC}(\hat{{\bf r}}_{A},\hat{{\bf r}}_{B},\hat{{\bf r}}_{C}) := \hat{V}_{AB}(|\hat{{\bf r}}_{A}-\hat{{\bf r}}_{B}|) + \hat{V}_{BC}(|\hat{{\bf r}}_{B}-\hat{{\bf r}}_{C}|) + \hat{V}_{CA}(|\hat{{\bf r}}_{C}-\hat{{\bf r}}_{A}|) \, . 
\end{eqnarray}
This generates time evolution of the total state according to
\begin{eqnarray} \label{tripartite_time-evolution_O-1}
\ket{\Psi_{ABC}^{(O)}(t)}_{ABC} := \exp\left(-\frac{i}{\hbar}\hat{H}_{ABC}^{(O)}t\right)\ket{\Psi_{ABC}^{(O)}(0)}_{ABC} \, , 
\end{eqnarray}
where $\ket{\Psi_{ABC}^{(O)}(0)}_{ABC} \equiv \ket{\Psi_{ABC}^{(O)}}_{ABC}$ and the latter is defined in Eqs. (\ref{|ABC>}).
(Here, we again assume that the initial state is separable, from $O$'s perspective, for simplicity.) 

However, there is, of course, no absolute measure of the energy of a system, since only relative energies are physically observable \cite{Rae:2002,Peskin&Schroeder:1995,QFT_Nutshell}. 
We must therefore construct three {\it relative} Hamiltonians, $\hat{\mathcal{H}}_{OBC}^{(A)}$, $\hat{\mathcal{H}}_{AOC}^{(B)}$ and $\hat{\mathcal{H}}_{ABO}^{(C)}$, for the QRFs $A$, $B$ and $C$. 
The eigenvalues of these operators represent, respectively, the possible measured values of the relative energies between (a) Alice's subsystem and the composite state of Bob and Charlie, (b) Bob's subsystem and the composite state of Charlie and Alice, and (c) Charlie's subsystem and the composite state of Alice and Bob. 
These are defined as
\begin{eqnarray} \label{relative_Hamiltonian_(A)OBC}
\hat{\mathcal{H}}_{OBC}^{(A)} := \frac{\hat{P}_{B}^2}{2m_B} + \frac{\hat{P}_{C}^2}{2m_C} + \hat{V}_{OB}(|\hat{{\bf X}}_{B}|) + \hat{V}_{CO}(|\hat{{\bf X}}_{C}|) + \hat{V}_{BC}(|\hat{{\bf X}}_{B}-\hat{{\bf X}}_{C}|) \, , 
\end{eqnarray}
where
\begin{eqnarray} \label{relative_Hamiltonian_(A)OBC*}
\hat{{\bf X}}_{B} &:=& \hat{{\bf x}}_{O} \otimes \hat{\mathbb{I}}_{B} \otimes \hat{\mathbb{I}}_{C} + \hat{\mathbb{I}}_{O} \otimes (\widehat{{\bf x}_{B}-{\bf x}_{O}})_{B} \otimes \hat{\mathbb{I}}_{C} \, , 
\nonumber\\
\hat{{\bf P}}_{B} &:=& \hat{{\bf p}}_{O} \otimes \hat{\mathbb{I}}_{B} \otimes \hat{\mathbb{I}}_{C} + \hat{\mathbb{I}}_{O} \otimes (\widehat{{\bf p}_{B}-{\bf p}_{O}})_{B} \otimes \hat{\mathbb{I}}_{C} \, , 
\nonumber\\
\hat{{\bf X}}_{C} &:=& \hat{{\bf x}}_{O} \otimes \hat{\mathbb{I}}_{B} \otimes \hat{\mathbb{I}}_{C} + \hat{\mathbb{I}}_{O} \otimes \hat{\mathbb{I}}_{B} \otimes (\widehat{{\bf x}_{C}-{\bf x}_{O}})_{C} \, , 
\nonumber\\
\hat{{\bf P}}_{C} &:=& \hat{{\bf p}}_{O} \otimes \hat{\mathbb{I}}_{B} \otimes \hat{\mathbb{I}}_{C} + \hat{\mathbb{I}}_{O} \otimes \hat{\mathbb{I}}_{B} \otimes (\widehat{{\bf p}_{C}-{\bf p}_{O}})_{C} \, ;
\end{eqnarray}
\begin{eqnarray} \label{relative_Hamiltonian_(B)AOC}
\hat{\mathcal{H}}_{AOC}^{(B)} := \frac{\hat{\Pi}_{C}^2}{2m_C} + \frac{\hat{\Pi}_{A}^2}{2m_A} + \hat{V}_{OC}(|\hat{{\bf Q}}_{C}|) + \hat{V}_{AO}(|\hat{{\bf Q}}_{A}|) + \hat{V}_{CA}(|\hat{{\bf Q}}_{C}-\hat{{\bf Q}}_{A}|) \, , 
\end{eqnarray}
where
\begin{eqnarray} \label{relative_Hamiltonian_(B)AOC*}
\hat{{\bf Q}}_{C} &:=& \hat{\mathbb{I}}_{A} \otimes \hat{{\bf q}}_{O} \otimes \hat{\mathbb{I}}_{C} + \hat{\mathbb{I}}_{A} \otimes \hat{\mathbb{I}}_{O} \otimes (\widehat{{\bf q}_{C}-{\bf q}_{O}})_{C} \, , 
\nonumber\\
\hat{\boldsymbol{\Pi}}_{C} &:=& \hat{\mathbb{I}}_{A} \otimes \hat{\boldsymbol{\pi}}_{O} \otimes \hat{\mathbb{I}}_{C} + \hat{\mathbb{I}}_{A} \otimes \hat{\mathbb{I}}_{O} \otimes (\widehat{\boldsymbol{\pi}_{C}-\boldsymbol{\pi}_{O}})_{C} \, , 
\nonumber\\
\hat{{\bf Q}}_{A} &:=& \hat{\mathbb{I}}_{A} \otimes \hat{{\bf q}}_{O} \otimes \hat{\mathbb{I}}_{C} + (\widehat{{\bf q}_{A}-{\bf q}_{O}})_{A} \otimes \hat{\mathbb{I}}_{O} \otimes \hat{\mathbb{I}}_{C} \, ,   
\nonumber\\
\hat{\boldsymbol{\Pi}}_{A} &:=& \hat{\mathbb{I}}_{A} \otimes \hat{\boldsymbol{\pi}}_{O} \otimes \hat{\mathbb{I}}_{C} + (\widehat{\boldsymbol{\pi}_{A}-\boldsymbol{\pi}_{O}})_{A} \otimes \hat{\mathbb{I}}_{O} \otimes \hat{\mathbb{I}}_{C} \, ; 
\end{eqnarray}
and
\begin{eqnarray} \label{relative_Hamiltonian_(C)ABO}
\hat{\mathcal{H}}_{ABO}^{(C)} := \frac{\hat{\mathlarger{\mathlarger{\mathlarger{\alpha}}}}_{A}^2}{2m_A} + \frac{\hat{\mathlarger{\mathlarger{\mathlarger{\alpha}}}}_{B}^2}{2m_B} + \hat{V}_{OA}(|\hat{{\bf A}}_{A}|) + \hat{V}_{BO}(|\hat{{\bf A}}_{B}|) + \hat{V}_{AB}(|\hat{{\bf A}}_{A}-\hat{{\bf A}}_{B}|) \, , 
\end{eqnarray}
where
\begin{eqnarray} \label{relative_Hamiltonian_(C)ABO*}
\hat{{\bf A}}_{A} &:=& \hat{\mathbb{I}}_{A} \otimes \hat{\mathbb{I}}_{B} \otimes \hat{{\bf a}}_{O} + (\widehat{{\bf a}_{A}-{\bf a}_{O}})_{A} \otimes \hat{\mathbb{I}}_{B} \otimes \hat{\mathbb{I}}_{O} \, , 
\nonumber\\
\hat{\mathlarger{\mathlarger{\boldsymbol{\alpha}}}}_{A} &:=& \hat{\mathbb{I}}_{A} \otimes \hat{\mathbb{I}}_{B} \otimes \hat{\boldsymbol{\alpha}}_{O} + (\widehat{\boldsymbol{\alpha}_{A}-\boldsymbol{\alpha}_{O}})_{A} \otimes \hat{\mathbb{I}}_{B} \otimes \hat{\mathbb{I}}_{O} \, , 
\nonumber\\
\hat{{\bf A}}_{B} &:=& \hat{\mathbb{I}}_{A} \otimes \hat{\mathbb{I}}_{B} \otimes \hat{{\bf a}}_{O} + \hat{\mathbb{I}}_{A} \otimes (\widehat{{\bf a}_{B}-{\bf a}_{O}})_{B} \otimes \hat{\mathbb{I}}_{O} \, , 
\nonumber\\
\hat{\mathlarger{\mathlarger{\boldsymbol{\alpha}}}}_{B} &:=& \hat{\mathbb{I}}_{A} \otimes \hat{\mathbb{I}}_{B} \otimes \hat{\boldsymbol{\alpha}}_{O} + \hat{\mathbb{I}}_{A} \otimes (\widehat{\boldsymbol{\alpha}_{B}-\boldsymbol{\alpha}_{O}})_{B} \otimes \hat{\mathbb{I}}_{O} \, .
\end{eqnarray}

Note that the definitions in Eqs. (\ref{relative_Hamiltonian_(A)OBC}), (\ref{relative_Hamiltonian_(B)AOC}) and (\ref{relative_Hamiltonian_(C)ABO}) suppose that the `observing' QRF is able to detect the presence of an interaction potential between itself and the `observed' subsystems. 
This is physically reasonable. 
For example, suppose that all three subsystems are charge-neutral and that the masses of the observed particles are small compared to the mass of the designated QRF. 
In this case, the gravitational interactions between the QRF and the two observed particles will dominate the dynamics of the tripartite system, while the interaction between the observed particles themselves may be considered negligible.
\footnote{The ability to account for physical interactions between the `observer' and the `observed' subsystems is, we believe, another advantage of our QRF model, over the relational models proposed by GCB and others \cite{Giacomini:2017zju,Vanrietvelde:2018pgb,Vanrietvelde:2018dit,Hohn:2018toe,Hohn:2018iwn,Krumm:2020fws,Ballesteros:2020lgl,QRF_Bell_Test:2021,Giacomini:2021gei,delaHamette:2021iwx,Cepollaro:2021ccc,Castro-Ruiz:2021vnq,AliAhmad:2021adn,Hoehn:2021flk,Carrozza:2021gju,delaHamette:2021oex,delaHamette:2021piz,Giacomini:2022hco,Overstreet:2022zgq,Kabel:2022cje,Apadula:2022pxk,Amelino-Camelia:2022dsj,Kabel:2023jve,Hoehn:2023ehz,Hoehn:2023axh,Wang:2023koz}. In these models, it is impossible to account for such interactions, since the QRF is not equipped with quantum degrees of freedom of its own. In our model, an interesting question then arises, namely, `what happens when these interactions generate entanglement, from the perspective of a classical observer, between the QRF and the observed subsystems?'. In this case, there is no clear distinction between the quantum-observer and the quantum-observed, even from the `objective' viewpoint defined by the classical background geometry. Throughout this work, we have considered only separable initial states, from the perspective of the CRF $O$, but we will consider this question in detail in a future analysis.} 
On the other hand, there is no physical mechanism by which the designated observer can measure the value of their own kinetic energy, relative to an arbitrary CRF that is defined by the classical background space. 
The kinetic terms in these Hamiltonians therefore contain only the relative momenta between physical subsystems. 

The three relative Hamiltonians generate three distinct {\it relative} time-evolutions, according to
\begin{eqnarray} \label{relative_Hamiltonian_(A)OBC_time-evolution}
&&\ket{\varPsi_{OBC}^{(A)}(t)}_{OBC} := \exp\left(-\frac{i}{\hbar}\hat{\mathcal{H}}_{OBC}^{(A)}t\right)\hat{\mathcal{P}}_{O \leftrightarrow A}\ket{\Psi_{ABC}^{(O)}(0)}_{ABC} 
\nonumber\\
&\neq& \ket{\Psi_{OBC}^{(A)}(t)}_{OBC} := \hat{\mathcal{P}}_{O \leftrightarrow A}\exp\left(-\frac{i}{\hbar}\hat{H}_{ABC}^{(O)}t\right)\ket{\Psi_{ABC}^{(O)}(0)}_{ABC} \, , 
\end{eqnarray}
\begin{eqnarray} \label{relative_Hamiltonian_(B)AOC_time-evolution}
&&\ket{\varPsi_{AOC}^{(B)}(t)}_{AOC} := \exp\left(-\frac{i}{\hbar}\hat{\mathcal{H}}_{AOC}^{(B)}t\right)\hat{\mathcal{P}}_{O \leftrightarrow B}\ket{\Psi_{ABC}^{(O)}(0)}_{ABC}
\nonumber\\
&\neq& \ket{\Psi_{AOC}^{(B)}(t)}_{AOC} := \hat{\mathcal{P}}_{O \leftrightarrow B}\exp\left(-\frac{i}{\hbar}\hat{H}_{ABC}^{(O)}t\right)\ket{\Psi_{ABC}^{(O)}(0)}_{ABC} \, , 
\end{eqnarray}
and
\begin{eqnarray} \label{relative_Hamiltonian_(C)ABO_time-evolution}
&&\ket{\varPsi_{ABO}^{(C)}(t)}_{ABO} := \exp\left(-\frac{i}{\hbar}\hat{\mathcal{H}}_{ABO}^{(C)}t\right)\hat{\mathcal{P}}_{O \leftrightarrow C}\ket{\Psi_{ABC}^{(O)}(0)}_{ABC}
\nonumber\\
&\neq& \ket{\Psi_{ABO}^{(C)}(t)}_{ABO} := \hat{\mathcal{P}}_{O \leftrightarrow C}\exp\left(-\frac{i}{\hbar}\hat{H}_{ABC}^{(O)}t\right)\ket{\Psi_{ABC}^{(O)}(0)}_{ABC} \, . 
\end{eqnarray} 
The expressions on the bottom lines of Eqs. (\ref{relative_Hamiltonian_(A)OBC_time-evolution}), (\ref{relative_Hamiltonian_(B)AOC_time-evolution}) and (\ref{relative_Hamiltonian_(C)ABO_time-evolution}) are equivalent to each other, and are each equivalent to the time-evolution given by Eq. (\ref{tripartite_Hamiltonian_O}), where it is understood that, when implementing the relevant parity swap, we must also implement the relevant coordinate transformation, given in Eqs. (\ref{CRF-O->QRF-A}), (\ref{CRF-O->QRF-B}) or (\ref{CRF-O->QRF-C}). 
Note, also, the subtle change in notation between the top and bottom lines, $\ket{\varPsi_{OBC}^{(A)}(t)}_{OBC} \neq \ket{\Psi_{OBC}^{(A)}(t)}_{OBC}$, etc., where the latter is defined as $\ket{\Psi_{OBC}^{(A)}(t)}_{OBC} := \hat{\mathcal{P}}_{O \leftrightarrow A}\ket{\Psi_{ABC}^{(O)}(t)}_{ABC}$. 
In other words, $ \ket{\Psi_{OBC}^{(A)}(t)}_{OBC}$ is equivalent to $\ket{\Psi_{ABC}^{(O)}(t)}_{ABC}$, since the two states differ by only a parity swap, plus the corresponding change of coordinates. 

By contrast, the three states denoted with the `varPsi' symbol, $\varPsi$, are {\it not} equivalent to $\ket{\Psi_{ABC}^{(O)}(t)}_{ABC}$. 
These represent the apparent time-evolutions of the tripartite state, as perceived by each of the three QRFs, Alice, Bob and Charlie. 
They encode only {\it relative} changes, which are physically observable (i.e., measurable), to the relevant designated observer. 
This description is clearly compatible with a Page-Wooters type scenario \cite{Page-Wooters:1983,Baumann:2019fbd,Hausmann:2023jpn,Marletto:2016gwv,Vedral:2022egu}, in which the wave function of the entire Universe exists in an energy eigenstate of the total, Universal Hamiltonian, but apparent time-evolution occurs between subsystems, due to their restricted access to information about the global state. 
The relation between the Page-Wooters model and the QRF model presented here is interesting, and worthy of further study, but a detailed analysis of this problem lies beyond the scope of the present work. 

\subsection{Single particles (wave functions) objectively exist} \label{Sec.4.2}

In our formalism, a collection of $N$ wave functions, $\left\{\psi_{I}\right\}_{I=0}^{N-1}$, can be associated with a system of $N$ quantum particles, whenever the state of the system is separable in some basis. 
Although, for entangled states, the notion of a single-particle wave function - that is, a single function of three quantum mechanical degrees of freedom - breaks down, the wave function of a general entangled state can still be expressed as a function of $3N$ degrees of freedom, $\Psi_{N}$. 
In terms of the linguistic shorthand introduced in Sec. \ref{Sec.2.2}, this is equivalent to `an $N$-wave function description of an $N$-particle state', and the fact that such a description is not only possible, but {\it necessary}, is a direct consequence of the objective existence of the classical background geometry \cite{Jones:1998,Stoica:2021owy}.

The key point is that, due to the objective existence of the classical spatial background, in which the wave functions of material particles are defined as complex-valued scalars \cite{Stoica:2021owy}, it is impossible for situations like the following to occur. 
Consider, for example, the three possible pairings of two particles out of a system of three, $(A, B)$, $(B, C)$ and $(C, A)$. 
In the GCB formalism \cite{Giacomini:2017zju}, the total state of the tripartite system is given by some functional, $F_{AB}[\psi\phi]$, from $C$'s perspective, by the functional $F_{BC}[\psi\phi]$, from $A$'s perspective, and by the functional $F_{CA}[\psi\phi]$, from $B$'s perspective. 
These therefore represent the {\it effective} wave functions of the pairs $(A, B)$, $(B, C)$ and $(C, A)$, respectively. 

Thus, even if there exists a basis in which one of these functionals is separable, e.g. $F_{AB}[\psi\phi] := \ket{\psi_{A}}_{A} \otimes \ket{\phi_{B}}_{B}$, we still cannot associate the function $\psi$ with the particle $A$, nor $\phi$ with the particle $B$, {\it objectively}, since `jumping' to a different QRF changes the labels on these functions, as well as the labels on the corresponding basis vectors. 
In our view, this leads to a paradoxical situation, in which the same two complex-valued functions of three variables describe, say, a proton and a neutron from an electron's perspective, a proton and an electron from a neutron's perspective, and an electron and a neutron from a proton's perspective. 

By contrast, in our formalism, it is always possible to prepare the initial state of such a composite system in the form $\ket{\Psi_{pne}}_{pne} = \ket{\psi_{p}}_{p} \otimes \ket{\psi_{n}}_{n} \otimes \ket{\psi_{e}}_{e}$, in some basis. 
Furthermore, although subsequent operations may generate entanglement between the subsystems, the labels associated with the individual functions, $(\psi_{p1},\psi_{p2},\dots)$, $(\psi_{n1},\psi_{n2},\dots)$ and $(\psi_{e1},\psi_{e2},\dots)$, which combine to make the total wave function of entangled state, are physically meaningful, since these {\it do not change} under CRF-to-QRF transitions, or under the transition from one QRF to another. 
One simply traces out the degrees of freedom associated with a particular subspace of the total Hilbert space, $\mathcal{H}_{p}$, $\mathcal{H}_{n}$ or $\mathcal{H}_{e}$, which is associated with the designated `observer', according to the procedure outlined in Sec. \ref{Sec.3}. 

\subsection{No ambiguities in transformations between QRFs} \label{Sec.4.3}

As explained in Sec. \ref{Sec.2.3}, there exists an ambiguity in the definition of the final state, when `jumping' from one QRF to another, in the GCB formalism \cite{Giacomini:2017zju}. 
There is, therefore, an inherent ambiguity in the definition of all QRF-to-QRF transitions, since the final state depends on which basis we choose to express the initial state in. 

In our formalism, however, there is only an {\it apparent} contradiction. 
In a bipartite system, in which the composite state of the two subsystems, $A$ and $B$, is both pure and separable from the perspective of a given CRF, $O$, Alice will perceive Bob's state as mixed. 
However, the mixed state she `sees' depends on the measurements she uses to probe Bob's state, with, for example, position and momentum measurements yielding different density matrices. 
But this is not a genuine ambiguity. 
In the limit $M_A \rightarrow \infty$, the two mixed states reduce to the same `objective' pure state, seen by $O$, and the measurement-sensitivity that Alice requires, in order to observe a difference between the two, is exactly the sensitivity she requires to resolve the non-Heisenberg terms in the GURs that characterise measurements of the relational variables. 
Thus, by resolving these terms, Alice may demonstrate that she, too, is quantum mechanical in nature, like Bob. 

In this sense, what appears at first sight to be a bug in our model, is in fact a feature, since it suggests experimental protocols by which competing definitions of the term `quantum reference frame' can be distinguished, empirically. 
Furthermore, we are able to describe both the initial and final states in a given QRF-to-QRF transition, as well as the corresponding position and momentum measurements, in whatever basis we choose, without ambiguity or contradiction. 
In short, our results are measurement-dependent, in an operational sense, but not basis-dependent, mathematically, like those presented in \cite{Giacomini:2017zju}. 
Our contention is that basis-dependent predictions, for the outcomes of physical measurements, do not make sense.

\subsection{Angular momentum and the spin degrees of freedom} \label{Sec.4.4} 

In this section, we again consider the bipartite quantum state, studied in Sec. \ref{Sec.3.2}, as the simplest nontrivial example of a QRF model. 
Our aim is to show that QRF-to-QRF transitions between states with different relative angular momenta, and different relative spins, can be rigorously and unambiguously treated within our proposed formalism. 
Our conclusions then generalise, naturally, to higher-order systems. 

We begin by notating that, were Alice and Bob classical systems, their angular momenta, relative to the coordinate origin of the CRF $O$, would be ${\bf L}_{A}^{(O)} = {\bf r}_{A} \times \boldsymbol{\kappa}_{A}$ and ${\bf L}_{B}^{(O)} = {\bf r}_{B} \times \boldsymbol{\kappa}_{B}$, respectively. The corresponding quantum operators are $\hat{{\bf L}}_{A}^{(O)} := (\hat{{\bf r}}_{A} \otimes \hat{\mathbb{I}}_{B}) \times (\hat{\boldsymbol{\kappa}}_{A} \otimes \hat{\mathbb{I}}_{B})$ and $\hat{{\bf L}}_{B}^{(O)} := (\hat{\mathbb{I}}_{A} \otimes \hat{{\bf r}}_{B}) \times (\hat{\mathbb{I}}_{A} \otimes \hat{\boldsymbol{\kappa}}_{B})$, where 
$\hat{\boldsymbol{\kappa}}_{A(B)}$ generate real space translations of Alice's (Bob's) state, relative to $O$, and $\hat{{\bf r}}_{A(B)}$ generate the corresponding translations in momentum space. 
The operators $\hat{{\bf L}}_{A}^{(O)}$ and $\hat{{\bf L}}_{B}^{(O)}$ then generate infinitesimal rotations of $A$ and $B$, about $O$, respectively.

The question therefore arises, what is the angular momentum of Alice, relative to Bob? 
For a classical system, this would clearly be ${\bf \mathfrak{L}}_{B}^{(A)} \equiv {\bf \mathfrak{L}}_{B} := {\bf x}_{B} \times {\bf p}_{B}$, in our chosen notation, where ${\bf x}_{B}$ and ${\bf p}_{B}$ denote the relative displacement and momentum of the two states \cite{Vallado:2001}. 
The corresponding quantum operator is
\begin{eqnarray} \label{L_B_bipartite}
\hat{{\bf \mathfrak{L}}}_{B} := \hat{{\bf X}}_{B} \times \hat{{\bf P}}_{B} \, , 
\end{eqnarray}
where the components of $\hat{{\bf X}}_{B}$ and $\hat{{\bf P}}_{B}$ are defined in Eqs. (\ref{X_{B}^{i}_bipartite}) and (\ref{P_{Bj}_bipartite}). 
Furthermore, since the definitions of $\hat{{\bf X}}_{B}$ and $\hat{{\bf P}}_{B}$ include the parity swap operation, $\hat{\mathcal{P}}_{O \leftrightarrow A}$, the eigenvalues of $\hat{{\bf \mathfrak{L}}}_{B}$ represent `the angular momentum of Bob', as perceived by Alice. 

To confirm that the definition (\ref{L_B_bipartite}) makes sense, we now show that $\hat{{\bf X}}_{B}$ and $\hat{{\bf P}}_{B}$ generate infinitesimal translations of Bob's state, relative to Alice's subsytem, in the relevant dual space. 
Using these results, it is straightforward to show that $\hat{{\bf \mathfrak{L}}}_{B}$ generates rotations of $B$'s state, relative to $A$, in both the position and momentum space representations. 
Thus, we have 
\begin{eqnarray} \label{P_B_bipartite_action_on_basis}
&{}&\exp\left[\frac{i}{2\hbar} \, \hat{{\bf P}}_{B} \, . \, {\bf a} \right] \, \ket{{\bf x}_{O}}_{O} \otimes \ket{{\bf x}_{B}-{\bf x}_{O}}_{B} 
\nonumber\\
&=& \exp\left[\frac{i}{2\hbar} \, \hat{{\bf \mathcal{P}}}_{B} \, . \, {\bf a} \right] \, \ket{{\bf x}_{O}}_{O} \otimes  \exp\left[\frac{i}{2\hbar} \, \hat{{\bf \mathcal{P}}}'_{B} \, . \, {\bf a} \right] \, \ket{{\bf x}_{B}-{\bf x}_{O}}_{B} 
\nonumber\\
&=& \ket{{\bf x}_{O}+{\bf a}/2}_{O} \otimes \ket{{\bf x}_{B}-{\bf x}_{O}+{\bf a}/2}_{B} \, , 
\end{eqnarray}
\begin{eqnarray} \label{X_B_bipartite_action_on_basis}
&{}&\exp\left[-\frac{i}{2\hbar} \, \boldsymbol{\alpha} \, . \, \hat{{\bf X}}_{B} \right] \, \ket{{\bf p}_{O}}_{O} \otimes \ket{{\bf p}_{B}-{\bf p}_{O}}_{B} 
\nonumber\\
&=& \exp\left[-\frac{i}{2\hbar} \, \boldsymbol{\alpha} \, . \, \hat{{\bf \mathcal{X}}}_{B} \right] \, \ket{{\bf p}_{O}}_{O} \otimes  \exp\left[-\frac{i}{2\hbar} \, \boldsymbol{\alpha} \, . \, \hat{{\bf \mathcal{X}}}'_{B} \right] \, \ket{{\bf p}_{B}-{\bf p}_{O}}_{B} 
\nonumber\\
&=&  \ket{{\bf p}_{O}+\boldsymbol{\alpha}/2}_{O} \otimes \ket{{\bf p}_{B}-{\bf p}_{O}+\boldsymbol{\alpha}/2}_{B} \, ,
\end{eqnarray}
which generates the transformations
\begin{eqnarray} \label{bipartite_X-translations}
{\bf x}_{O} \mapsto {\bf x}'_{O} = {\bf x}_{O} + {\bf a}/2  \, , \quad {\bf x}_{B} - {\bf x}_{O} \mapsto {\bf x}'_{B} - {\bf x}'_{O}  = ({\bf x}_{B} - {\bf x}_{O}) + {\bf a}/2  \, ,
\nonumber\\
{\bf x}_{B} \mapsto {\bf x}'_{B} = {\bf x}_{B} + {\bf a} \, , 
\end{eqnarray}
\begin{eqnarray} \label{bipartite_P-translations}
{\bf p}_{O} \mapsto {\bf p}'_{O} = {\bf p}_{O} + \boldsymbol{\alpha}/2  \, , \quad {\bf p}_{B} - {\bf p}_{O} \mapsto {\bf p}'_{B} - {\bf p}'_{O}  = ({\bf p}_{B} - {\bf p}_{O}) + \boldsymbol{\alpha}/2  \, ,
\nonumber\\
{\bf p}_{B} \mapsto {\bf p}'_{B} = {\bf p}_{B} + \boldsymbol{\alpha} \, , 
\end{eqnarray}
or, equivalently, 
\begin{eqnarray} \label{bipartite_X-translations*}
-{\bf r}_{A} \mapsto -{\bf r}'_{A} = -{\bf r}_{A} + {\bf a}/2 \, , \quad {\bf r}_{B} \mapsto {\bf r}'_{B} = {\bf r}_{B} + {\bf a}/2 \, , 
\nonumber\\
{\bf r}_{B} - {\bf r}_{A} \mapsto {\bf r}'_{B} - {\bf r}'_{A} = {\bf r}_{B} - {\bf r}_{A} + {\bf a} \, , 
\end{eqnarray}
\begin{eqnarray} \label{bipartite_P-translations*}
-\boldsymbol{\kappa}_{A} \mapsto -\boldsymbol{\kappa}'_{A} = -\boldsymbol{\kappa}_{A} + {\bf a}/2 \, , \quad \boldsymbol{\kappa}_{B} \mapsto \boldsymbol{\kappa}'_{B} 
= \boldsymbol{\kappa}_{B} + {\bf a}/2 \, , 
\nonumber\\
\boldsymbol{\kappa}_{B} - \boldsymbol{\kappa}_{A} \mapsto \boldsymbol{\kappa}'_{B} - \boldsymbol{\kappa}'_{A} = \boldsymbol{\kappa}_{B} - \boldsymbol{\kappa}_{A} + {\bf a} \, .
\end{eqnarray}
(See Fig. 8.)

\begin{figure}[h] \label{Fig.8}
\centering 
	\includegraphics[width=7cm]{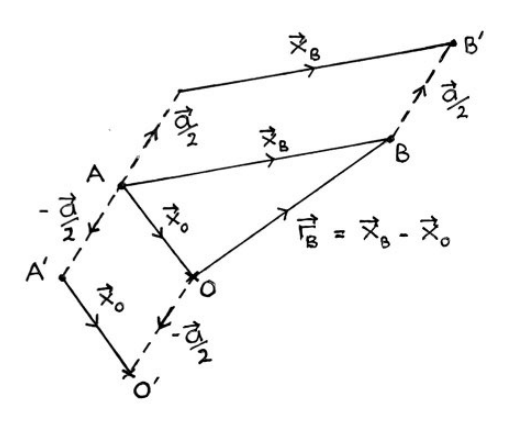}
\caption{The operator $\exp\left[\frac{i}{2\hbar} \, \hat{{\bf P}}_{B} \, . \, {\bf a} \right]$ translates ${\bf r}_{B}$ by ${\bf a}/2$ and $-{\bf r}_{A}$ by ${\bf a}/2$. The latter is equivalent to shifting ${\bf r}_{A}$ by 
$-{\bf a}/2$, so that the net result is a shift of $B$'s state, relative to $A$, by ${\bf a}$. The analogous diagram in momentum space corresponds to the action of the operator $\exp\left[-\frac{i}{2\hbar} \, \boldsymbol{\alpha} \, . \, \hat{{\bf X}}_{B} \right]$.}
\end{figure}

Note that, in Eqs. (\ref{P_B_bipartite_action_on_basis}) and (\ref{X_B_bipartite_action_on_basis}), the factor of $(2\hbar)^{-1}$ in the exponent is included due to the rescaled Heisenberg algebra, 
\begin{eqnarray} \label{bipartite_Heisenberg-1}
[\hat{X}_{B}^{i},\hat{P}_{Bj}] = 2i\hbar \, \delta^{i}{}_{j} \hat{\mathbb{I}}_{OB} \, , 
\end{eqnarray}
\begin{eqnarray} \label{bipartite_Heisenberg-2}
[\hat{X}_{B}^{i},\hat{X}_{B}^{j}] = 0 \, , 
\end{eqnarray}
\begin{eqnarray} \label{bipartite_Heisenberg-3}
[\hat{P}_{Bi},\hat{P}_{Bj}] = 0 \, , 
\end{eqnarray}
which, in turn, is generated by the algebra of the subcomponents $\left\{\hat{\mathcal{X}}_{B}^{i},\hat{\mathcal{X}}_{B}'^{i},\hat{\mathcal{P}}_{Bj},\hat{\mathcal{P}}'_{Bj}\right\}$,
\begin{eqnarray} \label{bipartite_subcomponents-1}
[\hat{\mathcal{X}}_{B}^{i},\hat{\mathcal{P}}_{Bj}] = i\hbar \delta^{i}{}_{j} \hat{\mathbb{I}}_{OB} \, , \quad [\hat{\mathcal{X}}_{B}'^{i},\hat{\mathcal{P}}'_{Bj}] = i\hbar \delta^{i}{}_{j} \hat{\mathbb{I}}_{OB} \, , 
\end{eqnarray}
\begin{eqnarray} \label{bipartite_subcomponents-2}
[\hat{\mathcal{X}}_{B}^{i},\hat{\mathcal{P}}'_{Bj}] = [\hat{\mathcal{X}}_{B}^{i},\hat{\mathcal{P}}'_{Bj}] = 0 \, , 
\end{eqnarray}
\begin{eqnarray} \label{bipartite_subcomponents-3}
[\hat{\mathcal{X}}_{B}^{i},\hat{\mathcal{X}}_{B}^{j}] = [\hat{\mathcal{X}}_{B}'^{i},\hat{\mathcal{X}}_{B}'^{j}] = 0 \, , 
\end{eqnarray}
\begin{eqnarray} \label{bipartite_subcomponents-4}
[\hat{\mathcal{X}}_{B}^{i},\hat{\mathcal{X}}_{B}'^{j}] = [\hat{\mathcal{X}}_{B}'^{i},\hat{\mathcal{X}}_{B}^{j}] = 0 \, , 
\end{eqnarray}
\begin{eqnarray} \label{bipartite_subcomponents-5}
[\hat{\mathcal{P}}_{Bi},\hat{\mathcal{P}}_{Bj}] = [\hat{\mathcal{P}}'_{Bi},\hat{\mathcal{P}}'_{Bj}] = 0 \, , 
\end{eqnarray}
\begin{eqnarray} \label{bipartite_subcomponents-6}
[\hat{\mathcal{P}}_{Bi},\hat{\mathcal{P}}'_{Bj}] = [\hat{\mathcal{P}}'_{Bi},\hat{\mathcal{P}}_{Bj}] = 0 \, . 
\end{eqnarray}

Equations (\ref{bipartite_subcomponents-1})-(\ref{bipartite_subcomponents-6}) define two commuting copies of the canonical Heisenberg algebra, the first of which acts nontrivially only on the subspace $\mathcal{H}_{O}^{(A)} := \hat{\mathcal{P}}_{A \rightarrow O}\mathcal{H}_{A} \equiv \mathcal{H}_{A}$ of the total Hilbert space of the bipartite system, while the second acts nontrivially only on the $\mathcal{H}_{B}$ subspace. 
However, as we will now show, the resulting algebra for the subcomponents of the relative angular momentum operators, $\left\{\hat{\mathfrak{L}}_{Bi}\right\}_{i=1}^{3}$, is not so simple. 
First, we define these via the relations
\begin{eqnarray} \label{bipartite_L_Bi}
\hat{\mathfrak{L}}_{Bi} := \epsilon_{ij}{}^{k}\hat{X}_{B}^{j}\hat{P}_{Bk} := \hat{\mathcal{L}}_{Bi} + \hat{\mathcal{L}}'_{Bi} + \hat{\mathbb{L}}_{Bi} \, , 
\end{eqnarray}
where 
\begin{eqnarray} \label{bipartite_L_B_subcomponents}
\hat{\mathcal{L}}_{Bi} := \epsilon_{ij}{}^{k}\hat{\mathcal{X}}_{B}^{j}\hat{\mathcal{P}}_{Bk} \, , \quad \hat{\mathcal{L}}'_{Bi} := \epsilon_{ij}{}^{k}\hat{\mathcal{X}}_{B}'^{j}\hat{\mathcal{P}}'_{Bk} \, , 
\quad
\hat{\mathbb{L}}_{Bi} := \epsilon_{ij}{}^{k}(\hat{\mathcal{X}}_{B}^{j}\hat{\mathcal{P}}'_{Bk} + \hat{\mathcal{X}}_{B}'^{j}\hat{\mathcal{P}}_{Bk}) \, .
\end{eqnarray}
Following our previous convention, the unprimed subcomponents $\hat{\mathcal{L}}_{Bi}$ act nontrivially only on the $\mathcal{H}_{O}^{(A)} := \hat{\mathcal{P}}_{A \rightarrow O}\mathcal{H}_{A} \equiv \mathcal{H}_{A}$ subspace, whereas the primed subcomponents $\hat{\mathcal{L}}'_{Bi}$ act nontrivially only on $\mathcal{H}_{B}$. 
In addition, there exists a third subcomponent, $\hat{\mathbb{L}}_{Bi}$, which acts nontrivially on both subspaces, and the resulting algebra for the subcomponents $\left\{\hat{\mathcal{L}}_{Bi},\hat{\mathcal{L}}'_{Bi},\hat{\mathbb{L}}_{Bi}\right\}$ is
\begin{eqnarray} \label{bipartite_L_Bi_subcomponents-1}
[\hat{\mathcal{L}}_{Bi},\hat{\mathcal{L}}_{Bj}] = i\hbar\epsilon_{ij}{}^{k} \, \hat{\mathcal{L}}_{Bk} \, , \quad [\hat{\mathcal{L}}'_{Bi},\hat{\mathcal{L}}'_{Bj}] = i\hbar\epsilon_{ij}{}^{k} \, \hat{\mathcal{L}}'_{Bk} \, , 
\end{eqnarray}
\begin{eqnarray} \label{bipartite_L_Bi_subcomponents-2}
[\hat{\mathcal{L}}_{Bi},\hat{\mathcal{L}}'_{Bj}] = [\hat{\mathcal{L}}'_{Bi},\hat{\mathcal{L}}_{Bj}] = 0 \, , 
\end{eqnarray}
\begin{eqnarray} \label{bipartite_L_Bi_subcomponents-3}
[\hat{\mathcal{L}}_{Bi},\hat{\mathbb{L}}_{Bj}] - [\hat{\mathcal{L}}_{Bj},\hat{\mathbb{L}}_{Bi}] = i\hbar\epsilon_{ij}{}^{k} \, \hat{\mathbb{L}}_{Bk} \, , 
\end{eqnarray}
\begin{eqnarray} \label{bipartite_L_Bi_subcomponents-4}
[\hat{\mathcal{L}}'_{Bi},\hat{\mathbb{L}}_{Bj}] - [\hat{\mathcal{L}}'_{Bj},\hat{\mathbb{L}}_{Bi}] = i\hbar\epsilon_{ij}{}^{k} \, \hat{\mathbb{L}}_{Bk} \, , 
\end{eqnarray}
\begin{eqnarray} \label{bipartite_L_Bi_subcomponents-5}
[\hat{\mathbb{L}}_{Bi},\hat{\mathbb{L}}_{Bj}] = i\hbar\epsilon_{ij}{}^{k} \, \hat{\mathcal{L}}_{Bk} + i\hbar\epsilon_{ij}{}^{l} \, \hat{\mathcal{L}}'_{Bl} \, .
\end{eqnarray}
Together, Eqs. (\ref{bipartite_L_Bi_subcomponents-1})-(\ref{bipartite_L_Bi_subcomponents-5}) yield
\begin{eqnarray} \label{bipartite_[L_Bi,L_Bj]}
[\hat{\mathfrak{L}}_{Bi},\hat{\mathfrak{L}}_{Bj}] = 2i\hbar \epsilon_{ij}{}^{k}\, \hat{\mathfrak{L}}_{Bk} \, ,
\end{eqnarray}
and, hence,
\begin{eqnarray} \label{bipartite_[L_B^2,L_Bj]}
[\hat{\mathfrak{L}}_{B}^2,\hat{\mathfrak{L}}_{Bi}] = 0 \, .
\end{eqnarray}
The corresponding uncertainty relations take the form
\begin{eqnarray} \label{bipartite_L_Bi_GUR-1}
(\Delta_{\Psi}\mathfrak{L}_{Bi})^2 \, (\Delta_{\Psi}\mathfrak{L}_{Bj})^2 \geq \dots \geq \hbar^2(\epsilon_{ij}{}^{k})^2 \, \langle\hat{\mathfrak{L}}_{Bk}\rangle_{\Psi}^2 \, ,  
\end{eqnarray}
where the absolute limit on the far right-hand side follows from the Schr{\" o}dinger-Robertson bound \cite{Rae:2002,Isham:1995}, in conjunction with (\ref{bipartite_[L_Bi,L_Bj]}), and the dots in the middle represent a sum of terms that are generically greater than or equal to this. 
There are $7^2 = 49$ such terms, due to the generalised uncertainties
\begin{eqnarray} \label{bipartite_L_Bi_GUR-2}
(\Delta_{\Psi}\mathfrak{L}_{Bi})^2 &=& (\Delta_{\Psi}\mathcal{L}_{Bi})^2 + (\Delta_{\Psi}\mathcal{L}'_{Bi})^2 + (\Delta_{\Psi}\mathbb{L}_{Bi})^2 
\nonumber\\
&+& {\rm cov}(\hat{\mathcal{L}}_{Bi},\hat{\mathbb{L}}_{Bi}) +  {\rm cov}(\hat{\mathbb{L}}_{Bi},\hat{\mathcal{L}}_{Bi})
\nonumber\\
&+& {\rm cov}(\hat{\mathcal{L}}'_{Bi},\hat{\mathbb{L}}_{Bi}) +  {\rm cov}(\hat{\mathbb{L}}_{Bi},\hat{\mathcal{L}}'_{Bi}) \, ,
\end{eqnarray}
where the subscript $\Psi$ again refers to the composite state of the bipartite system, $\ket{\Psi_{OB}^{(A)}}_{OB}$ (\ref{Psi_OB(A)}), and the covariance of two random variables is defined as ${\rm cov}(X,Y) := \langle XY \rangle - \langle X \rangle \langle Y \rangle$. 

Finally, we are able to demonstrate that the operator $\exp\left[\frac{i}{2\hbar} \, \theta \hat{\underline{n}} \, . \, \hat{{\bf \mathfrak{L}}}_{B}\right]$ generates rotations of Bob's state, by an angle $\theta$, about an arbitrary unit vector $\hat{\underline{n}}$ centred on the origin of Alice's `coordinate system'. 
We recall, also, that the latter is really a superposition of coordinate systems, all of which have the same relative orientation, as shown in Sec. \ref{Sec.3}. 
It is therefore sufficient to show that the operator $\exp\left[\frac{i}{2\hbar} \, \delta\theta_i \hat{\mathfrak{L}}_{Bi}\right]$ generates infinitesimal rotations of $B$'s state, {\it relative to each possible coordinate system in $A$'s superposition}, for all $i \in \left\{1,2,3\right\}$ $\equiv \left\{x,y,z\right\}$. 
Thus, we have
\begin{eqnarray} \label{bipartite_infinitesimal_rotations-1}
&{}&\exp\left[\frac{i}{2\hbar} \, \delta\theta_z \hat{{\bf \mathfrak{L}}}_{Bz}\right] \, \ket{{\bf x}_{O}}_{O} \otimes \ket{{\bf x}_{B}-{\bf x}_{O}}_{B} 
\nonumber\\
&=& \exp\left[\frac{i}{2\hbar} \, \delta\theta_z (\hat{{\bf \mathcal{L}}}_{Bz}+\hat{{\bf \mathcal{L}}}'_{Bz})\right] \, \exp\left[\frac{i}{2\hbar} \, \delta\theta_z \hat{{\bf \mathbb{L}}}_{Bz}\right] \, 
\ket{{\bf x}_{O}}_{O} \otimes \ket{{\bf x}_{B}-{\bf x}_{O}}_{B} 
\nonumber\\
&=& \exp\left[\frac{i}{2\hbar} \, \delta\theta_z \left(\hat{\mathcal{X}}_{B}\hat{\mathcal{P}}_{By} - \hat{\mathcal{Y}}_{B}\hat{\mathcal{P}}_{Bx} + \hat{\mathcal{X}}'_{B}\hat{\mathcal{P}}'_{By} - \hat{\mathcal{Y}}'_{B}\hat{\mathcal{P}}'_{Bx}\right)\right] 
\nonumber\\
&\times& \exp\left[\frac{i}{2\hbar} \, \delta\theta_z \left(\hat{\mathcal{X}}_{B}\hat{\mathcal{P}}'_{By} - \hat{\mathcal{Y}}_{B}\hat{\mathcal{P}}'_{Bx} + \hat{\mathcal{X}}'_{B}\hat{\mathcal{P}}_{By} - \hat{\mathcal{Y}}'_{B}\hat{\mathcal{P}}_{Bx}\right)\right]
\nonumber\\
&\times& \ket{x_{O}, \, y_{O}, \, z_{O}}_{O} \otimes \ket{x_{B}-x_{O}, \, y_{B}-y_{O}, \, z_{B}-z_{O}}_{B} 
\nonumber\\
&=& \exp\left[\frac{i}{2\hbar} \, \delta\theta_z \left(\hat{\mathcal{X}}_{B}\hat{\mathcal{P}}_{By} - \hat{\mathcal{Y}}_{B}\hat{\mathcal{P}}_{Bx} + \hat{\mathcal{X}}'_{B}\hat{\mathcal{P}}'_{By} - \hat{\mathcal{Y}}'_{B}\hat{\mathcal{P}}'_{Bx}\right)\right]
\nonumber\\
&\times& \bigg| x_{O} - \frac{\delta\theta_z}{2}(y_{B}-y_{O}), \, y_{O} + \frac{\delta\theta_z}{2}(x_{B}-x_{O}), \, x_{O} \bigg\rangle_{O} 
\nonumber\\
&\otimes&\bigg|(x_{B}-x_{O}) - \frac{\delta\theta_z}{2}y_{O}, \, (y_{B}-y_{O}) + \frac{\delta\theta_z}{2}x_{O}, \, z_{B}-z_{O}\bigg\rangle_{B}
\nonumber\\
&=& \bigg| x_{O} - \frac{\delta\theta_z}{2}y_{B}, \, y_{O} + \frac{\delta\theta_z}{2}x_{B}, \, z_{O} \bigg\rangle_{O} 
\nonumber\\
&\otimes&\bigg| (x_{B}-x_{O}) - \frac{\delta\theta_z}{2}y_{B}, \, (y_{B}-y_{O}) + \frac{\delta\theta_z}{2}x_{B}, \, z_{B}-z_{O}\bigg\rangle_{B} \, , 
\end{eqnarray}
plus an analogous expression in momentum space, to first order in $\delta\theta_z \ll 1$. 
Note that, in this derivation, we have used the fact that, although $[\hat{{\bf \mathcal{L}}}_{Bi},\hat{{\bf \mathbb{L}}}_{Bi}] \neq 0$ and $[\hat{{\bf \mathcal{L}}}'_{Bi},\hat{{\bf \mathbb{L}}}_{Bi}] \neq 0$, the relevant commutators for the application of the Baker-Campbell-Hausdorff formula \cite{Baker-Campbell-Hausdorff} are $[\delta\theta_z\hat{{\bf \mathcal{L}}}_{Bi},\delta\theta_z\hat{{\bf \mathbb{L}}}_{Bi}]$ and 
$[\delta\theta_z\hat{{\bf \mathcal{L}}}'_{Bi},\delta\theta_z\hat{{\bf \mathbb{L}}}_{Bi}]$. 
These are of order $(\delta\theta_z)^2$ and can therefore be neglected.
Equation (\ref{bipartite_infinitesimal_rotations-1}), plus its momentum space counterpart, yield the transformations
\begin{eqnarray} \label{bipartite_infinitesimal_rotations-2}
&&x_{O} \mapsto x'_{O} = x_{O} - \frac{\delta\theta_z}{2}y_{B} \, , \quad p_{Ox} \mapsto p'_{Ox} = p_{Ox} - \frac{\delta\theta_z}{2}p_{By} \, , 
\nonumber\\
&&y_{O} \mapsto y'_{O} = y_{O} + \frac{\delta\theta_z}{2}x_{B} \, , \quad p_{Oy} \mapsto p'_{Oy} = p_{Oy} + \frac{\delta\theta_z}{2}p_{Bx} \, , 
\nonumber\\
&&z_{O} \mapsto z'_{O} = z_{O} \, , \quad \quad \quad \quad \quad \, p_{Oz} \mapsto p'_{Oz} = p_{Oz} \, , 
\end{eqnarray}
and
\begin{eqnarray} \label{bipartite_infinitesimal_rotations-3}
&&x_{B}-x_{O} \mapsto x'_{B}-x'_{O} = x_{B}-x_{O} - \frac{\delta\theta_z}{2}y_{B} \, , 
\nonumber\\
&&y_{B}-y_{O} \mapsto y'_{B}-y'_{O} = y_{B}-y_{O} + \frac{\delta\theta_z}{2}x_{B} \, , 
\nonumber\\
&&z_{B}-z_{O} \mapsto z'_{B}-z'_{O} = z_{B}-z_{O} \, ;
\nonumber\\
&&p_{xB}-p_{xO} \mapsto p'_{Bx}-p'_{Ox} = p_{Bx}-p_{Ox} - \frac{\delta\theta_z}{2}p_{By} \, , 
\nonumber\\
&&p_{yB}-p_{yO} \mapsto p'_{By}-p'_{Oy} = p_{By}-p_{Oy} + \frac{\delta\theta_z}{2}p_{Bx} \, , 
\nonumber\\
&&p_{zB}-p_{zO} \mapsto p'_{Bz}-p'_{Oz} = p_{Bz}-p_{Oz} \, ,
\end{eqnarray}
giving
\begin{eqnarray} \label{bipartite_infinitesimal_rotations-4}
&&x_{B} \mapsto x'_{B} = x_{B} - \delta\theta_z \, y_{B} \, , \quad p_{Bx} \mapsto p'_{Bx} = p_{Bx} - \delta\theta_z \, p_{By} \, , 
\nonumber\\
&&y_{B} \mapsto y'_{B} = y_{B} + \delta\theta_z \, x_{B} \, , \quad p_{By} \mapsto p'_{By} = p_{By} + \delta\theta_z \, p_{By} \, , 
\nonumber\\
&&z_{B} \mapsto z'_{B} = z_{B} \, , \quad \quad \quad \quad \quad \, p_{Bz} \mapsto p'_{Bz} = p_{Bz} \, , 
\end{eqnarray}
and it is straightforward to show that similar expressions also hold for $\exp\left[\frac{i}{2\hbar} \, \delta\theta_y \hat{\mathfrak{L}}_{By}\right]$ and $\exp\left[\frac{i}{2\hbar} \, \delta\theta_x \hat{\mathfrak{L}}_{Bx}\right]$. 
This validates our interpretation of $\hat{{\bf \mathfrak{L}}}_{B}$ as the operator representing Bob's angular momentum, as seen by Alice, and the generator of `rotations' of his state, relative to hers, in both the position and momentum space representations. 
\footnote{Note that, by simply {\it interpreting} the coordinates ${\bf x}_{B}$ and ${\bf p}_{B}$ as relative coordinates, but leaving their mathematical forms, and the state spaces on which they act, unchanged from the canonical theory, the GCB formalism is unable to capture behaviour of this kind. Here, we have shown that, when both Alice and Bob exist in a superposition of position  (momentum) eigenstates, a `rotation of $A$ relative to $B$' really represents infinitely many rotations, of each possible set of axes in a superposition of coordinate systems, $\psi_{A}$ ($\tilde{\psi}_{A}$), with respect to each possible set of axes in second superposition of coordinate systems, $\phi_{B}$ ($\tilde{\phi}_{B}$). Physically, this situation is different to that in which $B$'s superposition of states is rotated around a fixed point in classical space, corresponding to the centre of classical-$A$'s fixed coordinate system. (See Fig. 9, for illustrations of rotation, with respect to both CRFs and QRFs.) If we wish to describe this using standard position, momentum and angular momentum operators, with their canonical mathematical forms unchanged, we must first construct the appropriate space of {\it effective} mixed states, as outlined in Sec. \ref{Sec.3}.}

Next, we consider the spin degrees of freedom within the QRF paradigm. 
In canonical QM, the Pauli matrices were originally discovered by imposing two conditions on the would-be spin operators; (a), that they be independent of the canonical phase space variables, position and momentum, as befits an `internal' property of a quantum state, and (b) that they satisfy the same algebra as the operators for orbital angular momentum \cite{Rae:2002}. 
Following the same logic here, we are inspired to search for a set of generalised spin operators, $\left\{\hat{S}_{Bi}\right\}_{i=1}^{3}$, which can each be split into the sum of three terms,
\begin{eqnarray} \label{bipartite_spin_ops-1}
\hat{S}_{Bi} := \hat{\mathcal{S}}_{Bi} + \hat{\mathcal{S}}'_{Bi} + \hat{\mathbb{S}}_{Bi} \, . 
\end{eqnarray}

The first term, $\hat{\mathcal{S}}_{Bi}$, should act nontrivially only on the Hilbert space corresponding to Alice's spin, after the parity swap $O \leftrightarrow A$, whereas the second term, $\hat{\mathcal{S}}'_{Bi}$, should act nontrivially only on Bob's spin space, giving
\begin{eqnarray} \label{bipartite_spin_ops-2}
\hat{\mathcal{S}}_{Bi} := \hat{s}_{Oi} \otimes \hat{\mathbb{I}}_{B} \, , \quad \hat{\mathcal{S}}'_{Bi} :=  \hat{\mathbb{I}}_{O} \otimes \hat{s}_{Bi} \, , 
\end{eqnarray}
where
\begin{eqnarray} \label{canonical_spin_ops_definition}
\hat{s}_{i} = \frac{\hbar}{2} \, \sigma_{i} \, , 
\end{eqnarray}
and $\left\{\sigma_{i}\right\}_{i=1}^{3}$ are the canonical Pauli matrices. 
These satisfy the ${\rm su}(2)$ Lie algebra, giving
\begin{eqnarray} \label{canonical_spin_ops_Lie_algebra}
[\sigma_{i},\sigma_{j}] = 2i \epsilon_{ij}{}^{k}\sigma_{k} \, , \quad [\hat{s}_{i},\hat{s}_{j}] = i\hbar \epsilon_{ij}{}^{k}\hat{s}_{k} \, ,
\end{eqnarray}
together with
\begin{eqnarray} \label{canonical_spin_ops_Casimir_operator-1}
[\sigma^2,\sigma_{i}] = 0 \, , \quad [\hat{s}^{2},\hat{s}_{i}] = 0 \, ,
\end{eqnarray}
where
\begin{eqnarray} \label{canonical_spin_ops_Casimir_operator-2}
\hat{s}^{2} = \sum_{i=1}^{3} \hat{s}_{i}^{2} = \left(\frac{\hbar}{2}\right)^2 \sum_{i=1}^{3} \sigma_{i}^{2} = \frac{\hbar^2}{4}\sigma^2 = \frac{3\hbar^2}{4} \, \hat{\mathbb{I}} 
\end{eqnarray}
is the Casimir operator, and we have used the fact that $\sigma_{i}^{2}=\hat{\mathbb{I}} $ for all $i$ \cite{Rae:2002,Jones:1998}. 
In the fundamental spin-$1/2$ representation the Pauli matrices also obey the Clifford algebra, 
\begin{eqnarray} \label{canonical_spin_ops_Clifford_algebra}
[\sigma_{i},\sigma_{j}]_{+} = 2\delta_{ij} \hat{\mathbb{I}} \, , \quad [\hat{s}_{i},\hat{s}_{j}]_{+} = \frac{\hbar^2}{2}\delta_{ij} \hat{\mathbb{I}} \, , 
\end{eqnarray}
where $[A,B]_{+} = AB + BA$ denotes the anti-commutator, and combining Eqs. (\ref{canonical_spin_ops_Lie_algebra}) and (\ref{canonical_spin_ops_Clifford_algebra}) gives
\begin{eqnarray} \label{canonical_spin_ops_Lie+Clifford_algebra}
\sigma_{i}\sigma_{j} = \delta_{ij} \hat{\mathbb{I}} + i \epsilon_{ij}{}^{k}\sigma_{k} \, , \quad \hat{s}_{i}\hat{s}_{j} = \left(\frac{\hbar}{2}\right)^2\delta_{ij} \hat{\mathbb{I}} + i\left(\frac{\hbar}{2}\right)\epsilon_{ij}{}^{k}\hat{s}_{k} \, .
\end{eqnarray}

Let us also recall that the canonical spin operators generate generalised rotations via the formula
\begin{eqnarray} \label{canonical_spin_ops_rotation}
\exp\left[\frac{i}{\hbar}\theta \, \hat{\underline{n}} \, . \, \hat{{\bf s}}\right] = \exp\left[\frac{i\theta}{2} \, \hat{\underline{n}} \, . \, \boldsymbol{\sigma}\right]
= \cos(\theta/2) \, \hat{\mathbb{I}} + i\sin(\theta/2) \, \hat{\underline{n}} \, . \, \boldsymbol{\sigma} \, , 
\end{eqnarray}
where we have used the fact that $(\hat{\underline{n}} \, . \, \boldsymbol{\sigma})^{2m} = \hat{\mathbb{I}}$, for $m \in \mathbb{Z}^{+}$ \cite{Chuang_Nielsen:2000}. 
Physically, the operator $\exp\left[\frac{i}{\hbar}\theta \, \hat{\underline{n}} \, . \, \hat{{\bf s}}\right]$ generates rotations of the spin-measurement axes, by the angle $\theta$, about the unit vector $\hat{\underline{n}}$, which is centred on the coordinate origin of $O$. 
The operators representing the total angular momentum, including both the orbital angular momentum and spin angular momentum components, may be constructed as
\begin{eqnarray} \label{canonical_J}
\hat{J}_{i} = \hat{L}_{i} + \hat{s}_{i} \, ,
\end{eqnarray}
and the operator $\exp\left[\frac{i}{\hbar}\theta \, \hat{\underline{n}} \, . \, \hat{{\bf J}}\right] = \exp\left[\frac{i}{\hbar}\theta \, \hat{\underline{n}} \, . \, (\hat{{\bf L}}+\hat{{\bf s}})\right]$ generates rotations of the axes along which both spin and orbital angular momentum are measured. 
These two sets of measurements therefore remain aligned. 

In our QRF formalism, the third subcomponent of the generalised spin operator (\ref{bipartite_spin_ops-1}), $\hat{\mathbb{S}}_{Bi}$, must therefore be constructed so that is satisfies the relevant terms in the algebra
\begin{eqnarray} \label{bipartite_S_Bi_subcomponents-1}
[\hat{\mathcal{S}}_{Bi},\hat{\mathcal{S}}_{Bj}] = i\hbar\epsilon_{ij}{}^{k} \, \hat{\mathcal{S}}_{Bk} \, , \quad [\hat{\mathcal{S}}'_{Bi},\hat{\mathcal{S}}'_{Bj}] = i\hbar\epsilon_{ij}{}^{k} \, \hat{\mathcal{S}}'_{Bk} \, , 
\end{eqnarray}
\begin{eqnarray} \label{bipartite_S_Bi_subcomponents-2}
[\hat{\mathcal{S}}_{Bi},\hat{\mathcal{S}}'_{Bj}] = [\hat{\mathcal{S}}'_{Bi},\hat{\mathcal{S}}_{Bj}] = 0 \, , 
\end{eqnarray}
\begin{eqnarray} \label{bipartite_S_Bi_subcomponents-3}
[\hat{\mathcal{S}}_{Bi},\hat{\mathbb{S}}_{Bj}] - [\hat{\mathcal{S}}_{Bj},\hat{\mathbb{S}}_{Bi}] = i\hbar\epsilon_{ij}{}^{k} \, \hat{\mathbb{S}}_{Bk} \, , 
\end{eqnarray}
\begin{eqnarray} \label{bipartite_S_Bi_subcomponents-4}
[\hat{\mathcal{S}}'_{Bi},\hat{\mathbb{S}}_{Bj}] - [\hat{\mathcal{S}}'_{Bj},\hat{\mathbb{S}}_{Bi}] = i\hbar\epsilon_{ij}{}^{k} \, \hat{\mathbb{S}}_{Bk} \, , 
\end{eqnarray}
\begin{eqnarray} \label{bipartite_S_Bi_subcomponents-5}
[\hat{\mathbb{S}}_{Bi},\hat{\mathbb{S}}_{Bj}] = i\hbar\epsilon_{ij}{}^{k} \, \hat{\mathcal{S}}_{Bk} + i\hbar\epsilon_{ij}{}^{l} \, \hat{\mathcal{S}}'_{Bl} \, , 
\end{eqnarray}
by analogy with Eqs. (\ref{bipartite_L_Bi_subcomponents-1})-(\ref{bipartite_L_Bi_subcomponents-5}). 
Assuming that Alice and Bob are both spin-$1/2$ particles, these requirements yield the unique definition 
\begin{eqnarray} \label{bipartite_spin_ops-3}
\hat{\mathbb{S}}_{Bi} := \frac{2}{\hbar}\epsilon_{i}{}^{jk} \, \hat{\mathcal{S}}_{Bj} \, \hat{\mathcal{S}}'_{Bk} := \frac{\hbar}{2}\epsilon_{i}{}^{jk} \, \sigma_{Oj} \otimes \sigma_{Bk} \, , 
\end{eqnarray}
where we have made use of both the Lie algebra and Clifford algebra properties of the fundamental representation, (\ref{canonical_spin_ops_Lie_algebra}) and (\ref{canonical_spin_ops_Clifford_algebra}), which also imply
\footnote{By making this assumption, we verify that our formalism holds for all fundamental particles that are {\it not} force-transmitting bosons, including all known leptons and quarks \cite{Peskin&Schroeder:1995,QFT_Nutshell}. It is clear that all such particles can, potentially, be accorded the status of  an `observer' within the QRF paradigm, but the status of force-transmitting bosons, which mediate interactions between them, is less clear. In a future study, we will analyse composite systems including spin-$0$, spin-$1/2$, spin-$1$, spin-$3/2$ and spin-$2$ particles, and will consider whether measures of the {\it relative} spin between these subsystems can be reasonably defined. However, such an analysis lies beyond the scope of the present work.}
\begin{eqnarray} \label{bipartite_spin_Clifford-1}
[\hat{\mathcal{S}}_{Bi},\hat{\mathcal{S}}_{Bj}]_{+} = \frac{\hbar^2}{2}\delta_{ij} \, \hat{\mathbb{I}}_{OB} \, , \quad [\hat{\mathcal{S}}'_{Bi},\hat{\mathcal{S}}'_{Bj}]_{+} = \frac{\hbar^2}{2}\delta_{ij} \, \hat{\mathbb{I}}_{OB} \, ,
\end{eqnarray}
\begin{eqnarray} \label{bipartite_spin_Clifford-2}
[\hat{\mathcal{S}}_{Bi},\hat{\mathbb{S}}_{Bj}]_{+} + [\hat{\mathcal{S}}_{Bj},\hat{\mathbb{S}}_{Bi}]_{+} = 0 \, , 
\end{eqnarray}
\begin{eqnarray} \label{bipartite_spin_Clifford-3}
[\hat{\mathcal{S}}'_{Bi},\hat{\mathbb{S}}_{Bj}]_{+} + [\hat{\mathcal{S}}'_{Bj},\hat{\mathbb{S}}_{Bi}]_{+} = 0 \, , 
\end{eqnarray}
\begin{eqnarray} \label{bipartite_spin_Clifford-4}
[\hat{\mathbb{S}}_{Bi},\hat{\mathbb{S}}_{Bj}]_{+} = \hbar^2\delta_{ij} \, \hat{\mathbb{I}}_{OB} - [\hat{\mathcal{S}}_{Bi},\hat{\mathcal{S}}'_{Bj}]_{+} - [\hat{\mathcal{S}}_{Bj},\hat{\mathcal{S}}'_{Bi}]_{+} \, . 
\end{eqnarray}

Next, we define the generalised Pauli matrices via the relation
\begin{eqnarray} \label{generalised_Pauli_matrices}
\hat{S}_{Bi} := \hbar \, \Sigma_{Bi} \, .
\end{eqnarray}
It follows immediately that these satisfy canonical-type Lie and Clifford algebras, yielding
\begin{eqnarray} \label{bipartite_[S_Bi,S_Bj]}
[\Sigma_{Bi},\Sigma_{Bj}] = 2i\epsilon_{ij}{}^{k} \, \Sigma_{Bk} \, , \quad [\hat{S}_{Bi},\hat{S}_{Bj}] = 2i\hbar\epsilon_{ij}{}^{k} \, \hat{S}_{Bk} 
\end{eqnarray}
and
\begin{eqnarray} \label{generalised_spin_ops_Clifford_algebra}
[\Sigma_{Bi},\Sigma_{Bj}]_{+} = 2\delta_{ij} \, \hat{\mathbb{I}}_{OB} \, , \quad [\hat{S}_{Bi},\hat{S}_{Bj}]_{+} = 2\hbar^2\delta_{ij} \, \hat{\mathbb{I}}_{OB} \, , 
\end{eqnarray}
so that 
\begin{eqnarray} \label{generalised_spin_ops_Lie+Clifford_algebra}
\Sigma_{Bi}\Sigma_{Bj} = \delta_{ij} \, \hat{\mathbb{I}}_{OB} + i\epsilon_{ij}{}^{k} \, \Sigma_{Bk} \, , \quad 
\hat{S}_{Bi}\hat{S}_{Bj} = \hbar^2\delta_{ij} \, \hat{\mathbb{I}}_{OB} + i\hbar\epsilon_{ij}{}^{k} \, \hat{S}_{Bk} \, . 
\end{eqnarray}
We then have
\begin{eqnarray} \label{generalised_spin_ops_Casimir_operator-1}
[\Sigma_{B}^{2},\Sigma_{Bi}] = 0 \, , \quad [\hat{S}_{B}^{2},\hat{S}_{Bi}] = 0 \, , 
\end{eqnarray}
where
\begin{eqnarray} \label{generalised_spin_ops_Casimir_operator-2}
\hat{S}_{B}^{2} = \sum_{i=1}^{3} \hat{S}_{Bi}^{2} = \hbar^2 \sum_{i=1}^{3} \Sigma_{Bi}^{2} = \hbar^2\Sigma_{B}^{2} = 3\hbar^2 \, \hat{\mathbb{I}}_{OB} \, ,
\end{eqnarray}
and where we have used the fact that $\Sigma_{Bi}^{2} = \hat{\mathbb{I}}_{OB}$, for all $i$.

Exponentiation of the generalised spin operators yields
\begin{eqnarray} \label{generalised_spin_ops_rotation}
\exp\left[\frac{i}{2\hbar}\theta \, \hat{\underline{n}} \, . \, \hat{{\bf S}}_{B}\right] = \exp\left[\frac{i\theta}{2} \, \hat{\underline{n}} \, . \, \boldsymbol{\Sigma}_{B}\right] 
= \cos(\theta/2) \, \hat{\mathbb{I}}_{OB} + i\sin(\theta/2) \, \hat{\underline{n}} \, . \, \boldsymbol{\Sigma} \, , 
\end{eqnarray}
where $(\hat{\underline{n}} \, . \, \boldsymbol{\Sigma}_{B})^{2m} = \hat{\mathbb{I}}_{OB}$ for $m \in \mathbb{Z}^{+}$. 
Note that, in Eqs. (\ref{generalised_spin_ops_rotation}), the factor of $(2\hbar)^{-1}$ arises due to the rescaled Lie algebra, given in Eqs. (\ref{bipartite_[S_Bi,S_Bj]}). 
Its inclusion in the exponential is consistent with the definition of the generalised Pauli matrices, Eqs. (\ref{generalised_Pauli_matrices}), and ensures that a rotation by $720^{{\rm o}}$ is needed to bring a generalised spin vector - that is, an arbitrary superposition of the eigenstates of $\hat{S}_{Bi}$ - back to its original state. 
The generalised uncertainty relations for the operators $\left\{\hat{S}_{Bi}\right\}_{i=1}^{3}$ are
\begin{eqnarray} \label{bipartite_S_Bi_GUR-1}
(\Delta_{\Psi}S_{Bi})^2 \, (\Delta_{\Psi}S_{Bj})^2 \geq \dots \geq \hbar^2(\epsilon_{ij}{}^{k})^2 \, \langle\hat{S}_{Bk}\rangle_{\Psi}^2 \, ,  
\end{eqnarray}
where
\begin{eqnarray} \label{bipartite_S_Bi_GUR-2}
(\Delta_{\Psi}S_{Bi})^2 &=& (\Delta_{\Psi}\mathcal{S}_{Bi})^2 + (\Delta_{\Psi}\mathcal{S}'_{Bi})^2 + (\Delta_{\Psi}\mathbb{S}_{Bi})^2 
\nonumber\\
&+& {\rm cov}(\hat{\mathcal{S}}_{Bi},\hat{\mathbb{S}}_{Bi}) +  {\rm cov}(\hat{\mathbb{S}}_{Bi},\hat{\mathcal{S}}_{Bi})
\nonumber\\
&+& {\rm cov}(\hat{\mathcal{S}}'_{Bi},\hat{\mathbb{S}}_{Bi}) +  {\rm cov}(\hat{\mathbb{S}}_{Bi},\hat{\mathcal{S}}'_{Bi}) \, ,
\end{eqnarray}
by complete analogy with Eqs. (\ref{bipartite_L_Bi_GUR-1}) and (\ref{bipartite_L_Bi_GUR-2}).

It is straightforward to find a unitary similarity transform such that 
\begin{eqnarray} \label{unitary_similarity_transform}
\hat{S}_{Bi} \mapsto \hat{U} \, \hat{S}_{Bi} \, \hat{U}^{-1} = \hbar \, (\sigma_{Oi} \oplus \sigma_{Bi}) \, , 
\end{eqnarray}
for all $i$, and its existence follows directly from the Lie group $-$ Lie algebra correspondence, which holds for all compact groups \cite{Hall:2015}. 
The generalised Pauli matrices are, therefore, reducible representations of the ${\rm SU}(2)$ group generators, equivalent to the direct sum of two fundamental representations, $\left\{\hat{U}\Sigma_{Bi}\hat{U}^{-1} = \sigma_{Oi} \oplus \sigma_{Bi}\right\}_{i=1}^{3}$. 
However, it would be incorrect to think that each of these acts, as a generator of ${\rm SU}(2)$, on an individual subspace of the bipartite system, $\mathcal{H}_{O}^{(A)}$ or $\mathcal{H}_{B}$. 
Instead, each irreducible representation acts in a distributed way, across both subspaces, so that the $\Sigma_{Bi}$ do {\it not} generate rotations of the spin-measurement axes used by the CRF $O$, relative to either Alice or Bob. 
The latter are generated by $\hat{\mathcal{S}}_{Bi}$ and $\hat{\mathcal{S}}'_{Bi}$, respectively, but the action of the third subcomponent, $\hat{\mathbb{S}}_{Bi}$, ensures that the total action of the operators $\hat{S}_{Bi} = \hbar\Sigma_{Bi}$ is subtler, and more complex. 

Nonetheless, the canonical-type expression for the operator $\exp\left[\frac{i\theta}{2} \, \hat{\underline{n}} \, . \, \boldsymbol{\Sigma}_{B}\right]$ (\ref{bipartite_S_Bi_GUR-1}) suggests that $\hat{{\bf S}}_{B} = \hbar\boldsymbol{\Sigma}_{B}$ does, in fact, rotate some set of spin-measurement axes, by an angle $\theta$, about the vector $\hat{\underline{n}}$. 
But what axes are these?
In fact, the generalised spin operators (\ref{bipartite_spin_ops-1}) generate rotations of each and every set of spin-measurement axes, in Alice's {\it superposition} of spin-measurement axes, relative to each and every set of spin-measurement axes in Bob's superposition. 
(See Figs. 9 and 10.)

We therefore interpret the eigenvalues of $\left\{\hat{S}_{Bi}\right\}_{i=1}^{3}$ as measures of the relative spin between the subsystems $A$ and $B$. 
Equivalently, these are possible the values of Bob's spin, measured by Alice, relative to her {\it superposition} of spin-measurement axes. 
These are coincident with the superposition of axes that Alice uses to determine the components of Bob's relative displacement, ${\bf x}_{B}$, and relative linear momentum, ${\bf p}_{B}$, as well as his relative angular momentum, $\bold{\mathfrak{L}}_{B}$. 
This scenario is illustrated in Fig. 10 and may be compared with the standard notion of rotation in canonical QM, depicted in Fig. 9. 
We stress, again, that the two scenarios are physically distinct and require different mathematical descriptions, which standard relational formalisms fail to capture \cite{Giacomini:2017zju,Vanrietvelde:2018pgb,Vanrietvelde:2018dit,Hohn:2018toe,Hohn:2018iwn,Krumm:2020fws,Ballesteros:2020lgl,QRF_Bell_Test:2021,Giacomini:2021gei,delaHamette:2021iwx,Cepollaro:2021ccc,Castro-Ruiz:2021vnq,AliAhmad:2021adn,Hoehn:2021flk,Carrozza:2021gju,delaHamette:2021oex,delaHamette:2021piz,Giacomini:2022hco,Overstreet:2022zgq,Kabel:2022cje,Apadula:2022pxk,Amelino-Camelia:2022dsj,Kabel:2023jve,Hoehn:2023ehz,Hoehn:2023axh,Wang:2023koz,Muller_Group_Website}. 
\footnote{See \cite{PhysRevA.70.032321,Hohn:2014ude,PhysRevA.95.022336,Giacomini:2018gxh,QRF_Qubits_NoGo:2021,Mikusch:2021kro,Baumann:2021urf,Piotrak:2023fsy}, for extensions of the GCB formalism to include QRFs with spin, as well as alternative approaches to this problem. To the best of our knowledge, no existing works arrive at the same conclusions reached here, namely, that the operator representing the relative angular momentum between two quantum subsystems should be defined as in Eqs. (\ref{bipartite_L_Bi})-(\ref{bipartite_L_B_subcomponents}), and that the operator representing the relative spin between these systems can be defined by requiring the algebraic consistency of the latter with the former.}

For the sake of compactness, we do not quote the explicit forms of the relative-spin operators here, but it is straightforward to show that each possesses 4 distinct eigenvectors, corresponding to 2 distinct eigenvalues. 
Each eigenvalue, $+\hbar$ and $-\hbar$, therefore possesses a 2-fold degeneracy. 
This is an important point: were some eigenvectors associated with different eigenvalues, Bob's subsystem would no longer appear, to Alice, as a spin-1/2 particle. 
For example, if at least some eigenvectors were associated with the eigenvalue zero, Bob's state would appear, instead, as a spin-1 particle. 
Although the eigenvalues $\pm \hbar$ are not exactly equal to the canonical values, $\pm \hbar/2$, there are still two of them, indicating binary spin `up' and spin `down' states.

To avoid confusion with the canonical spin vectors $\left\{\ket{\uparrow_{i}}_{O},\ket{\downarrow_{i}}_{O}\right\}$ and $\left\{\ket{\uparrow_{i}}_{B},\ket{\downarrow_{i}}_{B}\right\}$, we may denote the generalised eigenstates as $\left\{\ket{\uparrow_{i}}\rangle_{OB},\ket{\downarrow_{i}}\rangle_{OB};\ket{\uparrow'_{i}}\rangle_{OB},\ket{\downarrow'_{i}}\rangle_{OB}\right\}$.
In general, these correspond to entangled states of the canonical spins. 
The interested reader is referred to \cite{Lake:2019nmn,Lake:2020rwc,Lake:2021beh}, for further details. 

The above results then allow us to define the components of the total relative angular momentum operator, $\hat{{\bf \mathfrak{J}}}_{B}$, as 
\begin{eqnarray} \label{relative_J}
\hat{\mathfrak{J}}_{Bi} := \hat{\mathfrak{L}}_{Bi} + \hat{S}_{Bi} \, ,
\end{eqnarray}
or, equivalently, 
\begin{eqnarray} \label{relative_J_subcomponents-1}
\hat{\mathfrak{J}}_{Bi} := \hat{\mathcal{J}}_{Bi} + \hat{\mathcal{J}}'_{Bi} + \hat{\mathbb{J}}_{Bi} \, ,
\end{eqnarray}
where  
\begin{eqnarray} \label{relative_J_subcomponents-1}
\hat{\mathcal{J}}_{Bi} := \hat{\mathcal{L}}_{Bi} + \hat{\mathcal{S}}_{Bi} \, , \quad 
\hat{\mathcal{J}}'_{Bi} :=  \hat{\mathcal{L}}'_{Bi} + \hat{\mathcal{S}}'_{Bi} \, , \quad
\hat{\mathbb{J}}_{Bi} := \hat{\mathbb{L}}_{Bi}  + \hat{\mathbb{S}}_{Bi} \, .
\end{eqnarray}
The algebra of the subcomponents $\left\{\hat{\mathcal{J}}_{Bi},\hat{\mathcal{J}}'_{Bi},\hat{\mathbb{J}}_{Bi}\right\}$ is completely analogous to Eqs. (\ref{bipartite_L_Bi_subcomponents-1})-(\ref{bipartite_L_Bi_subcomponents-5}) and (\ref{bipartite_S_Bi_subcomponents-1})-(\ref{bipartite_S_Bi_subcomponents-5}), yielding the GURs
\begin{eqnarray} \label{bipartite_J_Bi_GUR-1}
(\Delta_{\Psi}\mathfrak{J}_{Bi})^2 \, (\Delta_{\Psi}\mathfrak{J}_{Bj})^2 \geq \dots \geq \hbar^2(\epsilon_{ij}{}^{k})^2 \, \langle\hat{\mathfrak{J}}_{Bk}\rangle_{\Psi}^2 \, ,  
\end{eqnarray}
where
\begin{eqnarray} \label{bipartite_J_Bi_GUR-2}
(\Delta_{\Psi}\mathfrak{J}_{Bi})^2 &=& (\Delta_{\Psi}\mathcal{J}_{Bi})^2 + (\Delta_{\Psi}\mathcal{J}'_{Bi})^2 + (\Delta_{\Psi}\mathbb{J}_{Bi})^2 
\nonumber\\
&+& {\rm cov}(\hat{\mathcal{J}}_{Bi},\hat{\mathbb{J}}_{Bi}) +  {\rm cov}(\hat{\mathbb{J}}_{Bi},\hat{\mathcal{J}}_{Bi})
\nonumber\\
&+& {\rm cov}(\hat{\mathcal{J}}'_{Bi},\hat{\mathbb{J}}_{Bi}) +  {\rm cov}(\hat{\mathbb{J}}_{Bi},\hat{\mathcal{J}}'_{Bi}) \, ,
\end{eqnarray}
as expected. 
The operator $\exp\left[\frac{i}{2\hbar}\theta \, \hat{\underline{n}} \, . \, \hat{{\bf \mathfrak{J}}}_{B}\right] = \exp\left[\frac{i}{2\hbar}\theta \, \hat{\underline{n}} \, . \, (\hat{{\bf \mathfrak{L}}}_{B}+\hat{{\bf S}}_{B})\right]$ rotates every possible set of axes in Alice's superposition of coordinate systems, relative to Bob's superposition of coordinate systems. 
In another example of (somewhat inaccurate) linguistic shorthand, we may say that it `rotates Bob's wave function, relative to Alice's wave function'. 
The more precise meaning of this operation is depicted, schematically, in Fig. 10. 
\begin{figure}[h]
        \centering
        \includegraphics[width=14cm]{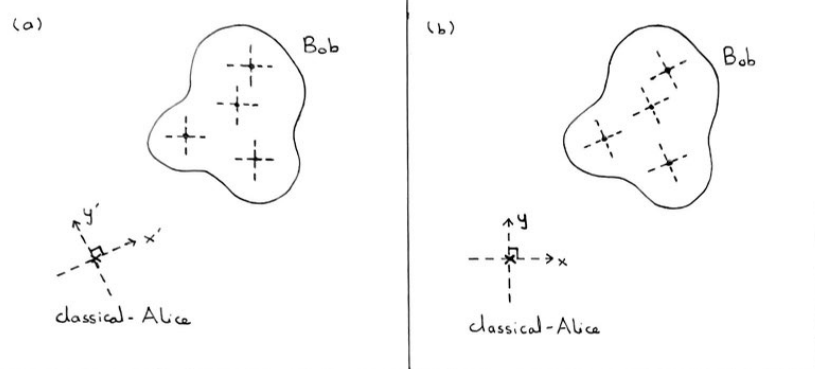}
        \caption{Exponentiating the total angular momentum operator of canonical QM, $\exp\left[i\theta \hat{\underline{n}/\hbar} \, . \, \hat{{\bf J}}\right]$, where $\hat{{\bf J}} = \hat{{\bf L}} + \hat{{\bf s}}$ and $\hat{{\bf L}}$, $\hat{{\bf s}}$ are the canonical orbital angular momentum and spin angular momentum operators, respectively, generates rotations of Bob's quantum state about the fixed origin of a classical reference frame. For the purposes of this diagram, we will call this CRF `classical Alice', instead of the usual `$O$'. This transformation may be viewed in two equivalent ways: (i) We may say that $\exp\left[i\theta \hat{\underline{n}/\hbar} \, . \, \hat{{\bf J}}\right]$ rotates Alice's classical coordinate system, relative to every possible coordinate system in Bob's superposition. This is depicted in Fig. 9(a). (ii) We may say that $\exp\left[i\theta \hat{\underline{n}/\hbar} \, . \, \hat{{\bf J}}\right]$ rotates every possible coordinate system in Bob's superposition, relative to classical-Alice's fixed coordinate system. This is depicted in Fig. 9(b). }
\end{figure}
%
%
\begin{figure}[h] 
        \centering 
        \includegraphics[width=14cm]{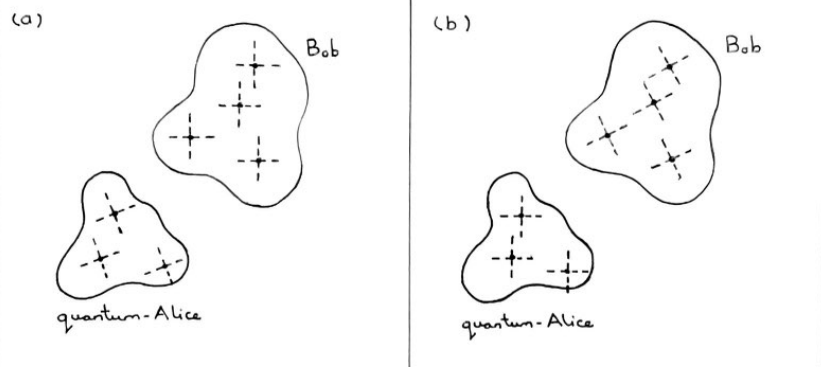}
\caption{In the QRF scenario, the situation is physically different to that presented in Fig. 9. Exponentiating the relative angular momentum operator, $\exp\left[i\theta \hat{\underline{n}/\hbar} \, . \, \hat{{\bf \mathfrak{J}}}_{B}\right]$, where $\hat{\mathfrak{J}}_{Bi} := \hat{\mathfrak{L}}_{Bi} + \hat{S}_{Bi}$ and $\hat{\mathfrak{L}}_{Bi}$, $\hat{S}_{Bi}$ denote the relative orbital angular momentum and relative spin operators, defined in Eqs. (\ref{bipartite_L_Bi})-(\ref{bipartite_L_B_subcomponents}) and (\ref{bipartite_spin_ops-1})-(\ref{bipartite_spin_ops-2}) plus (\ref{bipartite_spin_ops-3}), respectively, generates rotations of Bob's quantum state, about every possible coordinate system in a superposition of coordinates (a QRF). For the purposes of this diagram, we call this QRF `quantum Alice'. This transformation may also be viewed in two equivalent ways: (i) We may say that $\exp\left[i\theta \hat{\underline{n}/\hbar} \, . \, \hat{{\bf \mathfrak{J}}}_{B}\right]$ rotates every possible coordinate system in quantum-Alice's superposition of coordinates, relative to every possible coordinate system in Bob's superposition. This is depicted in Fig. 10(a). (ii) We may say that $\exp\left[i\theta \hat{\underline{n}/\hbar} \, . \, \hat{{\bf \mathfrak{J}}}_{B}\right]$ rotates every possible coordinate system in Bob's superposition of coordinates, relative to every possible coordinate system in quantum-Alice's superposition. This is depicted in Fig. 10(b). By simply interpreting the canonical operator $\hat{{\bf J}}_{B} := \hat{{\bf L}}_{B} + \hat{{\bf s}}_{B}$ as the relative angular momentum between two quantum subsystems, but leaving its mathematical form unchanged from the canonical theory, we believe that existing relational formalisms, including the GCB formalism \cite{Giacomini:2017zju}, fail to capture the important physical distinction between the two scenarios depicted in Figs. 9 and 10.}
\end{figure}

Finally, before concluding this subsection, we note that operators of the form (\ref{bipartite_L_Bi}) {\it must}, in fact, be included in any QRF model, including the GCB formalism and related models. 
However, in this case, their physical interpretation is problematic. 
For example, let us consider the tripartite state defined in \cite{Giacomini:2017zju}. 
Mathematically, this very much resembles our bipartite state, in the sense that it depends on only two wave functions, or, equivalently, $6$ quantum mechanical degrees of freedom. 
In this scenario, the orbital angular momentum of Bob's state, relative to Alice's, from Charlie's perspective, is evidently described by an operator with the same basic structure as (\ref{bipartite_L_Bi}), i.e., one that can be split into $3$ subcomponents obeying the algebra (\ref{bipartite_L_Bi_subcomponents-1})-(\ref{bipartite_L_Bi_subcomponents-5}). 
By contrast, both Bob's angular momentum and Alice's angular momentum, relative to Charlie's frame of reference, are described by canonical-type operators, $\hat{\bold{L}}_{B}^{(C)}$ and $\hat{\bold{L}}_{A}^{(C)}$, in this formalism. 
This is the case in our model only when $M_C \rightarrow \infty$, that is, in the limit where Charlie becomes, effectively, a classical observer. 
The mathematical distinction between these operators, in this limit, therefore corresponds to a real {\it physical} distinction between the observer and the observed. 
In the GCB model, however, there is no such distinction, as all three subsystems are supposed to be quantum mechanical in nature. 
It is, therefore, very difficult to explain the necessity of using operators with different mathematical structures, to describe apparently similar measurements, all of which, supposedly, represent measurements of {\it relative} variables.  

\section{The significance of QRFs for a future theory of quantum gravity} \label{Sec.5}

In this section, we show how the conceptual and mathematical machinery of the QRF formalism, developed in Secs. \ref{Sec.1}-\ref{Sec.4}, can be adapted and applied to the background geometry in which material quantum systems live. 
This is the only part of our work that goes beyond the basic postulates of canonical quantum theory, which are formulated for material particles inhabiting a classical three-dimensional Euclidean space, $\mathbb{E}^3$ \cite{Jones:1998,Giachetta:2011,Rae:2002,Isham:1995,Stoica:2021owy}. 

\subsection{Quantised spacetime as a QRF} \label{Sec.5.1}

Einstein's theory of general relativity equates the gravitational force with the curvature of the spacetime geometry \cite{Hobson:2006se}. 
It is therefore expected that the quantisation of the gravitational field should be equivalent to the quantisation of the spacetime background, in which material bodies live \cite{QG_Oriti:2009,QG_DeWitt:2011}.
In addition, dimensional analysis and a host of thought experiments \cite{Hossenfelder:2012jw,Garay:1994en}, plus a variety of specific approaches to the problem of quantum gravity, including string theory \cite{Kiritsis:2007zza} and loop quantum gravity \cite{LQG_Gambini_Pullin}, suggest that the Planck length, $l_{\rm Pl} \simeq \sqrt{\hbar G/c^3} \simeq 10^{-33}$ ${\rm cm}$, should act as a fundamental limit to the resolution of spacetime distances. 
Although the theoretical status of minimum momentum scale is less clear, it is noteworthy that, in a dark energy Universe, there exists an upper bound on the positional uncertainty of any particle, equal to radius of the de Sitter horizon, $l_{\rm dS} \simeq \sqrt{3/\Lambda} \simeq 10^{-28}$ ${\rm cm}$, where $\Lambda \simeq 10^{-56}$ ${\rm cm}^{-2}$ is the observed value of the cosmological constant \cite{Reiss1998,Perlmutter1999,Betoule:2014frx,Planck:2018vyg,Martin:2012bt,Spradlin:2001pw}. 
By the Heisenberg uncertainty principle (HUP), this immediately implies the existence of a minimum measurable momentum, of order $\sim \hbar\sqrt{\Lambda/3} \simeq 10^{-61}$ ${\rm kg \, ms^{-1}}$. 

In this section, we show that by promoting individual points in the classical spatial background, $\mathbb{E}^3$, to quantum superpositions of points, i.e., to non-material QRFs, we are able to re-derive some of the key phenomenological consequences of the existence of minimum length and / or momentum scales, in a new way. 
These include the most widely studied generalised uncertainty relations, considered in the phenomenological quantum gravity literature, namely, the generalised uncertainty principle (GUP) \cite{Maggiore:1993rv,Adler:1999bu,Scardigli:1999jh}, extended uncertainty principle (EUP) \cite{Bolen:2004sq,Park:2007az,Bambi:2007ty} and extended generalised uncertainty principle (EGUP) \cite{Kempf:1996ss,Hossenfelder:2012jw,Tawfik:2015rva,Tawfik:2014zca,Bosso:2023aht}.
The new derivations are then compared with standard derivations in the literature, based on modified commutation relations \cite{Kempf:1996ss,Hossenfelder:2012jw,Tawfik:2015rva,Tawfik:2014zca,Bosso:2023aht}, and several advantages of the new formalism are identified \cite{Lake:2018zeg,Lake:2019nmn,Lake:2020rwc,Lake:2023uoi}.
\footnote{The results presented in this section were, in fact, first derived in a series of previous papers \cite{Lake:2018zeg,Lake:2019nmn,Lake:2019oaz,Lake:2020rwc,Lake:2021beh,Lake:2020chb,Lake:2021gbu,Lake:2022hzr,Lake:2023lvh,Lake:2023uoi}. However, the analysis presented here, as well as in Sec. \ref{Sec.4.4}, goes beyond that provided in these works by clarifying the explicit connection between the QRF formalism developed in Secs. \ref{Sec.2}-\ref{Sec.4} and the GURs commonly studied in the quantum gravity literature. That such a connection may exist was hinted at, and discussed previously \cite{Lake:2018zeg,Lake:2019nmn,Lake:2020rwc}, but the exact correspondence between these models had not been clarified, until now.}

For later convenience, we define the Planck mass and length scales, and the de Sitter mass and length scales, as
\begin{eqnarray} \label{Planck_scales}
m_{\rm Pl} := \sqrt{\frac{\hbar c}{G}} \simeq 10^{-5} \, {\rm g} \, , \quad l_{\rm Pl} := \sqrt{\frac{\hbar G}{c^3}} \simeq 10^{-33} \, {\rm cm} \, , 
\end{eqnarray}
and
\begin{eqnarray} \label{dS_scales}
m_{\rm dS} := \frac{\hbar}{c}\sqrt{\frac{\Lambda}{3}} \simeq 10^{-66} \, {\rm g} \, , \quad l_{\rm dS} := \sqrt{\frac{3}{\Lambda}} \simeq 10^{28} \, {\rm cm} \, ,
\end{eqnarray}
respectively. 
The corresponding densities are
\begin{eqnarray} \label{rho_Lambda}
\rho_{\rm Pl} := \frac{3}{4\pi} \frac{c^5}{\hbar G^2} \simeq 10^{93} \, {\rm g \, . \, cm}^{-3} \, ,
\end{eqnarray} 
and
\begin{eqnarray} \label{rho_Lambda}
\rho_{\Lambda} := \frac{\Lambda c^2}{8\pi G} \simeq 10^{-29} \, {\rm g \, . \, cm}^{-3} \, ,
\end{eqnarray}   
where $\rho_{\Lambda}$ is approximately equal to the critical density of the present-day Universe, $\rho_{\Lambda} \simeq \rho_{\rm crit}(t_0)$ \cite{Hobson:2006se}. 

  
\subsection{GUP and EUP} \label{Sec.5.2}

When searching for phenomenological consequences of a quantised gravitational field, with or without a cosmological constant term, it is useful to note that the HUP can be `derived' in two different ways. 
The first `derivation' is heuristic, and is based on the so-called Heisenberg microscope thought experiment, in which an experimenter attempts to localise the position of a particle, by means of a scattering event, which inevitably disturbs its momentum \cite{Rae:2002}. 
The resulting trade-off between the position and momentum `uncertainties', $\Delta x$ and $\Delta p$, result in the relation
\begin{eqnarray} \label{HUP-1}
\Delta x^{i} \, \Delta p_j \gtrsim \frac{\hbar}{2} \delta^{i}{}_{j} \, .
\end{eqnarray}  
The inequality in Eqs. (\ref{HUP-1}) is imprecise because the associated `uncertainties' are not mathematically well defined. 
Alternatively, a more rigorous version of the HUP can be derived from abstract Hilbert space structure of canonical QM. 
By combining the Schr{\"o}dinger-Robertson inequality, 
\begin{eqnarray} \label{Robertson-Schrodinger-1}
\Delta_\psi O_1 \, \Delta_\psi O_2 \geq \frac{1}{2} |\braket{\psi | [\hat{O}_1,\hat{O}_2] | \psi}|  \, ,
\end{eqnarray}  
which holds for any two Hermitian operators, $\hat{O}_1$ and $\hat{O}_2$ \cite{Isham:1995}, with the canonical commutation relation, 
\begin{eqnarray} \label{[x,p]-1}
[\hat{x}^i,\hat{p}_j] = i\hbar\delta^{i}{}_{j} \ \hat{\mathbb{I}} \, ,
\end{eqnarray}
we obtain 
\begin{eqnarray} \label{HUP-2}
\Delta_\psi x^i \, \Delta_\psi p_j  \geq \frac{\hbar}{2} \delta^{i}{}_{j} \, .
\end{eqnarray}
The inequality in Eqs. (\ref{HUP-2}) is precise and the uncertainties $\Delta_\psi x^i$ and $\Delta_\psi p_j$ are well defined, as the standard deviations of the probability distributions $|\psi({\bf x})|^2$ and $|\tilde{\psi}({\bf p})|^2$, respectively. 
From here on, we use an appropriate subscript, attached to the symbol $\Delta$, to distinguish well-defined standard deviations from heuristic `uncertainties', introduced via thought experiments. 

Despite being heuristic, the thought experiment leading to Eqs. (\ref{HUP-1}) has its uses. 
For example, one can reanalyse this experimental setup, accounting also for the gravitational interaction between the observed particle and the probe particle. 
This leads to an additional non-Heisenberg term in the positional uncertainty, of order $\sim G\Delta p/c^3$ \cite{Maggiore:1993rv,Adler:1999bu,Scardigli:1999jh}, and, hence, to the GUP, 
\begin{eqnarray} \label{GUP-1}
\Delta x^i \gtrsim \frac{\hbar}{2\Delta p_j} \delta^{i}{}_{j} \left[1 + \alpha_0 \frac{2G}{\hbar c^3}(\Delta p_j)^2\right] \, ,
\end{eqnarray} 
where $\alpha_0$ is a numerical constant of order unity. 
The GUP implies the existence of a minimum resolvable position uncertainty, of the order of the Planck length, but no minimum momentum scale. 

Similarly, one can reanalyse the thought experiment, accounting for the presence of a background dark energy density (\ref{rho_Lambda}).
This yields an additional non-Heisenberg term in the momentum uncertainty, of order $\sim \hbar\sqrt{\Lambda/3}\Delta x$ \cite{Bolen:2004sq,Park:2007az,Bambi:2007ty}, giving rise to the EUP, 
\begin{eqnarray} \label{EUP-1}
\Delta p_j \gtrsim \frac{\hbar}{2\Delta x^i} \delta^{i}{}_{j} \left[1 + 2\eta_0 \Lambda (\Delta x^i)^2\right] \, , 
\end{eqnarray} 
where $\eta_0$ is of order unity. 
The EUP implies the existence of a minimum resolvable momentum uncertainty, of the order of the de Sitter momentum $\sim m_{\rm dS}c$, but no minimum length scale. 
Accounting for both canonical gravitational attraction, and the effects of repulsive dark energy, then leads to the EGUP, 
\begin{eqnarray} \label{EGUP-1}
\Delta x^i\Delta p_j \gtrsim \frac{\hbar}{2} \delta^{i}{}_{j} \left[1 + \alpha_0 \frac{2G}{\hbar c^3}(\Delta p_j)^2 + 2\eta_0\Lambda (\Delta x^i)^2\right] \, ,
\end{eqnarray} 
which implies the existence of both minimum length and momentum scales in nature. 

The great advantage of these thought experiment arguments is their model-independence. 
If we trust our intuition, regarding the way in which the gravitational field and dark energy density can interact with microscopic quantum systems, the EGUP offers a phenomenological guide, which, arguably, any would-be theory of quantum gravity, including a dark energy term, should be able to reproduce. 
It is also their greatest disadvantage. 
The `uncertainties' in Eqs. (\ref{GUP-1}), (\ref{EUP-1}) and (\ref{EGUP-1}) are heuristic, like those in Eqs. (\ref{HUP-1}), and it is not clear how to derive them, rigorously, as the standard deviations of probability distributions, obtained from a modified quantum formalism. 

Until recently, the only approach considered in the literature was to modify the canonical commutator (\ref{[x,p]-1}), such that $[\hat{x}^i,\hat{p}_j] = i\hbar \hat{F}^{i}{}_{j}(\hat{{\bf x}},\hat{{\bf p}})$ \cite{Kempf:1996ss,Hossenfelder:2012jw,Tawfik:2015rva,Tawfik:2014zca,Bosso:2023aht}. 
This immediately leads to modified uncertainty relations, due to the modification of the Schr{\"o}dinger-Robertson bound (\ref{Robertson-Schrodinger-1}). 
Specifically, setting $\hat{F}^{i}{}_{j}(\hat{{\bf x}},\hat{{\bf p}})  = 1 + \hat{p}_{j}^2 + \hat{x}^{i2}$ leads to an EGUP-type expression, though it may be shown that the resulting formalism exhibits various pathologies. 
These include, but are not limited to, the following \cite{Hossenfelder:2012jw,Tawfik:2015rva,Tawfik:2014zca,Lake:2020rwc,Lake:2023uoi}:
\begin{enumerate} \label{pathologies}

\item The generation of mass-dependent accelerations, even in the classical limit. 
Clearly, this violates any sensible definition of the equivalence principle (EP), in the quantum regime \cite{Lebed:2023edp,Zych:2018bmk,Pikovski:2011zk,Giacomini:2021aof,Giacomini:2020ahk,Marletto:2020agp,Zych:2015fka,Balsells:2023jdn}.

\item The fact that the background geometry, though not Euclidean, is still manifestly classical in nature. 
This an undesirable trait in a would-be model of quantum gravity, but, worse still, the exact metric of the underlying classical geometry is {\it unknown}. 
It cannot be inferred from the form of the uncertainty relations, alone, so that it is not clear `where in space' the eigenvalues of the modified position operators, $\left\{\hat{x}^{i}\right\}_{i=1}^{3}$, refer to \cite{Lake:2020rwc,Lake:2023uoi}. 
Similar remarks hold for $\left\{\hat{p}_{j}\right\}_{i=1}^{3}$, with respect to momentum space. 

\item The dependence of the position and momentum uncertainties on the {\it classical} frame of reference. 
For example, including a GUP-type term via the function $\hat{F}^{i}{}_{j}(\hat{{\bf p}}) = 1 + \hat{p}_{j}^2$ generates velocity-dependent positional uncertainties, whereas including an EUP-type term using $\hat{F}^{i}{}_{j}(\hat{{\bf x}}) = 1 +  \hat{x}^{i2}$ generates momentum uncertainties that depend of the shift-translation parameter \cite{Lake:2020rwc,Lake:2023uoi}.   

\item The so-called `soccer ball problem', which refers to the inability to define sensible multi-particle states, with energies exceeding the Planck-energy. 
This limit arises as an effective upper bound for the energies of {\it all} systems, including macroscopic bodies, due to the way in which modified commutators implement the Planck length as a minimum length scale \cite{Hossenfelder:2012jw,Tawfik:2015rva,Tawfik:2014zca,Lake:2019oaz,Lake:2020rwc,Lake:2023uoi}.

\end{enumerate}

Recently, an alternative basis for EGUP was developed, by a research team that includes the present authors, which neatly avoids these problems \cite{Lake:2018zeg,Lake:2019nmn,Lake:2019oaz,Lake:2020rwc,Lake:2021beh,Lake:2020chb,Lake:2021gbu,Lake:2022hzr,Lake:2023lvh,Lake:2023uoi}. 
In the remainder of this section, we review the basic formalism of this model and show how it is equivalent to applying the QRF formalism, developed in Secs. \ref{Sec.2}-\ref{Sec.4}, to spatial points in the background geometry, as opposed to point-like quantum particles in classical Euclidean space.  

To begin, we note that in canonical QM the eigenstate $\ket{{\bf x}}$ represents the localisation of a quantum particle at the point `${\bf x}$' in the background geometry. 
It does {\it not} correspond to a quantum state of this geometry, since the latter remains purely classical. 
To associate quantum degrees of freedom with the spatial background, therefore, we first `square' the quantum mechanical phase space (in the position space representation), via the map $\ket{{\bf x}} \mapsto \ket{{\bf x}} \otimes \ket{{\bf x'}}$. 
The new eigenstates $\ket{{\bf x'}}$ are interpreted, from here on, as representing the perfect localisation of a spatial `point' in the {\it quantum mechanical} background geometry. 
Generically, however, we expect such points to be delocalised, in the quantum gravity regime, over a region comparable to the Planck volume. 
We therefore implement the map
\begin{eqnarray} \label{smearing_map_x}
\hat{S}_{\bold{x}}: \ket{{\bf x}} \mapsto \ket{{\bf x}} \otimes \ket{g_{{\bf x}}} \, .
\end{eqnarray}
where
\begin{eqnarray} \label{g_x}
\ket{g_{{\bf x}}} = \int g({\bf x}'-{\bf x}) \ket{{\bf x}'} {\rm d}^{3}{\rm x}' \, , 
\end{eqnarray}
and $g(\bf{x}'-\bf{x})$ is any normalised function,   
\begin{eqnarray} \label{g_normalisation}
\int |g({\bf x}'-{\bf x})|^2 {\rm d}^{3}{\rm x}' = \int |g({\bf x}'-{\bf x})|^2 {\rm d}^{3}{\rm x} = 1 \, . 
\end{eqnarray}
For convenience, we will visualise $|g(\bold{x}'-\bold{x})|^2$ as a Planck-scale Gaussian distribution, 
\begin{eqnarray} \label{Planck-scale_Gaussian_x}
|g({\bf x}'-{\bf x})|^2 = \left(\frac{1}{\sqrt{2\pi}\sigma_g}\right)^3\exp\left[-\frac{({\bf x}'-{\bf x})^2}{2\sigma_g^2}\right] \, , 
\end{eqnarray}
where $\sigma_g \simeq l_{\rm Pl}$, from here on. 

The associated amplitude, $g(\bf{x}'-\bf{x})$, is interpreted as the quantum probability amplitude for the coherent transition $\bf{x} \leftrightarrow \bf{x}'$, where it is assumed that this transition also interchanges the amplitude(s) of any material-particle wave functions associated with the classical point `{\bf x}', $\psi(\bf{x}) \leftrightarrow \psi(\bf{x}')$ \cite{Lake:2018zeg,Lake:2019nmn,Lake:2019oaz,Lake:2020rwc,Lake:2021beh,Lake:2020chb,Lake:2021gbu,Lake:2022hzr,Lake:2023lvh,Lake:2023uoi}. 
By `smearing'  each spatial point in the classical Euclidean space, $\mathbb{E}^3$, we map the classical background geometry to a {\it superposition} of Euclidean geometries, which is formally equivalent to the Cartesian product $\mathbb{E}^3 \times \mathbb{E}^3$. 
Each geometry in the superposition is equivalent to every other geometry, except for the pair-wise exchange of at least two points, $\bf{x} \leftrightarrow \bf{x}'$. 
For isotropic smearing functions, including (\ref{Planck-scale_Gaussian_x}), the most probable position of each point, {\it relative} to all others, is precisely the classical value, ${\bf x}' = {\bf x}$, but fluctuations within a volume of order $\sim l_{\rm Pl}^3$ remain relatively likely. 
We now show how these fluctuations generate additional Planck-scale contributions to the positional uncertainties of material particles. 

Acting with $\hat{S}_{\bold{x}}$ (\ref{smearing_map_x}) on the canonical one-particle wave function $\ket{\psi} = \int \psi({\bf x})\ket{{\bf x}}{\rm d}^3{\rm x}$, which, we recall, `lives' in classical Euclidean space $\mathbb{E}^3$, we have $\hat{S}_{\bold{x}}:\ket{\psi} \mapsto \ket{\Psi'}$, where
\begin{eqnarray} \label{smearing_map_|psi>'}
\ket{\Psi'} := \int \psi(\bold{x}) \ket{\bold{x}}\ket{g_{{\bold{x}}}} {\rm d}^{3}{\rm x}
= \int\int \psi(\bold{x})g(\bold{x}'-{\bf x})\ket{\bold{x}}\ket{\bold{x}'} {\rm d}^{3}{\rm x} {\rm d}^{3}{\rm x}' \, .
\end{eqnarray}
This represents the wave function of a single particle that `lives' in the smeared superposition of geometries, $\mathbb{E}^3 \times \mathbb{E}^3$. 
For simplicity, we will analyse only smeared one-particle wave functions, in this paper. 
The interested reader is refereed to \cite{Lake:2018zeg} for a preliminary analysis of multi-particle states.

From (\ref{smearing_map_|psi>}), it is clear that, since an observed value of the particle's position, `${\bf x}'$', cannot determine which point(s) in the smeared superposition of geometries underwent the transition ${\bf x} \leftrightarrow {\bf x}'$, we must sum over all possibilities by integrating the joint probability distribution, $|\Psi({\bf x},{\bf x}')|^2 := |\psi({\bf x})|^2 |g({\bf x}'-{\bf x})|^2$, over ${\rm d}^3{\rm x}$. 
This yields 
\begin{eqnarray} \label{EQ_XPRIMEDENSITY}
\frac{{\rm d}^{3}P(\bold{x}' | \Psi)}{{\rm d}{\rm x}'^{3}} = \int |\Psi(\bold{x},\bold{x}')|^2 {\rm d}^3{\rm x} = |\psi|^2 * |g|^2(\bold{x}') \, ,
\end{eqnarray}
where the star again denotes a convolution. 
Thus, in this formalism, only primed degrees of freedom represent measurable quantities, whereas the unprimed degrees of freedom, inherited from canonical QM, are physically inaccessible \cite{Lake:2018zeg,Lake:2019nmn,Lake:2019oaz,Lake:2020rwc,Lake:2021beh,Lake:2020chb,Lake:2021gbu,Lake:2022hzr,Lake:2023lvh,Lake:2023uoi}. 
They are recovered only in the limit $|g(\bold{x}'-\bold{x})|^2 \rightarrow \delta^3(\bold{x}'-\bold{x})$ ($\sigma_g \rightarrow 0$), for which $\bold{x}' = \bold{x}$. 

However, we may apply exactly the same logic to the momentum space representation of canonical QM, yielding 
\begin{eqnarray} \label{smearing_map_p}
\hat{S}_{\bold{p}}: \ket{{\bf x}} \mapsto \ket{{\bf p}} \otimes \ket{g_{{\bf p}}} \, .
\end{eqnarray}
where
\begin{eqnarray} \label{g_p}
\ket{g_{{\bf p}}} = \int \tilde{g}({\bf p}'-{\bf p}) \ket{{\bf p}'} {\rm d}^{3}{\rm p}' \, , 
\end{eqnarray}
and 
\begin{eqnarray} \label{tilde_g_normalisation}
\int |\tilde{g}({\bf p}'-{\bf p})|^2 {\rm d}^{3}{\rm p}' = \int |\tilde{g}({\bf p}'-{\bf p})|^2 {\rm d}^{3}{\rm p} = 1 \, . 
\end{eqnarray}
Again, for simplicity, we may visualise $|\tilde{g}({\bf p}'-{\bf p})|^2$ as a Gaussian,
\begin{eqnarray} \label{Planck-scale_Gaussian_p}
|\tilde{g}({\bf p}'-{\bf p})|^2 = \left(\frac{1}{\sqrt{2\pi}\tilde{\sigma}_g}\right)^3\exp\left[-\frac{({\bf p}'-{\bf p})^2}{2\tilde{\sigma}_g^2}\right] \, , 
\end{eqnarray}
where $\tilde{\sigma}_g$ represents the minimum momentum space smearing and $\tilde{g}({\bf p}'-{\bf p})$ is the quantum probability amplitude for the transition ${\bf p} \leftrightarrow {\bf p}'$, $\tilde{\psi}(\bf{p}) \leftrightarrow \tilde{\psi}(\bf{p}')$.

Applying the map $\hat{S}_{\bold{p}}$ (\ref{smearing_map_p}) to the spectral representation of the canonical one-particle wave vector in the momentum basis, $\ket{\psi} = \int \tilde{\psi}({\bf p})\ket{{\bf p}}{\rm d}^3{\rm p}$, gives $\hat{S}_{\bold{p}}:\ket{\psi} \mapsto \ket{\Psi''}$, where 
\begin{eqnarray} \label{smearing_map_|psi>''}
\ket{\Psi''} := \int \psi(\bold{p}) \ket{\bold{p}}\ket{g_{{\bold{p}}}} {\rm d}^{3}{\rm p}
= \int\int \psi(\bold{p})g(\bold{p}'-{\bf p})\ket{\bold{p}}\ket{\bold{p}'} {\rm d}^{3}{\rm p} {\rm d}^{3}{\rm p}' \, . 
\end{eqnarray}
The joint probability distribution, associated with the measured value ${\bf p}'$, is $|\tilde{\Psi}(\bold{p},\bold{p}')|^2 := |\tilde{\psi}({\bf p})|^2|\tilde{g}({\bf p}'-{\bf p})|^2$, giving
\begin{eqnarray} \label{EQ_PPRIMEDENSITY}
\frac{{\rm d}^{3}P(\bold{p}' | \tilde{\Psi})}{{\rm d}{\rm p}'^{3}} = \int |\tilde{\Psi}(\bold{p},\bold{p}')|^2 {\rm d}^3{\rm x} = |\psi|^2 * |\tilde{g}|^2(\bold{p}') \, .
\end{eqnarray}
Here too, the unprimed degrees of freedom, which are inherited from canonical QM, are physically inaccessible and are recovered only in the limit $|\tilde{g}(\bold{p}'-\bold{p})|^2 \rightarrow \delta^3(\bold{p}'-\bold{p})$ ($\tilde{\sigma}_g \rightarrow 0$), for which $\bold{p}' = \bold{p}$.

Using our previous terminology, we may say that both $\hat{S}_{\bold{x}}$ (\ref{smearing_map_x}) and $\hat{S}_{\bold{p}}$ (\ref{smearing_map_p}) represent CRF-to-QRF transitions, and the physical process of `smearing' the background geometry is illustrated, heuristically, in Fig. 11. 
(See also Fig. 12, for a simplified, but phenomenologically {\it almost} equivalent picture). 
Yet again, we appear to have run into a similar problem to that faced by the GCB formalism \cite{Giacomini:2017zju}, namely, an ambiguity in the definition of such a transition, due to the fact that $\hat{S}_{\bold{x}}\ket{\psi} \neq \hat{S}_{\bold{p}}\ket{\psi}$ ($\ket{\Psi'} \neq \ket{\Psi''}$). 
Which choice is correct?

Yet again, the ambiguity is only apparent. 
We require the `objective' state of the composite quantum-particle-plus-quantum-background system to be independent of the chosen representation, so that we must define the momentum space representation of $\ket{\Psi'}$ as
\begin{eqnarray} \label{smearing_map_|psi>'*}
\ket{\Psi'} := \int\int \tilde{\psi}(\bold{p})\tilde{g}(\bold{p}'-{\bf p})\ket{\bold{p} \, \bold{p}'} {\rm d}^{3}{\rm p} {\rm d}^{3}{\rm p}' \, ,
\end{eqnarray}
where
\begin{eqnarray} \label{smeared_brakets-1}
&&\bra{\bold{x}}\braket{\bold{x}'|\bold{p} \, \bold{p}'} := \left(\frac{1}{\sqrt{2\pi\hbar\beta}}\right)^{6}\exp\left[\frac{i}{\hbar}\bold{p}.\bold{x} + \frac{i}{\beta}(\bold{p}'-\bold{p}).(\bold{x}'-\bold{x})\right] \, , 
\nonumber\\
&&\braket{\bold{p} \, \bold{p}'|\boldsymbol{\alpha} \, \boldsymbol{\alpha}'} := \delta^{3}(\bold{p}-\boldsymbol{\alpha}) \, \delta^{3}(\bold{p}'-\boldsymbol{\alpha}') \, , 
\end{eqnarray}
and the position space representation of $\ket{\Psi''}$ as
\begin{eqnarray} \label{smearing_map_|psi>''*}
\ket{\Psi''} := \int\int \psi(\bold{x})g(\bold{x}'-{\bf x})\ket{\bold{x} \, \bold{x}'} {\rm d}^{3}{\rm x} {\rm d}^{3}{\rm x}' \, ,
\end{eqnarray}
where
\begin{eqnarray} \label{smeared_brakets-2}
&&\bra{\bold{p}}\braket{\bold{p}'|\bold{x} \, \bold{x}'} := \left(\frac{1}{\sqrt{2\pi\hbar\beta}}\right)^{6}\exp\left[\frac{i}{\hbar}\bold{p}.\bold{x} + \frac{i}{\beta}(\bold{p}'-\bold{p}).(\bold{x}'-\bold{x})\right] \, , 
\nonumber\\
&&\braket{\bold{x} \, \bold{x}'|\bold{a} \, \bold{a}'} := \delta^{3}(\bold{p}-\bold{a}) \, \delta^{3}(\bold{p}'-\bold{a}') \, , 
\end{eqnarray}
together with
\begin{eqnarray} \label{smeared_Fourier_Transforms-1}
\tilde{\psi}(\bold{p}) := \int \psi(\bold{x}) \exp\left[-\frac{i}{\hbar}\bold{p}.\bold{x}\right]{\rm d}^{3}{\rm x} \, , 
\end{eqnarray}
as in canonical QM, and
\begin{eqnarray} \label{smeared_Fourier_Transforms-2}
\tilde{g}(\bold{p}'-\bold{p}) := \int g(\bold{x}'-\bold{x}) \exp\left[-\frac{i}{\beta}(\bold{p}'-\bold{p}).(\bold{x}'-\bold{x})\right]{\rm d}^{3}{\rm x}' \, .
\end{eqnarray}

The physical interpretation of the constant $\beta$, introduced in Eqs. (\ref{smeared_brakets-1}), (\ref{smeared_brakets-1}) and (\ref{smeared_Fourier_Transforms-2}), will be considered shortly. 
Before that, the careful reader will easily anticipate our next step. 
It is most convenient to work in the symmetric bases, for which we define the state
\begin{eqnarray} \label{smearing_map_|psi>}
\ket{\Psi} &:=& \int\int \psi(\bold{x})g(\bold{x}'-\bold{x})\ket{\bold{x}}\ket{\bold{x}'-\bold{x}}{\rm d}^{3}{\rm x}{\rm d}^{3}{\rm x}'
\nonumber\\
&:=& \int\int \tilde{\psi}(\bold{p})\tilde{g}(\bold{p}'-\bold{p})\ket{\bold{p}}\ket{\bold{p}'-\bold{p}}{\rm d}^{3}{\rm p}{\rm d}^{3}{\rm p}'
\nonumber\\
&:=& \ket{\psi} \otimes \ket{g} \, ,
\end{eqnarray}
which corresponds to the application of the map
\begin{eqnarray} \label{smearing_map_S}
\hat{S}: \ket{\psi} \mapsto \ket{\psi} \otimes \ket{g} \, . 
\end{eqnarray}
Note that, here, 
\begin{eqnarray} \label{|g>}
\ket{g} := \int g(\bold{x}'-\bold{x})\ket{\bold{x}'-\bold{x}}{\rm d}^{3}{\rm x}' := \int \tilde{g}(\bold{p}'-\bold{p})\ket{\bold{p}'-\bold{p}}{\rm d}^{3}{\rm p}' \, , 
\end{eqnarray}
which is not equivalent to either $\ket{g_{{\bf x}}}$ (\ref{g_x}) or $\ket{g_{{\bf p}}}$ (\ref{g_p}), corresponds to the quantum state of the {\it whole} background geometry, rather than the quantum state of a specific `point' in either physical space or momentum space. 

Next, we construct the generalised position and momentum operators,
\begin{eqnarray} \label{smeared_X}
\hat{X}^{i} &:=& \hat{\mathcal{X}}^{i} + \hat{\mathcal{X}}'^{i}
\nonumber\\
&:=& \int\int x^{i} \ket{\bold{x}}\bra{\bold{x}} \otimes \ket{\bold{x}'-\bold{x}}\bra{\bold{x}'-\bold{x}} {\rm d}^{3}{\rm x}{\rm d}^{3}{\rm x}' 
\nonumber\\
&+& \int\int (x'^{i}-x^{i}) \ket{\bold{x}}\bra{\bold{x}} \otimes \ket{\bold{x}'-\bold{x}}\bra{\bold{x}'-\bold{x}} {\rm d}^{3}{\rm x}{\rm d}^{3}{\rm x}' \, , 
\end{eqnarray}
and
\begin{eqnarray} \label{smeared_P}
\hat{P}_{j} &:=& \hat{\mathcal{P}}_{j} + \hat{\mathcal{P}}'_{j}
\nonumber\\
&:=& \int\int p_{j} \ket{\bold{p}}\bra{\bold{p}} \otimes \ket{\bold{p}'-\bold{p}}\bra{\bold{p}'-\bold{p}} {\rm d}^{3}{\rm p}{\rm d}^{3}{\rm p}' 
\nonumber\\
&+& \int\int (p'_{j}-p_{j}) \ket{\bold{p}}\bra{\bold{p}} \otimes \ket{\bold{p}'-\bold{p}}\bra{\bold{p}'-\bold{p}} {\rm d}^{3}{\rm p}{\rm d}^{3}{\rm p}' \, , 
\end{eqnarray}
whose eigenvalues are $x'^{i}$ and $p'_{j}$, respectively. 
Physically, these represent the possible measured values of the particle's position and momentum {\it relative} to the fluctuations of the quantum background geometry, in accordance with our interpretation of the latter as a non-material QRF. 
In this scenario, the sub-operators $\left\{\hat{\mathcal{X}}^{i},\hat{\mathcal{X}}'^{i};\hat{\mathcal{P}}_{j},\hat{\mathcal{P}}'_{j}\right\}$ do not represent observable quantities. 
We are restricted to performing measurements on material particles, whose motion is determined by a combination of canonical quantum diffusion and fluctuations of the quantum background geometry they inhabit, but there is no physical measurement that enables us to isolate either of these effects. 
Nonetheless, we can again use the position and momentum sub-operators to construct the useful unitaries
\begin{eqnarray} \label{smeared_unitaries}
\hat{\mathcal{U}}' := \exp\left[\frac{i}{\beta}\hat{\bold{\mathcal{P}}}'.\hat{\bold{\mathcal{X}}}\right] \, , \quad 
\hat{\mathcal{U}}'' := \exp\left[-\frac{i}{\beta}\hat{\bold{\mathcal{P}}}.\hat{\bold{\mathcal{X}}}'\right] \, .
\end{eqnarray}
These act on the entangled bases according to
\begin{eqnarray} \label{smeared_unitaries_bases}
\ket{\bold{x}}\ket{\bold{x}'}, \, \ket{\bold{p} \, \bold{p}'} \stackrel{\hat{\mathcal{U}}'^{\dagger}}{\longrightarrow} \ket{\bold{x}}\ket{\bold{x}'-\bold{x}}, \, \ket{\bold{p}}\ket{\bold{p}'-\bold{p}} 
\stackrel{\hat{\mathcal{U}}''^{\dagger}}{\longleftarrow} \ket{\bold{x} \, \bold{x}'}, \, \ket{\bold{p}}\ket{\bold{p}'} \, , 
\end{eqnarray}
and, hence, on the states $\ket{\Psi'}$ (\ref{smearing_map_|psi>'}), $\ket{\Psi''}$ (\ref{smearing_map_|psi>''}) and $\ket{\Psi}$ (\ref{smearing_map_|psi>}), according to
\begin{eqnarray} \label{smeared_unitaries_wave_functions}
\ket{\Psi'} \stackrel{\hat{\mathcal{U}}'^{\dagger}}{\longrightarrow} \ket{\Psi} \stackrel{\hat{\mathcal{U}}''^{\dagger}}{\longleftarrow} \ket{\Psi''} \, .
\end{eqnarray}
By analogy with our previous results, presented in Sec. \ref{Sec.3}, the unitary equivalence of these states indicates their {\it physical} equivalence, as representations of the canonical state $\ket{\psi}$, `viewed by' the smeared geometry $\ket{g}$ (\ref{|g>}), which acts as a non-material QRF. 

The different but equivalent representations of `non-material QRF states', $\ket{g_{\bold{x}}}$ (\ref{g_x}), $\ket{g_{\bold{p}}}$ (\ref{g_p}) and $\ket{g}$ (\ref{smearing_map_|psi>}), are also related according to
\begin{eqnarray} \label{smeared_unitaries_wave_g}
\ket{g_{\bold{x}}} \stackrel{\hat{\mathcal{U}}'^{\dagger}}{\longrightarrow} \ket{g} \stackrel{\hat{\mathcal{U}}''^{\dagger}}{\longleftarrow}\ket{g_{\bold{p}}} \, .
\end{eqnarray}
Hence, the smearing maps $\hat{S}_{\bold{x}}$ (\ref{smearing_map_x}), $\hat{S}_{\bold{p}}$ (\ref{smearing_map_p}) and $\hat{S}$ (\ref{smearing_map_S}) are related via 
\begin{eqnarray} \label{smeared_unitaries_maps}
\hat{\mathcal{U}}' \hat{S}_{\bold{x}} \, \hat{\mathcal{U}}'^{\dagger} = \hat{S} = \hat{\mathcal{U}}'' \hat{S}_{\bold{p}} \, \hat{\mathcal{U}}''^{\dagger} \, .
\end{eqnarray}
This again demonstrates the fact that the CRF-to-QRF transition, which represents the transition from classical Euclidean space $\mathbb{E}^3$ to the smeared superposition of geometries $\mathbb{E}^3 \times \mathbb{E}^3$, is, in fact, basis-independent, as required. 
In other words, we could just as easily define $\hat{S}_{\bold{x}}$ via its action on the momentum space eigenstates, $\ket{\bold{p}}$, and $\hat{S}_{\bold{p}}$ via its action on the position space eigenstates, $\ket{\bold{x}}$, and we may act with either operator on either representation of the canonical state $\ket{\psi}$. 
This indicates that, despite initial appearances to the contrary, $\hat{S}_{\bold{x}}$ and $\hat{S}_{\bold{p}}$ are not really basis-dependent operations, unlike the operators $\hat{S}_{x}^{(C \rightarrow A)}$ and $\hat{S}_{p}^{(C \rightarrow A)}$, considered in \cite{Giacomini:2017zju}.

It is straightforward to show that $\hat{X}^{i}$ and $\hat{P}_{j}$ generate GURs, via their action on the smeared-state $\ket{\Psi}$, according to 
\begin{eqnarray} \label{X_uncertainty}
(\Delta_\Psi X^{i})^2 &=& \braket{\Psi |(\hat{X}^{i})^{2}|\Psi} - \braket{\Psi|\hat{X}^{i}|\Psi}^2 
\nonumber\\
&=& (\Delta_{\Psi} \mathcal{X}^{i})^2 + (\Delta_{\Psi} \mathcal{X}'^{i})^2
\nonumber\\
&=& (\Delta_{\psi} x'^{i})^2 + (\Delta_{g}x'^{i})^2 \, ,
\end{eqnarray}
and
\begin{eqnarray} \label{P_uncertainty}
(\Delta_\Psi P_{j})^2 &=& \braket{\Psi |(\hat{P}_{j})^{2}|\Psi} - \braket{\Psi|\hat{P}_{j}|\Psi}^2 
\nonumber\\
&=& (\Delta_{\Psi} \mathcal{P}_{j})^2 + (\Delta_{\Psi} \mathcal{P}'_{j})^2
\nonumber\\
&=& (\Delta_{\psi} p'_{j})^2 + (\Delta_{g}p'_{j})^2 \, .
\end{eqnarray}
Alternatively, we may generate the same relations by acting with canonical-type position and momentum operators on the density matrices of the effective mixed states, corresponding to the probability distributions (\ref{EQ_XPRIMEDENSITY}) and (\ref{EQ_PPRIMEDENSITY}). 
These are constructed by complete analogy with the mixed states (\ref{Bob_effective_mixed_state_x}) and (\ref{Bob_effective_mixed_state_p}), obtained 
in Sec. \ref{Sec.3.2.6}.  

However, when treating the background geometry as a non-material QRF, we encounter a subtle issue that was not present in the canonical QRF model, developed in Secs. \ref{Sec.3}-\ref{Sec.4}, namely, that the canonical displacement vector experiences both `front-end' and `back-end' smearing. 
This is depicted, schematically, in Fig. 11, in which the the canonical displacement is labelled as ${\bf r}$. 
Strictly, every point along ${\bf r}$ is smeared into a Planck-width superposition of points, but, due to the vector-equivalence between curved lines and straight lines in flat Euclidean space \cite{Frankel:1997ec,Nakahara:2003nw}, only the smearing of the start and end points of the vector need concern us. 

Thus, the back-end smearing is simply the smearing of the fixed classical origin, $O$, which is mapped to a superposition of origins, $O$', as in the case of a canonical QRF depicted in Fig. 5. 
In addition to this, every {\it possible} end point of ${\bf r}$ is also smeared, when $|{\bf r}| > 0$, according to the map $\ket{\bf r} \mapsto \ket{\bf r} \otimes \ket{g_{\bf r}}$ (\ref{S_x}). 
The back-end smearing therefore has a double effect; in addition to mapping $\psi({\bf r})$ to $\psi({\bf r}-{\bf x}'_{O})$ (in the notation of Fig. 11), with probability $|g(-{\bf x}'_{O})|^2$, it also maps the origin of each and every smearing function, $g({\bf r}'-{\bf r})$, to $g({\bf r}'-{\bf r}-{\bf x}'_{O})$, for $|{\bf r}| > 0$. 
In other words, it maps the composite matter-plus-geometry wave function, $\Psi({\bf r},{\bf r}') := \psi({\bf r})g({\bf r}'-{\bf r})$, to the shifted composite matter-plus-geometry wave function, $\Psi({\bf r},{\bf r}';{\bf x}'_{O}) := \psi({\bf r}-{\bf x}'_{O})g({\bf r}'-{\bf r}-{\bf x}'_{O})$, with probability $|g({\bf x}'_{O})|^2$. 

This represents the perspective of the `observer', who would have been located at $O$, {\it if} fluctuations of the background geometry hadn't also smeared out his/her position, in addition to smearing out the canonical wave function of the observed particle, $\psi({\bf r})$.
The resulting density operator is
\begin{eqnarray} \label{rho_Psi-1}
\hat{\varrho}_{\Psi} &:=& \int \frac{{\rm d}^{3}P({\bf x}'_{O})}{{\rm d}^{3}{\rm x}'_{O}}\ket{\Psi({\bf x}_{O})}\bra{\Psi({\bf x}'_{O})} {\rm d}^{3}{\rm x}'_{O}
\nonumber\\
&:=& \int \frac{{\rm d}^{3}P({\bf x}'_{O})}{{\rm d}^{3}{\rm x}'_{O}}\ket{\psi({\bf x}'_{O})}\bra{\psi({\bf x}'_{O})} \otimes \ket{g({\bf x}'_{O})}\bra{g({\bf x}'_{O})} {\rm d}^{3}{\rm x}'_{O}
\end{eqnarray}
where 
\begin{eqnarray} \label{rho_Psi-2}
\ket{\Psi({\bf x}'_{O})} := \ket{\psi({\bf x}'_{O})}\ket{g({\bf x}'_{O})} := \int\int \psi({\bf r})g({\bf r}'-{\bf r}) \ket{{\bf r}+{\bf x}'_{O}}\ket{{\bf r}'-{\bf r}+{\bf x}'_{O}} {\rm d}^{3}{\rm r}{\rm d}^{3}{\rm r}' \, , 
\end{eqnarray}
and ${\rm d}^{3}P({\bf x}'_{O}) := |g(-{\bf x}'_{O})|^2{\rm d}^{3}{\rm x}'_{O}$. 

Equations (\ref{rho_Psi-1})-(\ref{rho_Psi-2}) give ${\rm d}^{3}P(\bold{x}'|\Psi) = |\psi|^2*|\mathfrak{g}|^2(\bold{x}'){\rm d}^{3}{\rm x}'$, where $|\mathfrak{g}|^2 = |g|^2*|g|^2$ and ${\bf x}' = {\bf r}'-{\bf x}'_{O}$ is the measured value of the displacement, which incorporates the effects of both the front-end and back-end smearing of the canonical displacement vector, as described above (see Fig. 11). 
Nonetheless, if $|g|^2$ is a Planck-width Gaussian, as assumed in Eq. (\ref{Planck-scale_Gaussian_x}), the only measurable difference between this scenario and the simplified picture shown in Fig. 12, in which only the front-end smearing of the canonical displacement is accounted for (for single-particle states), is a change in the effective smearing scale such that $\Delta_{g}x'^{i} = \sigma_g \simeq l_{\rm Pl} \rightarrow 2\sigma_g \simeq 2l_{\rm Pl}$ in the GUR (\ref{X_uncertainty}). 

For momentum measurements, the corresponding density operator is 
\begin{eqnarray} \label{rho_Psi-3}
\hat{\tilde{\varrho}}_{\Psi} &:=& \int \frac{{\rm d}^{3}P({\bf p}'_{O})}{{\rm d}^{3}{\rm p}'_{O}}\ket{\Psi({\bf p}_{O})}\bra{\Psi({\bf p}'_{O})} {\rm d}^{3}{\rm p}'_{O}
\nonumber\\
&:=& \int \frac{{\rm d}^{3}P({\bf p}'_{O})}{{\rm d}^{3}{\rm p}'_{O}}\ket{\psi({\bf p}'_{O})}\bra{\psi({\bf p}'_{O})} \otimes \ket{g({\bf p}'_{O})}\bra{g({\bf p}'_{O})} {\rm d}^{3}{\rm p}'_{O}
\end{eqnarray}
where 
\begin{eqnarray} \label{rho_Psi-4}
\ket{\Psi({\bf p}'_{O})} := \ket{\psi({\bf p}'_{O})}\ket{g({\bf p}'_{O})} := \int\int \tilde{\psi}(\boldsymbol{\kappa})\tilde{g}(\boldsymbol{\kappa}'-\boldsymbol{\kappa}) \ket{\boldsymbol{\kappa}+{\bf p}'_{O}}\ket{\boldsymbol{\kappa}'-\boldsymbol{\kappa}+{\bf p}'_{O}} {\rm d}^{3}{\rm \kappa}{\rm d}^{3}{\rm \kappa}' \, , 
\end{eqnarray}
and ${\rm d}^{3}P({\bf p}'_{O}) := |\tilde{g}(-{\bf p}'_{O})|^2{\rm d}^{3}{\rm p}'_{O}$. 
Again, the only measurable difference between the statistical predictions of Eqs. (\ref{rho_Psi-3})-(\ref{rho_Psi-4}) and those given by the pure-state density matrix $\hat{\rho}_{\Psi} := \ket{\Psi}\bra{\Psi}$, for single-particle states, is accounted for by substituting $\Delta_{g}p'_{j} = \tilde{\sigma}_g \simeq m_{\rm dS}c \rightarrow 2\tilde{\sigma}_g \simeq 2m_{\rm dS}c$ in the GUR (\ref{P_uncertainty}). 
Because both the Planck length and the de Sitter momentum are defined arbitrarily, up to numerical factors of order unity (\ref{Planck_scales})-(\ref{dS_scales}), these scenarios are phenomenologically equivalent, so that we may choose to neglect Eqs. (\ref{rho_Psi-1})-(\ref{rho_Psi-2}) and (\ref{rho_Psi-3})-(\ref{rho_Psi-4}), for simplicity, and to work directly with the GURs (\ref{X_uncertainty}) and (\ref{P_uncertainty}) in the following analysis.  

\begin{figure}[h] \label{Fig.11}
\centering 
	\includegraphics[width=11.25cm]{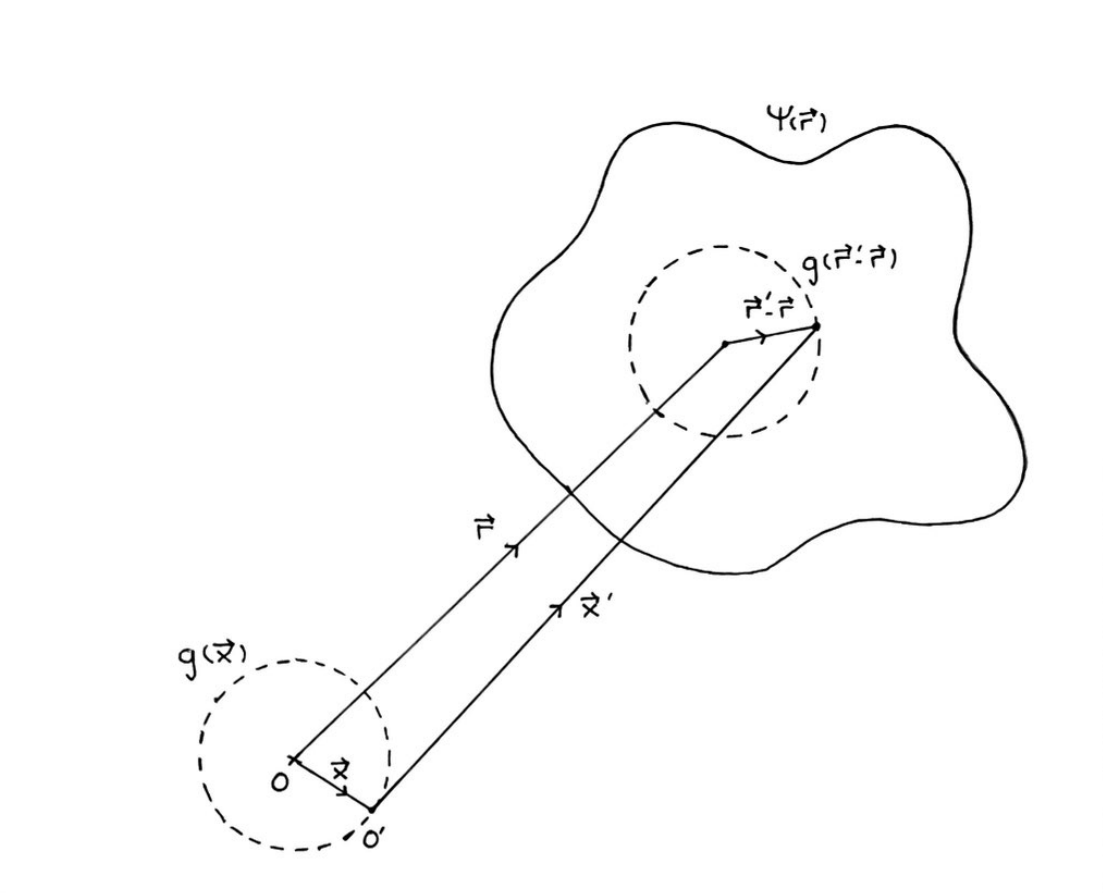}
\caption{The quantised spatial background as a non-material QRF: Strictly speaking, all points along the classical displacement vector $\bold{r}$ are `smeared' into Planck-width Gaussian amplitudes, $g(\bold{r}'-\bold{r})$, which affects both the origin and the end point of the displacement. For single-particle measurements, the resulting joint probability density is $|\Psi(\bold{r},\bold{r}';\bold{x}'_{O})|^2 := |\psi(\bold{r}-\bold{x}'_{O})|^2|g(\bold{r}'-\bold{r}-\bold{x}'_{O})|^2|g(-\bold{x}'_{O})|^2$, where $\bold{x} = -\bold{x}'_{O}$. This gives $\int |\Psi(\bold{r},\bold{r}';\bold{x}'_{O})|^2 {\rm d}^{3}{\rm r} = |\psi(-\bold{x}'_{O})|^2*|g(\bold{r}'-\bold{x}'_{O})|^2 \, . \, |g(-\bold{x}'_{O})|^2 \equiv |\psi({\bf x}'-{\bf r}')|^2*|g({\bf x}')|^2 \, . \, |g(\bold{x}'-\bold{r}')|^2$, where we have used the fact that ${\bf x}' = {\bf r}'-{\bf x}'_{O}$. Integrating over ${\bf r}'$, we then have ${\rm d}^{3}P(\bold{x}'|\Psi) = |\psi|^2*|\mathfrak{g}|^2(\bold{x}'){\rm d}^{3}{\rm x}'$, where $|\mathfrak{g}|^2 = |g|^2*|g|^2$. If $\sigma_g \simeq l_{\rm Pl}$ is the width of $|g|^2$, as in Eq. (\ref{Planck-scale_Gaussian_x}), then the width of $|\mathfrak{g}|^2$ is simply $2\sigma_g$, which is also of the order of the Planck length. Phenomenologically, this scenario is practically interchangeable with the simplified picture, depicted in Fig. 12, in which only the end point of the displacement vector is smeared. A completely analogous picture holds in momentum space, for which we interchange $\bold{r} \leftrightarrow \boldsymbol{\kappa}$, $\bold{r}' \leftrightarrow \boldsymbol{\kappa}'$, $\bold{x}'_{O} \leftrightarrow \bold{p}'_{O}$, $\bold{x}' \leftrightarrow \bold{p}'$ and $\psi \leftrightarrow \tilde{\psi}$, $g \leftrightarrow \tilde{g}$, $\Psi \leftrightarrow \tilde{\Psi}$, together with $\sigma_g \leftrightarrow \tilde{\sigma}_g$.}
\end{figure}

\begin{figure}[h] \label{Fig.12}
\centering 
	\includegraphics[width=11.25cm]{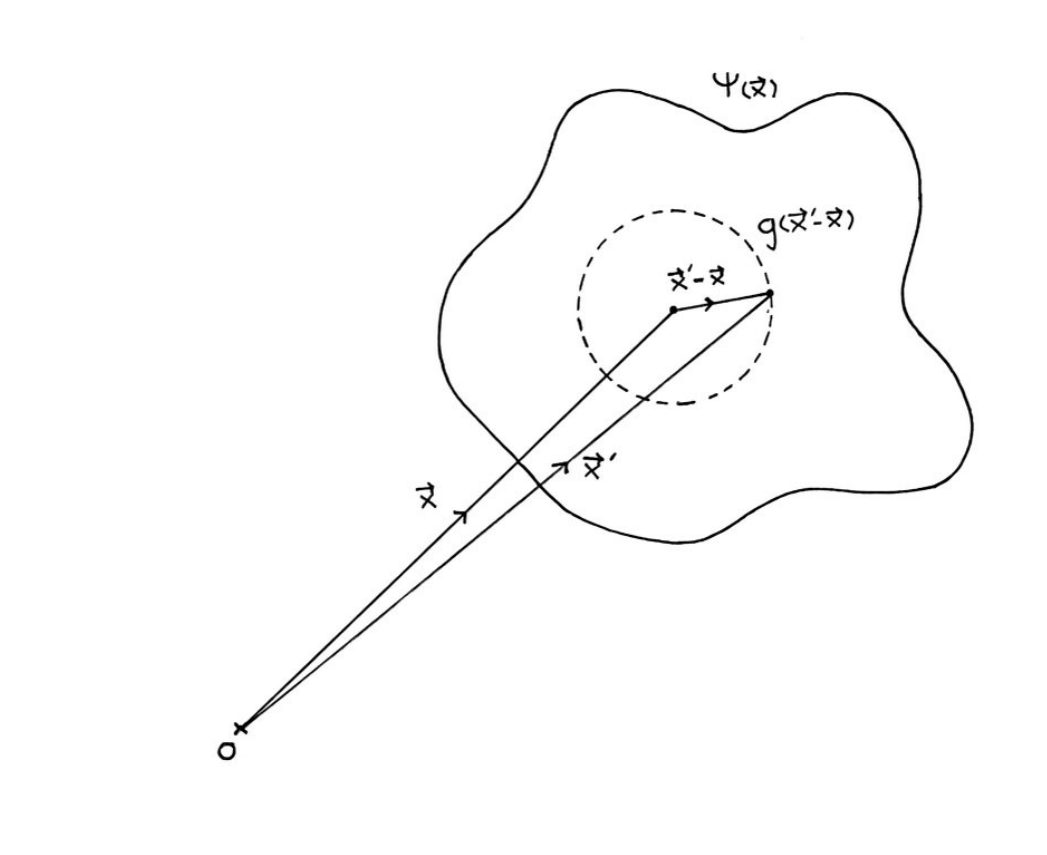}
\caption{The simplified picture of the quantised spatial background as a non-material QRF in which only the end point of the canonical displacement vector is smeared. For the sake of notational elegance, we relabel the canonical displacement as $\bold{x}$, rather than $\bold{r}$ (as in Fig. 11), and perform a similar notational change with respect to the measurable (primed) displacement, $\bold{r}'$, which we relabel as $\bold{x}'$. Analogous changes are affected in the momentum space picture, giving $\bold{r} \leftrightarrow \bold{x}$, $\bold{r}' \leftrightarrow \bold{x}'$, $\boldsymbol{\kappa} \leftrightarrow \bold{p}$ and $\boldsymbol{\kappa}' \leftrightarrow \bold{p}'$, overall. Phenomenologically, the simplified scenario depicted here is practically interchangeable with the more complicated, but physically accurate picture, shown in Fig. 11. For single-particle states the only measurable differences are expressed by simply interchanging $\sigma_g \leftrightarrow 2\sigma_g$ and $\tilde{\sigma}_g \leftrightarrow 2\tilde{\sigma}_g$ in Eqs. (\ref{X_uncertainty})-(\ref{P_uncertainty}). For multi-particle states, however, the situation is more complicated, since one may consider both simultaneous and non-simultaneous measurements performed on multiple subsystems. The corresponding analysis is similar to that performed in Secs. \ref{Sec.3.3}-\ref{Sec.3.4}, for multi-particle systems in canonical QM, but we here restrict our attention to smeared single-particle states, for simplicity. } 
\end{figure}

Clearly, both (\ref{X_uncertainty}) and (\ref{P_uncertainty}) are empirically indistinguishable from the uncertainty relation for the canonical pure state, $\ket{\psi}$, when both the position and momentum space smearing scales,
\begin{eqnarray} \label{sigmas}
\Delta_{g}x'^{i} = \sigma_{g} \, , \quad \Delta_{g} p'_{j} = \tilde{\sigma}_{g} \, , 
\end{eqnarray}
are below the thresholds of experimental sensitivity. 
These scales represent the minimum sensitivity required to resolve the non-canonical terms in the GURs derived from Eqs. (\ref{X_uncertainty})-(\ref{P_uncertainty}), which can be combined to give a pair of alternative, but informationally-equivalent expressions, namely
\begin{eqnarray} \label{GUR_X}
(\Delta_\Psi X^{i})^2 \, (\Delta_\Psi P_{j})^2 \geq \left(\frac{\hbar}{2}\right)^2(\delta^{i}{}_{j})^2 + (\Delta_\psi x'^{i})^2 \tilde{\sigma}_{g}^2 
+ \sigma_{g}^2\frac{(\hbar/2)^2}{(\Delta_\psi x'^{j})^2} + \sigma_{g}^2 \tilde{\sigma}_{g}^2 \, ,
\end{eqnarray}
and
\begin{eqnarray} \label{GUR_P}
(\Delta_\Psi X^{i})^2 \, (\Delta_\Psi P_{j})^2 \geq \left(\frac{\hbar}{2}\right)^2(\delta^{i}{}_{j})^2 + \sigma_{g}^2(\Delta_{\psi}p'_{j})^2 + \frac{(\hbar/2)^2}{(\Delta_{\psi}p'_{j})^2}\tilde{\sigma}_{g}^2 + \sigma_{g}^2 \tilde{\sigma}_{g}^2 \, ,
\end{eqnarray}
where we have made use of Eqs. (\ref{sigmas}), together with the relations
\begin{eqnarray} \label{g_uncertainty}
\Delta_{\psi}x'^{i}\Delta_{\psi}p'_{j} \geq \frac{\hbar}{2}\delta^{i}{}_{j} \, , \quad \Delta_{g}x'^{i}\Delta_{g}p'_{j} \geq \frac{\beta}{2}\delta^{i}{}_{j} \, .
\end{eqnarray}

The first of these is simply the HUP, expressed in terms of the (measurable) primed variables of the smeared space formalism, and the latter follows directly from the properties of the $\beta$-scaled Fourier transform (\ref{smeared_Fourier_Transforms-2}) \cite{pinsky2008introduction}. 
These relations hold independently and are useful for developing our intuition in the smeared-space background. 
Roughly speaking, the HUP is the correct uncertainty relation for material point-particles, inhabiting classical background geometries, whereas the new relation applies to quantised spatial `points', directly. 
However, as already noted, neither $\Delta_{\psi}x'^{i}$, $\Delta_{\psi}p'_{j}$, $\Delta_{g}x'^{i}$ nor $\Delta_{g}p'_{j}$ can be computed, individually, from the sets of measured values, $\left\{\bold{x}'\right\}_{-\infty}^{\infty}$ and $\left\{\bold{p}'\right\}_{-\infty}^{\infty}$.

Setting $i = j$ in the `geometric' uncertainty relation, employing the summation convention, and substituting from Eqs. (\ref{sigmas}), we obtain the definition of the constant $\beta$ as
\begin{eqnarray} \label{beta}
\beta := 2\sigma_{g}\tilde{\sigma}_{g} \, . 
\end{eqnarray}
We will consider the possible form of the dynamical equation for $\Psi(\bold{x},\bold{x}';t) := \psi(\bold{x};t)g(\bold{x}'-\bold{x};t)$, including the evolution of the geometric part, $g(\bold{x}'-\bold{x};t)$, later in this section, but, for now, we treat both $\sigma_g$ and $\tilde{\sigma}_g$ as fixed constants, also. 

Physically, $\beta$ represents the fundamental quantum of action associated with the wave function of the quantised background geometry \cite{Lake:2018zeg,Lake:2019nmn,Lake:2019oaz,Lake:2020rwc,Lake:2021beh,Lake:2020chb,Lake:2021gbu,Lake:2022hzr,Lake:2023lvh,Lake:2023uoi}. 
At first glance, this appears to be very bad news for our model. 
If the momentum space smearing scale is significantly less than the Planck momentum, $\tilde{\sigma}_g \ll m_{\rm Pl}c$, the geometric quantum of action is significantly less than Planck's constant, $\beta \ll \hbar$. 
This appears to contradict well-known no-go theorems, which state that it is impossible for the energies and momenta of different particle species to be quantised with respect to different Planck-type constants, in any theory with self-consistent interactions (see, for example, \cite{Sahoo:2004,Deser:2022lmi} and references therein). 
Such theorems imply, for example, that interactions between material particles with momenta $\bold{p} = \hbar\bold{k}$, and gravitons with momenta $\bold{p}' = \beta\bold{k}'$, cannot be consistently defined for $\beta \neq \hbar$. 
On the other hand, setting $\tilde{\sigma}_g \simeq m_{\rm Pl}c$ ($\beta = \hbar$) implies that the `minimum' momentum of any particle, in any frame of reference considered in our model, is of the order of the Planck momentum. 
This clearly violates both theoretical limits \cite{Hossenfelder:2012jw,Garay:1994en} and, more importantly, observational data.   

However, closer inspection shows that the no-go theorems presented in \cite{Sahoo:2004,Deser:2022lmi}, and related literature, do not contradict the results presented here. 
The fundamental brakets for the symmetrised bases $\left\{\ket{\bold{x}},\ket{\bold{x}'-\bold{x}};\ket{\bold{p}},\ket{\bold{p}'-\bold{p}}\right\}$, are
\begin{eqnarray} \label{smeared_brakets-3}
\braket{\bold{x}|\bold{p}} = \exp\left[\frac{i}{\hbar}\bold{p}.\bold{x}\right] \, , \quad \braket{\bold{x}'-\bold{x}|\bold{p}'-\bold{p}} = \exp\left[\frac{i}{\beta}(\bold{p}'-\bold{p}).(\bold{x}'-\bold{x})\right] \, .
\end{eqnarray}
These are equivalent to the relations
\begin{eqnarray} \label{smeared_dB-1}
\bold{p} = \hbar\bold{k} \, , \quad \bold{p}'-\bold{p} = \beta(\bold{k}'-\bold{k}) \, , 
\end{eqnarray}
which, together, combine to give
\begin{eqnarray} \label{smeared_dB-2}
\bold{p}' = \hbar\bold{k} + \beta(\bold{k}'-\bold{k}) \, . 
\end{eqnarray}
Heuristically, we may think of the non-canonical term in this relation as representing the additional momentum `kick', transferred to a material particle, due to quantum fluctuations of the background geometry in which the canonical de Broglie waves propagate. 
Most importantly, for our purposes, such a scenario is not ruled out, physically, since existing no-go theorems do not apply to modified de Broglie relations of the form (\ref{smeared_dB-2}) \cite{Sahoo:2004,Deser:2022lmi}. 

The key difference between this relation and those considered previously in the literature is the presence of the relative (i.e., relational) variables, $\bold{k}'-\bold{k}$. 
Such variables are the signature of the QRF paradigm and we note that, for $\beta \rightarrow \hbar$, we recover the QRF formalism of canonical quantum theory, outlined in Secs. \ref{Sec.3}-\ref{Sec.4}. 
In this limit, there is no distinction between matter and geometry and $g(\bold{x}'-\bold{x})$ may be interpreted, instead, as the wave function of a canonical quantum particle, inhabiting the classical background space.   

Next, we note that minmising the right-hand side of (\ref{GUR_X}) with respect to $\Delta_\psi x'^{i}$, and the right-hand side of (\ref{GUR_P}) with respect to $\Delta_\psi p'_{j}$, yields
\begin{equation} \label{EQ_CAN_DX_OPT}
(\Delta_\Psi X^{i})_{\mathrm{opt}} = \sqrt{\frac{\hbar}{2} \frac{\sigma_g}{\tilde{\sigma}_g} + \sigma_g^2} \, , \quad (\Delta_\Psi P_{j})_{\mathrm{opt}} = \sqrt{\frac{\hbar}{2} \frac{\tilde{\sigma}_g}{\sigma_g} + \tilde{\sigma}_g^2} \, ,
\end{equation}
and, hence,
\begin{eqnarray} \label{DXDP_opt}
\Delta_\Psi X^{i} \, \Delta_\Psi P_{j} \geq \frac{(\hbar + \beta)}{2} \, \delta^{i}{}_{j} \, .
\end{eqnarray}
The right-hand side of (\ref{DXDP_opt}) is the Schr{\" o}dinger-Robertson bound on the product of the generalised uncertainties, $\Delta_\Psi X^{i} \, \Delta_\Psi P_{j}$, and it is straightforward to show that $\hat{X}^{i}$ and $\hat{P}_{j}$ satisfy the rescaled Heisenberg algebra
\begin{equation} \label{[X,P]}
[\hat{X}^{i},\hat{P}_{j}] = i(\hbar + \beta)\delta^{i}{}_{j} \, {\bf\hat{\mathbb{I}}} \, , \quad [\hat{X}^{i},\hat{X}^{j}] = 0 \, , \quad [\hat{P}_{i},\hat{P}_{j}] = 0 \, . 
\end{equation}
This, in turn, follows from the algebra of the sub-operators $\left\{\hat{\mathcal{X}}^{i},\hat{\mathcal{X}}'^{i};\hat{\mathcal{P}}_{j},\hat{\mathcal{P}}'_{j}\right\}$,
\begin{eqnarray} \label{smeared_Heisenberg_subcomponents-1}
[\hat{\mathcal{X}}^{i},\hat{\mathcal{P}}_{j}] = i\hbar \delta^{i}{}_{j} \hat{\mathbb{I}} \, , \quad [\hat{\mathcal{X}}'^{i},\hat{\mathcal{P}}'_{j}] = i\beta \delta^{i}{}_{j} \hat{\mathbb{I}} \, , 
\end{eqnarray}
\begin{eqnarray} \label{smeared_Heisenberg_subcomponents-2}
[\hat{\mathcal{X}}^{i},\hat{\mathcal{P}}'_{j}] = [\hat{\mathcal{X}}^{i},\hat{\mathcal{P}}'_{j}] = 0 \, , 
\end{eqnarray}
\begin{eqnarray} \label{smeared_Heisenberg_subcomponents-3}
[\hat{\mathcal{X}}^{i},\hat{\mathcal{X}}^{j}] = [\hat{\mathcal{X}}'^{i},\hat{\mathcal{X}}'^{j}] = 0 \, , 
\end{eqnarray}
\begin{eqnarray} \label{smeared_Heisenberg_subcomponents-4}
[\hat{\mathcal{X}}^{i},\hat{\mathcal{X}}'^{j}] = [\hat{\mathcal{X}}'^{i},\hat{\mathcal{X}}^{j}] = 0 \, , 
\end{eqnarray}
\begin{eqnarray} \label{smeared_Heisenberg_subcomponents-5}
[\hat{\mathcal{P}}_{i},\hat{\mathcal{P}}_{j}] = [\hat{\mathcal{P}}'_{i},\hat{\mathcal{P}}'_{j}] = 0 \, , 
\end{eqnarray}
\begin{eqnarray} \label{smeared_Heisenberg_subcomponents-6}
[\hat{\mathcal{P}}_{i},\hat{\mathcal{P}}'_{j}] = [\hat{\mathcal{P}}'_{i},\hat{\mathcal{P}}_{j}] = 0 \, . 
\end{eqnarray}
This is clearly analogous to the position and momentum sub-operator algebra presented in Sec. \ref{Sec.4.4}, substituting $\beta \leftrightarrow \hbar$ as the quantum of action associated with the subspace of the QRF. 

Finally, we are now able to show that the most general uncertainty relation, satisfied by the smeared-space position and momentum operators, is of the form $\Delta_\Psi X^{i} \, \Delta_\Psi P_{j} \geq \dots \geq \frac{(\hbar + \beta)}{2} \, \delta^{i}{}_{j}$, where the sum of terms in the middle reduce to the bound on the right-hand side if and only if $\Delta_\Psi X^{i} = (\Delta_\Psi X^{i})_{\mathrm{opt}}$, $\Delta_\Psi P_{j} = (\Delta_\Psi P_{j})_{\mathrm{opt}}$ (\ref{EQ_CAN_DX_OPT}). 
Equations (\ref{GUR_X}), (\ref{GUR_P}) and (\ref{g_uncertainty}) can be combined, directly, to give
\begin{eqnarray} \label{smeared-spaceEGUP-1}
(\Delta_{\Psi} X^{i})^2 (\Delta_{\Psi} P_{j})^2 \geq \left(\frac{\hbar}{2}\right)^2(\delta^{i}{}_{j})^2 + \sigma_g^2(\Delta_{\Psi} P_{j})^2 + (\Delta_{\Psi} X^{i})^2\tilde{\sigma}_g^2 - \left(\frac{\beta}{2}\right)^2(\delta^{i}{}_{j})^2 \, .
\end{eqnarray}
Taking the square root and expanding to first order, then ignoring the subdominant term proportional to $(\beta/2)^2$, gives 
\begin{eqnarray} \label{smeared-spaceEGUP-2}
\Delta_{\Psi} X^{i} \Delta_{\Psi} P_{j} \gtrsim \frac{\hbar}{2}\delta^{i}{}_{j}\left[1 + \frac{2\sigma_g^2}{\hbar}(\Delta_{\Psi} P_{j})^2 + \frac{2\tilde{\sigma}_g^2}{\hbar}(\Delta_{\Psi} X^{i})^2\right] \, .
\end{eqnarray}
This expression is of EGUP form, but with the heuristic uncertainties $\Delta x^{i}$ and $\Delta p_{j}$ replaced by the well-defined standard deviations of the generalised probability distributions, $|\Psi(\bold{x},\bold{x}')|^2 := |\psi(\bold{x})|^2|g(\bold{x}'-\bold{x})|^2$ and $|\tilde{\Psi}(\bold{p},\bold{p}')|^2 := |\tilde{\psi}(\bold{p})|^2|\tilde{g}(\bold{p}'-\bold{p})|^2$, respectively. 
Furthermore, we have shown that this phenomenology can be obtained without introducing modified commutation relations, of non-Heisenberg type. 
For this reason, none of the pathologies afflicting conventional GUP and EUP models arise in our formalism \cite{Lake:2018zeg,Lake:2019nmn,Lake:2019oaz,Lake:2020rwc,Lake:2021beh,Lake:2020chb,Lake:2021gbu,Lake:2022hzr,Lake:2023lvh,Lake:2023uoi}, which, in addition, has a clear and intuitive physical basis. 

By comparing Eq. (\ref{smeared-spaceEGUP-2}) with Eq. (\ref{EGUP-1}), we can at last fix the values of both the position and the momentum space smearing scales, $\sigma_g$ and $\tilde{\sigma}_g$, to within numerical factors of order unity. 
Thus, setting 
\begin{eqnarray} \label{smearing_scales}
\sigma_g := \alpha_0 \, l_{\rm Pl} \, , \quad \tilde{\sigma}_g := 3\eta_0 \, m_{\rm dS}c \, , 
\end{eqnarray}
where $\alpha_0$, $\eta_0 \sim 1$, we have 
\begin{eqnarray} \label{smeared-spaceEGUP-3}
\Delta_{\Psi} X^{i} \Delta_{\Psi} P_{j} \gtrsim \frac{\hbar}{2}\delta^{i}{}_{j} \left[1 + \alpha_0\frac{2G}{\hbar c^3}(\Delta_{\Psi} P_{j})^2 + 2\eta_0 \Lambda(\Delta_{\Psi} X^{i})^2 \right] \, ,
\end{eqnarray}
as required. 
The corresponding value of $\beta$ is 
\begin{eqnarray} \label{beta_mag}
\beta = 2\hbar\sqrt{\frac{\rho_{\Lambda}}{\rho_{\rm Pl}}} \simeq \hbar \times 10^{-61} \, , 
\end{eqnarray}
and the the values of the position and momentum uncertainties that optimise the the most general GUR, Eq. (\ref{smeared-spaceEGUP-1}), are
\begin{eqnarray} \label{optimised_uncertainties}
&&(\Delta_\Psi X^{i})_{\mathrm{opt}} \simeq l_{\Lambda} := \sqrt{l_{\rm Pl}l_{\rm dS}} \simeq 0.1 \, {\rm mm} \, , 
\nonumber\\
&&(\Delta_\Psi P_{j})_{\mathrm{opt}} \simeq m_{\Lambda}c := \sqrt{m_{\rm Pl}m_{\rm dS}}c \simeq 10^{-3} \, {\rm eV/c} \, .
\end{eqnarray}
It is noteworthy that the corresponding energy density is of the order of the observed vacuum energy,
\begin{eqnarray} \label{optimised_energy_density}
(\rho_{\Psi})_{\rm opt} := \frac{3}{4\pi}\frac{(\Delta_\Psi P)_{\mathrm{opt}}c}{(\Delta_\Psi X)_{\mathrm{opt}}^3} \simeq \frac{3}{4\pi}\frac{m_{\Lambda}c^2}{l_{\Lambda}^3} \simeq \frac{\Lambda c^2}{8\pi G} =: \rho_{\Lambda} \, , 
\end{eqnarray}
where we have introduced the shorthand notation $(\Delta_\Psi X)^2 := \sum_{i=1}^{3}(\Delta_\Psi X^{i})^2$, $(\Delta_\Psi P) := \sum_{j=1}^{3}(\Delta_\Psi P_{j})^2$, and that the granularity of $\rho_{\Psi}$ on small scales, as implied by Eqs. (\ref{optimised_uncertainties}), may generate sub-millimetre oscillations in the effective gravitational field strength \cite{Perivolaropoulos:2016ucs,Antoniou:2017mhs}.

\subsection{Spacetime symmetries} \label{Sec.5.3}

In the remainder of this section, we investigate how the Planck-scale smearing of position space, in conjunction with the de Sitter-scale smearing of momentum space, impacts the symmetries that each space can be said to possess, in an operationally well-defined manner. 
Intuitively, we expect the Planck-scale fluctuations of physical space to destroy both the translation and rotational invariance of material systems, with respect to the background geometry they inhabit. 

Operationally, however, we expect the destruction of translation invariance to be manifest - that is, physically measurable - only when the magnitude of the shift parameter is of the order the smearing scale, $|\bold{a}| \simeq \sigma_g \simeq 10^{-33}$ cm. 
If one attempts, instead, to translate a material system through space by a macroscopic distance, or even by a distance comparable to the length scales of sub-atomic particles, $|\bold{a}| \simeq 10^{-12}$ m, the effect of such fluctuations should be negligible, and beyond the resolution of present-day experiments by many orders of magnitude.   

Likewise, if an experimenter attempts to boost the velocity of a material quantum system of mass $M$, with respect to the lab frame, by a constant amount $\bold{v} = \boldsymbol{\alpha}/M$, we expect such a procedure to be approximately successful only in the limit $|\boldsymbol{\alpha}| = M |\bold{v}| \gg \tilde{\sigma}_g \simeq 10^{-66}$ g. 
For $|\boldsymbol{\alpha}| \simeq 10^{-66}$ g, momentum space fluctuations - that is, fluctuations in the momentum of the material system, induced by Planck-scale fluctuations of the physical background geometry, in accordance with the `geometric' uncertainty relation $\Delta_{g}p'_{j} \geq \beta/(2\Delta_{g}x'^{i})$ - will tend to scatter the {\it intended} boost velocity in random directions, at different points in the canonical wave function $\psi(\bold{x})$ ($\tilde{\psi}(\bold{v})$).

To illustrate this point more clearly, we may consider a toy system comprised of a grid of evenly spaced, effectively {\it classical} point-particles, each with very large mass, $M \gtrsim m_{\rm Pl} \gg m_{\rm dS}$. 
The basic idea is to reduce both the canonical positional uncertainty and the canonical velocity uncertainty to near-zero, as outlined in Sec. \ref{Sec.3.2.6}, so that both can be considered negligible in the experiment. 
Next, we consider an experimenter constructing such a system in classical Euclidean space, before attempting to translate it by a fixed displacement vector, $\bold{a}$. 
Mathematically, this transformation is defined by the unitary operator 
\begin{eqnarray} \label{translation_op_canonical_QM}
\hat{U}(\bold{a})^{\dagger} := \exp\left[-\frac{i}{\hbar}\hat{\bold{p}}.\bold{a}\right] \, , 
\end{eqnarray}
and its action on the toy system is simple; a rigid translation of all masses in the grid, by the vector ${\bf a}$, which leaves the grid structure itself unchanged. 
For masses in a superposition of position eigenstates, that is, canonical wave vectors $\ket{\psi} = \int \psi(\bold{x})\ket{\bold{x}}{\rm d}^{3}{\rm x}$, $\hat{U}(\bold{a})^{\dagger}$ acts in a completely analogous manner, on each point `$\bold{x}$' in the superposition $\psi(\bold{x})$, giving
\begin{eqnarray} \label{translation_op_canonical_QM_action-1}
\hat{U}(\bold{a})^{\dagger} \ket{\psi} = \ket{\psi(\bold{a})} := \int \psi(\bold{x})\ket{\bold{x}-\bold{a}}{\rm d}^{3}{\rm x} \, , 
\end{eqnarray}
and the induced transformation of the wave function is
\begin{eqnarray} \label{translation_op_canonical_QM_action-2}
\hat{U}(\bold{a})^{\dagger}: \psi(\bold{x}) \mapsto \psi_{\bold{a}}(\bold{x}) := \psi(\bold{x}+\bold{a}) \, . 
\end{eqnarray}
From Eqs. (\ref{translation_op_canonical_QM_action-1})-(\ref{translation_op_canonical_QM_action-2}), we see that $\hat{U}(\bold{a})^{\dagger}$ successfully implements a rigid translation, by $\bold{a}$, of each point in the position space representation of the canonical wave function, $\psi(\bold{x})$. 

In the smeared superposition of geometries, the corresponding unitary operator is 
\begin{eqnarray} \label{translation_op_smeared_QM}
\hat{\mathcal{U}}(\bold{a})^{\dagger} := \exp\left[-\frac{i}{(\hbar+\beta)}\hat{\bold{P}}.\bold{a}\right] \, , 
\end{eqnarray}
where the individual components of $\hat{\bold{P}}$ are defined in Eqs. (\ref{smeared_P}). 
This represents an {\it attempt}, by our experimenter, to perform a rigid translation of the canonical one-particle wave function, $\psi(\bold{x})$, according to Eq. (\ref{translation_op_canonical_QM_action-2}), but the unavoidable Planck-scale fluctuations of the background geometry send this attempt awry. 
In this case, $\hat{\mathcal{U}}(\bold{a})^{\dagger}$ acts on the total smeared-state, $\ket{\Psi} := \ket{\psi} \otimes \ket{g}$ (\ref{smearing_map_|psi>}), giving
\begin{eqnarray} \label{translation_op_smeared_QM_action-1}
\hat{\mathcal{U}}(\bold{a})^{\dagger}\ket{\Psi} = \ket{\Psi({\bold{a}})} := \int\int \psi(\bold{x}) g(\bold{x}'-\bold{x}) \bigg|\bold{x} - \frac{\bold{a}}{1+\delta}\bigg\rangle \bigg|\bold{x}'-\bold{x}-\frac{\delta\bold{a}}{1+\delta}\bigg\rangle {\rm d}^{3}{\rm x}{\rm d}^{3}{\rm x}' \, , 
\end{eqnarray}
where the parameter $\delta$ is defined as 
\begin{eqnarray} \label{delta}
\delta := \beta/\hbar \simeq 10^{-61} \, . 
\end{eqnarray}

The induced transformation of the smeared-state wave function, $\Psi(\bold{x},\bold{x}') := \psi(\bold{x})g(\bold{x}'-\bold{x})$, is 
\begin{eqnarray} \label{translation_op_smeared_QM_action-2}
\hat{\mathcal{U}}(\bold{a})^{\dagger} : \Psi(\bold{x},\bold{x}') \mapsto \Psi_{\bold{a}}(\bold{x},\bold{x}') := \psi\left(\bold{x} + \frac{\bold{a}}{1+\delta}\right) g\left(\bold{x}'-\bold{x} + \frac{\delta \bold{a}}{1+\delta}\right) \, . 
\end{eqnarray}
From Eqs. (\ref{translation_op_smeared_QM_action-1})-(\ref{translation_op_smeared_QM_action-2}), we see that $\hat{\mathcal{U}}(\bold{a})^{\dagger}$ rigidly translates the canonical wave function, $\psi(\bold{x})$, by the vector $\bold{a}/(1+\delta)$, 
according to
\begin{eqnarray} \label{translation_op_smeared_QM_action-3}
\bold{x} \mapsto \bold{x} + \frac{\bold{a}}{1+\delta} \, , 
\end{eqnarray}
which, for $\delta \simeq 10^{-61}$ (\ref{delta}), is very close to $\bold{a}$. 
Nonetheless, the unprimed variable $\bold{x}$ is not an observable quantity in the smeared space formalism and we must consider, instead, the induced displacement of the measured position, $\bold{x}'$. 
Clearly, this is
\begin{eqnarray} \label{translation_op_smeared_QM_action-4}
\bold{x}' = \bold{x} + (\bold{x}' - \bold{x}) \mapsto \bold{x} + \frac{\bold{a}}{1+\delta} + \left(\bold{x}'-\bold{x} + \frac{\delta \bold{a}}{1+\delta}\right) = \bold{x}' + \bold{a} \, .
\end{eqnarray}

The action of $\hat{\mathcal{U}}(\bold{a})^{\dagger}$ on our toy system, consisting of a grid of heavy masses in the smeared superposition of geometries, is illustrated, schematically, in Fig. 13. 
The key point is that, although the measured positions of each mass, ${\bf x}'_1$, ${\bf x}'_2$, . . etc., are translated rigidly by ${\bf a}$, the transitions ${\bf x}_1 \leftrightarrow {\bf x}'_1$, ${\bf x}_2 \leftrightarrow {\bf x}'_2$, . . , are randomly determined by the probability distributions $g({\bf x}'_1-{\bf x}_1)$, $g({\bf x}'_2-{\bf x}_2)$, . . , so that the final displaced vectors, $\bold{x}'_1 + \bold{a}$, $\bold{x}'_2 + \bold{a}$, . . , are also randomly distributed. 
In the event that the random fluctuation ${\bf x}'_{I}-{\bf x}_{I}$ is exactly parallel, or anti-parallel, to the intended translation vector ${\bf a}$, this results in lengthening or shortening of the displacement by $\pm |{\bf x}'_{I}-{\bf x}_{I}|$, which only has a significant effect on the grid structure when $|{\bf a}| \simeq |{\bf x}'_{I}-{\bf x}_{I}| \sim \mathcal{O}(l_{\rm Pl})$. 
If, on the other hand, the random fluctuation ${\bf x}'_{I}-{\bf x}_{I}$ is exactly perpendicular to ${\bf a}$, this results in lengthening of the effective displacement according to $|{\bf a}| \rightarrow \sqrt{|{\bf a}|^2+|{\bf x}'_{I}-{\bf x}_{I}|^2}$, and an angular deflection by $\theta = {\rm tan}^{-1}(\pm |{\bf x}'_{I}-{\bf x}_{I}|/|{\bf a}|)$. 
Again, this is clearly only significant for $|{\bf a}| \simeq |{\bf x}'_{I}-{\bf x}_{I}| \sim \mathcal{O}(l_{\rm Pl})$, for which $\theta \simeq \pm 45^{o}$, and its effects are negligible in the limit $|{\bf a}| \gg |{\bf x}'_{I}-{\bf x}_{I}|$. 

\begin{figure}[h] \label{Fig.13}
\centering 
	\includegraphics[width=14cm]{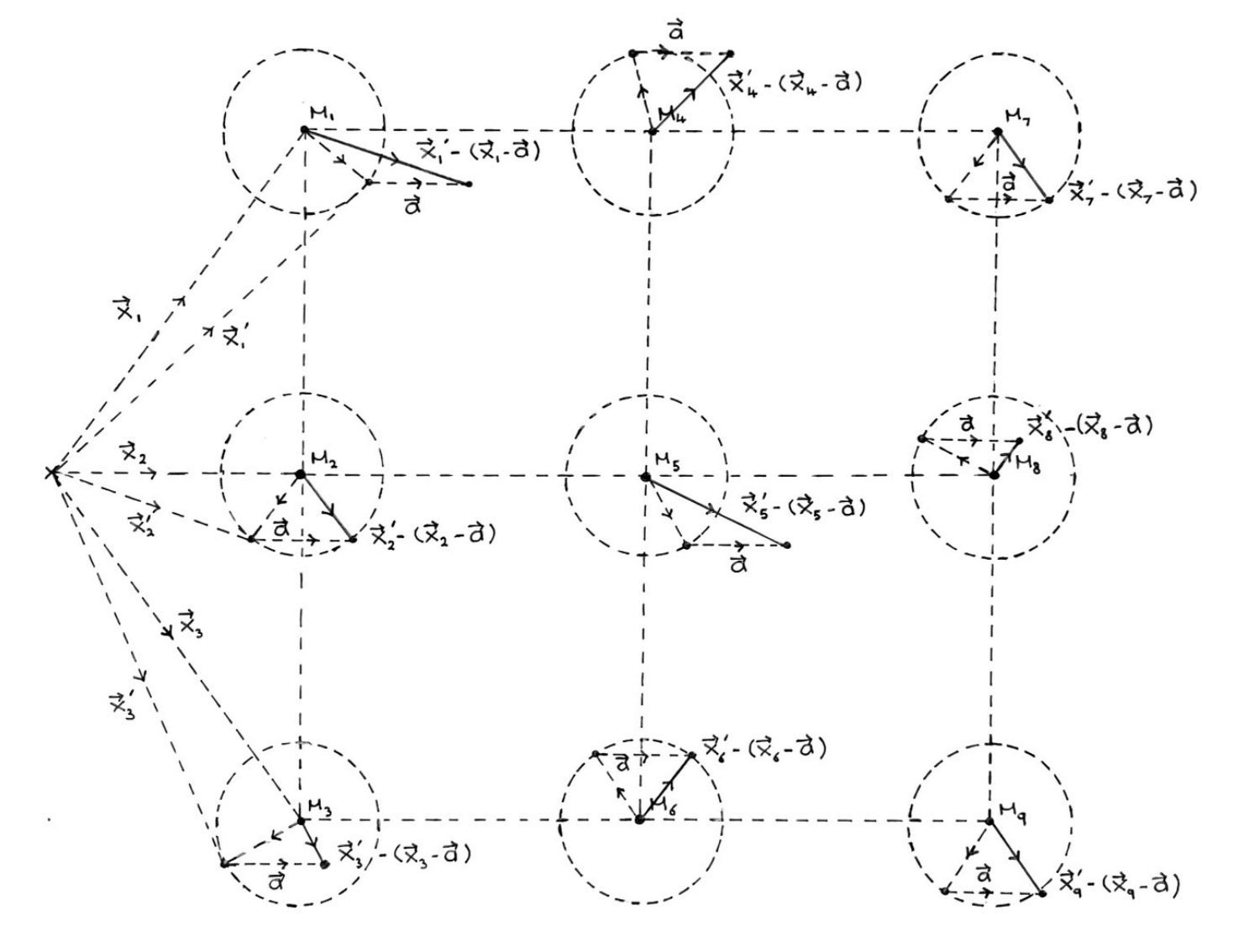}
\caption{The action of the operator $\hat{\mathcal{U}}(\bold{a})^{\dagger} := \exp\left[-\frac{i}{(\hbar+\beta)}\hat{\bold{P}}.\bold{a}\right]$ (\ref{translation_op_smeared_QM}) on a grid of massive point-particles, $M_1$, $M_2$ . . $M_N$, where $\hat{\bold{P}}$ (\ref{smeared_P}) measures the momentum of a material quantum system {\it relative} to the fluctuating quantum background geometry. Thus, $\hat{\bold{P}}$ also acts as the generator of translations, of the material quantum system, {\it relative} to the fluctuating background. We assume $M_I = M \gg m_{\rm Pl}$, for all $I \in \left\{1,2, \dots N\right\}$, so that both the canonical position and velocity uncertainties of each mass can be considered negligible in the experiment (see Sec. \ref{Sec.3.2}). For simplicity, we also neglect the Planck-scale smearing of the origin of the experimenter's coordinate system (see Fig. 12), where the latter is represented by a cross on the left-hand side of the diagram. Every mass in the grid is represented by a dot and the dashed circular line around each dot represents one standard deviation of the relevant smearing function, $g({\bf x}'_I-{\bf x}_I)$, which is of width $\sigma_g \simeq l_{\rm Pl}$. The would-be translation vector, ${\bf a}$, and the individual fluctuations, ${\bf x}'_{I}-{\bf x}_{I}$, are represented as dashed straight lines, whereas the resultant translations, ${\bf x}'_{I}-({\bf x}_{I}-{\bf a})$, which include the effects of the random fluctuations, are depicted as solid black lines.}
\end{figure}

Analogous results also hold, with respect to would-be Galilean velocity boosts, in the momentum space picture. 
Here, the relevant unitary operator, which generates momentum-boosts of the material quantum system, relative to the fluctuating background, is $\hat{\tilde{\mathcal{U}}}(\boldsymbol{\alpha})^{\dagger} := \exp\left[\frac{i}{(\hbar+\beta)}\boldsymbol{\alpha}.\hat{\bold{X}}\right]$, where $\hat{\bold{X}}$ is defined in Eq. (\ref{smeared_X}). 
Hence, the Planck-scale smearing of real space, together with the de Sitter scale smearing of momentum space, also destroys the rotational symmetry of both these spaces. 
Nonetheless, we expect rotational symmetry to hold, at least {\it approximately}, on angular scales in the range $\theta \gg \delta := \beta/\hbar \sim \mathcal{O}(10^{-61})$. 
An in depth analysis of this problem, and its implications for the existence of a minimum measurable angular momentum, of order $\beta$, lies beyond the scope of the present work, but we will return to this interesting question in a future study.
 
\subsection{Gauge symmetries} \label{Sec.5.4}

Given the intimate connection between gauge symmetries and spacetime symmetries, it should come as no surprise that our previous analysis suggests, strongly, that `smearing' the latter to include Planck-scale fluctuations also implies a concomitant smearing of gauge invariance. 
We stress, again, that this is equivalent to endowing the background spacetime with quantum degrees of freedom, or, in other words, to treating it formally as a non-material QRF.
The issue is subtle, however, and, although a detailed analysis lies outside the scope of the present work, we may make some preliminary remarks regarding its potential resolution. 
For concreteness, we restrict our attention to the most physically relevant and least hypothetical gauge group, namely, the Standard Model gauge group, $U(1) \times SU(2) \times SU(3)$ \cite{Peskin&Schroeder:1995,QFT_Nutshell}. 

First, let us consider the $U(1)$ subgroup. 
In the wave mechanics picture, our smearing procedure implies the map \cite{Lake:2018zeg,Lake:2019nmn}
\begin{eqnarray} \label{wave_mechanics_smearing}
\hat{p}_{i} = -i\hbar \frac{\partial}{\partial x^{i}} \mapsto \hat{P}_{i} := -i\hbar \frac{\partial}{\partial x^{i}}\Big|_{\bold{x}' = {\rm const.}} - i\beta \frac{\partial}{\partial (x'^{i}-x^{i})}\Big|_{\bold{x} = {\rm const.}} \, ,
\end{eqnarray}
for nonrelativistic momenta.
Potentially, this expression may be extended, in order to apply to smeared flat spacetime (smeared Minkowski space), by defining 
\begin{eqnarray} \label{wave_mechanics_smearing*}
\hat{P}_{\mu} := -i\hbar \frac{\partial}{\partial x^{\mu}}\Big|_{x' = {\rm const.}} - i\beta \frac{\partial}{\partial (x'^{\mu}-x^{\mu})}\Big|_{x = {\rm const.}} \, ,
\end{eqnarray}
where $\mu \in \left\{0,1,2,3\right\}$ and $x'^{\mu} = (x'^{0},x'^{i})$, $x^{\mu} = (x^{0},x^{i})$. 
This, in turn, may be extended to include some notion of smeared curved spacetime, by replacing the standard partial derivatives above with the covariant derivatives, $\nabla_{\mu} := \nabla_{\mu}|_{x' = {\rm const.}}$ and $\nabla'_{\mu} := \nabla'_{\mu}|_{x = {\rm const.}}$. 
(Here, the prime indicates that the derivative is to be taken with respect to the relative variable $(x'^{\mu}-x^{\mu})$, whereas the unprimed derivative is taken with respect to the canonical spacetime variable, $x^{\mu}$.)
However, in the fibre bundle approach to gauge theories \cite{Frankel:1997ec,Nakahara:2003nw}, $U(1)$ invariance of the kinetic term is incorporated by replacing standard covariant derivatives with gauge covariant derivatives, $\nabla_{\mu} \rightarrow \nabla_{\mu} - ie A_{\mu}(x)$, where $A_{\mu}(x) = (A_0(x),A_i(x))$ is the electromagnetic $4$-potential and $e$ is the coupling strength (charge) \cite{Peskin&Schroeder:1995,QFT_Nutshell}. 
This suggests an exchange of the form $A_{\mu}(x) \mapsto A_{\mu}(x')$, in the Planck-scale regime, where spacetime fluctuations are controlled by a $(3+1)$-dimensional quantum probability amplitude density, $g(x'-x)$. 
How, exactly, this should be implemented in quantum field theory, where the canonical quantum wave functions are `promoted' to field-operators \cite{Peskin&Schroeder:1995,QFT_Nutshell}, remains an open question, to be addressed in future work. 

Next, let us consider the $SU(2)$ subgroup. 
We begin by noting the close correspondence between the spin and orbital angular momentum operators, or, equivalently, between the generators of the groups $SU(2)$ and $SO(3)$ \cite{Jones:1998}. 
Strictly, analysis of the latter belongs in the previous subsection, but we include it here, for convenience.
In the nonrelativisitic smeared space formalism, reducible representations of the rotation generators represent measurements of the {\it relative} angular momentum, of a quantum particle, in relation to the quantum background geometry \cite{Lake:2019nmn,Lake:2020rwc,Lake:2021beh}. 
They are defined, formally, as $\hat{\mathfrak{L}}_{i} := \epsilon_{ij}{}^{k}\hat{X}^{j}\hat{P}_{k}$, where $\hat{X}^{j}$ and $\hat{P}_{k}$ are given by Eqs. (\ref{smeared_X}) and (\ref{smeared_P}), and may be split into the sum of three terms, $\hat{L}_{i} := \hat{\mathbb{L}}_{i} + \hat{\mathcal{L}}_{i}+\hat{\mathcal{L}}'_{i}$, where $\hat{\mathcal{L}}_{i} := \epsilon_{ij}{}^{k}\hat{\mathcal{X}}^{j}\hat{\mathcal{P}}_{k}$, $\hat{\mathcal{L}}'_{i} := \epsilon_{ij}{}^{k}\hat{\mathcal{X}}'^{j}\hat{\mathcal{P}}'_{k}$ and $\hat{\mathbb{L}}_{i} := \epsilon_{ij}{}^{k}(\hat{\mathcal{X}}'^{j}\hat{\mathcal{P}}_{k}+\hat{\mathcal{X}}^{j}\hat{\mathcal{P}}'_{k})$. 
The subcomponents $\left\{\hat{\mathbb{L}}_{i},\hat{\mathcal{L}}_{i},\hat{\mathcal{L}}'_{i}\right\}$ obey an algebra that is very similar to Eqs. (\ref{bipartite_L_Bi_subcomponents-1})-(\ref{bipartite_L_Bi_subcomponents-5}), but in which $\beta$ (\ref{beta}) replaces $\hbar$ as the quantum of action associated with the geometric part of the total Hilbert space of the matter-plus-geometry system \cite{Lake:2019nmn,Lake:2020rwc,Lake:2021beh}; 
\begin{eqnarray} \label{smeared_L_i_subcomponents-1}
[\hat{\mathcal{L}}_{i},\hat{\mathcal{L}}_{j}] = i\hbar\epsilon_{ij}{}^{k} \, \hat{\mathcal{L}}_{k} \, , \quad [\hat{\mathcal{L}}'_{i},\hat{\mathcal{L}}'_{j}] = i\beta\epsilon_{ij}{}^{k} \, \hat{\mathcal{L}}'_{k} \, , 
\end{eqnarray}
\begin{eqnarray} \label{smeared_L_i_subcomponents-2}
[\hat{\mathcal{L}}_{i},\hat{\mathcal{L}}'_{j}] = [\hat{\mathcal{L}}'_{i},\hat{\mathcal{L}}_{j}] = 0 \, , 
\end{eqnarray}
\begin{eqnarray} \label{smeared_L_i_subcomponents-3}
[\hat{\mathcal{L}}_{i},\hat{\mathbb{L}}_{j}] - [\hat{\mathcal{L}}_{j},\hat{\mathbb{L}}_{i}] = i\hbar\epsilon_{ij}{}^{k} \, \hat{\mathbb{L}}_{k} \, , 
\end{eqnarray}
\begin{eqnarray} \label{smeared_L_i_subcomponentss-4}
[\hat{\mathcal{L}}'_{i},\hat{\mathbb{L}}_{j}] - [\hat{\mathcal{L}}'_{j},\hat{\mathbb{L}}_{i}] = i\beta\epsilon_{ij}{}^{k} \, \hat{\mathbb{L}}_{k} \, , 
\end{eqnarray}
\begin{eqnarray} \label{smeared_L_i_subcomponents-5}
[\hat{\mathbb{L}}_{i},\hat{\mathbb{L}}_{j}] = i\hbar\epsilon_{ij}{}^{k} \, \hat{\mathcal{L}}_{k} + i\beta\epsilon_{ij}{}^{l} \, \hat{\mathcal{L}}'_{l} \, , 
\end{eqnarray}
We stress that, here, the $\hat{\mathfrak{L}}_{i}$ represent measurements performed on a {\it single} material particle, not on the two-particle state associated with the algebras considered in Sec. \ref{Sec.4.4}. 

To construct a model of smeared curved spacetime, as opposed to flat Euclidean space, with time as a parameter, we may consider rewriting the $\hat{\mathfrak{L}}_{i}$ operators such that
\begin{eqnarray} \label{rotation_generators_rewrite}
\hat{\mathfrak{L}}_{i} =: (\hbar+\beta)\epsilon_{ijk} \, \hat{\mathfrak{M}}^{jk} \, , 
\end{eqnarray}
and defining a new set of components, $\hat{\mathfrak{M}}^{0k}$, which represent generalised time-space rotations, i.e., Lorentz boosts {\it relative} to a fluctuating background, such that
\begin{eqnarray} \label{smeared_Lorentz_generators-1}
\hat{\mathfrak{M}}^{\mu\nu} :=  (\hbar+\beta)^{-1}(\hat{X} \times \hat{P})^{\mu\nu} \, ,
\end{eqnarray}
where the eigenvalues of the smeared $4$-vector operators, $\hat{X}^{\mu}$ and $\hat{P}_{\nu}$, are $x'^{\mu}$ and $p'_{\nu}$, respectively. 
As before, these generators should split into the sum of three terms, 
\begin{eqnarray} \label{smeared_Lorentz_generators-2}
\hat{\mathfrak{M}}^{\mu\nu} := \hat{\mathcal{M}}^{\mu\nu} + \hat{\mathcal{M}}'^{\mu\nu} + \hat{\mathbb{M}}^{\mu\nu} \, , 
\end{eqnarray}
where
\begin{eqnarray} \label{smeared_Lorentz_generators-3}
\hat{\mathcal{M}}^{\mu\nu} := \hbar^{-1}(\hat{\mathcal{X}} \times \hat{\mathcal{P}})^{\mu\nu} \, , \quad 
\hat{\mathcal{M}}'^{\mu\nu} := \beta^{-1}(\hat{\mathcal{X}'} \times \hat{\mathcal{P}}')^{\mu\nu} \, ,  \nonumber\\
\hat{\mathbb{M}}^{\mu\nu} := \sqrt{\hbar\beta}^{-1}[(\hat{\mathcal{X}'} \times \hat{\mathcal{P}})^{\mu\nu} + (\hat{\mathcal{X}} \times \hat{\mathcal{P}}')^{\mu\nu}] \, ,
\end{eqnarray}
and where the subcomponents of the relative $4$-momentum operator are defined in terms of the covariant derivatives, $\nabla_{\mu}$ and $\nabla'_{\mu}$, introduced above, as $\hat{P}_{\mu} := -i\hbar\nabla_{\mu}-i\beta\nabla'_{\mu}$. 
Detailed analysis is required to determine whether the definitions (\ref{smeared_Lorentz_generators-1})-(\ref{smeared_Lorentz_generators-3}) can, in fact, be self-consistently implemented in a generalised `smeared' quantum field theory, but, if they can, then we would expect the subcomponents $\left\{\hat{\mathcal{M}}^{\mu\nu},\hat{\mathcal{M}}'^{\mu\nu},\hat{\mathbb{M}}^{\mu\nu}\right\}$ to obey the algebra
\begin{eqnarray} \label{smeared_M_{munu}_subcomponents-1}
\left[\hat{\mathcal{M}}^{\mu\nu},\hat{\mathcal{M}}^{\rho\sigma}\right] &=& \frac{i}{1+\delta}(\eta^{\mu\rho}\hat{\mathcal{M}}^{\nu\sigma} - \eta^{\nu\rho}\hat{\mathcal{M}}^{\mu\sigma} - \eta^{\mu\sigma}\hat{\mathcal{M}}^{\nu\rho}+ \eta^{\nu\sigma}\hat{\mathcal{M}}^{\mu\rho}) \, , 
\nonumber\\ 
\left[\hat{\mathcal{M}}'^{\mu\nu},\hat{\mathcal{M}}'^{\rho\sigma}\right] &=& \frac{i\delta}{1+\delta}(\eta^{\mu\rho}\hat{\mathcal{M}}'^{\nu\sigma} - \eta^{\nu\rho}\hat{\mathcal{M}}'^{\mu\sigma} - \eta^{\mu\sigma}\hat{\mathcal{M}}'^{\nu\rho}+ \eta^{\nu\sigma}\hat{\mathcal{M}}'^{\mu\rho}) \, , 
\end{eqnarray}
\begin{eqnarray} \label{smeared_M_{munu}_subcomponents-2}
\left[\hat{\mathcal{M}}^{\mu\nu},\hat{\mathcal{M}}'^{\rho\sigma}\right] = \left[\hat{\mathcal{M}}'^{\mu\nu},\hat{\mathcal{M}}^{\rho\sigma}\right] = 0 \, , 
\end{eqnarray}
\begin{eqnarray} \label{smeared_M_{munu}_subcomponents-3}
\left[\hat{\mathcal{M}}^{\mu\nu},\hat{\mathbb{M}}^{\rho\sigma}\right] - \left[\hat{\mathcal{M}}^{\rho\sigma},\hat{\mathbb{M}}^{\mu\nu}\right] = \frac{i}{1+\delta}(\eta^{\mu\rho}\hat{\mathbb{M}}^{\nu\sigma} - \eta^{\nu\rho}\hat{\mathbb{M}}^{\mu\sigma} - \eta^{\mu\sigma}\hat{\mathbb{M}}^{\nu\rho}+ \eta^{\nu\sigma}\hat{\mathbb{M}}^{\mu\rho}) \, ,  
\end{eqnarray}
\begin{eqnarray} \label{smeared_M_{munu}_subcomponentss-4}
\left[\hat{\mathcal{M}}'^{\mu\nu},\hat{\mathbb{M}}^{\rho\sigma}\right] - \left[\hat{\mathcal{M}}'^{\rho\sigma},\hat{\mathbb{M}}^{\mu\nu}\right] = \frac{i\delta}{1+\delta}(\eta^{\mu\rho}\hat{\mathbb{M}}^{\nu\sigma} - \eta^{\nu\rho}\hat{\mathbb{M}}^{\mu\sigma} - \eta^{\mu\sigma}\hat{\mathbb{M}}^{\nu\rho}+ \eta^{\nu\sigma}\hat{\mathbb{M}}^{\mu\rho}) \, ,   
\end{eqnarray}
\begin{eqnarray} \label{smeared_M_{munu}_subcomponents-5}
\left[\hat{\mathbb{M}}^{\mu\nu},\hat{\mathbb{M}}^{\rho\sigma}\right] 
&=& \frac{i\delta}{1+\delta}(\eta^{\mu\rho}\hat{\mathcal{M}}^{\nu\sigma} - \eta^{\nu\rho}\hat{\mathcal{M}}^{\mu\sigma} - \eta^{\mu\sigma}\hat{\mathcal{M}}^{\nu\rho}+ \eta^{\nu\sigma}\hat{\mathcal{M}}^{\mu\rho}) 
\nonumber\\
&+& \frac{i}{1+\delta}(\eta^{\mu\rho}\hat{\mathcal{M}}'^{\nu\sigma} - \eta^{\nu\rho}\hat{\mathcal{M}}'^{\mu\sigma} - \eta^{\mu\sigma}\hat{\mathcal{M}}'^{\nu\rho}+ \eta^{\nu\sigma}\hat{\mathcal{M}}'^{\mu\rho}) \, , 
\end{eqnarray}
as the natural generalisation of Eqs. (\ref{smeared_L_i_subcomponents-1})-(\ref{smeared_L_i_subcomponents-5}).
\footnote{Strictly, in order to implement (\ref{smeared_M_{munu}_subcomponents-1})-(\ref{smeared_M_{munu}_subcomponents-5}) in smeared curved spacetime, $\eta_{\mu\nu} = {\rm diag}(-c^2,1,1,1)$ must be interpreted as the {\it locally} flat metric, which remains locally flat despite coherent transitions of the form $x \leftrightarrow x'$, which are controlled by the {\it local} spacetime smearing function $g(x'-x)$. Subtle issues like this require careful attention and we here sketch only the bare bones of a possible relativistic extension of our model.}

After these preliminary remarks concerning the generalisation of Lorentz invariance in a smeared superposition of geometries, we may now turn our attention to the issue at hand, namely, the generalisation of $SU(2)$ invariance. 
The spin analogue of the algebra (\ref{smeared_L_i_subcomponents-1})-(\ref{smeared_L_i_subcomponents-5}) is
\begin{eqnarray} \label{smeared_S_i_subcomponents-1}
[\hat{\mathcal{S}}_{i},\hat{\mathcal{S}}_{j}] = i\hbar\epsilon_{ij}{}^{k} \, \hat{\mathcal{S}}_{k} \, , \quad [\hat{\mathcal{S}}'_{i},\hat{\mathcal{S}}'_{j}] = i\beta\epsilon_{ij}{}^{k} \, \hat{\mathcal{S}}'_{k} \, , 
\end{eqnarray}
\begin{eqnarray} \label{smeared_S_i_subcomponents-2}
[\hat{\mathcal{S}}_{i},\hat{\mathcal{S}}'_{j}] = [\hat{\mathcal{S}}'_{i},\hat{\mathcal{S}}_{j}] = 0 \, , 
\end{eqnarray}
\begin{eqnarray} \label{smeared_S_i_subcomponents-3}
[\hat{\mathcal{S}}_{i},\hat{\mathbb{S}}_{j}] - [\hat{\mathcal{S}}_{j},\hat{\mathbb{S}}_{i}] = i\hbar\epsilon_{ij}{}^{k} \, \hat{\mathbb{S}}_{k} \, , 
\end{eqnarray}
\begin{eqnarray} \label{smeared_S_i_subcomponentss-4}
[\hat{\mathcal{S}}'_{i},\hat{\mathbb{S}}_{j}] - [\hat{\mathcal{S}}'_{j},\hat{\mathbb{S}}_{i}] = i\beta\epsilon_{ij}{}^{k} \, \hat{\mathbb{S}}_{k} \, , 
\end{eqnarray}
\begin{eqnarray} \label{smeared_S_i_subcomponents-5}
[\hat{\mathbb{S}}_{i},\hat{\mathbb{S}}_{j}] = i\hbar\epsilon_{ij}{}^{k} \, \hat{\mathcal{S}}_{k} + i\beta\epsilon_{ij}{}^{l} \, \hat{\mathcal{S}}'_{l} \, , 
\end{eqnarray}
where the subcomponents $\left\{\hat{\mathbb{S}}_{i},\hat{\mathcal{S}}_{i},\hat{\mathcal{S}}'_{i}\right\}$ of the relative spin operator, $\hat{S}_{i} := \hat{\mathcal{S}}_{i}+\hat{\mathcal{S}}'_{i}+\hat{\mathbb{S}}_{i}$, are defined as \cite{Lake:2019nmn,Lake:2020rwc,Lake:2021beh}
\begin{eqnarray} \label{smeared_S_i_subcomponents_definitions-1}
\hat{\mathcal{S}}_{i} := \frac{\hbar}{2} \, \sigma_{i} \otimes \hat{\mathbb{I}}_2 \, , \quad \hat{\mathcal{S}}'_{i} := \frac{\beta}{2} \, \hat{\mathbb{I}}_2 \otimes \sigma_{i} \, , \quad  
\end{eqnarray}
and
\begin{eqnarray} \label{smeared_S_i_subcomponents_definitions-2}
\hat{\mathbb{S}}_{i} := \frac{2}{\sqrt{\hbar\beta}}\epsilon_{i}{}^{jk} \, \hat{\mathcal{S}}_{j} \, \hat{\mathcal{S}}'_{k} = \frac{\sqrt{\hbar\beta}}{2}\epsilon_{i}{}^{jk} \, \sigma_{j} \otimes \sigma_{k} \, . 
\end{eqnarray}
We may then rewrite these as 
\begin{eqnarray} \label{spin_generators_rewrite}
\hat{S}_{i} =: \frac{(\hbar+\beta)}{2}\epsilon_{ijk} \, \hat{S}^{jk} \, , 
\end{eqnarray}
following standard methods \cite{Peskin&Schroeder:1995,QFT_Nutshell}, and define a new set of components, $\hat{S}^{0k}$, such that
\begin{eqnarray} \label{smeared_spin_generators-1}
\hat{S}^{\mu\nu} := \frac{(\hbar+\beta)}{2}^{-1}\epsilon_{\mu\nu}{}^{\rho} \, \hat{S}_{\rho} \, ,  
\end{eqnarray}
where $\hat{S}_{\rho} := (\hat{S}_{0},\hat{S}_{k})$ and $\hat{S}_{0} := \hat{\mathbb{I}}_2 \otimes \hat{\mathbb{I}}_2 = \hat{\mathbb{I}}_4$ is the four-dimensional identity operator. 
The new operators (\ref{smeared_spin_generators-1}) can be split into a sum of three subcomponents, $\hat{S}^{\mu\nu} := \hat{\mathcal{S}}^{\mu\nu} + \hat{\mathcal{S}}'^{\mu\nu} + \hat{\mathbb{S}}^{\mu\nu}$, where 
$\hat{\mathcal{S}}^{\mu\nu} := (\hbar/2)^{-1}\epsilon_{\mu\nu}{}^{\rho} \, \hat{\mathcal{S}}_{\rho}$, $\hat{\mathcal{S}}'^{\mu\nu} := (\beta/2)^{-1}\epsilon_{\mu\nu}{}^{\rho} \, \hat{\mathcal{S}}'_{\rho}$ and $\hat{\mathbb{S}}^{\mu\nu} := (\sqrt{\hbar\beta}/2)^{-1}\epsilon_{\mu\nu}{}^{\rho} \, \hat{\mathbb{S}}_{\rho}$, and the relativistic generalisation of Eqs. (\ref{smeared_S_i_subcomponents-1})-(\ref{smeared_S_i_subcomponents-5}) is then identical to (\ref{smeared_M_{munu}_subcomponents-1})-(\ref{smeared_M_{munu}_subcomponents-5}), but with $\hat{\mathcal{S}}^{\mu\nu} \leftrightarrow \hat{\mathcal{M}}^{\mu\nu}$, $\hat{\mathcal{S}}'^{\mu\nu} \leftrightarrow \hat{\mathcal{M}}'^{\mu\nu}$ and $\hat{\mathbb{S}}^{\mu\nu} \leftrightarrow \hat{\mathbb{M}}^{\mu\nu}$. 
By this reasoning, the generalised spin operators $\hat{S}^{\mu\nu}$ should also define a set of generalised Dirac matrices \cite{Peskin&Schroeder:1995,QFT_Nutshell}, $\left\{\Gamma^{\mu}\right\}_{\mu=0}^{3}$, via the relation
\begin{eqnarray} \label{Gamma_matrices-1}
\hat{S}^{\mu\nu} := \frac{i(\hbar+\beta)}{4}[\Gamma^{\mu},\Gamma^{\nu}]_{+} \, , 
\end{eqnarray}
and these, in turn, should split into subcomponents of their own. 
It is not unreasonable to imagine that the subcomponents of the $\Gamma^{\mu}$ matrices should obey anti-commutation relations that represent a relativistic generalisation of the smeared-space Clifford algebra \cite{Lake:2021beh}, 
\begin{eqnarray} \label{smeared_spin_Clifford-1}
[\hat{\mathcal{S}}_{i},\hat{\mathcal{S}}_{j}]_{+} = \frac{\hbar^2}{2}\delta_{ij} \, \hat{\mathbb{I}}_{4} \, , \quad [\hat{\mathcal{S}}'_{i},\hat{\mathcal{S}}'_{j}]_{+} = \frac{\beta^2}{2}\delta_{ij} \, \hat{\mathbb{I}}_{4} \, ,
\end{eqnarray}
\begin{eqnarray} \label{smeared_spin_Clifford-2}
[\hat{\mathcal{S}}_{i},\hat{\mathbb{S}}_{j}]_{+} + [\hat{\mathcal{S}}_{j},\hat{\mathbb{S}}_{i}]_{+} = 0 \, , 
\end{eqnarray}
\begin{eqnarray} \label{smeared_spin_Clifford-3}
[\hat{\mathcal{S}}'_{i},\hat{\mathbb{S}}_{j}]_{+} + [\hat{\mathcal{S}}'_{j},\hat{\mathbb{S}}_{i}]_{+} = 0 \, , 
\end{eqnarray}
\begin{eqnarray} \label{smeared_spin_Clifford-4}
[\hat{\mathbb{S}}_{i},\hat{\mathbb{S}}_{j}]_{+} = \hbar\beta\delta_{ij} \, \hat{\mathbb{I}}_{4} - [\hat{\mathcal{S}}_{i},\hat{\mathcal{S}}'_{j}]_{+} - [\hat{\mathcal{S}}_{j},\hat{\mathcal{S}}'_{i}]_{+} \, . 
\end{eqnarray}

In principle, the algebras sketched above may provide all the ingredients we need to construct a smeared spacetime generalisation of the canonical Dirac equation. 
Such an equation would describe the propagation of charged spin-$1/2$ fermions on a smeared superposition of curved spacetime geometries. 
Furthermore, it may be hoped that `squaring' this equation would lead to a smeared-space generalisation of the Klein-Gordon equation for the propagation of bosons, and that by imposing an appropriate group contraction we may, from there, arrive at a smeared-space generalisation of the nonrelativistic Pauli equation. 
Finally, we may neglect both spin and charge to obtain the smeared-space Schr{\" o}dinger equation, which we hypothesise will take the form  
\begin{eqnarray} \label{smeared_Schrodinger_equation}
\left[\frac{\hat{\bold{P}}^2}{2m} + \hat{V}(\hat{\bold{X}})\right]\Psi = i(\hbar + \beta)\frac{\partial\Psi}{\partial t} \, . 
\end{eqnarray}

This begs the question as to what the possible time-evolution of the smearing function $g$ might be? 
Assuming that $g$ is, indeed, evolvable in time, $g = g(\bold{x}'-\bold{x};t)$, we note that it may nonetheless be regarded as existing in an energy eigenstate, corresponding to the ground state energy of spacetime, under non-extreme conditions (i.e, far away from the singularities at the centres of black holes, or in the violent high-energy environment of the early Universe \cite{Hobson:2006se}). 
Wth this in mind, we note that $g$ takes the form of a three-dimensional Planck-width Gaussian, for all $t$, if it exists in the ground state defined by the `geometric' Schr{\" o}dinger equation  
\begin{eqnarray} \label{g_Schrodinger_equation}
\left[\frac{\hat{\bold{\mathcal{P}}}'^2}{2m_g} + \frac{1}{2}k \, \hat{\bold{\mathcal{X}}}'^2\right]g= i\beta\frac{\partial g}{\partial t} \, ,
\end{eqnarray}
where $\hat{\bold{\mathcal{X}}}'$, $\hat{\bold{\mathcal{P}}}'$ are defined in Eqs. (\ref{smeared_X})-(\ref{smeared_P}) and $m_g k \simeq \beta^2/\sigma_{g}^{4}$ ($\sigma_{g} \simeq l_{\rm Pl}$). 
Equation (\ref{g_Schrodinger_equation}) defines an infinite number of simple harmonic potentials, each one centred a different point `${\bf x}$' in the {\it classical} geometry. 
These potentials regulate the coherent transitions ${\bf x} \leftrightarrow {\bf x}'$, or, in other words, constrain the fluctuations of the delocalised `points' in the quantised background geometry, to be of the order of the Planck length. 
This is, perhaps, not as crazy as it may at first seem, since it is generally compatible with a field-theoretic description of the spacetime background \cite{EFT_Crowther:2016,Barcelo:2015bja} (see, also, \cite{EFT_limitations:2023} for an alternative perspective).
\footnote{We recall that even canonical quantum field theory can be described as ``treating the harmonic oscillator in ever-increasing levels of abstraction'', quote attributed to Sidney Coleman in a lecture at Harvard University  \cite{Sidney_Coleman:QFT_Quote} (unverified).} 
Interestingly, the associated energy density, $\rho_g := (3/4\pi)E_{0}/\sigma_g^3$, where $E_{0} = (3/2)\beta\omega$ is the ground-state energy, is also of the order of dark energy density, $\rho_g \simeq \rho_{\Lambda}$, if and only if $m_g \simeq m_{\rm Pl}$ and $\omega^2 := k/m_g \simeq \Lambda c^2$. 
However, it is not clear whether such a description is really appropriate, or whether it can emerge naturally from the nonrelativistic limit of the correct relativistic wave equation for the $(3+1)$-dimensional spacetime smearing function, $g(x'-x) \equiv g({\bf x}'-{\bf x},t'-t)$. 
These issues require careful attention and will be addressed in a future study.

Lastly, we consider the $SU(3)$ subgroup. 
This is the only subgroup of the Standard Model gauge group that we cannot, at this stage, make any detailed comments about, regarding its treatment in the smeared superposition of geometries. 
Based on the general ideas of our previous analysis, we can only say that we expect irreducible representations of the $SU(3)$ Lie algebra to be mapped to reducible representations. 
The generators of these representations should decompose into subcomponents, obeying a suitable smeared-space generalisation of the canonical $SU(3)$ algebra, and should be interpreted as the generators of colour-rotations, of the state of a material quantum system, {\it relative} to the quantum spacetime background. 
We recall that a canonical colour-rotation of the Standard Model gluon fields is described by a {\it spacetime-dependent} $SU(3)$ group element \cite{Peskin&Schroeder:1995,QFT_Nutshell}, which, therefore, must necessarily be affected by the smearing of the classical spacetime - from an infinite collection of CRFs, to an infinite collection of non-material QRFs, $\left\{\ket{g_{\bold{x}}}\right\}_{\bold{x} \in \mathbb{E}^3}$.
        
\section{Discussion} \label{Sec.6}

There is no distinction, mathematically, between relational and non-relational coordinates in classical mechanics. 
Formally, it makes little difference whether they represent the displacements and momenta of material particles, relative to an idealised `external' observer, or relative to another material particle; or even to a virtual dynamical entity like the centre-of-mass. 
The only meaningful differences between coordinates of these kinds are whether or not they are ignorable, in the sense outlined in Sec. \ref{Sec.1} \cite{Classical_Mechanics_Kibble}, and whether they can be eliminated from the classical Hamiltonian by means of constraints, as in \cite{Vanrietvelde:2018dit} and related works. 

In this work we have argued that, by contrast, there {\it is} a meaningful distinction between relational and non-relational coordinates, in the quantum regime. 
Put simply, coordinates describing the displacements or momenta of quantum systems, relative to an idealised CRF, are represented by operators that depend on three quantum mechanical degrees of freedom, 
$\left\{\hat{{\bf r}}_{I}\right\}_{I=0}^{N-1}$ or $\left\{\hat{\boldsymbol{\kappa}}_{I}\right\}_{I=0}^{N-1}$.
By contrast, those describing relative displacements or momenta between two quantum subsystems, each of which exists in a superposition of position (momentum) eigenstates, relative to the classical background geometry, must be represented, instead, by operators that depend on six quantum degrees of freedom, $\left\{\hat{{\bf X}}_{I}\right\}_{I=1}^{N-1}$ or  $\left\{\hat{{\bf P}}_{I}\right\}_{I=1}^{N-1}$.

The simple physical reason for this is that we require six degrees of freedom, or, equivalently, `two wave functions' (see Sec. \ref{Sec.1}), to describe the superpositions of both the `observer' and the `observed' particles. 
Thus, by subjecting classical relational variables to canonical quantisation of the form
\begin{eqnarray} \label{canonical_quantisation_relational_variables}
{\bf x}_{I} \mapsto \hat{{\bf x}}_{I} := \int {\bf x}_{I} \ket{{\bf x}_{I}}\bra{{\bf x}_{I}}_{I} {\rm d}^3{\rm x}_{I} \, , \quad {\bf p}_{I} \mapsto \hat{{\bf p}}_{I} := \int {\bf p}_{I} \ket{{\bf p}_{I}}\bra{{\bf p}_{I}}_{I} {\rm d}^3{\rm p}_{I} \, , 
\end{eqnarray}
existing relational formalisms, including the widely used GCB formalism \cite{Giacomini:2017zju}, effectively treat these coordinates as {\it non-relational}, while continuing to interpret them as {\it relational}. 
Ultimately, this leads such models to describe systems of $N$ quantum particles using $3(N-1)$, or fewer \cite{Giacomini:2017zju,Vanrietvelde:2018pgb,Vanrietvelde:2018dit,Hohn:2018toe,Hohn:2018iwn,Krumm:2020fws,Ballesteros:2020lgl,QRF_Bell_Test:2021,Giacomini:2021gei,delaHamette:2021iwx,Cepollaro:2021ccc,Castro-Ruiz:2021vnq,AliAhmad:2021adn,Hoehn:2021flk,Carrozza:2021gju,delaHamette:2021oex,delaHamette:2021piz,Giacomini:2022hco,Overstreet:2022zgq,Kabel:2022cje,Apadula:2022pxk,Amelino-Camelia:2022dsj,Kabel:2023jve,Hoehn:2023ehz,Hoehn:2023axh,Wang:2023koz}, quantum mechanical degrees of freedom. 
This is simply too few to describe the superpositions of both the observing QRF and the remaining $N-1$ subsystems it is supposed to observe. 

We contend, therefore, that what such formalisms {\it really} describe is the view, from a QRF embodied as a material quantum system, of $N-2$ quantum subsystems, plus one effectively classical system, capable of acting as a CRF. 
The latter is seen, by a genuinely quantum system in a superposition of position (momentum) eigenstates, as existing in an equivalent superposition of position (momentum) eigenstates, as illustrated in Fig. 1. 
In this scenario, the two `superpositions' - that is, the genuine superposition of the QRF and the {\it apparent} superposition of the CRF - are related via a simple coordinate transformation. 
This can be represented, mathematically, by a unitary operator - as in, for example, \cite{Giacomini:2017zju,Muller_Group_Website} - without the need to introduce additional quantum mechanical degrees of freedom. 

Thus, in our model, an arbitrary CRF $O$, whether it be represented by a massive, macroscopic observer, or by an abstract spatial point in the classical background geometry, is assigned an {\it effective} Hilbert space, $\mathcal{H}_{O} := \hat{\mathcal{P}}_{A \leftrightarrow O}\mathcal{H}_{A}$, by a given QRF, $A$. 
The corresponding state of the total system, as viewed by $O$, $\ket{\Psi_{ABC\dots}^{(O)}}_{ABC\dots}$ , is mapped according to
\begin{eqnarray} \label{O<-->A}
\ket{\Psi_{ABC\dots}^{(O)}}_{ABC\dots} \mapsto \ket{\Psi_{OBC\dots}^{(A)}}_{OBC\dots} := \hat{\mathcal{P}}_{A \leftrightarrow O} \, \ket{\Psi_{ABC\dots}^{(O)}}_{ABC\dots} \, , 
\end{eqnarray}
where $\hat{\mathcal{P}}_{A \leftrightarrow O}$ is the parity swap operator. 
To describe relational degrees of freedom, one must act on this state with operators of the form
\begin{eqnarray} \label{non-canonical_quantisation_relational_variables}
&&{\bf x}_{I} \mapsto \hat{{\bf X}}_{I} := ({\bf r}_{A})_{O} \otimes \hat{\mathbb{I}}_{BC\dots} + \hat{\mathbb{I}}_{O} \otimes (\widehat{{\bf r}_{I}-{\bf r}_{A}})_{BC\dots} \, , 
\nonumber\\
&&{\bf p}_{I} \mapsto \hat{{\bf P}}_{I} := (\boldsymbol{\kappa}_{A})_{O} \otimes \hat{\mathbb{I}}_{BC\dots} + \hat{\mathbb{I}}_{O} \otimes (\widehat{\boldsymbol{\kappa}_{I}-\boldsymbol{\kappa}_{A}})_{BC\dots} \, , 
\end{eqnarray}
for $I \in \left\{B,C,\dots\right\}$ ($I \neq A$). 
The eigenvalues of these operators are, truly, the desired relational quantities, 
\begin{eqnarray} \label{true_relational_variables}
{\bf x}_{I} := {\bf r}_{I} - {\bf r}_{A} \, , \quad {\bf p}_{I} := \boldsymbol{\kappa}_{I} - \boldsymbol{\kappa}_{A} \, . 
\end{eqnarray}

More complicated operators, representing other relational variables, can then be constructed in the usual way, from combinations of these fundamental, conjugate, relational variables; $F({\bf x}_{I},{\bf p}_{I}) \mapsto \hat{F}(\hat{{\bf X}}_{I},\hat{{\bf P}}_{I})$ (up to ordering ambiguities \cite{Isham:1995}). 
In this work, we analysed the structure of the generalised (relative) orbital angular momentum operator, $\hat{{\bf \mathfrak{L}}}_{I} := \hat{{\bf X}}_{I} \times \hat{{\bf P}}_{I}$, and determined the algebraic structure of its subcomponents. 
Using this, we defined a measure of the relative spin, $\hat{{\bf S}}_{I}$, and constructed the total relative angular momentum operator, $\hat{{\bf \mathfrak{J}}}_{I} := \hat{{\bf \mathfrak{L}}}_{I} + \hat{{\bf S}}_{I}$, which, we believe, has not been presented thus far in the QRF literature. 

An important consequence of our analysis is the demonstration that, when both the `observing' and the `observed' quantum systems exist in superpositions of position (momentum) eigenstates, this leads, unavoidably, to generalised uncertainty relations for the relevant relational variables. 
This phenomenology {\it may} be recovered from the action of canonical-type operators $\left\{\hat{{\bf x}}_{I},\hat{{\bf p}}_{I}\right\}_{I=1}^{N-1}$ (\ref{canonical_quantisation_relational_variables}), and functions thereof, $\hat{F}(\hat{{\bf x}}_{I},\hat{{\bf p}}_{I})$, but we must first construct the effective mixed states that these operators act on. 
This is achieved by performing a partial trace over $\mathcal{H}_{O}$ - the subspace of the total Hilbert space associated with the `observing' QRF - after projecting the `objective' global state of the $N$-partite system onto an appropriate set of eigenstates. 
These must correspond to genuinely {\it relational} variables, in the sense described above, and must be chosen in such a way as to implement the required canonical coordinate transformation between the arbitrary CRF, $O$, and the chosen QRF, $A$, as outlined in Sec. \ref{Sec.3}. 

This operation renders certain information, about the total state of the composite system, inaccessible to the observing QRF subsystem, but does not remove it entirely from the description of physical reality. 
It is, for example, still possible for $A$ to detect her own quantum mechanical nature, by means of sufficiently accurate measurements of both $\hat{{\bf x}}_{I}$ and $\hat{{\bf p}}_{I}$. 
These must be accurate enough to resolve the non-canonical (non-Heisenberg) terms in the relevant GURs for position and momentum measurements, Eqs. (\ref{indep_uncertainties_O_AB**})-(\ref{sigmas_OAB}). 
This suggests various experimental protocols that are capable of distinguishing between the definition of a QRF proposed in \cite{Giacomini:2017zju}, and that proposed here, as discussed in Secs. \ref{Sec.3.2}-\ref{Sec.3.3}. 

We stress that a primary distinction between the two formalisms is that, whereas QRF-to-QRF transitions are represented by unitary operators in \cite{Giacomini:2017zju}, and related works \cite{Vanrietvelde:2018pgb,Vanrietvelde:2018dit,Hohn:2018toe,Hohn:2018iwn,Krumm:2020fws,Ballesteros:2020lgl,QRF_Bell_Test:2021,Giacomini:2021gei,delaHamette:2021iwx,Cepollaro:2021ccc,Castro-Ruiz:2021vnq,AliAhmad:2021adn,Hoehn:2021flk,Carrozza:2021gju,delaHamette:2021oex,delaHamette:2021piz,Giacomini:2022hco,Overstreet:2022zgq,Kabel:2022cje,Apadula:2022pxk,Amelino-Camelia:2022dsj,Kabel:2023jve,Hoehn:2023ehz,Hoehn:2023axh,Wang:2023koz}, this is impossible in our model. 
Performing partial traces over the subspaces associated with two different observers, for example, $\mathcal{H}_{O} := \hat{\mathcal{P}}_{A \leftrightarrow O}\mathcal{H}_{A}$ and $\mathcal{H}'_{O} := \hat{\mathcal{P}}_{B \leftrightarrow O}\mathcal{H}_{B}$, renders different information about the total state of the system inaccessible. 
Since there is, a priori, no relationship between the wave functions of the subsystems $A$ and $B$, the transition QRF $A$ $\rightarrow$ QRF $B$ is, necessarily, {\it non-unitary}. 

In the final part of this work, Sec. \ref{Sec.5}, we applied the QRF formalism developed in earlier sections, for canonical QM, to the problem of quantum gravity, or, more specifically, quantum {\it geometry}. 
Treating the background geometry as a non-material QRF, by endowing it with additional quantum mechanical degrees of freedom, we were able to re-derive the most widely studied GURs in the quantum gravity literature, namely, the generalised uncertainty principle (GUP)\cite{Maggiore:1993rv,Adler:1999bu,Scardigli:1999jh}, extended uncertainty principle (EUP), and extended generalised uncertainty principle (EGUP) \cite{Bolen:2004sq,Park:2007az,Bambi:2007ty}, in a completely new way. 
The mathematical structure underlying these relations, in our model, is completely different to that used in the bulk of the existing literature \cite{Hossenfelder:2012jw,Tawfik:2015rva,Tawfik:2014zca}, over the last 30 years \cite{Kempf:1996ss,Bosso:2023aht}, and neatly avoids the pathologies inherent in the standard modified-commutator approach \cite{Lake:2020rwc,Lake:2023uoi}. 

In this sense, our work represents a unification of the GUR and QRF paradigms, which, within the formalism presented, are seen to be `flip sides of the same quantum coin'. 
Such generalised (non-Heisenberg) uncertainty relations are an inescapable consequence of the measurement of truly {\it relational} variables, that is, of the {\it relative} values of properties such as position, momentum, angular momentum, energy and spin, between one quantum system and another. 
They emerge, necessarily, because each subsystem exists in superposition of the relevant eigenstates, with reference to the background geometry - whether quantum or classical - which material quantum systems inhabit \cite{Jones:1998,Stoica:2021owy}. 

\subsection{Conclusions} \label{Sec.6.1}

In this section, we present a point-by-point summary of the most important conclusions and interesting features of our work. 
We also present a point-by-point summary of its main limitations, and indicate, briefly, how these may be overcome in future analyses. 

Summary of conclusions:
\begin{enumerate}

\item The use of the Born rule, to construct the probability ${\rm d}^{3}P({\bf x}|\psi) = |\psi({\bf x})|^2{\rm d}^{3}{\rm x}$, implies that the wave function $\psi({\bf x})$ is defined as a complex-valued scalar in an `objective' (i.e., classical) Euclidean background geometry. 
Hence, the latter exists, at least formally, in canonical QM \cite{Jones:1998,Stoica:2021owy}. 

\item The Erlangen program in geometry demonstrates the formal equivalence between classical Euclidean space and the group of Euclidean transformations \cite{ErlangenProgram_Klein_1872,ErlangenProgram_EMS_2015,Zuber:2013rha,Kisil:2010,Horwood:2004uh,ErlangenProgram_Encycolpedia_Srpinger,Lev:2020igj,Goenner:2015}. 
For material systems within this space, the corresponding momentum space in also Euclidean, and the resulting group of transformations for these systems is the Galilean group \cite{Jones:1998}. 

\item From 2, it follows that, in {\it any} theory in which the classical background exists, classical frames of reference (CRFs) also exist. 
In Euclidean space, CRFs are related by Galilean transformations, which can be represented by unitary operators in canonical QM \cite{Jones:1998}.

\item Conclusions 1-3 are not new, but our analysis demonstrates their relevance to so-called `relational' quantum theories \cite{Rovelli:1995fv,Zych:2018nao,Hoehn:2019fsy,Hoehn:2020epv,Robson:2023hux}, in which only the relative values of variables, such as position, momentum, angular momentum, energy and spin, between different quantum subsystems, are considered physical. 

\item Our key observation is that relational theories, such as \cite{Giacomini:2017zju} and related works \cite{Vanrietvelde:2018pgb,Vanrietvelde:2018dit,Hohn:2018toe,Hohn:2018iwn,Krumm:2020fws,Ballesteros:2020lgl,QRF_Bell_Test:2021,Giacomini:2021gei,delaHamette:2021iwx,Cepollaro:2021ccc,Castro-Ruiz:2021vnq,AliAhmad:2021adn,Hoehn:2021flk,Carrozza:2021gju,delaHamette:2021oex,delaHamette:2021piz,Giacomini:2022hco,Overstreet:2022zgq,Kabel:2022cje,Apadula:2022pxk,Amelino-Camelia:2022dsj,Kabel:2023jve,Hoehn:2023ehz,Hoehn:2023axh,Wang:2023koz}, must also make use of the Born rule, ${\rm d}^{3}P({\bf x}|\psi) = |\psi({\bf x})|^2{\rm d}^{3}{\rm x}$, where the volume element ${\rm d}^{3}{\rm x}$ represents an infinitesimal volume of {\it classical} Euclidean space \cite{Frankel:1997ec,Nakahara:2003nw}. 
This remains the case, even when ${\bf x}$ represents the {\it relative} displacement between two quantum subsystems. 

\item We contend, therefore, that even relational models must assume the existence of the classical Euclidean background, in order to make sense of, and apply, the Born rule for position measurements. 

\item It follows, from 6, that wave functions must exist `objectively' - that is, as quantum superpositions of position eigenstates, `${\bf x}$', relative to an arbitrary CRF defined by the classical background space \cite{ErlangenProgram_Klein_1872,ErlangenProgram_EMS_2015,Zuber:2013rha,Kisil:2010,Horwood:2004uh,ErlangenProgram_Encycolpedia_Srpinger,Lev:2020igj,Goenner:2015} - even in relational theories such as \cite{Giacomini:2017zju,Vanrietvelde:2018pgb,Vanrietvelde:2018dit,Hohn:2018toe,Hohn:2018iwn,Krumm:2020fws,Ballesteros:2020lgl,QRF_Bell_Test:2021,Giacomini:2021gei,delaHamette:2021iwx,Cepollaro:2021ccc,Castro-Ruiz:2021vnq,AliAhmad:2021adn,Hoehn:2021flk,Carrozza:2021gju,delaHamette:2021oex,delaHamette:2021piz,Giacomini:2022hco,Overstreet:2022zgq,Kabel:2022cje,Apadula:2022pxk,Amelino-Camelia:2022dsj,Kabel:2023jve,Hoehn:2023ehz,Hoehn:2023axh,Wang:2023koz}. 

\item From points 6 and 7 - that is, from the objective existence of wave functions in the classical background geometry - it follows that a system of $N$ quantum particles must be described by a total of $3N$ quantum mechanical degrees of freedom. 
Or, in other words, that we require one wave function to describe each particle. 

\item A key prediction of relational quantum theories, including the GCB formalism \cite{Giacomini:2017zju}, is that a system of $N$ quantum particles can be described by a maximum of $3(N-1)$ degrees of freedom. 
Ultimately, this follows from the assumption that {\it all} non-relational degrees of freedom are unphysical, and can therefore be neglected \cite{Muller_Group_Website}.

\item Conclusion 9 clearly contradicts conclusion 8. 

\item Since conclusion 8 follows from the necessity of utilising the Born rule, even in relational theories, and, ultimately, from the objective existence of the wave function in the classical background geometry, whereas conclusion 9 follows from the assumption that {\it all} non-relational degrees of freedom can be neglected, even in the quantum regime, these two assumptions are {\it incompatible}.

\item This indicates an inconsistency in standard relational models, such as \cite{Giacomini:2017zju,Vanrietvelde:2018pgb,Vanrietvelde:2018dit,Hohn:2018toe,Hohn:2018iwn,Krumm:2020fws,Ballesteros:2020lgl,QRF_Bell_Test:2021,Giacomini:2021gei,delaHamette:2021iwx,Cepollaro:2021ccc,Castro-Ruiz:2021vnq,AliAhmad:2021adn,Hoehn:2021flk,Carrozza:2021gju,delaHamette:2021oex,delaHamette:2021piz,Giacomini:2022hco,Overstreet:2022zgq,Kabel:2022cje,Apadula:2022pxk,Amelino-Camelia:2022dsj,Kabel:2023jve,Hoehn:2023ehz,Hoehn:2023axh,Wang:2023koz}, which must make use of the first assumption, as well as the second, in order to predict the probabilities associated with measurements of relative position. 

\item Closer analysis reveals the origin of the error, which leads to conclusion 9. 
In classical mechanics, non-relational degrees of freedom are truly ignorable, in the sense defined in \cite{Classical_Mechanics_Kibble}, since they have no measurable physical consequences. 
However, in the quantum regime, this is no longer the case. 

\item It is straightforward to show that, because both the `observing' and the `observed' quantum systems exist as superpositions of position eigenstates, relative to an arbitrary CRF $O$, defined by the classical background, any measure of their {\it relative} displacement must depend on a superposition of 6 position variables. 
For example, `the displacement of $A$ relative to $B$' is defined as ${\bf x}_{B} := {\bf r}_{B}-{\bf r}_{A}$, where $\left\{r_{A}^{i}\right\}_{i=1}^{3}$ and $\left\{r_{B}^{i}\right\}_{i=1}^{3}$ denote the components of $A$'s position and $B$'s position, respectively, relative to the origin of $O$'s coordinate system. 
These displacements exist in a superposition of states, $\Psi_{AB}({\bf r}_{A},{\bf r}_{B})$, and it is not possible to neglect 3 of the 6 quantum mechanical degrees of freedom, without discarding relevant physical information. 
The arguments above clearly generalise to $N$-particle systems, and to measurements of relative momenta. 

\item Following conclusion 14, we may then ask, {\it what are the physical consequences of the degrees of freedom that relational theories neglect?} 
Our analysis, presented in Secs. \ref{Sec.3}-\ref{Sec.4} of the text, shows that combining the quantum fluctuations of the `observer' with those of the `observed' quantum system generates generalised uncertainty relations (GURs) of non-Heisenberg type. 
The Heisenberg uncertainty relation is then recovered, only approximately, in the limit that the quantum fluctuations of the observer can be neglected, for all practical purposes. 

\item The generic existence of GURs, for measurements of {\it relative} variables, provided that wave functions exist objectively in the classical background geometry, suggests experimental protocols by which the correct definition of the term `quantum reference frame' can be established, empirically, once and for all. 

\item Since there is, a priori, no relation between the wave functions of the observing and the observed subsystems, it follows that QRF-to-QRF transitions cannot be represented by unitary operators. 
Such operators are information-preserving but different information, about the state of the total system of which they form a part, is (in)accessible to different observers.  

\item By applying a modified version of our QRF formalism to the background geometry itself - that is, by treating the background space as a non-material QRF, endowed with its own quantum mechanical degrees of freedom - we are able to recover the three most widely studied GURs in the phenomenological quantum gravity literature, the generalised uncertainty principle (GUP), extended uncertainty principle (EUP), and extended generalised uncertainty principle (EGUP) \cite{Maggiore:1993rv,Adler:1999bu,Scardigli:1999jh,Bolen:2004sq,Park:2007az,Bambi:2007ty,Kempf:1996ss,Hossenfelder:2012jw,Tawfik:2015rva,Tawfik:2014zca,Bosso:2023aht}. 

\item This analysis shows that the QRF and GUR paradigms may be united into a single formalism, which may have profound implications for a future theory of quantum gravity. 
We speculate that, just as the classical spacetime background may be regarded as an infinite collection of non-material CRFs, the quantised spacetime background of the quantum gravity regime may be regarded, instead, as an infinite collection of non-material QRFs. 
If true, this implies that a future research program should focus on developing a quantum generalisation of the Erlangen program in classical geometry \cite{ErlangenProgram_Klein_1872,ErlangenProgram_EMS_2015,Zuber:2013rha,Kisil:2010,Horwood:2004uh,ErlangenProgram_Encycolpedia_Srpinger,Lev:2020igj,Goenner:2015}.

\item Conclusions 18 and 19 are subject to an important caveat. 
The extension of our QRF formalism, which was developed for canonical quantum systems in Secs. \ref{Sec.3}-\ref{Sec.4}, to include the background space as a non-material QRF (as in Sec. \ref{Sec.5}), {\it requires} the existence of a new quantum of action, $\beta \ll \hbar$, in addition to Planck's constant. 
This is associated with {\it relative} fluctuations, of material quantum systems, in relation to the quantum background geometry. 

\item Interestingly, the generalised de Broglie relations that this new quantum of action implies, ${\bf p}'=\hbar{\bf k} + \beta({\bf k}'-{\bf k})$ (\ref{smeared_dB-2}), where ${\bf p}'$ is the momentum of a material quantum particle {\it relative} to fluctuations of the background, do not contradict existing no-go theorems which suggest the uniqueness of $\hbar$ \cite{Sahoo:2004,Deser:2022lmi}. 
Roughly speaking, the condition $\beta \ll \hbar$ indicates that spacetime is `somewhat more classical than matter', in the sense that its quantum features become apparent at much smaller scales of action, which is clearly in accordance with physical observations. 

\item To the best of our knowledge, no existing theories of quantum gravity make use of a second Planck-type constant, but ours requires it, in order to recover the observed vacuum (dark energy) density \cite{Reiss1998,Perlmutter1999,Betoule:2014frx,Planck:2018vyg}. 
If there is even a slim chance that the standard assumption is incorrect - namely, that the energy and momentum of gravitational waves is quantised in the same way as that of material quantum systems, according to the canonical de Broglie relations, $E = \hbar\omega$ and ${\bf p}=\hbar{\bf k}$ - then this possibility should be thoroughly investigated.  

\end{enumerate}

Limitations of the present analysis:
\begin{enumerate}

\item Throughout the present work, we considered only composite quantum systems, $ABC \dots $, whose initial states were separable, from the perspective of an arbitrary CRF $O$. 
We then considered CRF-to-QRF transitions which allow us to `jump' from $O$'s viewpoint to that of a given quantum subsystem, $A$, $B$, $C$, . . , etc. 
However, in order to give a full description of such transitions, we must also consider the case in the which the subsystems of the composite state are entangled, even from the perspective of a classical observer. 
Conceptually, this scenario is extremely interesting, as it implies that there is no clear distinction between the observer and the observed, even from the `objective' viewpoint of the classical background geometry. 
It is unclear, at present, how the predictions of our model might change in this scenario.

\item Although we defined a measure of the relative angular momentum between two quantum subsystems, $A$ and $B$, as the generator of relative rotations $\hat{\mathfrak{L}}_{Bi}$ (\ref{bipartite_L_Bi}), we did not determine the explicit form of this operator's eigenstates. 
Likewise, we did not determine the eigenstates of $\hat{\mathfrak{L}}_{B}^2$, or those of the {\it relative} Hamiltonian, whose eigenvalues are ($E_{B}-E_{A}$), where  $E_{B}$ and $E_{A}$ are the energies of each subsystem, as measured by the CRF $O$. 
In short, we did not determine the simultaneous eigenstates of these {\it relational} operators, $\ket{\mathfrak{n} \, \mathfrak{l} \, \mathfrak{m}}$, which ought to represent the appropriate QRF-generalisation of the canonical eigenstates, $\ket{n \, l \, m}$, seen by $O$. 
This task must be completed as a matter of urgency. 
We may then use the states $\ket{\mathfrak{n} \, \mathfrak{l} \, \mathfrak{m}}$ to construct the density matrix of the effective mixed state, seen by Bob, when he performs measurements of Alice's energy, total angular momentum, and magnetic angular momentum, relative to his own state. 
This analysis of the energy-angular momentum basis of the bipartite state should then be generalised to $N$-particle systems. 

\item By analogy with the measure of relative angular momentum, we defined a measure of relative spin between Alice and Bob's states, $\hat{S}_{Bi}$ (\ref{bipartite_spin_ops-1}), but we did not investigate its properties in detail. 
In particular, we found that the composite $AB$ system may exist in one of two {\it distinct} eigenstates, of $\hat{S}_{Bi}$ and $\hat{S}_{B}^2$, each of which appear, to Bob, to represent the unique spin `up' state of Alice's subsystem, $\left\{\ket{\uparrow_{i}}\rangle,\ket{\uparrow'_{i}}\rangle\right\}$. 
Likewise, there exist two distinct eigenstates of the total $AB$ system which each appear, to Bob, to represent Alice's unique spin `down' state, $\left\{\ket{\downarrow_{i}}\rangle,\ket{\downarrow'_{i}}\rangle\right\}$. 
We may therefore ask if there is {\it any} physical measurement, that Bob can perform, which is capable of distinguishing between $\ket{\uparrow_{i}}\rangle$ and $\ket{\uparrow'_{i}}\rangle$ or $\ket{\downarrow_{i}}\rangle$ and $\ket{\downarrow'_{i}}\rangle$? 
This question is clearly relevant to the significance and interpretation of the Pauli exclusion principle \cite{Rae:2002}, in the QRF paradigm, and how, if at all, it may be extended to a relational perspective. 

\item We gave a broad outline of the conceptual basis of experimental protocols, which, if implemented, should enable the competing definitions of the terms `quantum reference frame', given in \cite{Giacomini:2017zju,Vanrietvelde:2018pgb,Vanrietvelde:2018dit,Hohn:2018toe,Hohn:2018iwn,Krumm:2020fws,Ballesteros:2020lgl,QRF_Bell_Test:2021,Giacomini:2021gei,delaHamette:2021iwx,Cepollaro:2021ccc,Castro-Ruiz:2021vnq,AliAhmad:2021adn,Hoehn:2021flk,Carrozza:2021gju,delaHamette:2021oex,delaHamette:2021piz,Giacomini:2022hco,Overstreet:2022zgq,Kabel:2022cje,Apadula:2022pxk,Amelino-Camelia:2022dsj,Kabel:2023jve,Hoehn:2023ehz,Hoehn:2023axh,Wang:2023koz} and presented here, to be distinguished empirically. 
However, we did not describe such protocols in detail, give realistic estimates of the likely sources of error, or assess the ability to implement them using current or near-future technology. 
Future theoretical analyses should work closely with experimental groups, to determine whether such experiments are viable within a reasonable time frame.   

\end{enumerate}

\subsection{Prospects for future work} \label{Sec.6.2}

As is often the case with scientific endeavours, our work begs more questions than it answers. 
Here, we consider some of the important unanswered questions, raised in the course of our analysis, and suggest directions for future work in this field. 
Possible extensions of our current model fall broadly into two categories. 
First, extensions of the canonical QRF formalism, presented in Secs. \ref{Sec.3}-\ref{Sec.4}, to include non-relativistic gravity, and the relativistic regime, including both flat and curved spacetimes. 
Second, extensions of the smeared-space formalism,  presented in Sec. \ref{Sec.5}, in which the background geometry is also endowed with quantum mechanical degrees of freedom. 
In the list below, we do not explicitly distinguish between these two, unless absolutely necessary. 
Thus, when we speak of a `special-relativistic extension' of our model, for example, this should be understood as implying two distinct but related extensions; the first treats relativistic QRFs as material systems inhabiting classical Minkowski space and the second seeks to generalise the concept of Minkowski space itself, to include `relativistic' non-material QRFs. 
Ultimately, our aim is to combine both these perspectives, to create a realistic model of QRFs, embodied as material quantum systems in a quantum background geometry. 

\begin{enumerate}

\item The model should be extended to the special-relativisitc regime of flat spacetime, which must, necessarily, include some definition of a `temporal reference frame' \cite{Castro-Ruiz:2019nnl,Henderson:2020zax,Debski:2022tfm,Lie:2023rjx}. 
The inclusion of dark energy, or a cosmological constant term, may then be achieved by replacing Poincar{\' e} group invariance with de Sitter group invariance \cite{Ibragimov:2015}.

\item In a different direction, our model should also be extended to include non-relativistic (Newtonian) gravity. 
This may be achieved in at least two ways; (i) by treating the Newtonian gravitational potential as an external field, existing in a flat Euclidean background - as Newton himself first conceived it - or, (ii) as a manifestation of a curved Riemannian geometry, as in Newton-Cartan theory \cite{Hansen:2018ofj,Banerjee:2018gqz}. 
Inclusion of the cosmological constant term, in the non-relativistic regime, may then be achieved via an appropriate group contraction, which breaks local de Sitter group invariance to local invariance under the Newton-Hooke group \cite{Newton-Hooke:book,Guo:2004yq,Schurmann:2018yuz}.

\item If extensions 1 and 2 can be performed, successfully, they may be regarded as different low-energy limits of the appropriate general-relativistic extension of our QRF model. 
This gives us two points of departure, from which to work towards a theory describing material-QRFs in a smeared superposition of curved spacetime geometries. 
Whatever results are obtained, these should be compared with existing models of spacetime superpositions \cite{Ford:1997zb,Hossenfelder:2012qg,Foo:2021exb,Foo:2022dnz,Foo:2023vbr}.

\item Clearly, our work has relevance for the gravity-matter entanglement hypothesis, proposed in \cite{Kay:2007rx,Kay:2018mxr,PhysRevA.98.052312}, and, more generally, for the possibility of observing the gravitationally-induced entanglement in material quantum systems. 
This subject has received a lot of attention in the recent literature \cite{Marletto:2017kzi,Feng:2022hfv,Neppoleon:2021pue,Marletto:2018fzw,Marletto:2018lsb,Fragkos:2022tbm,Guff:2021mfw,Chevalier:2021xyw,Ma:2021rve,Bosso:2016ycv,Altamirano:2016fas,Rosi:2017ieh,Wood:2019ymk,Verma:2021rbh,Bose:2017nin,Christodoulou:2018cmk,Belenchia:2018szb,Carney:2021yfw}. 
However, at present, it is not clear, exactly, what the implications of our model are for existing predictions of this phenomenon. 

\item The smeared-space model, developed in Sec. \ref{Sec.5}, strongly suggests that the quantised background geometry exhibits $SU(2)$ symmetry, as also suggested by other authors, based on the spinor representation of Minkowski space \cite{Woit-SU(2):2023}. 
Interestingly, if this is really the case, then a material field with particle-like excitations of mass $m_{\Lambda} \simeq 10^{-3} \, {\rm eV/c^2}$ (\ref{optimised_uncertainties}), and with corresponding position and momentum uncertainties, $(\Delta_\Psi X)_{\mathrm{opt}} \simeq \hbar/(m_{\Lambda}c) \simeq 0.1 \, {\rm mm}$ and $(\Delta_\Psi P)_{\mathrm{opt}} \simeq m_{\Lambda}c \simeq 10^{-3} \, {\rm eV/c}$, must optimise the smeared-space GUR (\ref{smeared-spaceEGUP-1}), from which the EGUP (\ref{smeared-spaceEGUP-2}) is derived as an approximation. 
The energy density of such a field-configuration is comparable to the dark energy density, $\rho_{\Lambda} := \Lambda c^2/(8\pi G) \simeq (3/4\pi)m_{\Lambda} c^2/l_{\Lambda}^3 \simeq 10^{-29}$ ${\rm g \, . \, cm^{-3}}$, which suggests a model of (fermionic) particulate dark energy. 
Such models were originally put forward in \cite{Burikham:2015nma,Burikham:2015sro,Lake:2017ync,Lake:2017uzd} and similar models have been proposed, by other authors, in \cite{Hashiba:2018hth} and \cite{Chu:2023mqi,Chu:2023agv}. 
The emergence of repulsive inter-particle forces, in fermion gas systems, is described in \cite{Keles:2023can}. 
Over cosmological distances, such forces may be capable of mimicking the action of a constant dark energy density \cite{Lake-Jain-Paterek-in-progress}.

\item The existence of a second Planck-type constant, associated with fluctuations of the quantum background geometry, which is treated formally as a non-material QRF, should have profound implications for the quantisation of both the gravitational potential and gravitational waves \cite{Carlip:2008zf,Mauro:2015,Marletto:2017pjr,Belenchia:2019gcc,Rydving:2021qua,Vedral:2022znc}, or, more generally, for the physics of `gravitons' \cite{Pauli-Fierz:1938,Ford:1994cr,Ford:1996qc,Dyson:2013hbl,Rothman:2006fp,Lieu:2017lzh,Parikh:2020nrd,Parikh:2020fhy,Parikh:2020kfh,Kanno:2020usf,Kanno:2021gpt,Cho:2021gvg,Tobar:2023ksi,Carney:2023nzz,Trenggana:2023ysu,Sen:2023ksj}. 
Taking our spacetime-as-a-QRF model seriously requires us to re-evaluate recent estimates regarding the viability of single-graviton detection \cite{Tobar:2023ksi}, as well as thresholds on the detectability of the `noise of gravitons' at future gravitational wave detectors \cite{Parikh:2020nrd,Parikh:2020fhy,Parikh:2020kfh}. 
We stress, again, that our approach evades existing no-go theorems \cite{Sahoo:2004,Deser:2022lmi}, which suggest the uniqueness of $\hbar$ for both gravitons and material particles \cite{Lake:2020rwc}.

\item The conventional approach to linearised quantum gravity, which goes right back to the early work of Pauli and Fierz \cite{Pauli-Fierz:1938}, is to expand the spacetime metric to first order as $g_{\mu\nu}(x) \simeq \eta_{\mu\nu} + \delta g_{\mu\nu}(x)$, where $\eta_{\mu\nu}$ is the flat Minkowski metric and $\delta g_{\mu\nu}(x)$ denotes a perturbation, then quantise the latter; $\delta g_{\mu\nu}(x) \mapsto \widehat{\delta g}_{\mu\nu}(x)$. 
Formally, the condition $\eta^{\mu\sigma}\widehat{\delta g}_{\sigma\nu}(x) = \widehat{\delta g}^{\mu}{}_{\nu}(x)$ indicates that $\widehat{\delta g}_{\mu\nu}(x)$ is treated as a {\it material} quantum field, living in the flat spacetime background. 
Hence, it is conventionally quantised by assuming that the standard de Broglie relations, $E = \hbar\omega$ and ${\bf p} = \hbar {\bf k}$, hold for each plane wave in the flat-space Fourier decomposition. 
The smeared-space quantisation of the background geometry works differently. 
Rather than quantising the perturbation, but neglecting to quantise the flat leading-order term in the metric expansion, we have, in this analysis, quantised {\it only} the leading order term, while neglecting the perturbation. 
(The latter is to be included in future analyses, based on steps 2-3 outlined above.) 
This leads to the modified de Broglie relations ${\bf p}'=\hbar{\bf k} + \beta({\bf k}'-{\bf k})$ (\ref{smeared_dB-2}) in which the non-canonical term ${\bf p}'-{\bf p}=\beta({\bf k}'-{\bf k})$ is induced by quantum fluctuations of the background geometry, which nonetheless remains {\it flat} \cite{Lake:2019nmn}. 
This analysis strongly suggests that perturbative terms should, likewise, be quantised non-canonically, using both relative variables and the `geometric' quantum of action, $\beta \ll \hbar$ \cite{Lake:2020rwc}. 
With this in mind, it is interesting to note a recent analysis, which claims that existing LIGO data already rules out standard $E = \hbar\omega$ gravitons \cite{Lieu:2017lzh}. 
Though by no means conclusive, our work indicates that such conclusions deserve serious scrutiny, and we urge the quantum gravity research community to be open minded, regarding the possibility of a `non-Planckian' spacetime.

\item A related question concerns the definition of material particles in the quantum gravity regime. 
In canonical quantum field theory, particle-like excitations are identified with irreducible representations of the Poincar{\' e} group, according to the Wigner classification \cite{Jones:1998,Straumann:2008kq}. 
The physical reason for this, as we have stressed throughout the current work, is that canonical quantum fields exist, as material objects, within the classical spacetime background \cite{Stoica:2021owy}. 
Hence, these fields, and their particle-like excitations, obey classical Poincar{\' e} symmetry. 
Put bluntly, in {\it any} model of quantised spacetime, our existing descriptions of material quantum systems go to hell-in-a-handbasket, since the formal identification between classical symmetries and the geometry in which material systems are embedded, as specified by the Erlangen program \cite{ErlangenProgram_Klein_1872,ErlangenProgram_EMS_2015,Zuber:2013rha,Kisil:2010,Horwood:2004uh,ErlangenProgram_Encycolpedia_Srpinger,Lev:2020igj,Goenner:2015}, no longer applies. 
\footnote{Surprisingly, this problem has received very little attention in the quantum gravity literature; see \cite{Scaria:2002yb,Donnelly:2020xgu,Cianfrani:2021tiy,Emelyanov:2021auc} for notable exceptions.} 
It is therefore unclear what role, if any, the {\it mathematics} of symmetry should play in our most fundamental descriptions of nature. 
Is the spacetime geometry, and its equivalent description in terms of local Poincar{\' e} symmetries \cite{Jones:1998,Straumann:2008kq}, an emergent phenomenon? 
Does it arise, approximately, from an effective coarse-graining over the symmetry{\it less} states of quantum gravity, as suggested by some authors \cite{Emergent_properties}? 
Or are symmetries embedded at the heart of even {\it pre}-geometric theories? 
Our current work suggests the latter. 
In Sec. \ref{Sec.5.3}, we showed how classical spacetime symmetries can be broken {\it operationally}, by quantum fluctuations of the background geometry, even when the operators corresponding to particle-properties such as linear momentum, angular momentum, and spin, are identified with the usual group generators, (e.g., with translation operators in the canonical the Heisenberg algebra, or with the generators of the canonical Lie algebras, $\mathfrak{so(3)}$ and $\mathfrak{su(3)}$). 
The key point is that, in this case, such properties are truly {\it relational}, and, as such, are their operators are identified with {\it reducible} representations of the relevant generators. 
The reducible representations, however, are not arbitrary; they must decompose into subcomponents that satisfy the generalised algebras presented in Secs. \ref{Sec.5.3}-\ref{Sec.5.4}. 
This suggests an interesting way forward for our future research, namely, that we may aim to develop a mathematical theory of quantum spacetime, based on classical symmetries, which, nonetheless, describes the operational breaking of these symmetries in the quantum gravity regime.   

\item Finally, our results may have implications for a veritable miscellany of research directions in contemporary theoretic physics, including gravitational decoherence and the quantum-to-classical transition \cite{Penrose_Gravity_Collapse,Pikovski:2013qwa,KTM:2014,Bonder:2015hja,Pikovski:2015dka,Vedral:2020dnh,Layton:2023wdo}, the existence of regular black holes \cite{Hayward:2005gi}, and the sharpness, or lack thereof, of black hole horizons \cite{Barrow:2020tzx,Lake:2022hzr}. 
The latter has implications for the information loss paradox \cite{Mathur:2009hf}, as well as for holographic theories \cite{Bousso:2002ju}, and we may aim to extended our model to include `non-commutative reference frames', that is, QRFs defined in a noncommutative background geometry \cite{Connes:1994,Bertolami:2015yga,Marletto:2018bcv}. 

\end{enumerate}

\section*{Acknowledgments}

We thank Flaminia Giacomini,  {\u C}aslav Brukner, Tomasz Paterek and Shi-Dong Liang for helpful discussions. 
We are also grateful to Flaminia Giacomini, Esteban Castro-Ruiz and {\u C}aslav Brukner for their previous work, and, quite simply, for having the guts to ask difficult questions. 
Without their pioneering efforts, our own work in this field would, most likely, never have existed. 

\section*{Funding}

This work was supported by the Grant of Scientific and Technological Projection of Guangdong Province (China), no. 2021A1515010036.



\end{document}